%% file: nanograv_12y_wb.tex
\documentclass[twocolumn,appendixfloats]{aastex63}

\usepackage{natbib}
\usepackage{amsmath}
\usepackage{verbatim}
\usepackage{afterpage}

\newcommand{\fiveyr}{NG5}        
\newcommand{\nineyr}{NG9}        
\newcommand{\elevenyr}{NG11}     
\newcommand{\twelveyr}{NG12.5}     

\newcommand{\nbin}{n_{\textrm{bin}}}
\newcommand{\nchan}{n_{\textrm{chan}}}
\newcommand{\phio}{\phi_\circ}
\newcommand{\dm}{\textrm{DM}}
\newcommand{\nuphi}{\nu_{\phi_\circ}}
\newcommand{\an}{a_{n}}
\newcommand{\dmunits}{\textrm{cm}^{-3}~\textrm{pc}}
\newcommand{\given}[1][]{\:#1\vert\:}

\NewPageAfterKeywords

\begin{document}

\title{The NANOGrav 12.5-year Data Set: Wideband Timing of 47 Millisecond Pulsars}
\shorttitle{The NANOGrav 12.5-year Wideband Data Set}
\shortauthors{Alam et al.}

\input{text/authors}

\collaboration{1000}{The NANOGrav Collaboration}
\noaffiliation

\correspondingauthor{Timothy T. Pennucci}
\email{tim.pennucci@nanograv.org}

\begin{abstract}

\input{text/abstract}

\end{abstract}



\section{Introduction}
\label{sec:intro}
\input{text/sec_intro}

\section{Observations}
\label{sec:obs}
\input{text/sec_obs}

\section{Construction of the Wideband\\Data Set}
\label{sec:wb}
\input{text/sec_wb}
\vspace{1in}
\section{Results \& Discussion}
\label{sec:results}
\input{text/sec_results}

\section{Summary \& Conclusions}
\label{sec:conclusion}
\input{text/sec_conclusion}


\input{text/sec_acknowledgement}

\facilities{Arecibo, GBT}

\software{\texttt{ENTERPRISE} \citep{enterprise}, \texttt{libstempo} \citep{libstempo}, \texttt{matplotlib} \citep{matplotlib}, \texttt{nanopipe} \citep{nanopipe}, \texttt{PSRCHIVE} \citep{vS11}, \texttt{PTMCMC} \citep{ptmcmc}, \texttt{PulsePortraiture} \citep{pulseportraiture}, \texttt{PyPulse} \citep{pypulse}, \texttt{Tempo} \citep{tempo}, \texttt{Tempo2} \citep{tempo2}, \texttt{tempo\_utils}\footnote{\href{https://github.com/demorest/tempo\_utils}{https://github.com/demorest/tempo\_utils}}}
\\\\
\input{text/sec_contribution}

\appendix

\section{Low S/N Wideband TOAs}
\label{sec:low_snr}
\input{text/sec_low_snr}


\section{Wideband Timing Likelihood}
\label{sec:wb_like}
\input{text/sec_wb_likelihood}

\section{Timing Residuals \& DM Variations}
\label{sec:resid}
\input{text/sec_resid}

\input{text/fig_summary}  

\clearpage

\bibliographystyle{aasjournal}
\bibliography{nano12y_wb}

\end{document}

%% file: text/authors.tex
\author[0000-0003-2449-5426]{Md F. Alam}
\affiliation{Department of Physics and Astronomy, Franklin \& Marshall College, P.O. Box 3003, Lancaster, PA 17604, USA}
\author{Zaven Arzoumanian}
\affiliation{X-Ray Astrophysics Laboratory, NASA Goddard Space Flight Center, Code 662, Greenbelt, MD 20771, USA}
\author[0000-0003-2745-753X]{Paul T. Baker}
\affiliation{Department of Physics and Astronomy, Widener University, One University Place, Chester, PA 19013, USA}
\author[0000-0003-4046-884X]{Harsha Blumer}
\affiliation{Department of Physics and Astronomy, West Virginia University, P.O. Box 6315, Morgantown, WV 26506, USA}
\affiliation{Center for Gravitational Waves and Cosmology, West Virginia University, Chestnut Ridge Research Building, Morgantown, WV 26505, USA}
\author{Keith E. Bohler}
\affiliation{Center for Advanced Radio Astronomy, University of Texas-Rio Grande Valley, Brownsville, TX 78520, USA}
\author{Adam Brazier}
\affiliation{Cornell Center for Astrophysics and Planetary Science and Department of Astronomy, Cornell University, Ithaca, NY 14853, USA}
\author[0000-0003-3053-6538]{Paul R. Brook}
\affiliation{Department of Physics and Astronomy, West Virginia University, P.O. Box 6315, Morgantown, WV 26506, USA}
\affiliation{Center for Gravitational Waves and Cosmology, West Virginia University, Chestnut Ridge Research Building, Morgantown, WV 26505, USA}
\author[0000-0003-4052-7838]{Sarah Burke-Spolaor}
\affiliation{Department of Physics and Astronomy, West Virginia University, P.O. Box 6315, Morgantown, WV 26506, USA}
\affiliation{Center for Gravitational Waves and Cosmology, West Virginia University, Chestnut Ridge Research Building, Morgantown, WV 26505, USA}
\author{Keeisi Caballero}
\affiliation{Center for Advanced Radio Astronomy, University of Texas-Rio Grande Valley, Brownsville, TX 78520, USA}
\author{Richard S. Camuccio}
\affiliation{Center for Advanced Radio Astronomy, University of Texas-Rio Grande Valley, Brownsville, TX 78520, USA}
\author{Rachel L. Chamberlain}
\affiliation{Department of Physics and Astronomy, Franklin \& Marshall College, P.O. Box 3003, Lancaster, PA 17604, USA}
\author[0000-0002-2878-1502]{Shami Chatterjee}
\affiliation{Cornell Center for Astrophysics and Planetary Science and Department of Astronomy, Cornell University, Ithaca, NY 14853, USA}
\author[0000-0002-4049-1882]{James M. Cordes}
\affiliation{Cornell Center for Astrophysics and Planetary Science and Department of Astronomy, Cornell University, Ithaca, NY 14853, USA}
\author[0000-0002-7435-0869]{Neil J. Cornish}
\affiliation{Department of Physics, Montana State University, Bozeman, MT 59717, USA}
\author[0000-0002-2578-0360]{Fronefield Crawford}
\affiliation{Department of Physics and Astronomy, Franklin \& Marshall College, P.O. Box 3003, Lancaster, PA 17604, USA}
\author[0000-0002-6039-692X]{H. Thankful Cromartie}
\affiliation{Cornell Center for Astrophysics and Planetary Science and Department of Astronomy, Cornell University, Ithaca, NY 14853, USA}
\author[0000-0002-2185-1790]{Megan E. DeCesar}
\affiliation{Department of Physics, Lafayette College, Easton, PA 18042, USA}
\affiliation{George Mason University, Fairfax, VA 22030, resident at U.S. Naval Research Laboratory, Washington, D.C. 20375, USA}
\author[0000-0002-6664-965X]{Paul B. Demorest}
\affiliation{National Radio Astronomy Observatory, 1003 Lopezville Rd., Socorro, NM 87801, USA}
\author[0000-0001-8885-6388]{Timothy Dolch}
\affiliation{Department of Physics, Hillsdale College, 33 E. College Street, Hillsdale, Michigan 49242, USA}
\author{Justin A. Ellis}
\affiliation{Infinia ML, 202 Rigsbee Avenue, Durham NC, 27701}
\author[0000-0002-2223-1235]{Robert D. Ferdman}
\affiliation{School of Chemistry, University of East Anglia, Norwich, NR4 7TJ, United Kingdom}
\author{Elizabeth C. Ferrara}
\affiliation{NASA Goddard Space Flight Center, Greenbelt, MD 20771, USA}
\author[0000-0001-5645-5336]{William Fiore}
\affiliation{Center for Gravitation, Cosmology and Astrophysics, Department of Physics, University of Wisconsin-Milwaukee,\\ P.O. Box 413, Milwaukee, WI 53201, USA}
\affiliation{Department of Physics and Astronomy, West Virginia University, P.O. Box 6315, Morgantown, WV 26506, USA}
\affiliation{Center for Gravitational Waves and Cosmology, West Virginia University, Chestnut Ridge Research Building, Morgantown, WV 26505, USA}
\author[0000-0001-8384-5049]{Emmanuel Fonseca}
\affiliation{Department of Physics, McGill University, 3600  University St., Montreal, QC H3A 2T8, Canada}
\author{Yhamil Garcia}
\affiliation{Center for Advanced Radio Astronomy, University of Texas-Rio Grande Valley, Brownsville, TX 78520, USA}
\author[0000-0001-6166-9646]{Nathan Garver-Daniels}
\affiliation{Department of Physics and Astronomy, West Virginia University, P.O. Box 6315, Morgantown, WV 26506, USA}
\affiliation{Center for Gravitational Waves and Cosmology, West Virginia University, Chestnut Ridge Research Building, Morgantown, WV 26505, USA}
\author[0000-0001-8158-638X]{Peter A. Gentile}
\affiliation{Department of Physics and Astronomy, West Virginia University, P.O. Box 6315, Morgantown, WV 26506, USA}
\affiliation{Center for Gravitational Waves and Cosmology, West Virginia University, Chestnut Ridge Research Building, Morgantown, WV 26505, USA}
\author[0000-0003-1884-348X]{Deborah C. Good}
\affiliation{Department of Physics and Astronomy, University of British Columbia, 6224 Agricultural Road, Vancouver, BC V6T 1Z1, Canada}
\author[0000-0001-6609-2997]{Jordan A. Gusdorff}
\affiliation{Department of Physics, Lafayette College, Easton, PA 18042, USA}
\author[0000-0003-2289-2575]{Daniel Halmrast}
\affiliation{Department of Physics, Hillsdale College, 33 E. College Street, Hillsdale, Michigan 49242, USA}
\affiliation{Department of Mathematics, University of California, Santa Barbara, CA 93106, USA}
\author[0000-0003-2742-3321]{Jeffrey S. Hazboun}
\affiliation{University of Washington Bothell, 18115 Campus Way NE, Bothell, WA 98011, USA}
\author{Kristina Islo}
\affiliation{Center for Gravitation, Cosmology and Astrophysics, Department of Physics, University of Wisconsin-Milwaukee,\\ P.O. Box 413, Milwaukee, WI 53201, USA}
\author[0000-0003-1082-2342]{Ross J. Jennings}
\affiliation{Cornell Center for Astrophysics and Planetary Science and Department of Astronomy, Cornell University, Ithaca, NY 14853, USA}
\author[0000-0002-4188-6827]{Cody Jessup}
\affiliation{Department of Physics, Hillsdale College, 33 E. College Street, Hillsdale, Michigan 49242, USA}
\affiliation{Department of Physics, Montana State University, Bozeman, MT 59717, USA}
\author[0000-0001-6607-3710]{Megan L. Jones}
\affiliation{Center for Gravitation, Cosmology and Astrophysics, Department of Physics, University of Wisconsin-Milwaukee,\\ P.O. Box 413, Milwaukee, WI 53201, USA}
\author[0000-0002-3654-980X]{Andrew R. Kaiser}
\affiliation{Department of Physics and Astronomy, West Virginia University, P.O. Box 6315, Morgantown, WV 26506, USA}
\affiliation{Center for Gravitational Waves and Cosmology, West Virginia University, Chestnut Ridge Research Building, Morgantown, WV 26505, USA}
\author[0000-0001-6295-2881]{David L. Kaplan}
\affiliation{Center for Gravitation, Cosmology and Astrophysics, Department of Physics, University of Wisconsin-Milwaukee,\\ P.O. Box 413, Milwaukee, WI 53201, USA}
\author[0000-0002-6625-6450]{Luke Zoltan Kelley}
\affiliation{Center for Interdisciplinary Exploration and Research in Astrophysics (CIERA), Northwestern University, Evanston, IL 60208}
\author[0000-0003-0123-7600]{Joey Shapiro Key}
\affiliation{University of Washington Bothell, 18115 Campus Way NE, Bothell, WA 98011, USA}
\author[0000-0003-0721-651X]{Michael T. Lam}
\affiliation{School of Physics and Astronomy, Rochester Institute of Technology, Rochester, NY 14623, USA}
\affiliation{Laboratory for Multiwavelength Astrophysics, Rochester Institute of Technology, Rochester, NY 14623, USA}
\author{T. Joseph W. Lazio}
\affiliation{Jet Propulsion Laboratory, California Institute of Technology, 4800 Oak Grove Drive, Pasadena, CA 91109, USA}
\author[0000-0003-1301-966X]{Duncan R. Lorimer}
\affiliation{Department of Physics and Astronomy, West Virginia University, P.O. Box 6315, Morgantown, WV 26506, USA}
\affiliation{Center for Gravitational Waves and Cosmology, West Virginia University, Chestnut Ridge Research Building, Morgantown, WV 26505, USA}
\author{Jing Luo}
\affiliation{Department of Astronomy \& Astrophysics, University of Toronto, 50 Saint George Street, Toronto, ON M5S 3H4, Canada}
\author[0000-0001-5229-7430]{Ryan S. Lynch}
\affiliation{Green Bank Observatory, P.O. Box 2, Green Bank, WV 24944, USA}
\author[0000-0003-2285-0404]{Dustin R. Madison}
\altaffiliation{NANOGrav Physics Frontiers Center Postdoctoral Fellow}
\affiliation{Department of Physics and Astronomy, West Virginia University, P.O. Box 6315, Morgantown, WV 26506, USA}
\affiliation{Center for Gravitational Waves and Cosmology, West Virginia University, Chestnut Ridge Research Building, Morgantown, WV 26505, USA}
\author{Kaleb Maraccini}
\affiliation{Center for Gravitation, Cosmology and Astrophysics, Department of Physics, University of Wisconsin-Milwaukee,\\ P.O. Box 413, Milwaukee, WI 53201, USA}
\author[0000-0001-7697-7422]{Maura A. McLaughlin}
\affiliation{Department of Physics and Astronomy, West Virginia University, P.O. Box 6315, Morgantown, WV 26506, USA}
\affiliation{Center for Gravitational Waves and Cosmology, West Virginia University, Chestnut Ridge Research Building, Morgantown, WV 26505, USA}
\author[0000-0002-4307-1322]{Chiara M. F. Mingarelli}
\affiliation{Center for Computational Astrophysics, Flatiron Institute, 162 5th Avenue, New York, New York, 10010, USA}
\affiliation{Department of Physics, University of Connecticut, 196 Auditorium Road, U-3046, Storrs, CT 06269-3046, USA}
\author[0000-0002-3616-5160]{Cherry Ng}
\affiliation{Dunlap Institute for Astronomy and Astrophysics, University of Toronto, 50 St. George St., Toronto, ON M5S 3H4, Canada}
\author{Benjamin M. X. Nguyen}
\affiliation{Department of Physics and Astronomy, Franklin \& Marshall College, P.O. Box 3003, Lancaster, PA 17604, USA}
\author[0000-0002-6709-2566]{David J. Nice}
\affiliation{Department of Physics, Lafayette College, Easton, PA 18042, USA}
\author[0000-0001-5465-2889]{Timothy T. Pennucci}
\altaffiliation{NANOGrav Physics Frontiers Center Postdoctoral Fellow}
\affiliation{National Radio Astronomy Observatory, 520 Edgemont Road, Charlottesville, VA 22903, USA}
\affiliation{Institute of Physics, E\"{o}tv\"{o}s Lor\'{a}nd University, P\'{a}zm\'{a}ny P. s. 1/A, 1117 Budapest, Hungary}
\author[0000-0002-8826-1285]{Nihan S. Pol}
\affiliation{Department of Physics and Astronomy, West Virginia University, P.O. Box 6315, Morgantown, WV 26506, USA}
\affiliation{Center for Gravitational Waves and Cosmology, West Virginia University, Chestnut Ridge Research Building, Morgantown, WV 26505, USA}
\author[0000-0002-4709-6236]{Joshua Ramette}
\affiliation{Department of Physics, Hillsdale College, 33 E. College Street, Hillsdale, Michigan 49242, USA}
\affiliation{Department of Physics, Massachusetts Institute of Technology, 77 Massachusetts Avenue, Cambridge, MA 02139-4307}
\author[0000-0001-5799-9714]{Scott M. Ransom}
\affiliation{National Radio Astronomy Observatory, 520 Edgemont Road, Charlottesville, VA 22903, USA}
\author[0000-0002-5297-5278]{Paul S. Ray}
\affiliation{Space Science Division, Naval Research Laboratory, Washington, DC 20375-5352, USA}
\author[0000-0002-7283-1124]{Brent J. Shapiro-Albert}
\affiliation{Department of Physics and Astronomy, West Virginia University, P.O. Box 6315, Morgantown, WV 26506, USA}
\affiliation{Center for Gravitational Waves and Cosmology, West Virginia University, Chestnut Ridge Research Building, Morgantown, WV 26505, USA}
\author[0000-0002-7778-2990]{Xavier Siemens}
\affiliation{Department of Physics, Oregon State University, Corvallis, OR 97331, USA}
\affiliation{Center for Gravitation, Cosmology and Astrophysics, Department of Physics, University of Wisconsin-Milwaukee,\\ P.O. Box 413, Milwaukee, WI 53201, USA}
\author[0000-0003-1407-6607]{Joseph Simon}
\affiliation{Jet Propulsion Laboratory, California Institute of Technology, 4800 Oak Grove Drive, Pasadena, CA 91109, USA}
\author[0000-0002-6730-3298]{Ren\'{e}e Spiewak}
\affiliation{Centre for Astrophysics and Supercomputing, Swinburne University of Technology, P.O. Box 218, Hawthorn, Victoria 3122, Australia}
\author[0000-0001-9784-8670]{Ingrid H. Stairs}
\affiliation{Department of Physics and Astronomy, University of British Columbia, 6224 Agricultural Road, Vancouver, BC V6T 1Z1, Canada}
\author[0000-0002-1797-3277]{Daniel R. Stinebring}
\affiliation{Department of Physics and Astronomy, Oberlin College, Oberlin, OH 44074, USA}
\author[0000-0002-7261-594X]{Kevin Stovall}
\affiliation{National Radio Astronomy Observatory, 1003 Lopezville Rd., Socorro, NM 87801, USA}
\author[0000-0002-1075-3837]{Joseph K. Swiggum}
\altaffiliation{NANOGrav Physics Frontiers Center Postdoctoral Fellow}
\affiliation{Department of Physics, Lafayette College, Easton, PA 18042, USA}
\author[0000-0003-0264-1453]{Stephen R. Taylor}
\affiliation{Department of Physics and Astronomy, Vanderbilt University, 2301 Vanderbilt Place, Nashville, TN 37235, USA}
\author[0000-0002-4657-9826]{Michael Tripepi}
\affiliation{Department of Physics, Hillsdale College, 33 E. College Street, Hillsdale, Michigan 49242, USA}
\affiliation{Department of Physics, The Ohio State University, Columbus, OH 43210, USA}
\author[0000-0002-4162-0033]{Michele Vallisneri}
\affiliation{Jet Propulsion Laboratory, California Institute of Technology, 4800 Oak Grove Drive, Pasadena, CA 91109, USA}
\author[0000-0003-4700-9072]{Sarah J. Vigeland}
\affiliation{Center for Gravitation, Cosmology and Astrophysics, Department of Physics, University of Wisconsin-Milwaukee,\\ P.O. Box 413, Milwaukee, WI 53201, USA}
\author[0000-0002-6020-9274]{Caitlin A. Witt}
\affiliation{Department of Physics and Astronomy, West Virginia University, P.O. Box 6315, Morgantown, WV 26506, USA}
\affiliation{Center for Gravitational Waves and Cosmology, West Virginia University, Chestnut Ridge Research Building, Morgantown, WV 26505, USA}
\author[0000-0001-5105-4058]{Weiwei Zhu}
\affiliation{National Astronomical Observatories, Chinese Academy of Science, 20A Datun Road, Chaoyang District, Beijing 100012, China}

%% file: text/abstract.tex

We present a new analysis of the profile data from the 47 millisecond pulsars comprising the 12.5-year data set of the North American Nanohertz Observatory for Gravitational Waves (NANOGrav), which is presented in a parallel paper \citep[][\twelveyr]{Alam20a}.
Our reprocessing is performed using ``wideband'' timing methods, which use frequency-dependent template profiles, simultaneous time-of-arrival (TOA) and dispersion measure (DM) measurements from broadband observations, and novel analysis techniques.
In particular, the wideband DM measurements are used to constrain the DM portion of the timing model.
We compare the ensemble timing results to \twelveyr\ by examining the timing residuals, timing models, and noise model components.
There is a remarkable level of agreement across all metrics considered. 
Our best-timed pulsars produce encouragingly similar results to those from \twelveyr.
In certain cases, such as high-DM pulsars with profile broadening, or sources that are weak and scintillating, wideband timing techniques prove to be beneficial, leading to more precise timing model parameters by $10-15$\%.
The high-precision, multi-band measurements of several pulsars indicate frequency-dependent DMs.
Compared to the narrowband analysis in \twelveyr, the TOA volume is reduced by a factor of 33, which may ultimately facilitate computational speed-ups for complex pulsar timing array analyses.
This first wideband pulsar timing data set is a stepping stone, and its consistent results with \twelveyr\ assure us that such data sets are appropriate for gravitational wave analyses.

%% file: text/sec_intro.tex

Pulsar timing arrays (PTAs) are poised to make the first detection of nanohertz gravitational waves (GWs) through the decades-long monitoring of dozens of millisecond pulsars (MSPs) \citep{Taylor2016,RosadoSesanaGair:2015}.
Current PTA experiments include the North American Nanohertz Observatory for Gravitational Waves \citep[NANOGrav\footnote{Please visit our website at \href{http://nanograv.org}{nanograv.org}.},][]{Alam20a,Cordes19,Ransom19}, the Parkes Pulsar Timing Array in Australia \citep[PPTA,][]{Kerr20,Hobbs13}, the European Pulsar Timing Array \citep[EPTA,][]{Desvignes16,Kramer13}, and newly established PTA efforts in India \citep{Susobhanan20,Joshi18} and China \citep{Hobbs19,Lee16}.
Together, the PTA collaborations work together under the umbrella venture called the International Pulsar Timing Array \citep[IPTA, ][]{Perera19,Manchester13b}.
Several other key science projects on premier radio telescopes, such as the MeerTime project with the MeerKAT telescope \citep{Bailes16} and the CHIME/Pulsar collaboration with the eponymous CHIME telescope \citep{Ng17}, will soon contribute to the ensemble PTA effort.
Furthermore, planned telescopes like the DSA-2000 \citep{Hallinan19} and the ngVLA \citep{Mckinnon19} will significantly broaden the impacts of PTA science.

The raw data collected by the PTA observations in all of the above often take the form of light curves, called pulse profiles, which map the average radio flux density to the rotational phase of the neutron star as a function of time, frequency, and polarization.
Pulsar timing methods in general obtain pulse times-of-arrival (TOAs) by cross-correlating these data profiles with a template profile \citep[e.g.,][]{Lommen13}.
A timing model of the neutron star's rotation is fit to the observed TOAs and predicts future rotations of the neutron star \citep[e.g., see Chapter~8 of][]{L&K05}.
TOA measurements have historically been and will continue to be the fundamental timing quantities of interest until other methods become more commonly implemented, such as those that produce timing model solutions by examining the profile data directly \citep{Lentati17a,Lentati15a}.

Along with the anticipation of GW detection are the expectations that the number of MSPs that comprise the array and the bandwidth of PTA observations will increase.
In particular, for NANOGrav, we project to have $>$100~MSPs timed by the middle of the decade and to be using an ultra-wideband receiver (between $\sim$0.7--4.0~GHz) at at least one of our facilities in the near future \citep[see][]{Ransom19}.
Indeed, large fractional-bandwidth receivers have already been deployed by the PPTA \citep{Hobbs20} and the EPTA \citep{Beacon} for high-precision pulsar timing, and large-fractional bandwidth or multi-band sub-arraying capabilities are either planned or are implemented in all of the aforementioned efforts.


The PTA detection of low-frequency GWs requires both high-cadence pulsar timing in addition to as many long pulsar data sets as possible \citep{bs+19,Lam18c,Siemens2013,Burt2011}.
The combination of more MSPs and increased bandwidth presents PTAs with an ever-increasing, and perhaps intractable, number of TOAs that need to be analyzed.
This problem is compounded not just by the long-term nature of PTAs, but also by increasing the rate of observation, as is the case for the roughly daily cadence of observations by CHIME/Pulsar, which will later be integrated into NANOGrav data sets.
As demonstrative examples: there are almost two-and-a-half times the number of TOAs for a single NANOGrav pulsar in our most recent data set than there are in the entire first NANOGrav data set, and, depending on the exact observations and processing protocol, CHIME/Pulsar by itself may double our current TOA volume after a single year of collecting data.
The absolute number of TOAs, as well as the number of timing model parameters (including parameterizations of the noise), play a determining role in the time it takes to perform GW analyses of PTA data \citep[][]{vHV15,vhV14,Ellis13a,Lentati13b}.
Advanced data analysis techniques to handle this deluge of TOAs will need to be significantly improved if we want to avoid delays on the numerous science deliverables offered by PTAs \citep{Cordes19,Fonseca19,Goulding:2019,Kelley19,Lynch19,Mingarelli2019,Siemens19,Stinebring19,Taylor19}.

A naive suggestion is to frequency-average the profiles, which would reduce the number of TOAs by factors of dozens.
However, maintaining frequency resolution in MSP timing observations is required when observing with even moderate fractional bandwidths for at least three reasons: (1) inter-observational changes in the dispersive delay due to the homogeneous, ionized interstellar medium (ISM) may be measurable and need to be modeled as part of the timing model, (2) the profile shape may change as a function of frequency, which will blunt the timing accuracy and precision if unmodeled, and (3) the effects of diffractive scintillation, particularly in combination with (2), may need to be accounted for.
The dispersive delay is proportional to the column density of free electrons along the line of sight, which is called the dispersion measure (DM), and the measurement and accommodation of DM changes is an outstanding problem in high-precision pulsar timing \citep{Jones17,Lam2016,Lee14,Keith13,Lentati13}.
Additionally, as bandwidths grow, more subtle effects arising from the inhomogeneity in the ISM become more prominent in pulsar timing; these effects include profile broadening \citep{Geyer17,lkd+17,Geyer16,Levin16}, non-dispersive delays \citep{Lam2018,FosterCordes90}, and frequency-dependent DMs \citep{Lam19,Donner19,Cordes16}.

Current methods to handle these issues grew mostly out of historical practices and do not address the TOA volume issue.
For instance, in the NANOGrav 5-year data set \citep[][hereafter NG5]{Demorest13}, we used individual phase offset parameters between frequency channels (called ``JUMP'' parameters) to account for frequency-dependent profile shapes that were evident even in the data from our narrower bandwidth data acquisition backends, ASP and GASP.
A simpler model was employed in the three subsequent data sets, the\\9-, 11-, and 12.5-year data sets (hereafter referred to as \nineyr\ \citep{Arzoumanian2015b}, \elevenyr\ \citep{Arzoumanian2018a}, and \twelveyr\ \citep{Alam20a}, respectively), in which a polynomial is fit to the average frequency-dependent TOAs as a function of log-frequency.
This model, parameterized by ``FD'' (frequency-dependent) parameters, was necessitated by the adoption of the PUPPI and GUPPI backends, which are capable of processing bandwidths that are wider by more than an order of magnitude.
However, no direct modeling of the evolving pulse profile shapes is performed to make those data sets.

\citet{PDR14} and \citet{Liu14} contemporaneously provided the beginnings of a new solution, which conveniently addresses profile evolution, ISM variations, and the TOA volume problem in one framework, referred to as ``wideband timing''.
The basic idea is to use a combination of a frequency-dependent profile model with an augmented TOA measurement algorithm to produce two measurements irrespective of the frequency resolution of the profile data: one TOA and one DM.
The usage of wideband TOAs and their associated DM measurements requires special attention and new techniques, which are detailed later.
For this reason, up until now, there has been no published, large-scale application of wideband timing for PTAs or other projects, although early, proof-of-concept demonstrations on \nineyr\ can be found in \citet{PennucciPhDT}.

In \twelveyr, we presented our 12.5-year data set, the creation and timing analyses of which use subbanded (i.e., per frequency channel) TOAs; we refer to that data set and its analysis with the moniker ``narrowband'' (NB).
Here we present new analyses of the same pulse profile data for the same 47 MSPs, reduced into the form of wideband TOAs with DM measurements and associated timing models, and refer to it as the ``wideband'' (WB) data set.
As we demonstrate, this first-ever wideband data set yields consistent timing results, and is publicly available in parallel with the narrowband data set\footnote{Please visit \href{https://data.nanograv.org/}{data.nanograv.org} for access to all of NANOGrav's data sets.  Specifically, the 12.5-year data set analyzed here is the ``v4'' version.  The data set presented here has the permanent DOI 10.5281/zenodo.4312887.}.

The structure of this paper is as follows.
In Section~\ref{sec:obs}, we briefly summarize the observations, but refer the reader to \twelveyr\ for the full description.
In Section~\ref{sec:wb}, we detail the generation of the wideband data set, including frequency-dependent template profile modeling, TOA measurement, and data set curation.
In Section~\ref{sec:results}, we present the ensemble results, which are largely consistent with those from \twelveyr, in a concise, comparative format; we also examine particular results from several individual pulsars.
In Section~\ref{sec:conclusion}, we summarize the discussion and comment on the future and on-going development of wideband timing for NANOGrav and other purposes.
Appendix~\ref{sec:low_snr} contains an analysis of wideband TOAs in the low signal-to-noise ratio (S/N) limit.
Appendix~\ref{sec:wb_like} describes the revised pulsar timing likelihood with which we analyze each pulsar's data set.
Appendix~\ref{sec:resid} contains the timing residuals and dispersion measure variations for all pulsars, from both data sets for ease of comparison.
We direct the reader to \twelveyr\ for discussions on new astrophysical results arising from the 12.5-year data set.
Furthermore, the results from searching the 12.5-year narrowband data set for a stochastic background of GWs have been reported in \citet{Arzoumanian20}, and a similar analysis of the wideband data set will be presented elsewhere.

%% file: text/sec_obs.tex

The observations comprising the NANOGrav 12.5-year data set were collected between July 2004 and June 2017, with timing baselines for individual pulsars in the range of 2.3 to 12.9~years.
Of the 47 MSPs presented here, 17 of them have been observed since the original \fiveyr\ data set, we added 20 more in \nineyr, 9 more in \elevenyr\ (with one \nineyr\ source, J1949$+$3106, removed), and 2 MSPs have been added for the present data set: J1946$+$3417, and J2322$+$2057.

All data were collected either at the 305-m Arecibo Observatory (AO), or the 100-m Robert C. Byrd Green Bank Telescope (GBT).
Any pulsar that is visible with the more sensitive AO dish is observed there, otherwise we observe it with the GBT.
Arecibo was used to observe 26 sources, while 23 sources have data from the GBT.
We regularly observe J1713$+$0747 and B1937$+$21 (a.k.a. J1939$+$2134) with both facilities.

Most pulsars are observed once every 3--4~weeks, with six sources being observed weekly: J0030$+$0451, J1640$+$2224, J1713$+$0747, J2043$+$1711, and J2317$+$1439 at AO since 2015, and J1713+0747 and J1909$-$3744 with the GBT since 2013.

All pulsars are observed with receivers in two widely separated frequency bands during each epoch in order to measure propagation effects from the ISM, including variations in the DM.
At Arecibo, these frequency bands are two of three possible receivers centered around 430~MHz ($\sim$70~cm), 1.4~GHz ($\sim$20~cm, ``L-band''), and 2.1~GHz ($\sim$15~cm, ``S-band''); the use of the 327~MHz ($\sim$90~cm) receiver for one source, J2317$+$1439, has been discontinued since the end of 2013.
At the GBT, all sources are observed with the 820~MHz ($\sim$35~cm) and 1.4~GHz receivers.  
The receiver turret at Arecibo accommodates back-to-back observations on the same day, defining one observational epoch, whereas mechanical and logistical factors demand that the two observations comprising a single epoch be separated by a few ($\sim$3) days at the GBT.

Between approximately 2010 and 2012 we transitioned from the 64~MHz bandwidth capable ASP and GASP data acquisition backend instruments at Arecibo and the GBT, respectively \citep{DemorestPhDT}, to the 800~MHz bandwidth capable PUPPI and GUPPI instruments \citep{Ford2010,DuPlain2008}.
Details of these instruments, their coverage of the receivers' bandwidth, and the transition can be found in \nineyr.
However, since the observed frequency ranges are of relevance to this work, we list them in Table~\ref{tab:observing_systems}, adopted from Table~1 of \nineyr.

\input{tables/tab_obssystem_table}

Our procedures for flux and polarization calibration, as well as for excision of radio frequency interference (RFI) are unchanged from \elevenyr.
Although dual polarization measurements are made, only the total intensity information is used in the timing analyses of either data set.

The profile data used to measure TOAs in both the narrowband and wideband data sets have $\nbin = 2048$ rotational phase bins and are time-averaged to have subintegration times up to 30~minutes or 2.5\% of the orbital period for binary pulsars, whichever is shorter.
The ASP and GASP data are left at their native 4~MHz frequency channel resolution, whereas the PUPPI and GUPPI data are frequency-averaged to have channel bandwidths in the range 1.5--12.5~MHz, depending on the frequency range observed.

These final, folded, calibrated, reduced profile data sets represent the same starting place for both the narrowband and wideband analyses.
Further details about the observations, their calibration, and data reduction can be found in \twelveyr\ and the earlier data set papers.

However, one new development in the preparation of these profiles that is important to highlight in the context of Section~\ref{subsec:prof_evol} is the correction of artifact images due to imperfect sampling of the pulsar signal.
To summarize the details found in \twelveyr, PUPPI and GUPPI use interleaved analog-to-digital converters (ADCs) that have slightly unbalanced gains and that do not sample exactly out of phase with one another.
If uncorrected, a very low amplitude band-flipped copy of the signal remains in the data, which corrupts the modeling of profile evolution for pulsars with certain combinations of spin period, DM, and S/N.
Following \citet{Kurosawa01}, the PUPPI and GUPPI profile data for each receiver were corrected for these artifact images using a routine implemented in the pulsar data reduction package \texttt{PSRCHIVE} \citep{Hotan04} as part of \twelveyr.
Some of the profiles for certain PUPPI observations could not be corrected; the TOAs obtained from these observations come with an additional metadata flag (see Table~\ref{tab:toa_flags}).

The timing baselines and observational coverage in the form of multi-frequency epochs for each pulsar are shown in Figure~\ref{fig:epochs}.
An analogous figure is presented in \twelveyr, but there are small differences in the exact epochs, as will be detailed in Section~\ref{subsec:wb_cleaning}.

\input{text/fig_epochs_12y}

%% file: tables/tab_obssystem_table.tex

\begin{deluxetable*}{@{\extracolsep{0pt}}cccccccccc}[ht!]
\centerwidetable
\tabletypesize{\footnotesize}
\tablewidth{0pt}
\tablecaption{Observing Frequencies and Bandwidths\tablenotemark{\footnotesize{a}}\label{tab:observing_systems}}
\tablehead{\\[-5pt] & \multicolumn{9}{c}{Backends} \\[0pt] \cline{2-10}\\[-13pt] \colhead{} & \multicolumn{4}{c}{ASP/GASP} & & \multicolumn{4}{c}{PUPPI/GUPPI} \\[-2pt]
           \cline{2-5}\cline{7-10}
           \rule{0pt}{1pt}
           Telescope      &                             & Frequency                         &  Usable                    &  $\Delta$DM &&                            & Frequency                         &   Usable                 &  $\Delta$DM \\
           Receiver       &  Data Span\tablenotemark{\tiny{b}} &        Range\tablenotemark{\tiny{c}} & Bandwidth\tablenotemark{\tiny{d}} & Delay\tablenotemark{\tiny{e}} && Data Span\tablenotemark{\tiny{b}} &        Range\tablenotemark{\tiny{c}} & Bandwidth\tablenotemark{\tiny{d}} & Delay\tablenotemark{\tiny{e}} \\
                          &            &     [MHz]        &  [MHz]    & [$\mu$s] &&           &    [MHz]         &   [MHz] & [$\mu$s]
           }
\startdata
\multicolumn{10}{l}{\phn Arecibo} \\[0pt]
\hline
\rule[0pt]{0pt}{11pt}%
  327   &  $2005.0-2012.0$ & $315-339$   &  34    & 2.86 && $2012.2-2017.5$ &  $302-352$        &  \phn 50    & 6.00 \\
  430   &  $2005.0-2012.3$ & $422-442$   &  20    & 1.03 && $2012.2-2017.5$ &  $421-445$        &  \phn 24    & 1.23 \\
  L-wide &  $2004.9-2012.3$ & $1380-1444$ &  64    & 0.09 && $2012.2-2017.5$ & $1147-1765$           &      603    & 0.91\\
  S-wide &  $2004.9-2012.6$ & $2316-2380$ &  64    & 0.02 && $2012.2-2017.5$ & \tablenotemark{\tiny{\phantom{f}}}$1700-2404$\tablenotemark{\tiny{f}}  &   460  & 0.36 \\[0pt]
\hline
\multicolumn{1}{c}{\rule[-3pt]{0pt}{11pt}GBT} & \multicolumn{9}{c}{} \\[0pt]
\hline
\rule[0pt]{0pt}{11pt}%
  Rcvr\_800 &  $2004.6-2011.0$ & $822-866$   &  64    & 0.30 && $2010.2-2017.5$ &  $722-919$       &   186  & 1.52 \\
  Rcvr1\_2 &  $2004.6-2010.8$ & $1386-1434$ &  48    & 0.07 && $2010.2-2017.5$ &  $1151-1885$      &   642  & 0.98
\enddata
\tablenotetext{a}{Table reproduced and modified from \nineyr.\vspace{-0.1in}}
\tablenotetext{b}{Dates of instrument use.  Observation dates of individual pulsars vary; see Figure~\ref{fig:epochs}.\vspace{-0.1in}}
\tablenotetext{c}{Typical values; some observations differed. Some frequencies unusable due to radio frequency interference.\vspace{-0.1in}}
\tablenotetext{d}{Nominal values after excluding narrow subbands with radio frequency interference.\vspace{-0.1in}}
\tablenotetext{e}{Representative dispersive delay between profiles at the extrema frequencies listed in the Frequency Range column induced by a $\Delta$DM$ = 5 \times 10^{-4}$~cm$^{-3}$~pc, which is approximately the median uncertainty across all wideband DM measurements in the data set; for scale, 1~$\mu$s~$\sim$~1~phase~bin for a 2~ms pulsar with our configuration of $\nbin = 2048$.\vspace{-0.1in}}
\tablenotetext{f}{Non-contiguous usable bands at $1700-1880$ and $2050-2404$~MHz.\vspace{-0.1in}}
\end{deluxetable*}

%% file: text/fig_epochs_12y.tex
\begin{figure*}[htb!]
\begin{center}
\includegraphics[width=6.0in]{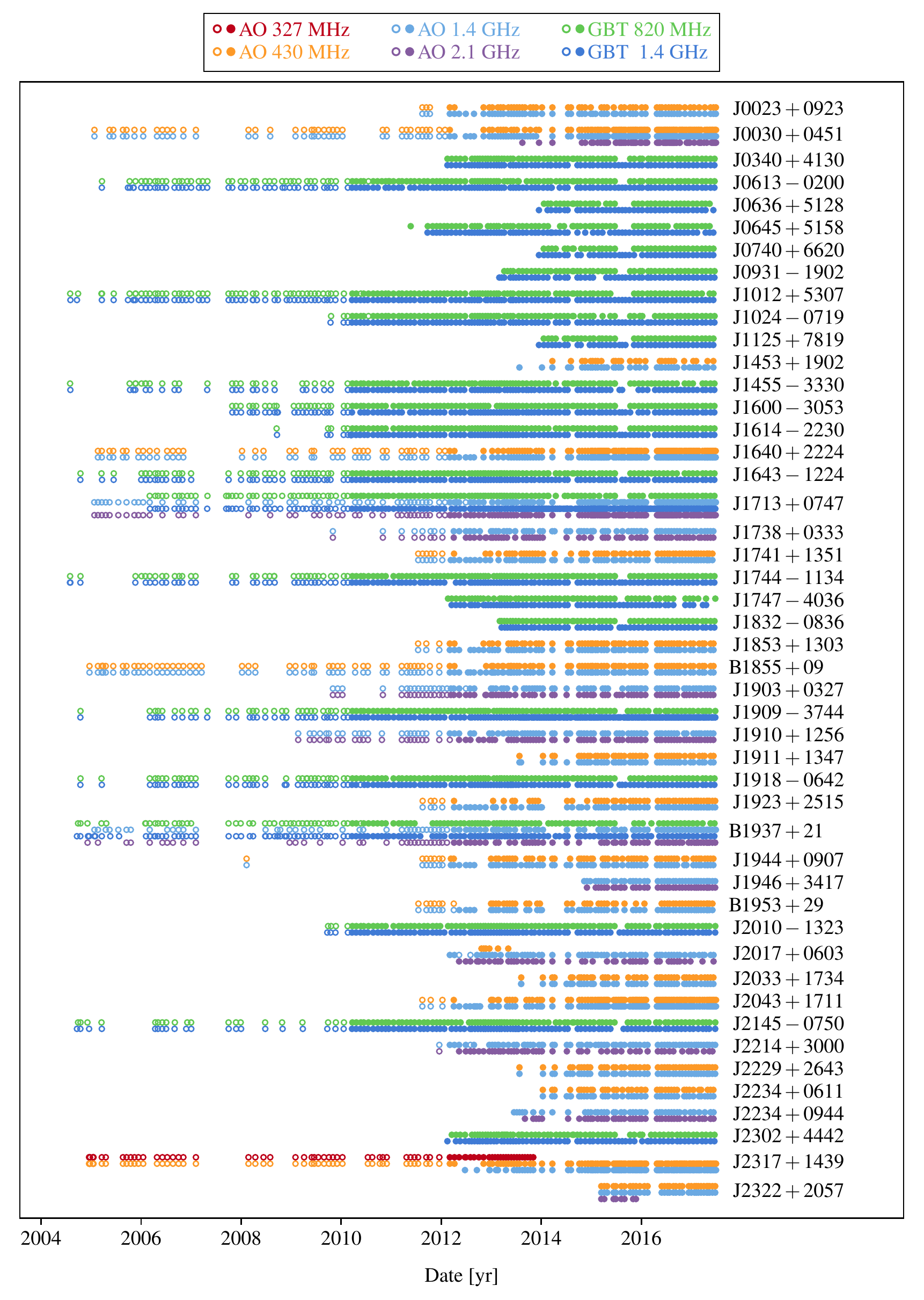}
\caption{\label{fig:epochs}
Epochs of all observations in the data set.
The color of each marker indicates the radio frequency band and observatory, as listed in the legend at the top; these colors are also used in Figures~\ref{fig:wb_toa_dm_err}, \ref{fig:wn_compare}, and \ref{fig:wb_low_snr}, and in the pulsar-specific plots of Appendix~\ref{sec:resid}.
The backend data acquisition system is indicated by marker type: open circles are ASP or GASP, whereas filled circles are PUPPI or GUPPI.
}
\end{center}
\end{figure*}

%% file: text/sec_wb.tex

\subsection{Overview}
\label{subsec:wideband_overview}

The measurement of TOAs from pulsar data with a large instantaneous bandwidth was first developed in \citet{Liu14} and \citet{PDR14}, and further explored in \citet{PennucciPhDT} and \citet{Pennucci19}.
We refer the reader to those works for details and here briefly summarize the important points. 

A single narrowband TOA corresponds to the time of arrival of a pulse profile observed in a single frequency channel\footnote{Another similar protocol used in the pulsar timing community is to produce band-averaged TOAs, in which the detected profiles are summed over the observing bandwidth, creating a single profile from which to extract the TOA.} (sometimes referred to as a ``subband''); in contrast, a single wideband measurement is composed of both the time of arrival of a pulse at some reference frequency and an estimate of the dispersion measure at the time of observation.
The difference can be conceptualized thusly: narrowband TOAs from a single subintegration are like the individual, scattered measurements around a linear relationship, whereas the fitted intercept and slope to this relationship are like the wideband TOA and DM, respectively.
The log-likelihood function for the wideband measurements is reproduced in Section~\ref{subsec:wb_toa_lnlike}.

The second important difference in the new wideband data set is not fundamental to the measurement of the TOA.
Heretofore we have used a single, frequency-independent template profile for each receiver band to generate narrowband TOAs and have used FD parameters \citep{Arzoumanian2015b} to account for constant phase offsets originating from the mismatch between the template and the evolving shape of the profiles.
For the measurement of wideband TOAs, we explicitly account for pulse profile evolution by using a high-fidelity, noise-free, frequency-dependent model for each receiver band.
See Section~\ref{subsec:prof_evol} for a brief description of how these models are created.

Although the narrowband and wideband data sets were developed in parallel, the established techniques in preparing the former allowed us to use some information from its final products to facilitate the production of the latter.
In particular, some of the curating performed, including flagging bad epochs, as well as the initial timing, was borrowed from the narrowband analysis.
In this way, the wideband data set is not completely independent, as is detailed in Sections~\ref{subsec:wb_cleaning}~\&~\ref{subsec:timing}.

It is important to underscore that the wideband data set for each pulsar is composed of TOAs that are paired with estimates of the instantaneous DM.
What makes the analysis of the wideband data set truly unique is that these DM estimates inform the portion of the timing model that accounts for DM variability (for our analyses, this is ``DMX''; see Section~\ref{subsec:timing}).
In Section~\ref{subsec:timing}, we describe our approach, with greater detail in Appendix~\ref{sec:wb_like}; the results are examined in Section~\ref{sec:results}.

Publicly available code\footnote{\url{https://github.com/pennucci/PulsePortraiture}} is used for both the generation of frequency-dependent templates and the measurement of the wideband TOAs \citep{pulseportraiture}.
\\
\subsection{Wideband TOA Log-Likelihood Function}
\label{subsec:wb_toa_lnlike}

All of our narrowband and wideband TOAs are measured using what is now referred to as the ``Fourier phase-gradient shift algorithm'' \citep[][historically known as ``FFTFIT'']{Taylor92}, which makes use of the Fourier shift theorem to achieve a phase offset precision much better than a single rotational phase bin, and which is computationally efficient by virtue of avoiding the time-domain cross-correlation calculation between the data and template pulse profiles.
We use a similar notation as Appendix~B of \nineyr, but see also \citet{DemorestPhDT} and \citet{PDR14} for details of what follows.
The time-domain model has the assumed form
\begin{equation}
\label{eqn:toa_signal}
    D(\nu, \varphi) = B(\nu) + a(\nu)\,T(\nu, \varphi-\phi(\nu)) + N(\nu),
\end{equation}
That is, for each subintegration in an observation, we assume that the data profiles $D$ as a function of rotational phase $\varphi$ and frequency $\nu$ can be described by a template $T$ that is shifted in phase by $\phi$ and scaled in amplitude by $a$, with added Gaussian-distributed phase-independent noise $N$; the term $B$ represents the bandpass shape.
After discretizing these quantities, taking the discrete Fourier transform (DFT), making use of the Fourier shift theorem, and rearranging terms, we can reformulate Equation~\ref{eqn:toa_signal} into our TOA log-likelihood,
\begin{equation}
\label{eqn:toa_lnlike}
    \chi^{2} = \sum_{n,k} \frac{\lvert d_{nk} - \an t_{nk}e^{-2\pi ik\phi_n}\rvert^{2}}{\sigma^2_n}.
\end{equation}
In Equation~\ref{eqn:toa_lnlike}, the integer index $k$ is the Fourier frequency (conjugate to rotational phase or time), $t_{nk}$ is the DFT of the template profile for the frequency channel indexed by $n$ (with frequency center $\nu_n$), $\an$ is the scaling amplitude parameter for the template, $\phi_n$ is the phase offset for the template, and $d_{nk}$ is the DFT of the data profile for frequency channel $n$, which has the corresponding Fourier-domain noise level $\sigma^2_n$\,\footnote{$\sigma^2_n$ is the noise for either the real or imaginary part, and is larger than its (real) time-domain counterpart by a factor of $\nbin/2$.}.

For conventional TOAs, the optimization of this function takes place on an individual channel basis, in which case there is no index $n$ in Equation~\ref{eqn:toa_lnlike} over which a summation occurs.
Moreover, for our narrowband TOAs, $t_{nk}$ is not a function of $n$; that is, profile evolution is not accounted for by changing the shape of the template across a single receiver's frequency band.
Instead, in \twelveyr, a single template profile is used for each receiver band and constant phase offsets arising from the mismatch between the template shape and the evolving pulse shape are accounted for via FD parameters in the narrowband timing models.

The crucial difference in wideband TOAs is that the phase offsets $\phi_n$ in Equation~\ref{eqn:toa_lnlike} are constrained to follow the cold-plasma dispersion law, proportional to $\nu^{-2}$:
\begin{equation}
\label{eqn:wb_constraint}
    \phi_n(\nu_n) = \phio + \frac{K\times\dm}{P_s}\Big(\nu_n^{-2}-\nuphi^{-2}\Big),
\end{equation}
where $P_s$ is the instantaneous spin period of the pulsar, $K$ is the dispersion constant (a combination of fundamental physical constants approximately equal to $4.148808 \times 10^3$~MHz$^2$~cm$^3$~pc$^{-1}$~s), $\dm$ is the dispersion measure, and $\phio$ is the phase offset at reference frequency $\nuphi$.
Equation~\ref{eqn:toa_lnlike} can be recast using the maximum-likelihood values of $\an$ and rewritten as a function of only the two parameters $\phio$ and DM (see \citet{PDR14}), which can then be readily optimized numerically.
We calculate the parameter uncertainties using the Fisher matrix and choose $\nuphi$ such that there is zero covariance between the DM and $\phio$, the latter of which is directly related to the TOA.

Additional terms to the wideband TOA log-likelihood are currently being explored (Pennucci et al., \textit{in prep.}), which include accounting for pulse broadening from multi-path propagation through the turbulent ISM (i.e., ``scattering'') in a similar fashion to \citet{lkd+17}, as well incorporating a higher-order delay term besides $\nu^{-2}$, the motivation for which are discrete ISM ``events'' \citep{Lam2018}.
The low-frequency, high-cadence capabilities offered by CHIME/Pulsar will make tracking the interstellar weather in this way an exciting endeavor, following in the footsteps of studies like \citet{Ramachandran06} and \citet{Driessen19} (long-term ISM tracking of B1937$+$21 and the Crab pulsar, respectively).

\subsection{Frequency-dependent Template Profiles}
\label{subsec:prof_evol}

The evolving template $t_{nk}$ in Equation~\ref{eqn:toa_lnlike} can be freely chosen, and in this work we employ the modeling method from \citet{Pennucci19}, which describes a generalized, frequency-dependent version of our usual protocol for making template profiles.
In contrast, to make the conventional noise-free templates used in \twelveyr\ for narrowband TOA measurement, all profiles for each combination of pulsar and receiver are averaged together to build a single, high S/N mean profile, which is then smoothed.
We direct the reader to \citet{Pennucci19} for details, but we summarize its novel procedure as follows.

An analogous averaging of the data for each combination of pulsar and receiver is performed, but frequency resolution is maintained to arrive at a high S/N mean ``portrait'' (a collection of nominally aligned mean pulse profiles across a contiguous frequency band); only the PUPPI and GUPPI data were averaged for this purpose.
A principal component analysis is performed on the average portrait, and the most significant, highest S/N eigenvectors (and mean profile) are smoothed to become noise-free basis functions (``eigenprofiles'').
The mean-profile-subtracted profiles from the average portrait are projected onto each of the eigenprofiles, producing a set of coefficients for each.
These coefficients are simultaneously fit to a slowly varying spline function that is parameterized by frequency and encapsulates the evolution of the pulse profile shape.

In this manner, a template profile $T$ at any frequency $\nu$ can be constructed by evaluating the $n_{\textrm{\scriptsize{eig}}}$ coefficient spline functions $B_i$ at $\nu$, linearly combining the eigenprofiles $\hat{e}_i$ using these coefficients, and adding the result to the mean profile $\widetilde{p}$,
\begin{equation}
\label{eqn:Tprof}
    T(\nu) = \sum^{n_{\textrm{\scriptsize{eig}}}}_{i=1} B_{i}(\nu) \, \hat{e}_i ~ + ~ \widetilde{p}.
\end{equation}

%

In summary, a single model for generating high-fidelity, noise-free template profiles is composed of the smoothed mean profile, the smoothed basis eigenprofiles, and a function to describe the profile evolution curve in that basis.

These models were made for each combination of pulsar and receiver, and then used to measure wideband TOAs according to Equation~\ref{eqn:toa_lnlike}.
The modeling procedure attempts to guess the true, unknown profile alignment by starting with the same Occam assumption used in the narrowband analysis: there is no profile evolution, neither in the shape nor alignment of the profiles.
This assumption is used to initially align and average the profile data by using the fixed, mean profile shape as a reference for the alignment.
After iteratively aligning and averaging the profile data and then creating a model, it should not come as a surprise that the absolute, average DM measured in each receiver band will differ slightly.
We minimize this difference by measuring the weighted-mean DM offset relative to the DM measured in the lowest frequency band.
The DM offset was then applied as a rotation proportional to $\nu^{-2}$ to the average portrait, the profile evolution model was recreated, and the TOAs were remeasured; this process was iterated a total of three times.
The reference DM choice was made relative to the lowest frequency band because, except in the cases of Arecibo pulsars observed only at L- and S-bands, this will be a frequency band with lower fractional bandwidth than L-band, but from which reasonably precise DM measurements are made.
This choice gave better modeling results than rotating the averaged low frequency data relative to the L-band alignment, which may be ambiguous due to profile evolution.
For the other sources, S-band generally does not give precise DM measurements, and so L-band is used as the reference.
See Section~\ref{subsec:model_results} for more discussion on this topic.

The initial set of wideband TOAs used in the timing and noise analyses were measured with these DM-aligned models, and instrumental time offsets were applied to TOAs from ASP and GASP profiles, as detailed in Appendix~A of \nineyr.
Metadata in the TOA files take the form of ``flags'', which get appended to each TOA line in the files.
A number of new TOA flags have been added to aid wideband timing analyses, and a few of the usual TOA flags have different meanings from their narrowband TOA counterparts; these are listed in the top portion of Table~\ref{tab:toa_flags}.

The choice of DM alignment in wideband profile models is analogous to the ambiguity of absolute phase between TOAs measured with different template profiles in the narrowband analysis.
Those constant phase offsets are modeled in the timing model with so-called ``JUMP'' parameters and are also present in the wideband analysis.
Our fiducial DM alignment is an attempt at getting the simplest profile evolution models, but a new, analogous timing model parameter is necessary when using multi-band DM measurements as data for the timing model.
To this end, we implemented ``DMJUMP'' parameters for wideband timing in the extended likelihood introduced in Section~\ref{subsec:timing}.
Appendix~\ref{sec:wb_like} contains details about how these parameters influence the timing model.

\input{tables/tab_toa_flags}

\subsection{Cleaning \& Curating the Wideband Data Set}
\label{subsec:wb_cleaning}

The narrowband data set was prepared in advance of the wideband data set, and as a part of its creation we kept track of bad observations that were corrupted by instrumentation or calibration issues, or were so affected by RFI that we excised them outright (250 of 11,178 observations).  
These observations (which are included in the narrowband data set as commented TOAs with the flag \texttt{-cut badepoch}) were simply not introduced into the wideband pipeline.
There are also a small number of observations (36) for which data were taken on a pulsar using a different receiver than usual, often for testing purposes (these are included in the narrowband data set as commented TOAs with the flag \texttt{-cut orphaned}).
These data are generally not sufficient to create good profile evolution models, and would add very few degrees of freedom; we similarly excluded them from the wideband analysis at the start.

There was one other additional step in curating the profile data set used to make wideband TOAs.
Upon finishing the modeling procedure described in Section~\ref{subsec:prof_evol}, we calculated goodness-of-fit statistics for each profile in the data set based on its predicted pulse shape from the corresponding model.
Profiles in a given subintegration were zero-weighted if their goodness-of-fit exceeded a threshold ($\chi^2_{\textrm{reduced}} > 1.25$), which was empirically determined after examining the distributions for each combination of pulsar and receiver.

For most combinations, the number of discarded profiles in this manner was of order a few percent.
After zero-weighting these profiles, the data were re-averaged and the profile evolution models were recreated.
This step was necessary because, as with the ADC artifact mentioned in Section~\ref{sec:obs}, unmitigated RFI can corrupt the modeling procedure.
More general RFI-flagging techniques based on template-matching using the wideband profile models are in development within NANOGrav and elsewhere (MeerTime collaboration, private communication).
Such techniques could potentially identify irregularities in the profiles, be it from RFI or other sources, earlier in the reduction pipeline.

The remainder of the cleaning of the wideband data set was performed on the measured TOAs; any TOAs ``cut'' from further analysis were given one of the flags listed in Table~\ref{tab:toa_flags}, but are included as commented TOAs in the publicly available text files.
Most of the cuts described in the table have counterparts in the preparation of the narrowband data set, and we refer the reader to \twelveyr\ for details beyond those offered in the table and those that follow.

The S/N threshold used for the wideband TOAs was set at $25$, compared to the value of $8$ used for narrowband TOAs.
The main reason for this was empirical and related to the fact that the estimated S/N for wideband TOAs is subject to significant bias in the low S/N regime, favoring a higher threshold than is naively derived.
We justify this choice in Appendix~\ref{sec:low_snr}.

Note that in \elevenyr\ and \twelveyr\ a numerical TOA outlier analysis is performed \citep{Vallisneri2017}.
Some of the narrowband TOAs identified in this way are from profiles corrupted by RFI or instrumental problems that were not otherwise identified.
Our goodness-of-fit filter of the profile data described earlier served a similar purpose, and no separate outlier TOA analysis was performed.
We found that after filtering the profiles in this way and thresholding the TOAs based on the S/N cutoff of $25$, the initial timing results were remarkably clean; there were only a handful of additional TOAs that were culled based on a large timing residual ($>100~\mu$s) or were otherwise identified by eye (see Table~\ref{tab:toa_flags}).

\input{tables/tab_ntoa_nchan}

Overall, despite the procedural differences in preparing the two data sets, the quality control for the wideband data set resulted in $\sim$~16\% more profiles used for TOA measurement, as can be seen in Table~\ref{tab:ntoa_nchan}.
This difference is largely due to the inclusion of low S/N ratio profiles that are discarded in the narrowband data set (see Appendix~\ref{sec:low_snr}); as such, it is unsurprising that these additional data in general do not carry a proportionally large impact on the timing results, as will be shown.
However, see Section~\ref{subsec:pulsar_results} for specific examples.

After curation, the resulting wideband data set has 12,598 TOAs, corresponding to 480,474 profiles; this is compared to the 415,122 TOAs in the narrowband data set, a factor of $\sim$~33 larger in TOA volume, which will only grow as the ASP and GASP TOAs become a fractionally smaller subset of the entire data set, and as new wideband facilities and receivers come into use.  
Note, however, that the overall wideband data set volume is only a factor of $33/2 \sim 16$ smaller, after including the DM measurements in the analysis.

\begin{table*}[th!]
    \caption{Basic Pulsar Parameters and TOA Statistics\label{tab:toa_summary}}
\centering
\input{tables/tab_toa_summary}

\end{table*}

\input{text/fig_wb_toa_dm_err}

A summary of the TOA uncertainties are presented in two forms.
First, the median uncertainties are listed, along with other basic pulsar parameters, in Table~\ref{tab:toa_summary}.
There is an analogous table in \twelveyr\ for the narrowband TOAs; in both cases, the uncertainties have been scaled to estimate the median TOA uncertainty from a 1800~second observation of the pulsar with 100~MHz of bandwidth.
Overall, the values are comparable to their narrowband counterparts, but differences may be attributable to any of: unmodeled profile evolution in the narrowband data set, the inclusion of very low S/N profiles in the wideband data set, the additional fit parameter (DM) in the wideband measurement, or other subtle discrepancies.
Second, in Figure~\ref{fig:wb_toa_dm_err} we graphically present the ``raw'' median TOA and DM uncertainties with central intervals covering the central 68\% of the distribution, ranking pulsars by their median PUPPI or GUPPI L-band TOA uncertainty.
We use ``raw'' to mean that these are the formal, estimated uncertainties from the template-matching procedure, which do not include any other sources of uncertainty and are not scaled in any way.
It is obvious from this plot that, depending on the pulsar, the improvement in raw TOA precision after moving from ASP and GASP to PUPPI and GUPPI is a factor of 2--3 or more in many cases, but the DM precision improves by an order of magnitude or more in all receiver bands except 327 and 430~MHz.
This improvement is due to the increase in bandwidth covered by PUPPI and GUPPI (see Table~\ref{tab:observing_systems}).

\subsection{Obtaining Timing Solutions}
\label{subsec:timing}

We used the 12.5-year data set results from \twelveyr\ as initial timing solutions instead of deriving completely new timing results from the extended baselines of the 11-year data set.
This was done in part to facilitate comparisons and in part to reduce the need for redundant analyses.
Specifically, any new spin, astrometric, or binary timing model parameters found to be significant in \twelveyr\ were retained, but FD parameters were removed, as were the parameters that describe the DM model, called DMX.

DMX is a piecewise-constant characterization of DM variability that is part of the timing model.
Simpler models of DM variability, such as low-order polynomials, do not describe the data well, but more advanced models, such as those that use a stochastic description of variability \citep[e.g., as a Gaussian process,][]{Lentati13}, are currently being investigated.
The criteria for dividing up the TOAs into DMX epochs defined by Modified Julian Dates (MJDs) can be found in \twelveyr.
For each DMX epoch, a DM is measured based on the $\nu^{-2}$ dependence of the TOAs that fall within the epoch, and all of these DMX model parameters are measured simultaneously with the fit for the rest of the timing model.

If we were to ignore the wideband DM measurements, the wideband TOA data set would be significantly hampered in the following ways.
There are a large number of DMX epochs which contain data from a single receiver.
In the cases where such an epoch has a single wideband TOA (instead of the dozens of analogous narrowband TOAs), the corresponding single DMX parameter removes the single degree of freedom, artificially zeroing out the timing residual for this epoch.
If there are a few wideband TOAs from the same receiver band in such an epoch, they will have similar reference frequencies, and so the DMX parameter will be poorly constrained and perhaps biased.
Finally, even for the majority of DMX epochs for which there are multi-frequency wideband TOAs from dual receiver observations, DMX only has access to the TOAs, their uncertainties, and reference frequencies.
That is, the information about the dispersive delays across the individual receiver bands (captured by the wideband DM measurements, or, equivalently, the multi-frequency TOAs in the narrowband data set) is lost, and DMX only sees the dispersive delay between the bands.
As can be seen in most pulsars' DM and DMX time series (see Appendix~\ref{sec:resid}), the wideband TOAs and their inter-band dispersive delay carry more weight in the DMX model than do the intra-band delays characterized by their corresponding wideband DM measurements.
How much more so depends on the pulsar and receiver bands in question, but it is important to highlight that disregarding the DM data is not a viable option for analyzing this data set.
Indeed, we attempted several such analyses, which yielded significantly worse results in many pulsars.

Therefore, not only was it appropriate, but it was also necessary to expand the likelihood used to fit our timing models so that the wideband DM measurements inform the DM model.
In effect, in the new likelihood, the wideband DM measurements influence the timing model as prior information on the DMX values.
Each of the TOAs falling within a DMX epoch have a corresponding DM measurement; the weighted average of these measurements is used as the mean of a Gaussian prior on the DMX value for that epoch, while the standard error of the weighted average is the prior distribution's standard deviation.
The details of this new likelihood and its implementation in the pulsar timing software packages \texttt{Tempo} \citep{tempo} and \texttt{ENTERPRISE} \citep{enterprise} can be found in Appendix~\ref{sec:wb_like}.

\afterpage{\clearpage}

The timing models from \twelveyr\ were first refit with \texttt{Tempo} using the wideband TOAs only, omitting the DM measurements, to setup the DMX epochs and to get initial DMX values.
Including the DM measurements at this point sometimes resulted in poor timing results because there is currently no way to fit the DMJUMP parameters simultaneously with the timing model within \texttt{Tempo}.
It is at this stage that TOAs were excluded from further analysis if they did not meet the frequency ratio criterion described in Table~\ref{tab:toa_flags} or if the entire epoch was removed based on a new analysis performed in \twelveyr\ (also mentioned in Table~\ref{tab:toa_flags}).

The wideband TOAs, DMs, and timing models were then subject to a Bayesian analysis with \texttt{ENTERPRISE} using the new wideband likelihood.
This analysis optimizes the probability of the observed data by characterizing the noise in the timing residuals, which has both white and red components, much in the same way as in \twelveyr, \elevenyr, and \nineyr, with a few important differences:

\textit{No ECORR --}
There is one parameter in the standard white noise model that is not used in the wideband analyses.
This parameter, called ECORR, accounts for the (assumed 100\%) correlation between multi-frequency TOAs taken at the same time and is used in the narrowband analyses of \nineyr, \elevenyr, and \twelveyr\ \citep{Arzoumanian2014,Arzoumanian2015b}.
Since wideband TOAs effectively consolidate the many narrowband TOAs into one, any physical effects contributing to this parameter (such as pulse jitter or ISM effects; see Section~\ref{subsubsec:wn_compare}) would be absorbed by the standard EQUAD noise parameter, which is added in quadrature to the measured TOA uncertainty \citep{Edwards06,Lentati14}.
Alternatively, any effects contributing to ECORR in the narrowband analysis may be modeled by a larger and shallower red noise process in the wideband analysis.
A comparison of the detected excess white noise in the two data sets is presented in Section~\ref{subsubsec:wn_compare}.

\textit{DMEFAC \& DMJUMP --}
Two additional parameters are needed in the new wideband likelihood.
The first, which we call ``DMEFAC'', is analogous to the standard TOA EFAC: it is a factor that scales the estimated wideband DM measurement uncertainty.
In a similar fashion to the other white noise parameters, a DMEFAC is assigned for each combination of receiver and backend in each pulsar's noise model.
The second was introduced in Section~\ref{subsec:prof_evol}, which we call ``DMJUMP''.
This parameter is analogous to standard JUMP parameters, but instead of modeling an achromatic phase offset between TOAs measured in different receiver bands, DMJUMP is a DM offset between wideband DMs measured in different bands.
These parameters account for the differences in alignment between profile evolution models in disparate bands, and amount to making a choice for the absolute DM.
It is important to stress that this ambiguity in absolute DM, as well as the offsets in DMs measured in disparate bands, exist also in the narrowband analyses; in \twelveyr, the choice of having fixed templates in each band, coupled with using FD parameters to account for constant TOA biases as a function of frequency, amount to addressing the analogous problems.
We assign one DMJUMP parameter per receiver in each pulsar's timing model, since the profile evolution models are independent of backend.
It may seem that we should use one less DMJUMP parameter than there are receivers in each pulsar's analysis, as is done for standard phase JUMP parameters.
However, because the DMX model is separately informed by the TOAs, it is not an overdetermined problem.
This fact is borne out by examining the posterior chains; although we see that the DMJUMP parameters are often highly covariant, they are not completely degenerate.
We used a uniform prior distribution on DMJUMP parameters in the range $[-0.01,0.01]~\dmunits$; virtually all of the values are $\lvert\textrm{DMJUMP}\rvert < 0.004~\dmunits$.

\textit{White noise priors --}
In the analyses of all of our other data sets, we have used large, uniform priors on EFAC between 0.1 and 10.0.
EFAC was originally implemented to account for instances when the profile data poorly matched the template profile in the TOA fit, which would underestimate the TOA uncertainty.
In the present analysis, we expect EFAC to be near 1.0 because we are using evolving profile templates and have carefully excised RFI at a number of stages in the pipeline.
We have found that allowing extreme EFAC values can inadvertently over- or down-weight subsets of the data when it is not justified.
One reason for this is that there is a larger amount of covariance between EFAC and EQUAD parameters in the wideband analysis because the formal TOA uncertainties (of which there are far fewer) are more homoscedastic; EFAC and EQUAD parameters can only be differentiated if there is variance in the uncertainties.
Equation~\ref{eqn:N_matrix} describes how EFAC and EQUAD parameters are related and affect the TOA measurement uncertainty.
Therefore, we used a Gaussian prior on all EFAC parameters with a mean of 1.0 and standard deviation of 0.25; for similar reasons, we applied the same prior to DMEFAC parameters.
This choice is further justified in Appendix~\ref{sec:low_snr}, where we show that the estimated TOA and DM uncertainties based on calculating the Fisher matrix of Equation~\ref{eqn:toa_lnlike} are accurate down to very low S/N.
It should also be noted that these uncertainties, being based on the Fisher information matrix, are equal to the Cram\'{e}r-Rao lower bound, which motivates the continued use of EFAC parameters.
We use the same prior on EQUAD parameters as is used for both EQUAD and ECORR in \twelveyr, which is a uniform distribution on log$_{10}$(EQUAD~[s])~$\in [-8.5, -5.0]$.
Due to our use of non-uniform priors for EFAC and DMEFAC parameters, we refer to all point estimates from the noise modeling as maximum a posteriori (MAP) values, instead of maximum-likelihood values.

\textit{Red noise priors --}
We use the exact same red noise model and priors as in \twelveyr, but because the determination of red noise significance differs slightly from \elevenyr\ and \nineyr, and because it will be relevant in the discussion of results, we summarize it here.
The red noise is assumed to be a stationary Gaussian process, which we parameterize with a power-law power spectral density $P$ of the form
\begin{equation}
    \label{eqn:red_pl}
    P(f_m) = A^2_{\small \textrm{red}} \left(\frac{f_m}{1~\textrm{yr}^{-1}}\right)^{\gamma_{\small \textrm{red}}},
\end{equation}
where $A_{\small \textrm{red}}$ is the amplitude of the red noise at a frequency of 1\,yr$^{-1}$ in units of $\mu$s~yr$^{1/2}$, and $\gamma_{\small \textrm{red}}$ is the spectral index.
The spectrum is evaluated at thirty linearly spaced frequencies $f_m$ indexed by $m$, incremented by 1/$T_{\textrm{span}}$, where $T_{\textrm{span}}$ is the span of the pulsar's data set.
The prior on the red noise amplitude is uniform on log$_{10}$($A_{\small \textrm{red}}$~[yr$^{3/2}$])~$\in [-20, -12]$, whereas the prior on the red noise index has been constrained in both 12.5-year analyses to be uniform on $\gamma \in [-7, -1.2]$.
A pulsar is deemed to have ``significant red noise'' in these analyses if the Savage-Dickey density ratio \citep[a proxy for the Bayes factor,][]{dickey1971} estimated from the posterior distribution of log$_{10}$($A_{\small \textrm{red}}$) is greater than one hundred.
Very low-index red noise is thought to primarily arise from imperfect modeling of various effects from the ISM \citep{ShannonCordes2017} and will be covariant with the white noise parameters.
Including shallow red noise instead of modeling it with only white noise parameters will not significantly change the timing model.
The analyses here and in \twelveyr\ are only indicative of the presence of red noise, which may or may not be wholly intrinsic to the pulsar; a comparison of the red noise models is presented in Section~\ref{subsubsec:rn_compare}.
Advanced noise modeling of the 11- and 12.5-year data sets, in which we explore bespoke models for each pulsar specifically in the context of GW analyses, is underway and will be presented elsewhere (Simon et al. \textit{in prep.}).

Upon completion of the noise analysis, following the same protocol as in \twelveyr, the MAP noise model is included as fixed parameters in the timing model, which is re-optimized using the generalized least squares implementation of \texttt{Tempo}, now using the augmented, wideband likelihood.
The large majority of the reduced chi-squared (goodness-of-fit) values fall between 0.9 and 1.1, with a few larger values.
Some of these are to be expected because the additional DM data may not be particularly informative, or they may not be modeled well by DMX (e.g., see Section~\ref{subsec:DM_nu}).
As in \twelveyr, we examined the significance of adding and removing various timing model parameters, but after finding no strong evidence favoring change, we kept the identical set of timing model parameters for ease of comparison.
The differences with respect to crossing the significance threshold for including or excluding parameters are marginal, and in several cases are a function of the difference in red noise model (see Section~\ref{subsubsec:rn_compare}).

The timing models are summarized in Table~\ref{tab:timing_model_summary}, which also lists the Bayes factor, $B$, indicating the significance of red noise.
There is an analogous table in \twelveyr\ containing the results from the analyses of the narrowband data set.
As mentioned in Table~\ref{tab:toa_flags}, we removed ASP and GASP TOAs that were taken simultaneously with concurrent PUPPI or GUPPI observations from the final TOA data sets.
The final timing models with noise parameters, curated wideband TOAs, and related auxiliary files are the furnished products comprising this data release.
We present the timing residuals and DM time series for these data in Appendix~\ref{sec:resid}, which includes visual comparisons with the counterpart averaged residuals and DMX models from \twelveyr.

\begin{table*}[th!]
\centering
\caption{Summary of Timing Model Fits\label{tab:timing_model_summary}}
\input{tables/tab_timing_model_summary}

\end{table*}

%% file: tables/tab_toa_flags.tex
\begin{deluxetable*}{p{2.8cm}p{5.5cm}p{9.0cm}}[p!]
\centerwidetable
\tabletypesize{\footnotesize}
\tablewidth{0pt}
\tablecaption{Wideband TOA Flags\label{tab:toa_flags}}
\tablehead{\colhead{Flag} & \colhead{Meaning} & \colhead{Notes}}
  \startdata
  \texttt{-pp\_dm} \textit{value} & Dispersion Measure [cm$^{-3}$~pc] & \textit{value} is the wideband DM estimate from Equation~\ref{eqn:toa_lnlike} associated with the TOA. \\
  \texttt{-pp\_dme} \textit{value} & Dispersion Measure Uncertainty [cm$^{-3}$~pc] & \textit{value} is the estimated 1$\sigma$ uncertainty on the DM estimate. \\
  \texttt{-nch} \textit{value} & Number of Channels & \textit{value} is the integer number of frequency channels ($\nchan$); this in contrast to the \texttt{-nch} flag in the narrowband data set, which is the number of channels averaged together from the original, raw data. \\
  \texttt{-nchx} \textit{value} & Number of Channels Used & \textit{value} is the integer number of \textit{non-zero-weighted} frequency channels in the associated subintegration used in the wideband TOA fit. \\
  \texttt{-chbw} \textit{value} & Channel Bandwidth [MHz] & \textit{value} = bandwidth / $\nchan$. The total bandwidth can be recovered from this number and $\nchan$. \\
  \texttt{-bw} \textit{value} & Effective Bandwidth [MHz] & \textit{value} is the difference between the highest and lowest channels' center frequencies used in the wideband TOA fit; this is in contrast to the \texttt{-bw} flag in the narrowband data set, which is the bandwidth for each TOA, i.e., the channel bandwidth provided here by the \texttt{-chbw} flag. \\
  \texttt{-fratio} \textit{value} & Frequency Ratio & \textit{value} is the ratio of the highest and lowest channels' center frequencies; in combination with the effective bandwidth, this value can be used to recover the two frequencies. \\
  \texttt{-snr} \textit{value} & TOA Signal-to-Noise Ratio (S/N) & Similar to the conventional TOA flag, but calculated using Equation~\ref{eqn:wb_toa_snr}. \\
  \texttt{-gof} \textit{value} & TOA Goodness-of-Fit ($\chi^2_{\textrm{reduced}}$) & Similar to the conventional TOA flag, but calculated using Equation~\ref{eqn:toa_lnlike} and the relevant number of degrees of freedom. \\
  \texttt{-flux} \textit{value} & Flux Density [mJy] & Analogous to the \texttt{-flux} flag in the narrowband data set, \textit{value} is the estimated mean flux density for the subintegration (see Section~\ref{subsec:port_results}). \\
  \texttt{-fluxe} \textit{value} & Flux Density Uncertainty [mJy] & Analogous to the \texttt{-fluxe} flag in the narrowband data set, \textit{value} is the estimated 1$\sigma$ uncertainty on the flux density. \\
  \texttt{-flux\_ref\_freq} \textit{value} & Flux Density Reference Frequency [MHz] & \textit{value} is the reference frequency for the mean flux density estimate. \\
  \texttt{-img uncorr} & Incomplete Artifact Image Correction & Some of the profiles in this subintegration did not undergo removal of the ADC artifact image (see Section~\ref{sec:obs} and \twelveyr). \\
  \tableline
  \multicolumn{3}{c}{Flags Indicating a Removed TOA} \\
  \tableline
  \texttt{-cut dmx} & The ratio of maximum to minimum frequencies observed in a DMX epoch $\nu_{max}/\nu_{min} < 1.1$. & $\nu_{max}$ and $\nu_{min}$ are calculated from \texttt{-bw} and \texttt{-fratio} flags here, but correspond to individual TOA reference frequencies in the narrowband data set.  This cut is based on the minimum and maximum frequencies across all TOAs in a DMX bin. (968) \\
  \texttt{-cut simul} & Identifies an ASP/GASP TOA acquired at the same time as a PUPPI/GUPPI TOA. & These TOAs represent duplicate information and were removed at the very last stage of analysis. (576) \\
  \texttt{-cut snr} & The TOA does not meet a signal-to-noise ratio threshold. & TOAs for which \texttt{-snr} has \textit{value} $< 25$; for the narrowband TOAs, the threshold is $8$ (see Appendix~\ref{sec:low_snr}). (500) \\
  \texttt{-cut epochdrop} & Entire epoch removed based on an epoch-by-epoch removal analysis. & Epochs identified by this analysis in the narrowband data set are removed also in the wideband data set; see \twelveyr\ for details. (68) \\
  \texttt{-cut one} & The subintegration only has one frequency channel. & TOAs for which \texttt{-nchx} has a value of 1; a DM cannot be estimated from this observation. (33) \\
  \texttt{-cut manual} & An outlier determined by manual inspection. & In most cases, the TOA's corresponding profile data is corrupted by instrumentation or RFI. These were identified independently from the narrowband TOAs that have the same flag. (29) \\
  \texttt{-cut cull} & The TOA had a large residual in the initial timing analysis. & We used the \texttt{Tempo} utility program \texttt{cull} to identify TOAs that had a residual $> 100$~$\mu$s.  These outliers were confirmed by human inspection to have an issue. (6) \\
  \enddata
  \tablecomments{The \texttt{-cut} flags are ordered here by how many such wideband TOAs were removed from the analyses (numbers in parentheses).  All cut TOAs are provided as commented-out TOAs in the ASCII-text TOA files; excluding these, there are 12,598 wideband TOAs in the data set.  See \twelveyr\ for additional information on TOA flags.  Other flags and the TOA format we use (``IPTA'') are conventional and are not listed nor explained here.} 
  \end{deluxetable*}

%% file: tables/tab_ntoa_nchan.tex
\begin{deluxetable}{crrrr}[ht!]
  \tabletypesize{\footnotesize}
  \tablewidth{\columnwidth}
  \tablecaption{Data Volume Comparison\label{tab:ntoa_nchan}}
  \tablehead{\colhead{Source} & \colhead{\# TOAs} & \colhead{\# Prof.} & \colhead{\# TOAs} & \colhead{Diff.} \\[-8pt] \colhead{} & \colhead{\hspace*{0pt}\hfill (WB)} & \colhead{\hspace*{0pt}\hfill (WB)} & \colhead{\hspace*{0pt}\hfill (NB)} & \colhead{\hspace*{0pt}\hfill [\%]}}
    \startdata
      J0023$+$0923  &  589  &  17846  &  12516  &  $43$ \\
      J0030$+$0451  &  488  &  12607  &  12543  &  $1$ \\
      J0340$+$4130  &  164  &  9092  &  8069  &  $13$ \\
      J0613$-$0200  &  360  &  13683  &  13201  &  $4$ \\
      J0636$+$5128  &  711  &  38309  &  21374  &  $79$ \\
      J0645$+$5158  &  217  &  11800  &  7893  &  $50$ \\
      J0740$+$6620  &  86  &  4679  &  3328  &  $41$ \\
      J0931$-$1902  &  123  &  6712  &  3712  &  $81$ \\
      J1012$+$5307  &  554  &  21334  &  19307  &  $11$ \\
      J1024$-$0719  &  230  &  12206  &  9792  &  $25$ \\
      J1125$+$7819  &  108  &  5853  &  4821  &  $21$ \\
      J1453$+$1902  &  68  &  2148  &  1555  &  $38$ \\
      J1455$-$3330  &  282  &  11996  &  8408  &  $43$ \\
      J1600$-$3053  &  313  &  14345  &  14374  &  $0$ \\
      J1614$-$2230  &  275  &  13433  &  12775  &  $5$ \\
      J1640$+$2224  &  418  &  10078  &  9256  &  $9$ \\
      J1643$-$1224  &  319  &  12786  &  12798  &  $0$ \\
      J1713$+$0747  &  1012  &  36501  &  37698  &  $-3$ \\
      J1738$+$0333  &  269  &  9542  &  6977  &  $37$ \\
      J1741$+$1351  &  147  &  4255  &  3845  &  $11$ \\
      J1744$-$1134  &  347  &  14106  &  13380  &  $5$ \\
      J1747$-$4036  &  151  &  8096  &  7572  &  $7$ \\
      J1832$-$0836  &  120  &  6630  &  5364  &  $24$ \\
      J1853$+$1303  &  134  &  3968  &  3544  &  $12$ \\
      B1855$+$09\phantom{....}  &  313  &  6340  &  6464  &  $-2$ \\
      J1903$+$0327  &  156  &  4893  &  4854  &  $1$ \\
      J1909$-$3744  &  550  &  24329  &  22633  &  $8$ \\
      J1910$+$1256  &  172  &  5392  &  5012  &  $8$ \\
      J1911$+$1347  &  88  &  2621  &  2625  &  $0$ \\
      J1918$-$0642  &  379  &  15000  &  13675  &  $10$ \\
      J1923$+$2515  &  119  &  3588  &  3009  &  $19$ \\
      B1937$+$21\phantom{....}  &  525  &  16067  &  17024  &  $-6$ \\
      J1944$+$0907  &  138  &  3923  &  3931  &  $0$ \\
      J1946$+$3417  &  78  &  3013  &  3016  &  $0$ \\
      B1953$+$29\phantom{....}  &  119  &  3395  &  3421  &  $-1$ \\
      J2010$-$1323  &  278  &  14360  &  13306  &  $8$ \\
      J2017$+$0603  &  127  &  4856  &  2986  &  $63$ \\
      J2033$+$1734  &  90  &  2720  &  2691  &  $1$ \\
      J2043$+$1711  &  316  &  9505  &  5624  &  $69$ \\
      J2145$-$0750  &  313  &  14332  &  13961  &  $3$ \\
      J2214$+$3000  &  233  &  9143  &  6269  &  $46$ \\
      J2229$+$2643  &  97  &  2853  &  2442  &  $17$ \\
      J2234$+$0611  &  88  &  2720  &  2475  &  $10$ \\
      J2234$+$0944  &  175  &  6584  &  5892  &  $12$ \\
      J2302$+$4442  &  174  &  9602  &  7833  &  $23$ \\
      J2317$+$1439  &  505  &  10733  &  9784  &  $10$ \\
      J2322$+$2057  &  80  &  2500  &  2093  &  $19$ \\
      \hline
      Total  & 12598  &  480474  &  415122  &  $16$ \\
    \enddata
  \tablecomments{\scriptsize
  The last column shows the difference between the number of profiles used in measuring all wideband TOAs (third and second columns, respectively) and the number of TOAs in the narrowband data set (fourth column), expressed as an integer-rounded percentage of the latter value.
  If the two data sets contained identical profiles, there would be an exact one-to-one correspondence between columns three and four.
  Note that the wideband data set has an equal number of DM measurements as wideband TOAs.
  The TOA numbers shown here do not include those with a \texttt{-cut} flag, which are included as part of the data release.
  }
\end{deluxetable}

%% file: tables/tab_toa_summary.tex
{\footnotesize%
\begin{tabular}{@{\extracolsep{0pt}}crrrrrlrlrlrlrlr}
\hline
\hline
Source & \multicolumn{1}{c}{$P$}  & \multicolumn{1}{c}{$dP/dt$}      & \multicolumn{1}{c}{DM}           & \multicolumn{1}{c}{$P_b$}
         & \multicolumn{10}{c}{Median scaled TOA uncertainty$^a$ [$\mu$s] / \# of epochs}
         & \multicolumn{1}{c}{Span} \\ \cline{6-15}
       & \multicolumn{1}{c}{[ms]} & \multicolumn{1}{c}{[$10^{-20}$]} & \multicolumn{1}{c}{[pc~cm$^{-3}$]} & \multicolumn{1}{c}{[d]}
         & \multicolumn{2}{c}{327~MHz}
         & \multicolumn{2}{c}{430~MHz}
         & \multicolumn{2}{c}{820~MHz}
         & \multicolumn{2}{c}{1.4~GHz}
         & \multicolumn{2}{c}{2.1~GHz}
         & \multicolumn{1}{c}{[yr]} \\
\hline
J0023$+$0923 & 3.05 & 1.14 & 14.3 & 0.1& \multicolumn{2}{c}{-} & 0.041 & 58 & \multicolumn{2}{c}{-} & 0.550 & 65 & \multicolumn{2}{c}{-} & 5.9\\
J0030$+$0451 & 4.87 & 1.02 & 4.3 & -& \multicolumn{2}{c}{-} & 0.193 & 174 & \multicolumn{2}{c}{-} & 0.368 & 187 & 0.998 & 70 & 12.4\\
J0340$+$4130 & 3.30 & 0.70 & 49.6 & -& \multicolumn{2}{c}{-} & \multicolumn{2}{c}{-} & 0.799 & 69 & 1.992 & 71 & \multicolumn{2}{c}{-} & 5.3\\
J0613$-$0200 & 3.06 & 0.96 & 38.8 & 1.2& \multicolumn{2}{c}{-} & \multicolumn{2}{c}{-} & 0.100 & 134 & 0.432 & 135 & \multicolumn{2}{c}{-} & 12.2\\
J0636$+$5128 & 2.87 & 0.34 & 11.1 & 0.1& \multicolumn{2}{c}{-} & \multicolumn{2}{c}{-} & 0.264 & 39 & 0.650 & 42 & \multicolumn{2}{c}{-} & 3.5\\
J0645$+$5158 & 8.85 & 0.49 & 18.2 & -& \multicolumn{2}{c}{-} & \multicolumn{2}{c}{-} & 0.388 & 66 & 1.050 & 74 & \multicolumn{2}{c}{-} & 6.1\\
J0740$+$6620 & 2.89 & 1.22 & 15.0 & 4.8& \multicolumn{2}{c}{-} & \multicolumn{2}{c}{-} & 0.545 & 38 & 0.583 & 41 & \multicolumn{2}{c}{-} & 3.5\\
J0931$-$1902 & 4.64 & 0.36 & 41.5 & -& \multicolumn{2}{c}{-} & \multicolumn{2}{c}{-} & 0.940 & 51 & 2.065 & 51 & \multicolumn{2}{c}{-} & 4.3\\
J1012$+$5307 & 5.26 & 1.71 & 9.0 & 0.6& \multicolumn{2}{c}{-} & \multicolumn{2}{c}{-} & 0.343 & 135 & 0.538 & 142 & \multicolumn{2}{c}{-} & 12.9\\
J1024$-$0719 & 5.16 & 1.86 & 6.5 & -& \multicolumn{2}{c}{-} & \multicolumn{2}{c}{-} & 0.564 & 89 & 0.851 & 94 & \multicolumn{2}{c}{-} & 7.7\\
J1125$+$7819 & 4.20 & 0.69 & 12.0 & 15.4& \multicolumn{2}{c}{-} & \multicolumn{2}{c}{-} & 0.663 & 40 & 1.713 & 42 & \multicolumn{2}{c}{-} & 3.5\\
J1453$+$1902 & 5.79 & 1.17 & 14.1 & -& \multicolumn{2}{c}{-} & 0.988 & 28 & \multicolumn{2}{c}{-} & 2.494 & 40 & \multicolumn{2}{c}{-} & 3.9\\
J1455$-$3330 & 7.99 & 2.43 & 13.6 & 76.2& \multicolumn{2}{c}{-} & \multicolumn{2}{c}{-} & 1.126 & 113 & 1.886 & 113 & \multicolumn{2}{c}{-} & 12.9\\
J1600$-$3053 & 3.60 & 0.95 & 52.3 & 14.3& \multicolumn{2}{c}{-} & \multicolumn{2}{c}{-} & 0.253 & 113 & 0.200 & 115 & \multicolumn{2}{c}{-} & 9.6\\
J1614$-$2230 & 3.15 & 0.96 & 34.5 & 8.7& \multicolumn{2}{c}{-} & \multicolumn{2}{c}{-} & 0.332 & 96 & 0.482 & 107 & \multicolumn{2}{c}{-} & 8.8\\
J1640$+$2224 & 3.16 & 0.28 & 18.5 & 175.5& \multicolumn{2}{c}{-} & 0.033 & 177 & \multicolumn{2}{c}{-} & 0.260 & 186 & \multicolumn{2}{c}{-} & 12.3\\
J1643$-$1224 & 4.62 & 1.85 & 62.3 & 147.0& \multicolumn{2}{c}{-} & \multicolumn{2}{c}{-} & 0.270 & 130 & 0.460 & 129 & \multicolumn{2}{c}{-} & 12.7\\
J1713$+$0747 & 4.57 & 0.85 & 15.9 & 67.8& \multicolumn{2}{c}{-} & \multicolumn{2}{c}{-} & 0.097 & 129 & 0.043 & 450 & 0.041 & 185 & 12.4\\
J1738$+$0333 & 5.85 & 2.41 & 33.8 & 0.4& \multicolumn{2}{c}{-} & \multicolumn{2}{c}{-} & \multicolumn{2}{c}{-} & 0.374 & 70 & 1.119 & 64 & 7.6\\
J1741$+$1351 & 3.75 & 3.02 & 24.2 & 16.3& \multicolumn{2}{c}{-} & 0.100 & 63 & \multicolumn{2}{c}{-} & 0.233 & 73 & \multicolumn{2}{c}{-} & 5.9\\
J1744$-$1134 & 4.07 & 0.89 & 3.1 & -& \multicolumn{2}{c}{-} & \multicolumn{2}{c}{-} & 0.107 & 128 & 0.198 & 126 & \multicolumn{2}{c}{-} & 12.9\\
J1747$-$4036 & 1.65 & 1.31 & 153.0 & -& \multicolumn{2}{c}{-} & \multicolumn{2}{c}{-} & 0.983 & 62 & 1.155 & 65 & \multicolumn{2}{c}{-} & 5.3\\
J1832$-$0836 & 2.72 & 0.83 & 28.2 & -& \multicolumn{2}{c}{-} & \multicolumn{2}{c}{-} & 0.608 & 53 & 0.450 & 53 & \multicolumn{2}{c}{-} & 4.3\\
J1853$+$1303 & 4.09 & 0.87 & 30.6 & 115.7& \multicolumn{2}{c}{-} & 0.278 & 64 & \multicolumn{2}{c}{-} & 0.378 & 70 & \multicolumn{2}{c}{-} & 5.9\\
B1855$+$09\phantom{....} & 5.36 & 1.78 & 13.3 & 12.3& \multicolumn{2}{c}{-} & 0.195 & 116 & \multicolumn{2}{c}{-} & 0.128 & 123 & \multicolumn{2}{c}{-} & 12.5\\
J1903$+$0327 & 2.15 & 1.88 & 297.5 & 95.2& \multicolumn{2}{c}{-} & \multicolumn{2}{c}{-} & \multicolumn{2}{c}{-} & 0.400 & 75 & 0.470 & 78 & 7.6\\
J1909$-$3744 & 2.95 & 1.40 & 10.4 & 1.5& \multicolumn{2}{c}{-} & \multicolumn{2}{c}{-} & 0.040 & 125 & 0.086 & 267 & \multicolumn{2}{c}{-} & 12.7\\
J1910$+$1256 & 4.98 & 0.97 & 38.1 & 58.5& \multicolumn{2}{c}{-} & \multicolumn{2}{c}{-} & \multicolumn{2}{c}{-} & 0.251 & 83 & 0.555 & 84 & 8.3\\
J1911$+$1347 & 4.63 & 1.69 & 31.0 & -& \multicolumn{2}{c}{-} & 0.586 & 42 & \multicolumn{2}{c}{-} & 0.109 & 46 & \multicolumn{2}{c}{-} & 3.9\\
J1918$-$0642 & 7.65 & 2.57 & 26.5 & 10.9& \multicolumn{2}{c}{-} & \multicolumn{2}{c}{-} & 0.358 & 126 & 0.605 & 128 & \multicolumn{2}{c}{-} & 12.7\\
J1923$+$2515 & 3.79 & 0.96 & 18.9 & -& \multicolumn{2}{c}{-} & 0.184 & 53 & \multicolumn{2}{c}{-} & 0.665 & 66 & \multicolumn{2}{c}{-} & 5.8\\
B1937$+$21\phantom{....} & 1.56 & 10.51 & 71.1 & -& \multicolumn{2}{c}{-} & \multicolumn{2}{c}{-} & 0.006 & 125 & 0.010 & 228 & 0.011 & 85 & 12.8\\
J1944$+$0907 & 5.19 & 1.73 & 24.4 & -& \multicolumn{2}{c}{-} & 0.242 & 62 & \multicolumn{2}{c}{-} & 0.495 & 72 & \multicolumn{2}{c}{-} & 9.3\\
J1946$+$3417 & 3.17 & 0.32 & 110.2 & 27.0& \multicolumn{2}{c}{-} & \multicolumn{2}{c}{-} & \multicolumn{2}{c}{-} & 0.365 & 40 & 0.510 & 38 & 2.6\\
B1953$+$29\phantom{....} & 6.13 & 2.97 & 104.5 & 117.3& \multicolumn{2}{c}{-} & 0.251 & 54 & \multicolumn{2}{c}{-} & 0.753 & 65 & \multicolumn{2}{c}{-} & 5.9\\
J2010$-$1323 & 5.22 & 0.48 & 22.2 & -& \multicolumn{2}{c}{-} & \multicolumn{2}{c}{-} & 0.344 & 94 & 0.685 & 96 & \multicolumn{2}{c}{-} & 7.8\\
J2017$+$0603 & 2.90 & 0.80 & 23.9 & 2.2& \multicolumn{2}{c}{-} & 0.199 & 6 & \multicolumn{2}{c}{-} & 0.327 & 67 & 0.765 & 46 & 5.3\\
J2033$+$1734 & 5.95 & 1.11 & 25.1 & 56.3& \multicolumn{2}{c}{-} & 0.189 & 40 & \multicolumn{2}{c}{-} & 0.901 & 46 & \multicolumn{2}{c}{-} & 3.8\\
J2043$+$1711 & 2.38 & 0.52 & 20.8 & 1.5& \multicolumn{2}{c}{-} & 0.058 & 132 & \multicolumn{2}{c}{-} & 0.385 & 148 & \multicolumn{2}{c}{-} & 5.9\\
J2145$-$0750 & 16.05 & 2.98 & 9.0 & 6.8& \multicolumn{2}{c}{-} & \multicolumn{2}{c}{-} & 0.190 & 111 & 0.485 & 115 & \multicolumn{2}{c}{-} & 12.8\\
J2214$+$3000 & 3.12 & 1.47 & 22.5 & 0.4& \multicolumn{2}{c}{-} & \multicolumn{2}{c}{-} & \multicolumn{2}{c}{-} & 0.560 & 71 & 1.491 & 50 & 5.5\\
J2229$+$2643 & 2.98 & 0.15 & 22.7 & 93.0& \multicolumn{2}{c}{-} & 0.229 & 46 & \multicolumn{2}{c}{-} & 0.750 & 48 & \multicolumn{2}{c}{-} & 3.9\\
J2234$+$0611 & 3.58 & 1.20 & 10.8 & 32.0& \multicolumn{2}{c}{-} & 0.342 & 39 & \multicolumn{2}{c}{-} & 0.189 & 44 & \multicolumn{2}{c}{-} & 3.4\\
J2234$+$0944 & 3.63 & 2.01 & 17.8 & 0.4& \multicolumn{2}{c}{-} & \multicolumn{2}{c}{-} & \multicolumn{2}{c}{-} & 0.214 & 45 & 0.617 & 44 & 4.0\\
J2302$+$4442 & 5.19 & 1.39 & 13.8 & 125.9& \multicolumn{2}{c}{-} & \multicolumn{2}{c}{-} & 0.967 & 69 & 1.996 & 68 & \multicolumn{2}{c}{-} & 5.3\\
J2317$+$1439 & 3.45 & 0.24 & 21.9 & 2.5& 0.081 & 78 & 0.056 & 186 & \multicolumn{2}{c}{-} & 0.409 & 141 & \multicolumn{2}{c}{-} & 12.5\\
J2322$+$2057 & 4.81 & 0.97 & 13.4 & -& \multicolumn{2}{c}{-} & 0.217 & 33 & \multicolumn{2}{c}{-} & 0.952 & 33 & 1.717 & 8 & 2.3\\
\hline
\multicolumn{5}{r}{Nominal scaling factor$^b$ (ASP/GASP)}
  & \multicolumn{2}{c}{0.6}
  & \multicolumn{2}{c}{0.4}
  & \multicolumn{2}{c}{0.8}
  & \multicolumn{2}{c}{0.8}
  & \multicolumn{2}{c}{0.8}
  & \\
\multicolumn{5}{r}{Nominal scaling factor$^b$ (GUPPI/PUPPI)}
  & \multicolumn{2}{c}{0.7}
  & \multicolumn{2}{c}{0.5}
  & \multicolumn{2}{c}{1.4}
  & \multicolumn{2}{c}{2.5}
  & \multicolumn{2}{c}{2.1}
  & \\
\hline
\end{tabular}

\vspace{0.5em}

{$^a$ For this table, the original TOA uncertainties were scaled by their
bandwidth-time product $\left( \frac{\Delta \nu}{100~\mathrm{MHz}}
\frac{\tau}{1800~\mathrm{s}} \right)^{1/2}$ to remove variation due to
different instrument bandwidths and integration time.}

\vspace{0.5em}

{$^b$ TOA uncertainties can be rescaled to the nominal full instrumental
bandwidth listed in Table~\ref{tab:observing_systems} by dividing by these
scaling factors.}

}

%% file: text/fig_wb_toa_dm_err.tex
\begin{figure*}[htb!]
\begin{center}
\includegraphics[width=2\columnwidth]{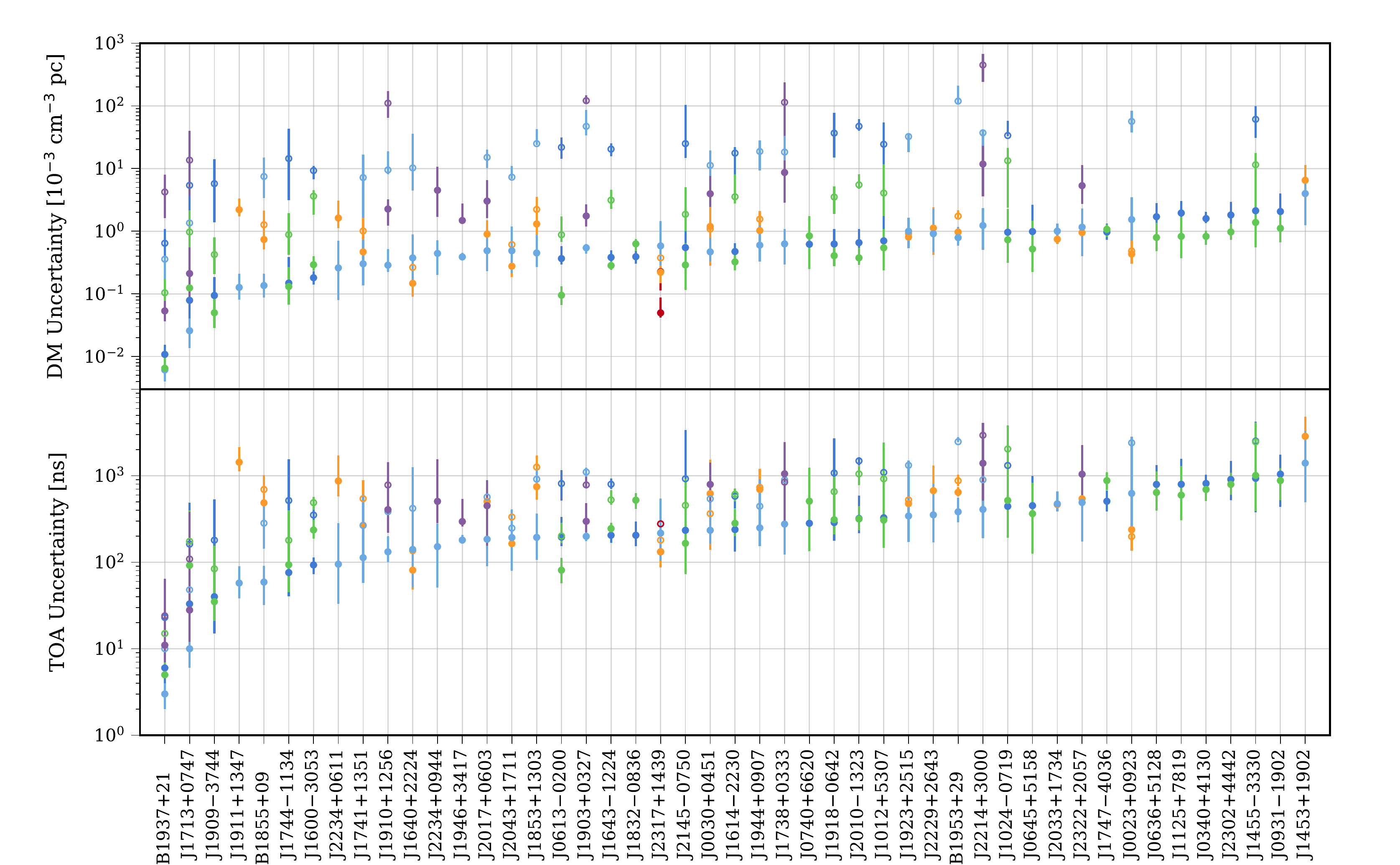}
\caption{\label{fig:wb_toa_dm_err}
The median raw wideband TOA and DM measurement uncertainties with central 68\% intervals.
Pulsars are ordered by their median PUPPI or GUPPI L-band (1.4~GHz) TOA uncertainties.
The dramatic increase in DM precision after moving from the ASP and GASP backends (open cirlces) to the PUPPI and GUPPI backends (filled circles) is evident.
The colors indicate the receiver as in Figure~\ref{fig:epochs}: 327~MHz (red), 430~MHz (orange), 820~MHz (green), 1.4~GHz (lighter blue for AO, darker blue for the GBT), 2.1~GHz (purple).
}
\end{center}
\end{figure*}

%% file: tables/tab_timing_model_summary.tex
{\footnotesize
\begin{tabular}{@{\extracolsep{6pt}}crrrrrrccccrc}
\hline
\hline
Source
  & \# of TOAs
  & \multicolumn{5}{c}{\# of Fit Parameters$^a$}
  & \multicolumn{2}{c}{RMS$^b$ [$\mu$s]}
  & \multicolumn{3}{c}{Red Noise$^c$}
  & Figure \\ \cline{3-7} \cline{8-9} \cline{10-12}

  &
  & S
  & A
  & B
  & DM
  & J
  & Full
  & White
  & $A_{\mathrm{red}}$
  & $\gamma_{\mathrm{red}}$
  & log$_{10}B$
  & \\
\hline
J0023$+$0923 & 589 & 3 & 5 & 9 & 64 & 1 & 0.288  & - & - & - & $>$2$^d$  & \ref{fig:summary-J0023+0923} \\
J0030$+$0451 & 488 & 3 & 5 & 0 & 190 & 2 & 2.868  & 0.200  & 0.006 & $-$5.3 & $>$2\phantom{$^x$}  & \ref{fig:summary-J0030+0451} \\
J0340$+$4130 & 164 & 3 & 5 & 0 & 75 & 1 & 0.449  & - & - & - & $-$0.20\phantom{$^x$}  & \ref{fig:summary-J0340+4130} \\
J0613$-$0200 & 360 & 3 & 5 & 8 & 139 & 1 & 0.188  & - & - & - & 1.19$^e$  & \ref{fig:summary-J0613-0200} \\
J0636$+$5128 & 711 & 3 & 5 & 6 & 44 & 1 & 0.596  & - & - & - & $-$0.07\phantom{$^x$}  & \ref{fig:summary-J0636+5128} \\
J0645$+$5158 & 217 & 3 & 5 & 0 & 79 & 1 & 0.196  & - & - & - & $-$0.17\phantom{$^x$}  & \ref{fig:summary-J0645+5158} \\
J0740$+$6620 & 86 & 3 & 5 & 7 & 44 & 1 & 0.106  & - & - & - & $-$0.15\phantom{$^x$}  & \ref{fig:summary-J0740+6620} \\
J0931$-$1902 & 123 & 3 & 5 & 0 & 57 & 1 & 0.424  & - & - & - & $-$0.16\phantom{$^x$}  & \ref{fig:summary-J0931-1902} \\
J1012$+$5307 & 554 & 3 & 5 & 6 & 142 & 1 & 0.891  & 0.209  & 0.472 & $-$1.3 & $>$2\phantom{$^x$}  & \ref{fig:summary-J1012+5307} \\
J1024$-$0719 & 230 & 4 & 5 & 0 & 100 & 1 & 0.240  & - & - & - & 0.36\phantom{$^x$}  & \ref{fig:summary-J1024-0719} \\
J1125$+$7819 & 108 & 3 & 5 & 5 & 43 & 1 & 0.614  & - & - & - & $-$0.09\phantom{$^x$}  & \ref{fig:summary-J1125+7819} \\
J1453$+$1902 & 68 & 3 & 5 & 0 & 39 & 1 & 0.798  & - & - & - & $-$0.10\phantom{$^x$}  & \ref{fig:summary-J1453+1902} \\
J1455$-$3330 & 282 & 3 & 5 & 6 & 120 & 1 & 0.544  & - & - & - & $-$0.13\phantom{$^x$}  & \ref{fig:summary-J1455-3330} \\
J1600$-$3053 & 313 & 3 & 5 & 8 & 128 & 1 & 0.213  & - & - & - & $-$0.09\phantom{$^x$}  & \ref{fig:summary-J1600-3053} \\
J1614$-$2230 & 275 & 3 & 5 & 8 & 114 & 1 & 0.175  & - & - & - & $-$0.23\phantom{$^x$}  & \ref{fig:summary-J1614-2230} \\
J1640$+$2224 & 418 & 3 & 5 & 8 & 185 & 1 & 0.142  & - & - & - & $-$0.16\phantom{$^x$}  & \ref{fig:summary-J1640+2224} \\
J1643$-$1224 & 319 & 3 & 5 & 6 & 140 & 1 & 2.385  & 0.292  & 1.815 & $-$1.2 & $>$2\phantom{$^x$}  & \ref{fig:summary-J1643-1224} \\
J1713$+$0747 & 1012 & 3 & 5 & 8 & 362 & 3 & 0.097  & 0.081  & 0.022 & $-$1.6 & $>$2\phantom{$^x$}  & \ref{fig:summary-J1713+0747} \\
J1738$+$0333 & 269 & 3 & 5 & 5 & 77 & 1 & 0.272  & - & - & - & $-$0.18\phantom{$^x$}  & \ref{fig:summary-J1738+0333} \\
J1741$+$1351 & 147 & 3 & 5 & 8 & 73 & 1 & 0.148  & - & - & - & $-$0.10\phantom{$^x$}  & \ref{fig:summary-J1741+1351} \\
J1744$-$1134 & 347 & 3 & 5 & 0 & 134 & 1 & 0.721  & 0.278  & 0.220 & $-$1.7 & $>$2\phantom{$^x$}  & \ref{fig:summary-J1744-1134} \\
J1747$-$4036 & 151 & 3 & 5 & 0 & 71 & 1 & 6.722  & 0.767  & 0.518 & $-$3.8 & $>$2\phantom{$^x$}  & \ref{fig:summary-J1747-4036} \\
J1832$-$0836 & 120 & 3 & 5 & 0 & 58 & 1 & 0.195  & - & - & - & $-$0.04\phantom{$^x$}  & \ref{fig:summary-J1832-0836} \\
J1853$+$1303 & 134 & 3 & 5 & 8 & 70 & 1 & 0.322  & 0.092  & 0.139 & $-$1.9 & $>$2\phantom{$^x$}  & \ref{fig:summary-J1853+1303} \\
B1855$+$09\phantom{....} & 313 & 3 & 5 & 7 & 123 & 1 & 1.387  & 0.322  & 0.045 & $-$3.4 & $>$2\phantom{$^x$}  & \ref{fig:summary-B1855+09} \\
J1903$+$0327 & 156 & 3 & 5 & 8 & 82 & 1 & 2.962  & 0.394  & 1.238 & $-$2.2 & $>$2\phantom{$^x$}  & \ref{fig:summary-J1903+0327} \\
J1909$-$3744 & 550 & 3 & 5 & 9 & 221 & 1 & 0.337  & 0.058  & 0.025 & $-$2.8 & $>$2\phantom{$^x$}  & \ref{fig:summary-J1909-3744} \\
J1910$+$1256 & 172 & 3 & 5 & 6 & 89 & 1 & 0.399  & - & - & - & $-$0.16\phantom{$^x$}  & \ref{fig:summary-J1910+1256} \\
J1911$+$1347 & 88 & 3 & 5 & 0 & 46 & 1 & 0.115  & - & - & - & 0.12\phantom{$^x$}  & \ref{fig:summary-J1911+1347} \\
J1918$-$0642 & 379 & 3 & 5 & 7 & 133 & 1 & 0.296  & - & - & - & $-$0.15\phantom{$^x$}  & \ref{fig:summary-J1918-0642} \\
J1923$+$2515 & 119 & 3 & 5 & 0 & 66 & 1 & 0.237  & - & - & - & $-$0.16\phantom{$^x$}  & \ref{fig:summary-J1923+2515} \\
B1937$+$21\phantom{....} & 525 & 3 & 5 & 0 & 207 & 3 & 2.243  & 0.099  & 0.087 & $-$3.5 & $>$2\phantom{$^x$}  & \ref{fig:summary-B1937+21} \\
J1944$+$0907 & 138 & 3 & 5 & 0 & 72 & 1 & 0.375  & - & - & - & $-$0.12\phantom{$^x$}  & \ref{fig:summary-J1944+0907} \\
J1946$+$3417 & 78 & 3 & 5 & 8 & 41 & 1 & 0.143  & - & - & - & $-$0.10\phantom{$^x$}  & \ref{fig:summary-J1946+3417} \\
B1953$+$29\phantom{....} & 119 & 3 & 5 & 6 & 65 & 1 & 0.394  & - & - & - & 0.83\phantom{$^x$}  & \ref{fig:summary-B1953+29} \\
J2010$-$1323 & 278 & 3 & 5 & 0 & 108 & 1 & 0.250  & - & - & - & $-$0.22\phantom{$^x$}  & \ref{fig:summary-J2010-1323} \\
J2017$+$0603 & 127 & 3 & 5 & 7 & 74 & 2 & 0.097  & - & - & - & $-$0.20\phantom{$^x$}  & \ref{fig:summary-J2017+0603} \\
J2033$+$1734 & 90 & 3 & 5 & 5 & 46 & 1 & 0.520  & - & - & - & $-$0.12\phantom{$^x$}  & \ref{fig:summary-J2033+1734} \\
J2043$+$1711 & 316 & 3 & 5 & 7 & 148 & 1 & 0.122  & - & - & - & 1.70\phantom{$^x$}  & \ref{fig:summary-J2043+1711} \\
J2145$-$0750 & 313 & 3 & 5 & 7 & 123 & 1 & 0.812  & 0.274  & 0.438 & $-$1.6 & $>$2\phantom{$^x$}  & \ref{fig:summary-J2145-0750} \\
J2214$+$3000 & 233 & 3 & 5 & 5 & 77 & 1 & 0.419  & - & - & - & 0.21\phantom{$^x$}  & \ref{fig:summary-J2214+3000} \\
J2229$+$2643 & 97 & 3 & 5 & 6 & 48 & 1 & 0.196  & - & - & - & $-$0.16\phantom{$^x$}  & \ref{fig:summary-J2229+2643} \\
J2234$+$0611 & 88 & 3 & 5 & 7 & 44 & 1 & 0.035  & - & - & - & $-$0.15\phantom{$^x$}  & \ref{fig:summary-J2234+0611} \\
J2234$+$0944 & 175 & 3 & 5 & 5 & 51 & 1 & 0.165  & - & - & - & $>$2$^d$  & \ref{fig:summary-J2234+0944} \\
J2302$+$4442 & 174 & 3 & 5 & 7 & 75 & 1 & 0.693  & - & - & - & $-$0.13\phantom{$^x$}  & \ref{fig:summary-J2302+4442} \\
J2317$+$1439 & 505 & 3 & 5 & 6 & 209 & 2 & 5.416  & 0.204  & 0.001 & $-$6.0 & $>$2\phantom{$^x$}  & \ref{fig:summary-J2317+1439} \\
J2322$+$2057 & 80 & 3 & 5 & 0 & 33 & 2 & 0.237  & - & - & - & $-$0.10\phantom{$^x$}  & \ref{fig:summary-J2322+2057} \\
\hline
\end{tabular}

\vspace{0.25em}

{$^a$ Fit parameters: S=spin; B=binary; A=astrometry; DM=dispersion measure; J=phase jump (and an equal number of DM jumps).}

\vspace{0.25em}

{$^b$ Weighted root-mean-square of post-fit timing residuals.
For sources with red noise, the ``Full'' RMS value includes the red noise contribution, while the ``White'' RMS does not.}

\vspace{0.25em}

{$^c$ Maximum-likelihood red noise parameters: $A_{\mathrm{red}}$ = amplitude of red noise power spectral density at $f$=1~yr$^{-1}$ with units $\mu$s yr$^{1/2}$;\\$\gamma_{\mathrm{red}}$ = spectral index; $B$ = Bayes factor (``$>$2'' indicates a Bayes factor larger than our threshold log$_{10}$B~$>$~2, but which could not be estimated using the Savage-Dickey ratio).}

\vspace{0.25em}

{$^d$ See text for additional details on this source.}

\vspace{0.25em}

{$^e$ This source has significant red noise in the analysis of the narrowband data set.}

}

%% file: text/sec_results.tex

\subsection{Average Portraits \& Flux Density Measurements}
\label{subsec:port_results}

A by-product of the profile evolution modeling procedure is a calibrated high S/N average portrait with a nominal profile alignment and full polarization information.
The polarization portraits contain a wealth of information and are of interest to model in their own right; their models could potentially be used to improve the TOA measurement in cases of significant polarization.
For sufficiently polarized, large bandwidth, high S/N data, the rotation measure (RM) could be measured as part of the wideband TOA measurement.
Such a development would combine the techniques summarized in Section~\ref{subsec:prof_evol} with those from \citet{vS06}, \cite{vS13}, and \citet{Oslowski13}, and is an active field of research.

We also estimated the phase- and frequency-averaged flux density for each of our PUPPI and GUPPI TOA measurements; ASP and GASP data were excluded because the profile data from which TOA measurements were made had been rescaled from their original flux calibration (see \nineyr\ for details).
The two main assumptions that go into the estimate and its formal, statistical uncertainty are that the profile evolution model sufficiently describes the data (i.e., no model error) and that it has a correct baseline of zero flux density; all phases contribute to the measurement.
The frequency-averaged flux density and uncertainty are calculated from the weighted-mean of the phase-averaged flux densities.
Since the scaling parameters $a_n$ enter the calculation in the same way as for the S/N estimate, the flux density estimates may contain similar biases (see Appendix~\ref{sec:low_snr}).
The relevant flags for these measurements are listed in Table~\ref{tab:toa_flags}, including a reference frequency for the flux density estimate.
No additional sources of uncertainties are considered, and the interpretation of these measurements should be treated with caution.
%

\subsection{Profile Evolution Models}
\label{subsec:model_results}

We find that for the majority of our pulsars, the profile evolution model for a given receiver band requires a single eigenprofile (62 of 102 pulsar-receiver combinations), which can be thought of as the gradient of the mean profile.
Most of the remainder required two (20 of 102) or zero (13 of 102; i.e., those data are consistent with a constant, non-evolving profile).
The few cases in which more than three basis eigenprofiles are used to describe profile evolution arise in two very high S/N pulsars (3 of 102 have three, the remaining 4 cases have more).  
B1937$+$21 shows spectral leakage from the overlapping, finite-attenuation filters used to subband the data\footnote{We note that a better choice of filter appears to drastically improve this situation \citep{Bailes20}.}, which results in the increased number of eigenprofiles in three of its models, and the imperfect correction of the ADC artifact image described in Section~\ref{sec:obs} has the same consequence for one model for J1713$+$0747.
Removing the perhaps spurious eigenprofiles for these pulsars does not appear to significantly change the timing results in Section~\ref{sec:results}, so we leave them for completeness.
Furthermore, these two pulsars are observed with both observatories at L-band, and we find that the first two eigenprofiles (which contribute the most to profile evolution) are qualitatively the same between the models from each receiver.

Profile broadening from scattering in the ISM or other drastic, intrinsic profile evolution may be responsible for second and third eigenprofiles in the cases where either of those are detected.
However, ``incorrect'' profile alignment with respect to a constant rotation proportional to $\nu^{-2}$ (corresponding to a small, constant DM offset, generally not larger than, but at most a few times $\sim$10$^{-3}~\dmunits$) may also be the culprit for additional eigenprofiles.

It is important to highlight that this subtle issue exists in the narrowband analysis as well; the implicit assumption there is perhaps the most parsimonious one, that the profile shape does not evolve with frequency and that the profiles are aligned in phase.
The choice of profile alignment sets the value of the absolute DMs measured and will not have an effect on the timing analyses, though a detailed study of this question is beyond the scope of this paper.
More interesting questions about disentangling profile evolution from ISM variations and possible magnetospheric effects are still open \citep{Hassall12}.
A possible future development in the context of the present work is to take a similarly parsimonious approach and simultaneously model profile evolution across all observed bands while minimizing the number of significant eigenprofiles as a function of dispersive rotation.
Furthermore, the underlying physical description of the observed profile evolution also warrants its own investigation.

One might expect a correlation between the total number of eigenprofiles for each pulsar and the number of FD parameters in the timing models from \twelveyr.
We see a rough correspondence between these two numbers, but its interpretation is dubious.
For example, the FD parameters for B1855$+$09 (a.k.a. J1857$+$0943) from \twelveyr\ account for an approximate 20~$\mu$s delay across the profiles in its 430~MHz band, purportedly from unmodeled frequency evolution of the profile shape.
Careful inspection reveals that its 430~MHz profiles show no evidence for profile evolution, neither in the number of significant eigenprofiles (zero), nor in the profile residuals after subtracting the model, nor by direct comparison of the profiles, whereas there is prominent profile evolution across the L-wide bandwidth.
Even though the 430~MHz band is a factor of three lower in frequency than L-wide, the latter's narrowband TOAs will be more influential in DM estimation.
This can be understood by the much larger fractional bandwidth of the L-wide receiver (see Table~\ref{tab:observing_systems}): although the dispersive delay across both receiver bands is comparable, the median raw wideband TOA uncertainty from L-wide is an order of magnitude more precise, and its median raw wideband DM uncertainty is $\sim$5 times smaller (see Figure~\ref{fig:wb_toa_dm_err}).
The spurious FD prediction may arise from the interplay between the relative weighting of the L-band and 430~MHz data in the DMX model, the covariance between FD parameters and DMX values, or perhaps something more interesting; most likely, the FD parameters are filling in for the role of DMJUMP, as mentioned in Section~\ref{subsec:timing}.
The details are beyond the scope of this paper and are under investigation elsewhere. 

\subsection{Frequency-dependent DMs}
\label{subsec:DM_nu}

For a handful of our highest DM pulsars, the DM time series from each frequency band appear significantly different from one another.
These trends are apparent in the panels second from the top in Appendix~\ref{sec:resid} for pulsars J1600$-$3053, J1643$-$1224, J1747$-$4036, and J1903$+$0327 (Figures \ref{fig:summary-J1600-3053}, \ref{fig:summary-J1643-1224}, \ref{fig:summary-J1747-4036}, and \ref{fig:summary-J1903+0327}, with DMs $\sim$~52.3, 62.3, 153.0, and 297.5~$\dmunits$, respectively).
It is also readily apparent in these panels, and in many other pulsars' DM time series, that the DM measurements are only significant after the switchover from the older generation of backend instruments (ASP and GASP) to the newer ones (PUPPI and GUPPI) due to their ability to process a larger bandwidth in real time (see Table~\ref{tab:observing_systems}).

All four of these pulsars have clear pulse broadening in the form of frequency-dependent tails on the trailing edges of their profile components.
To estimate the amount of scattering present in these pulsars, we decomposed their concatenated average portraits into a small number of fixed Gaussian components and an evolving one-sided exponential function \citep{PDR14}.
In this way we estimated the scattering timescale $\tau$ at 1400~MHz for each of these four pulsars to be $\tau_{\textrm{1400}} \sim$~26, 52, 22, and 130~$\mu$s, respectively.

If the scattering timescale is changing with time and is not accounted for in the TOA measurement, the wideband DM measurements will be biased similarly as a function of time.
As mentioned in Section~\ref{subsec:wb_toa_lnlike}, a forthcoming publication will present extensions to the wideband TOA measurement that will be better able to segregate time-variable profile broadening from classical DM variations (Pennucci et al. \textit{in prep.}).
The scattering timescale scales more steeply with frequency than does the dispersive delay (approximately, $\tau \propto \nu^{-4}$), and therefore the wideband DMs measured at lower frequencies will incur a greater bias, since the centroids of scattered pulse components shift by a greater amount.
However, one expects that these biases, even if they are different in magnitude, will be correlated in time.
Conditioned on that assumption, it is difficult to explain the DM time series of these pulsars arising solely from time-variable scattering.
In all four instances, there are periods of correlation \textit{and} anti-correlation between the DM time series measured in each frequency band.

This sort of behavior is, however, predicted by the phenomenon of ``frequency-dependent DM'' \citep{Cordes16}, and very similar behavior has been seen in at least one other (canonical) pulsar \citep{Lam19,Donner19}, although earlier indications existed in B1937$+$21 \citep{DemorestPhDT,Ramachandran06,Cordes90} and in sparse multi-frequency measurements of the highest DM pulsar \citep{Pennucci15}.
The dispersion measure is defined as the path integral of the free-electron density sampled by a propagating electromagnetic wave.
Due to the refractive nature of the ISM, the path will vary as a function of the frequency of the wave, and due to the density inhomogeneities in the ISM, the integrated density -- the DM -- will therefore also be a function of frequency.
However, these differences are expected to be small, with root-mean-square (RMS) values typically $\ll 10^{-3}~\dmunits$, and thus only high-precision observations (e.g., bright MSPs, or bright low-frequency sources) of high-DM pulsars over long periods of time are expected to convincingly show this phenomenon.

To substantiate the claim that the DM trends seen in these four pulsars may arise from this peculiar ISM effect, we can calculate the predicted RMS difference between DMs measured at a fiducial frequency $\nu$ and a lower frequency $\nu'$, $\sigma_{\overline{\textrm{DM}}}(\nu,\nu')$, using Equations~12 and 15 of \citet{Cordes16}.
Using our rough scattering timescales to estimate the scintillation bandwidths at $\nu$, and using the appropriate frequencies for each pulsar, we find $\sigma_{\overline{\textrm{DM}}}(\nu,\nu') \approx 2,~4,~2,~\textrm{and~} 3~\times~10^{-3}~\dmunits$ for J1600$-$3053, J1643$-$1224, J1747$-$4036, and J1903$+$0327, respectively.
These values are all within a factor of $\sim$~2--3 of the RMS differences measured in the observed DM time series: 0.6, 1.7, 2.8, and 5.9~$\times~10^{-3}~\dmunits$, respectively, where we only considered the PUPPI and GUPPI data for these measurements.
Given that this quick assessment involves the assumptions that the density inhomogeneities in the ISM are Kolmogorov in nature, and that the scattering occurs in a single thin screen, we find this level of agreement suggestive.
A more in depth analysis is beyond the scope of this work, but these results indicate that long-term timing of high-DM MSPs in the context of PTA experiments offer a unique opportunity to study this phenomenon, as well as time-variable scattering; the low-frequency, high-cadence observations of CHIME/Pulsar are especially promising in these areas. 


In the two largest DM pulsars (J1747$-$4036 and J1903$+$0327), there are obvious chromatic trends in the timing residuals from \twelveyr\ that are ameliorated in the wideband analysis.
The narrowband noise analyses compensate for this by having larger white noise parameters and slightly larger, shallower red noise, which helps to explain the timing improvements seen in the wideband data set.
Similarly, because the ISM effects appear as apparently chromatic DM measurements in the wideband data set, the DMEFAC parameters are larger than expected ($\sim$1.5$-$2.0).
That is, the boilerplate DMX model may not be good representation of these data, even with DMEFAC and DMJUMP parameters, and more advanced DM and noise models are required.

In addition to these four pulsars, there are four more in our sample that have a DM $\gtrsim$~50~$\dmunits$: J0340$+$4130, B1937$+$21, J1946$+$3417, and B1953$+$29 (a.k.a. J1955$+$2908; Figures \ref{fig:summary-J0340+4130}, \ref{fig:summary-B1937+21}, \ref{fig:summary-J1946+3417}, and \ref{fig:summary-B1953+29}, with DMs $\sim$~49.6, 71.1, 110.2, and 104.5~$\dmunits$, respectively).
None of their DM time series show the clear chromatic trends seen in the other four, but they all have some amount of additional variance that inflates their DMEFAC parameters.
Using the measured scintillation parameters from \citet{Levin16} for the three lower DM pulsars and repeating similar calculations as above, we find that the RMS differences predicted from \citet{Cordes16} are much smaller ($\sim$ an order of magnitude or more) than what is seen in the data.
We could not find a published value for J1946$+$3417, so we estimated its scattering timescale by modeling its profile with Gaussian components in the same fashion as the first four pulsars and find $\tau_{\textrm{1400}} \sim 64~\mu$s.
The predicted and observed RMS DM differences are again similar, $\sim$~4 and 2~$\times~10^{-3}~\dmunits$, respectively, and so frequency dependent DM effects could also be playing a role here.

A couple of other well-timed pulsars with intermediate DM values show the same kind of extra DM variance (e.g., J0613$-$0200; Figure~\ref{fig:summary-J0613-0200}, DM $\sim$~38.8~$\dmunits$, and see also Section~\ref{subsubsec:rn_compare}).
Neither frequency-dependent dispersion nor time-variable scattering (in the form of profile broadening) appears to be playing a role here or in the three pulsars mentioned above.
Another subtle effect may be at play in some of these pulsars, which is a manifestation of short-timescale variations referred to as ``pulse jitter''.
Pulse jitter arises due to the fact that any finite collection of real single pulses will produce a mean profile with a slightly different shape and location between realizations, despite the long-term stability of the average profile \citep{Helfand75}.
For broadband observations that are significantly influenced by pulse jitter, the wideband DM estimate will be biased (cf. Parthasarathy et al. \textit{in prep.}).
Depending on the frequency dependence of pulse jitter \citep{Lam19b}, the bias may also be strongly frequency dependent.
Alternatively, the ``finite scintle effect'', in which the relevant time-variable scattering effects occur in the strong diffractive regime \citep{CordShan10,Cordes90}, can have the same effect as pulse jitter and similarly bias the DM measurements.
An investigation of the observed DM variance in some of our pulsars is beyond the scope of this paper, but it will play a role in considerations of future wideband data sets.

\subsection{Comparison of Collective Timing Results}
\label{subsec:timing_results}

In this section we give an overview of the wideband timing results by assessing the overall characteristics in comparison with the those found in \twelveyr\ and addressing a few pulsars individually.
Besides those already mentioned, and those that will be included in Section~\ref{subsec:pulsar_results}, specific interesting, astrophysical results from the 12.5-year data set can be found in \twelveyr; in particular, these include new or improved astrometric and binary timing model parameters.
The timing residuals from both data sets are presented in Appendix~\ref{sec:resid}.

\subsubsection{Timing Model Parameters}
\label{subsubsec:param_compare}

\input{text/fig_param_compare}

As mentioned in Section~\ref{subsec:timing}, the set of spin, astrometric, and binary timing model parameters used in our analyses is identical to that in \twelveyr; the phrase ``timing model parameters'' used for the remainder of the text refers to this collection, excluding DMX parameters, which are compared separately.
The ensemble of differences in these parameters is shown in Figure~\ref{fig:param_compare}.
The differences between parameter values are plotted, where each difference has been normalized by the parameter uncertainty from \twelveyr\ ($\sigma_{\textrm{NB}}$), and the ``error bar'' on each difference has a length equal to the ratio of parameter uncertainties, with the uncertainty from \twelveyr\ in the denominator (i.e., $\sigma_{\textrm{WB}}/\sigma_{\textrm{NB}}$).
Such a convention allows us to discuss the relative differences we see in parameters and address their consistency without having to reference their absolute units.
In the discussions that follow, we suppress the subscript on $\sigma_{\textrm{NB}}$ and use the standalone symbol $\sigma$ to refer to the units of these normalized differences.
JUMP parameters are not included in Figure~\ref{fig:param_compare}, as they are not meaningful, and parameters that reference an epoch were excluded if the epochs differed.
Generally, this was not the case; 526 of 549 total parameters (96\%), not including JUMPs or DMX parameters, were directly compared.

At a glance, we see that the timing model parameters are in very good agreement, almost entirely $<2\sigma$ different (99\% of the parameters), and with very similar parameter uncertainties.
The cases in which red noise is detected, or is detected in only one analysis, can be harder to interpret (these parameters are semi-transparent in Figure~\ref{fig:param_compare}); due to covariance with the MAP red noise model, especially if the red noise is shallow, the parameter uncertainties can differ by a large factor.
Only 1 of the 526 parameters (0.2\%) is $>5\sigma$ away, while in total 4 (0.8\%) are $>3\sigma$ away; for context, if these were independent experiments and we interpreted these differences as random samples from the unit normal distribution, we would ``expect'' $\sim$~0 deviations $>5\sigma$ and $\sim$~1 deviation $>3\sigma$.
The only parameter larger than $5\sigma$ different is a known peculiarity in our data set (see Section~\ref{subsubsec:J1640+2224}).
Two of the remaining three parameters with a difference larger than $3\sigma$ belong to J2234$+$0611, but also have larger uncertainties by $\sim$30\% (see Section~\ref{subsubsec:J2234+0611}).
The last differing parameter is the parallax measurement of the black widow pulsar J2234$+$0944 (see Section~\ref{subsubsec:J2234+0944}).
Nevertheless, the parameters agree remarkably well across the board, even in cases where red noise is detected.

\subsubsection{DMX Parameters}
\label{subsubsec:dmx_compare}

\input{text/fig_dmx_compare}

We compare the mean-subtracted DMX model parameters in Figure~\ref{fig:dmx_compare}, which has the same presentation as Figure~\ref{fig:param_compare}.
Of the 4,685 differences, 13 (0.3\%) are $>5\sigma$ away ($\sim$~0 ``expected''), a total of 104 (2.2\%) are $>3\sigma$ away ($\sim$~6 ``expected''), and 94\% agree to better than $2\sigma$.
B1937$+$21 is responsible for 35 of the differences $>3\sigma$, which are due to the scatter in its DM measurements and the large influence they have on the DMX model (see Sections~\ref{subsec:DM_nu}~\&~\ref{subsubsec:B1937+21}).
Another 37 of these belong to the combination of J1713$+$0747 (see below), J1903$+$0327 (Section~\ref{subsec:DM_nu}), and J2234$+$0944 (Section~\ref{subsubsec:J2234+0944}).
The remaining 32 differences are distributed among 11 pulsars for which we have no particular suspicions.

Besides the influence of differing red noise models that was already mentioned, the FD parameters in the narrowband data set are covariant with all DMX parameters, which makes the interpretation of the uncertainty ratios in Figure~\ref{fig:dmx_compare} difficult. 
Nevertheless, 92\% of the DMX uncertainties agree to within a factor of 1.5, 99\% agree to within a factor of two, and the median DMX uncertainty for each pulsar is comparable between the data sets.

The number of DMX parameters often differs slightly between the data sets by one, two, or three parameters; there are 10, 6, and 4 such instances, respectively, with 26 pulsars having the same number of DMX parameters.
The discrepancies in the number of DMX epochs arise from the slight differences in curating the data sets (Section~\ref{subsec:wb_cleaning}).
The exception to this is J1713$+$0747, where we opted to use a higher density of DMX bins during and after the second dip in its DM time series \citep{Lam2018}, resulting in 37 additional DMX values; a similar binning exception is made for the same reason in \twelveyr.
However, this means the relevant DMX epochs in \twelveyr\ average over a greater span of time and will be biased in comparison to their counterparts here.
The DMX time series are plotted in the topmost panels of the figures in Appendix~\ref{sec:resid}.
In most instances, the DMX parameters from \twelveyr\ are hidden by those from the present analysis, demonstrating their close agreement.

\subsubsection{Excess White Noise}
\label{subsubsec:wn_compare}

\input{text/fig_wn_compare}

A pulse TOA is a proxy for the moment a fixed point of longitude on the neutron star passes over the line of sight.
There are a number of sources of uncertainty that obfuscate and bias the determination of this moment in time, even if the formal arrival time of the pulse can be very precisely determined.
These additional uncertainties introduce either time-uncorrelated (white) scatter or time-correlated (red) trends into the timing residuals, and can originate from a wide variety of sources local to the observatory, Earth, the solar system, the pulsar, or the intervening ISM.
For thorough reviews of the sources of these uncertainties, we direct the reader to \citet{Verbiest18} and \citet{CordShan10}.
Here, we compare the excess white noise seen in both data sets, followed by the red noise in Section~\ref{subsubsec:rn_compare}.

The formal TOA uncertainties are scaled in both analyses by EFAC parameters, which do not have a straightforward interpretation with respect to physical, excess noise; nominally, EFAC parameters account for misestimation of the system noise level or template matching errors.
Comparing the EFAC parameters is not very enlightening, particularly because the narrowband analysis uses fixed, non-evolving templates and uses a much broader prior on EFAC.
The wideband analysis also uses DMEFAC parameters, which can absorb some excess noise that might be modeled by EFAC in the narrowband analysis.
As mentioned in Section~\ref{subsec:timing}, there is a difference between the two analyses in how white noise is modeled in the timing residuals.
EQUAD and ECORR parameters capture the additional variance in the narrowband analysis, but because ECORR accounts for fluctuations that are completely correlated for simultaneously obtained measurements (i.e., narrowband TOAs), it cannot be differentiated from EQUAD in the wideband analysis and is therefore left out of the noise model.

To effectively compare the white noise, we plot the MAP EQUAD parameters from our analyses against the quadrature sum of the corresponding maximum-likelihood EQUAD and ECORR parameters in Figure~\ref{fig:wn_compare}.
In the figure, points appear more opaque in proportion to how constrained the posterior distribution of the parameter is.
There is a clear correspondence over two orders of magnitude in the white noise parameters, suggesting that both analyses see very similar white noise.

Although the integrated pulse profile shapes of MSPs are secularly stable \citep[][with some exceptions, e.g., \citet{slk+16}]{Brook18}, they vary minutely (indeed) on short timescales due pulse jitter (see Section~\ref{subsec:DM_nu}). 
Pulse jitter contributes additional uncertainty to the TOA and is expected to manifest in ECORR parameters, though the measured ECORR values exceed the predicted level of jitter \citep{Lam16}.
Jitter is thought to be weakly or modestly dependent on frequency \citep{Shannon14,Lam19b}, and its effects can only be reduced by longer integration times or actively accounting for shape change \citep{Oslowski11}.
On the other hand, the various ISM effects that can contribute to EQUAD have a stronger (and mostly pulsar-independent) frequency dependence.
From this perspective, analyzing narrowband TOAs may help to discriminate between sources of excess white noise, although using evolving profile templates would be an improvement to the overall approach.
In this way, both forms of analysis may contribute to arriving at the best results for a given pulsar.
For example, if some of our MSPs have large white noise because of time-variable scattering, then the pulse broadening can be included as part of the wideband TOA measurement.

\subsubsection{RMS Timing Residual}
\label{subsubsec:rms_compare}

\input{text/fig_rms_compare}

A second metric for gauging the overall level of noise is the RMS timing residual (see Table~\ref{tab:timing_model_summary}).
In Figure~\ref{fig:rms_compare} we compare the RMS values between the two analyses, taking care to use the averaged residuals from the narrowband data set and the whitened set of residuals whenever red noise was detected in either of the analyses.
Almost no pulsars differ by more than a factor of 1.5, with a number of exceptions explained as part of Section~\ref{subsec:pulsar_results}, and half of them agree to within a factor of 1.1.
However, the RMS residual from either analysis can be very sensitive to the exact noise model, which is fixed in the final optimization of the timing model.
The noise analyses explore the logarithm of the EQUAD, ECORR, and red noise amplitude parameters, so small statistical deviations in the best-fit parameters arising from the Monte Carlo analysis can lead to rather different RMS values.
The RMS should be thought of as a random variable, whose variance is influenced by the posterior distributions of the noise parameters.
This is true even when the noise model parameters are constrained, and it underscores the need for advanced noise modeling techniques.
Nevertheless, it is encouraging that 31 of the 47 pulsars show some amount of improvement and that all but five pulsars have RMS residuals no more than $\sim10\%$ larger than their narrowband counterparts.




\subsubsection{Detection of Red Noise}
\label{subsubsec:rn_compare}

\input{text/fig_rn_compare}

A final, and perhaps most crucial, litmus test for the wideband analyses is the detection of red noise in individual pulsars.
Obviously, the presence of red noise in the wideband data set (or lack thereof), in relation to what is seen in the narrowband data set, guides our expectations of full-scale GW analyses, which heretofore have only been vetted on our narrowband data sets.
We introduced the red noise model in Section~\ref{subsec:timing}; there are additional details in Appendix~\ref{sec:wb_like}, \nineyr, \elevenyr, and \twelveyr.
Here we discuss our findings in contrast to those from the narrowband analysis.

In Figure~\ref{fig:rn_compare} we show the significantly detected power-law red noise in our analyses compared to those from \twelveyr.
We again find the level of agreement between the data sets reassuring.
Recall from Section~\ref{subsec:timing} that a pulsar is deemed to have ``significant red noise'' if the estimated Bayes factor is above one hundred (see Table~\ref{tab:timing_model_summary} for Bayes factors).
Thirteen pulsars have detected red noise in both analyses, one pulsar has significant red noise detected in just the narrowband analysis (J0613$-$0200), and two black widow pulsars, which are not shown in the plot, are treated differently and not discussed further here (J0023$+$0923 and J2234$+$0944; see Section~\ref{subsubsec:J2234+0944}).
Ten of these pulsars (plus J1713$+$0747) had detected red noise in \elevenyr; J1744$-$1134, J1853$+$1303, and J2317$+$1439 are new detections in \twelveyr, which are all have significant red noise here.

It is thought that unmitigated ISM effects can manifest as shallow-spectrum red noise \citep[][]{ShannonCordes2017,CordShan10,FosterCordes90,Rickett90}, which we indicate in Figure~\ref{fig:rn_compare} for $\gamma_{\textrm{\scriptsize red}} > -3$.
J0613$-$0200's DM is in the top third of our sample ($\sim$ 38.8~$\dmunits$, respectively), and has fairly shallow red noise in its narrowband analyses.
Red noise is only marginally favored in its wideband analysis, as indicated by the Bayes factors of $\sim$~15 in Table~\ref{tab:timing_model_summary}.
When red noise is included in the wideband analysis (the dashed-dotted lines in Figure~\ref{fig:rn_compare}), the MAP model has the same index, a slightly smaller amplitude, and similar white noise parameters, than in \twelveyr.
Without red noise, the corresponding wideband white noise EQUAD parameters are statistically unchanged.
This suggests that the wideband analysis might be able to mitigate some of the ISM-induced red noise.


Intrinsic spin noise in pulsars has been modeled in the literature as a random walk in phase, frequency, or frequency derivative, with corresponding power spectral indices of $-2$, $-4$, and $-6$, respectively, as well as arising from chaotic behavior \citep[e.g.,][]{Harding90}.
The lighter gray region in Figure~\ref{fig:rn_compare} represents the best fit index ($\gamma_{\textrm{\scriptsize spin}} = -4.46 \pm 0.16$) for timing noise seen across pulsars of all types from \citet{Lam17a}, consistent with a mixture of random walks \citep[e.g.,][]{DAlessandro95,CordesDowns85}.
The scatter in this best fit relation, however, is large enough to essentially cover the range of observed spectra.
It is therefore difficult to interpret the spread of red noise we have detected, particularly because we suspect that some of the pulsars with shallow red noise are dominated by contributions from the ISM, whereas others may have a mix of contributions.
Coexisting with the red noise intrinsic to the pulsar and that from the ISM, there is a contribution from the background of stochastic, low-frequency GWs, which is thought to have a steep power-law index ($\gamma_{\textrm{\scriptsize GWB}} = -13/3$; \citet{Jaffe2003,Phinney2001}), indicated by a dotted vertical line in the figure.
For scale, the dashed vertical line indicates the 95\% upper limit on the amplitude of the GW background from analyzing the 11-year data set \citep{Arzoumanian2018b}.
A more recent search for the stochastic GW background in the 12.5-year narrowband data set is presented in \citet{Arzoumanian20}.



\subsection{Additional Discussion of Individual Pulsars}
\label{subsec:pulsar_results}

The results from a number of pulsars, some of which have been previously mentioned, deserve additional comments, caveats, or emphasis, which we detail here.
In addition, for the simple purpose of highlighting one example of generally good, comprehensive agreement with the narrowband results, and one example of where perhaps wideband timing did not prove beneficial, we direct the reader to J0931$-$1902 and J1910$+$1256, respectively.

\subsubsection{J0931$-$1902}
\label{subsubsec:J0931-1902}

J0931$-$1902 has the distinction of having the largest fractional difference between the number of pulse profiles used in the wideband analysis and the number of TOAs in the narrowband analysis; its wideband data set makes use of 81\% more profiles (see Table~\ref{tab:ntoa_nchan}).
Again, this difference arises because of the S/N threshold used in the narrowband analysis; because this pulsar is fairly weak and scintillates, a large number of its low S/N profiles get individually discarded in the narrowband analysis, even though they combine to yield useful wideband TOAs.
J0931$-$1902 is the second worst pulsar in our data set in terms of raw L-band timing precision (see Figure~\ref{fig:wb_toa_dm_err}), but is somewhere in the middle in terms of RMS ($\sim$~440~ns in both data sets).
There is absolutely nothing else different about its results from the narrowband analysis -- except that its timing model parameters are all $\sim$~15\% more precise in the wideband analysis.
At least two other pulsars show this level of improvement that is most likely attributable to a similar explanation -- J0340$+$4130 and J0740$+$6620 -- although their differences in data volume are not extreme.
These improvements underscore the benefit of using the wideband TOA approach for salvaging all information contained in less bright or scintillating pulsars.

\subsubsection{J1640$+$2224}
\label{subsubsec:J1640+2224}

The difference in J1640$+$2224's ecliptic longitude is the lone culprit referred to earlier for being very different ($\sim$~6$\sigma$) from its counterpart in the narrowband analysis.
However, this is a known anomaly to us, albeit of unknown origin; we have previously compared timing results from different timing software using the exact same data sets, and J1640$+$2224's ecliptic longitude was the single outlier to be significantly different (see also the comparison between \texttt{Tempo} and \texttt{PINT} \citep{PINT} in \twelveyr).
The published position from Very Long Baseline Interferometry \citep{Vigeland18} is not precise enough to discern between the two measurements.
However, it should be noted that the value from \twelveyr\ is better than 1$\sigma$ consistent with the extrapolated value from \elevenyr, whereas the value from the wideband analysis is $\sim$~2$\sigma$ consistent with the extrapolated value from \nineyr.
\citet{Fonseca2016} followed up on \nineyr\ and suspected that J1640$+$2224 is a massive neutron star \citep[see also][]{Deng20}; the improvements on the mass measurements will be presented elsewhere.

\subsubsection{J1643$-$1224}
\label{subsubsec:J1643-1224}

We have already discussed J1643$-$1224 at some length in Section~\ref{subsec:DM_nu}.
It is worth emphasizing, though, that some of the complexity and chromatic dependence seen in the DM measurements and timing residuals of this pulsar almost certainly arise from the fact that it lies directly behind the HII region Sh 2-27 associated with $\zeta$-Ophiuchi \citep{Ocker20}.
This association may also be responsible for a protracted decrease in its flux density \citep{Maitia03}.
In addition to the confounding factors of the ISM, at least one intrinsic profile shape change event is thought to have occurred in this pulsar around February 2015 \citep{slk+16}.
Although we see the corresponding discrete perturbation in J1643$-$1224's timing residuals at this time, the follow-up analysis by \citet{Brook18} on our 11-year data set argues that ISM effects cannot be ruled out.
The $\sim$~45\% improvement in its RMS timing residual seen in Figure~\ref{fig:rms_compare} is almost certainly a result of the mitigation of the chromatic structure in its residuals; see the discussion at the end of Section~\ref{subsec:DM_nu}.

\subsubsection{J1747$-$4036}
\label{subsubsec:J1747-4036}

Similarly, we have already discussed J1747$-$4036 in Section~\ref{subsec:DM_nu}.
J1747$-$4036 also stands out in Figure~\ref{fig:rms_compare}, with the same level of improvement in RMS residual as J1643$-$2224 due to the mitigation of chromatic structure in the timing residuals.

\subsubsection{J1910$+$1256}
\label{subsubsec:J1910+1256}

J1910$+$1256 has a significantly worse RMS wideband timing residual in Figure~\ref{fig:rms_compare}, by just over a factor of two.
We find a significant L-band EQUAD detected in the wideband analysis, whereas the posterior distributions for the L-band EQUAD and ECORR are consistent with upper-limits.
Interestingly, the white noise parameters for S-band are significantly measured in both analysis and are of similar amplitude.
J1910$+$1256 is in the top ten pulsars by raw L-band timing precision (Figure~\ref{fig:wb_toa_dm_err}), with the median L-band PUPPI TOA having a precision just above 100~ns; the MAP PUPPI L-band EQUAD is more than three times larger.
The source of this discrepancy has not been determined but despite the difference, the timing model parameters are no more than $\sim$~10\% worse than their narrowband counterparts.

\subsubsection{B1937$+$21}
\label{subsubsec:B1937+21}

B1937$+$21 (a.k.a. J1939$+$2134) presents a special set of challenges for the wideband analysis, being the brightest pulsar in the data set with the smallest formal measurement uncertainties by a considerable margin (see Figure~\ref{fig:wb_toa_dm_err}).
As mentioned in Section~\ref{subsec:prof_evol}, its profile modeling is contaminated by spectral leakage because it is so bright, although we do not believe this meaningfully affects the timing results.
As mentioned in Section~\ref{subsec:DM_nu}, it has a substantial amount of scatter in its wideband DM measurements once the long-term trend is removed; this results in the highest DMEFAC parameters in the data set as well as the worst goodness-of-fit value for its timing model, due to the additional contribution from the DM model.
The restrictive Gaussian prior (see Section~\ref{subsec:timing}) inhibits the DMEFAC parameters from taking even larger values, which would encapsulate more of the variance in the DM time series.
Relaxing the prior is not physically motivated, and so these results direct us to implement an additional DM model parameter in future analyses, one that is analogous to the standard EQUAD parameter.
Given that both pulse jitter and variable diffractive interference in the ISM (i.e., the ``finite scintle effect'') play a roll in the observations of this pulsar \citep{Lam19b}, it is feasible that both effects serve to bias the wideband DM estimates, resulting in extra variance in the DM time series.
Despite the additional variance in the wideband DM time series, the astrometric timing model parameters are in very good agreement ($<$~1$\sigma$) with similar uncertainties.

\subsubsection{J1946$+$3417}
\label{subsubsec:J1946+3417}

J1946$+$3417 is one of the two new pulsars in this data set, which has already been discussed in Section~\ref{subsec:DM_nu} due to it having the third largest DM in the data set ($\sim$110.2~$\dmunits$).
It has the distinction of showing the single largest difference in RMS in either direction, seen in Figure~\ref{fig:rms_compare}; the wideband RMS timing residual is a factor of three smaller.
Both analyses examine the same amount of data and the wideband raw timing precision is $\sim$~10\% better (see Table~\ref{tab:toa_summary} and the equivalent table in \twelveyr).
The Bayes factor for red noise in the narrowband analysis is $\sim$~59, whereas it is not at all favored in the wideband analysis.
The preferred red noise model is large and shallow, and as a result of it not being included, the narrowband white noise parameters are larger than their wideband counterparts.
The timing model parameters agree to $\le 1\sigma$, but with $\sim$~10\% larger uncertainties in the wideband analysis.
It should also be noted that J1946$+$3417 is an astrophysically interesting source, as it is one of the few eccentric binary MSPs in the field and also contains a massive neutron star \citep[][all of which also make note of J2234$+$0611]{Barr17,Jiang15,Antoniadis14,Freire14}.

%

\subsubsection{J2043$+$1711}
\label{subsubsec:J2043+1711}

J2043$+$1711 has the highest sub-threshold Bayes factor in Table~\ref{tab:timing_model_summary}, $B\sim50$.
In repeated analyses, the statistic $B$ was noisy enough to sometimes cross our significance threshold.
The narrowband analysis also favors red noise, but with a lower Bayes factor~$\sim$~26.
The difference between the analyses may arise from the amount of data examined.
As can be seen in Table~\ref{tab:ntoa_nchan}, J2043$+$1711's wideband data set is $\sim$~70\% larger than its narrowband counterpart, the third largest difference, which is due to its scintillation characteristics combined with the S/N ratio cut off in the narrowband analysis.
J2043$+$1711 has a fairly low DM ($\sim$20.8~$\dmunits$), a timing baseline of six years and, importantly, it has been included in our high-cadence observations at Arecibo since 2015, which has increased its data volume by $\sim$~70\% since the 11-year data set.
The narrow features in its profile enable this pulsar to be timed very precisely when it is detected (see Tables~\ref{tab:toa_summary}~\&~\ref{tab:timing_model_summary}, and Figure~\ref{fig:wb_toa_dm_err}), and so we expect the emerging red noise in this pulsar to be significantly detected in the near future.

\subsubsection{J2234$+$0611}
\label{subsubsec:J2234+0611}

At face value, J2234$+$0611 is the best timed pulsar in the data set: the narrowband and wideband timing RMS values are $\sim$~60 and 35~ns, respectively (Figure~\ref{fig:rms_compare}), with no preference for red noise in either analysis.
This is partially due to it only having a timing baseline 3.4~years in length, although it is in the top ten pulsars by raw L-band timing precision.
Both analyses detect excess white noise, although the wideband analysis measures a significantly larger EQUAD in the 430~MHz band; this results in an overweighting of the wideband L-band data, which may explain the significantly smaller RMS value.
We make special mention of this pulsar also because it stands out for its level of disagreement in its timing model parameters with \twelveyr, as mentioned in Section~\ref{subsubsec:param_compare}.
All of its wideband timing model parameters have larger uncertainties by $\sim$~30$-$40\%, but no other pulsar shows quite this level of disagreement.
Its ecliptic latitude and parallax measurements are $\sim$~3.5$\sigma$ different from their narrowband analysis counterparts.
The fact that J2234$+$0611 has a relatively short timing baseline but has significantly measured secular binary parameters is a testament to its timing precision.
In fact, additional modeling of its binary orbit is necessary, which was carried out in \citet{Stovall2019} with an additional 1.5~years of data, most of which were NANOGrav observations collected beyond the cutoff of the present data set.
Along with the Shapiro delay and annual orbital parallax, \citet{Stovall2019} were able to determine the 3-D orbital geometry of the binary.
We are confident that the discrepancies seen here will be resolved with the implementation of the \citet{Stovall2019} timing solution in future data sets.

\subsubsection{J2234$+$0944}
\label{subsubsec:J2234+0944}

J2234$+$0944 was previously mentioned in Sections~\ref{subsubsec:param_compare}, \ref{subsubsec:dmx_compare}, and \ref{subsubsec:rn_compare}.
This pulsar is one of four black widow pulsars in the data set (along with J0023$+$0923, J0636$+$5128, and J2214$+$3000), and one of two (along with J2214$+$3000) that do not show orbital or secular variability according to \citet{BakNielsen20}, who studied three of these systems over $\sim~8$~year baselines (J0636$+$5128 was not part of their study).
This pulsar has very significantly detected ``red noise'' in the wideband analysis, but no indication of it in the narrowband analysis, according to the Bayes factors in Table~\ref{tab:timing_model_summary} and its analog in \twelveyr.
However, this ``red noise'' is specious; the preferred model is extremely shallow and the power-law fit to the function in Equation~\ref{eqn:red_pl} is dominated by the frequencies higher than 1\,yr$^{-1}$, reflecting the short timing baseline ($T_\textrm{span} \sim 4.0$~yr).
Based on this reasoning, we exclude the red noise component from J2234$+$0944's analysis; we similarly excluded the ``red noise'' seen in J0023$+$0923 for the same reason.
The origin of the excess noise seen in the wideband data set is not known; it could be a sign of variability \citep{Torres17}, but the findings of \citet{BakNielsen20} refute this.
Interestingly, the black widow J2214$+$3000 had a very similar issue in \elevenyr, when it had a similar timing baseline, which was resolved with the additional data in this data set and a reparameterization of its timing model.
Finally, \twelveyr\ reports that the parallax measurement is no longer significant, in contrast to \elevenyr; this loss of significance in \twelveyr\ is marginal and not the case in the wideband analysis.
Despite being $\sim$~3$\sigma$ different from \twelveyr, J2234$+$0944's parallax is significantly measured and is $< 1\sigma$ consistent with the value from \elevenyr.

%% file: text/fig_param_compare.tex
\begin{figure}[ht!]
\begin{center}
\includegraphics[width=\columnwidth]{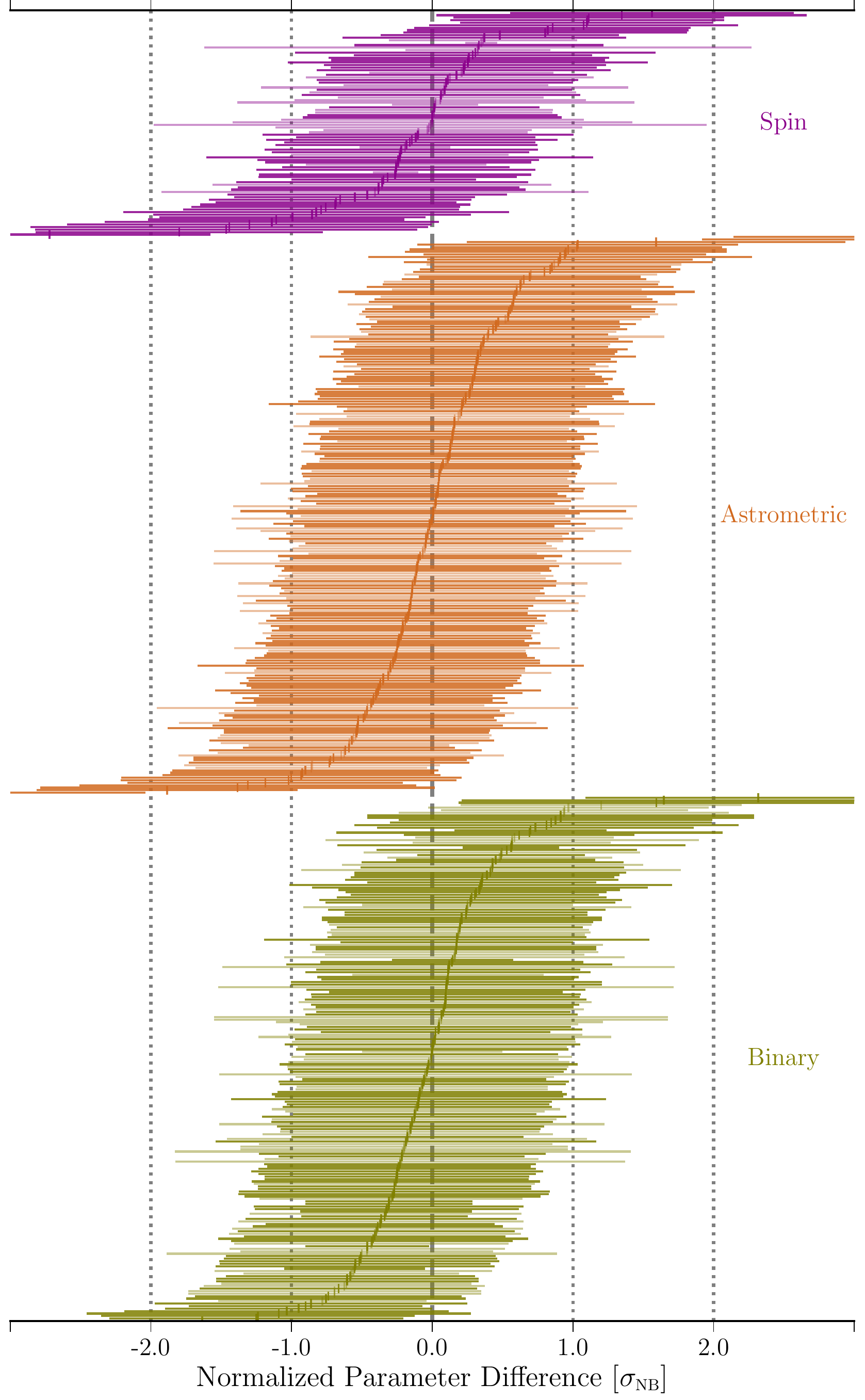}
\caption{\label{fig:param_compare}
Snapshot comparison of 526 timing model parameters measured in the two data sets, divided into three main groups, each ordered by the normalized difference in parameter value.
The parameter differences have been normalized by their uncertainties from \twelveyr\ ($\sigma_{\textrm{NB}} \equiv \sigma$), and ``error bars'' have a length $= \sigma_{\textrm{WB}}/\sigma_{\textrm{NB}}$.
The more transparent points are parameters from timing models containing red noise in at least one analysis; due to covariance with, and small differences in, the MAP red noise model (see Figure~\ref{fig:rn_compare}), these differences may be harder to interpret.
}
\end{center}
\end{figure}

%% file: text/fig_dmx_compare.tex
\begin{figure}[ht!]
\begin{center}
\includegraphics[width=\columnwidth]{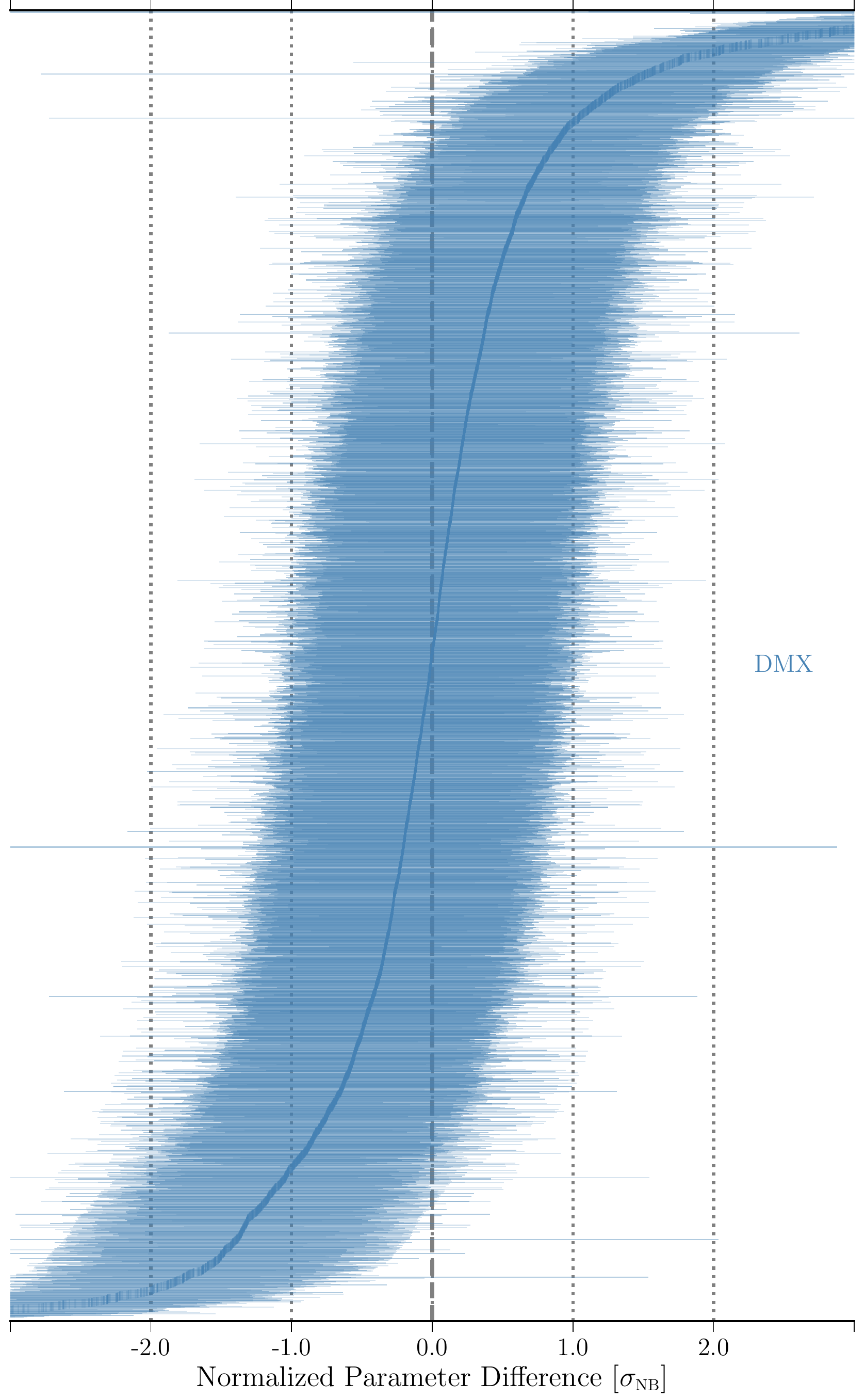}
\caption{\label{fig:dmx_compare}
Snapshot comparison of 4,685 DMX model parameters measured in the two data sets, presented in the same manner as Figure~\ref{fig:param_compare}. 
In addition to the effect of differing red noise models between the analyses, the covariance present between DMX parameters and the FD parameters in the narrowband analysis may also skew the ratio of parameter uncertainties.
}
\end{center}
\end{figure}

%% file: text/fig_wn_compare.tex
\begin{figure}[t!]
\begin{center}
\includegraphics[width=\columnwidth]{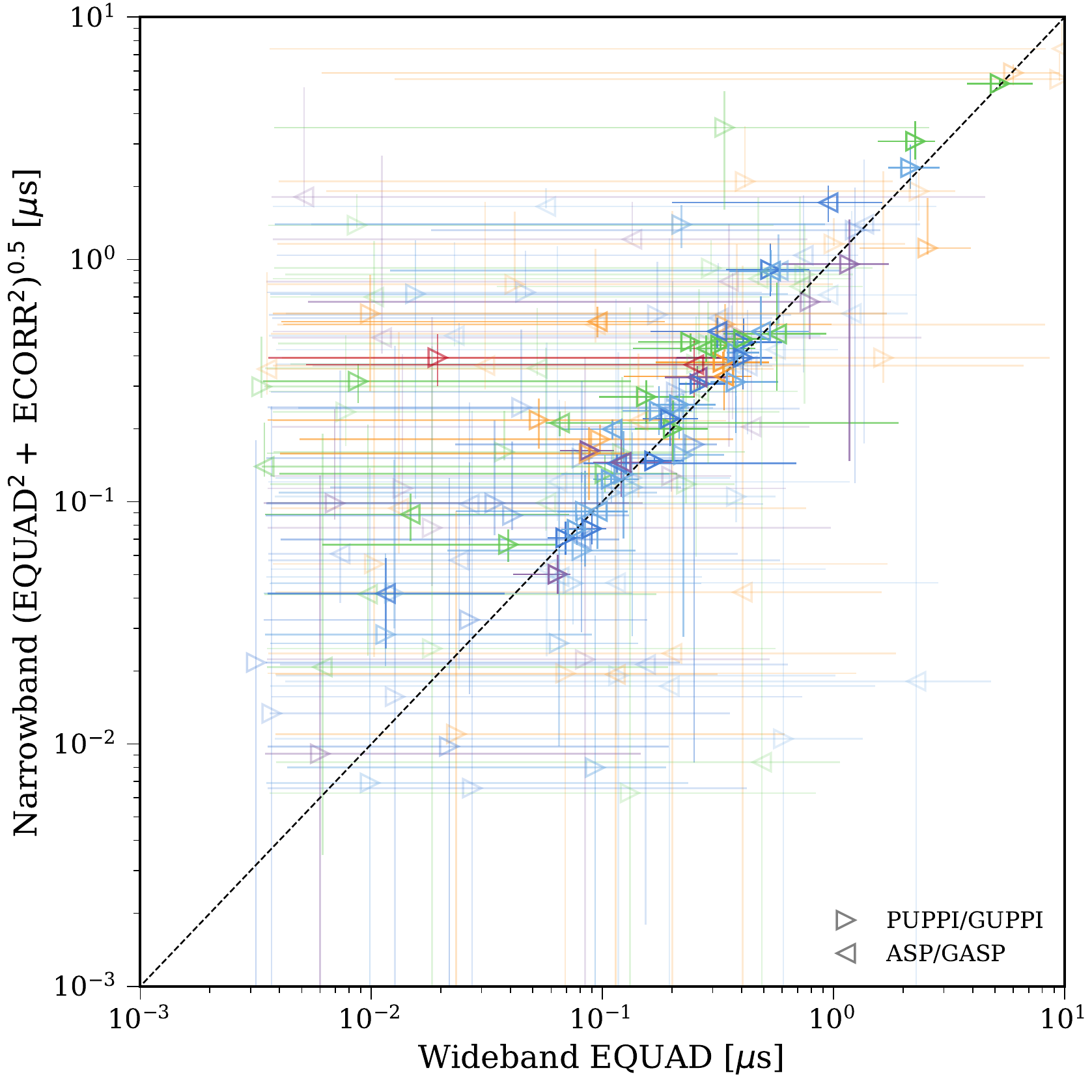}
\caption{\label{fig:wn_compare}
Comparison of excess white noise seen in the two data sets.
Each symbol demarcates a single combination of pulsar, receiver, and backend instrument; the symbol direction indicates the backend, whereas the colors indicate the receiver as in Figure~\ref{fig:epochs}: 327~MHz (red), 430~MHz (orange), 820~MHz (green), 1.4~GHz (lighter blue for AO, darker blue for the GBT), 2.1~GHz (purple).
As ECORR parameters are not part of the wideband noise model, the quadrature sum of maximum-likelihood estimates for the narrowband EQUAD and ECORR parameters is plotted against the corresponding wideband MAP EQUAD estimate.
The central 95\% interval of each parameter's posterior is shown as a superimposed horizontal or vertical line (the smaller of the EQUAD and ECORR intervals was chosen for each narrowband value).
The transparency of the symbols is a proxy for the mutual significance of the parameter; the smaller the combined intervals, the more opaque the marker.
In this way, a clear correlation is brought out, suggesting similar white noise is seen in both analyses.
}
\end{center}
\end{figure}

%% file: text/fig_rms_compare.tex
\begin{figure}[t!]
\begin{center}
\includegraphics[width=\columnwidth]{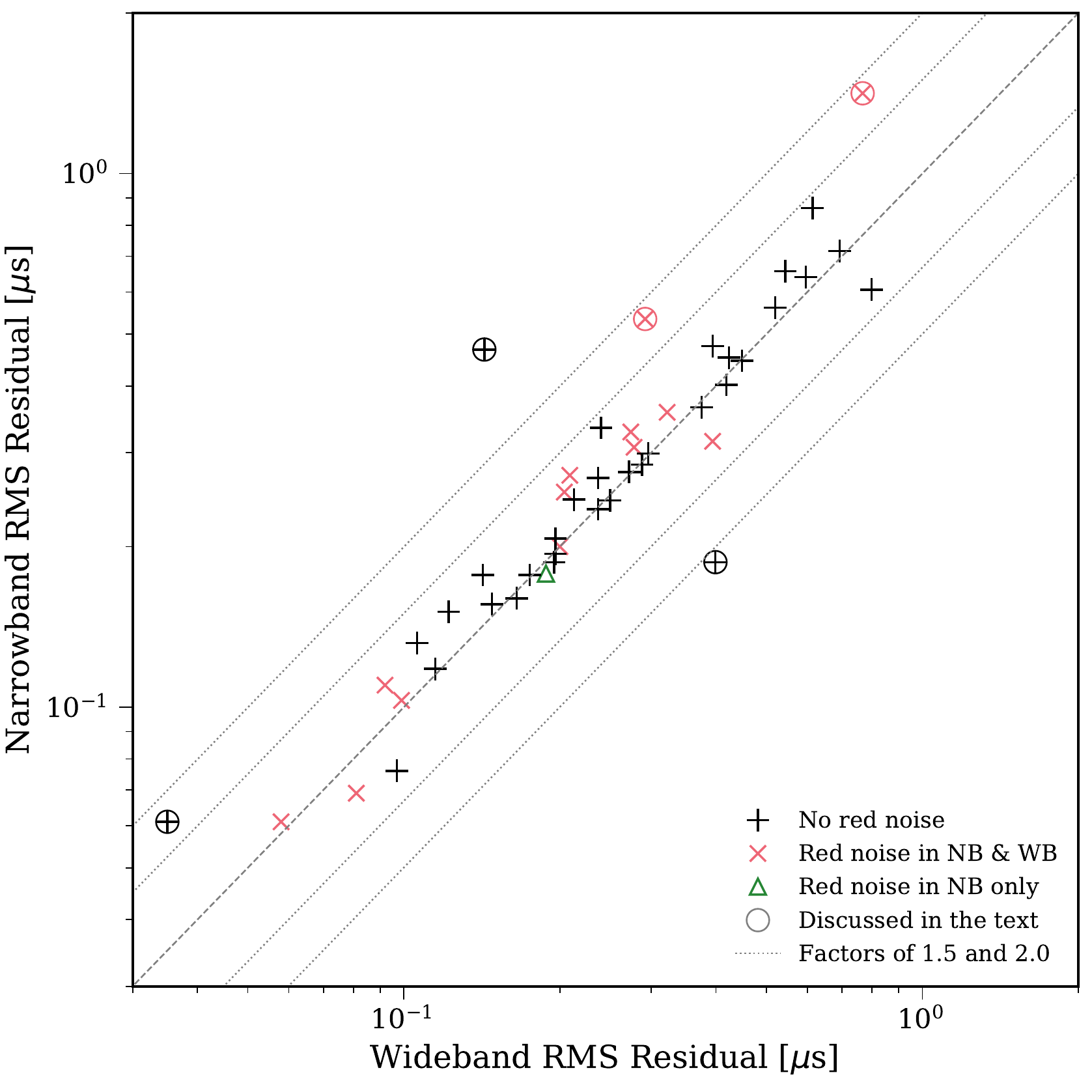}
\caption{\label{fig:rms_compare}
Comparison of the timing residuals' weighted root-mean-square (RMS) values between the two data sets.
The RMS of the ``whitened'' residual is plotted in all applicable cases; the values are given in Table~\ref{tab:timing_model_summary}.
Encircled pulsars are addressed as part of Section~\ref{subsec:pulsar_results}; in increasing order of their wideband timing RMS, these are J2234$+$0611, J1946$+$3417, J1643$-$1224, J1910$+$1256, and J1747$-$4036.
}
\end{center}
\end{figure}

%% file: text/fig_rn_compare.tex
\begin{figure*}[htb!]
\begin{center}
\includegraphics[width=2\columnwidth]{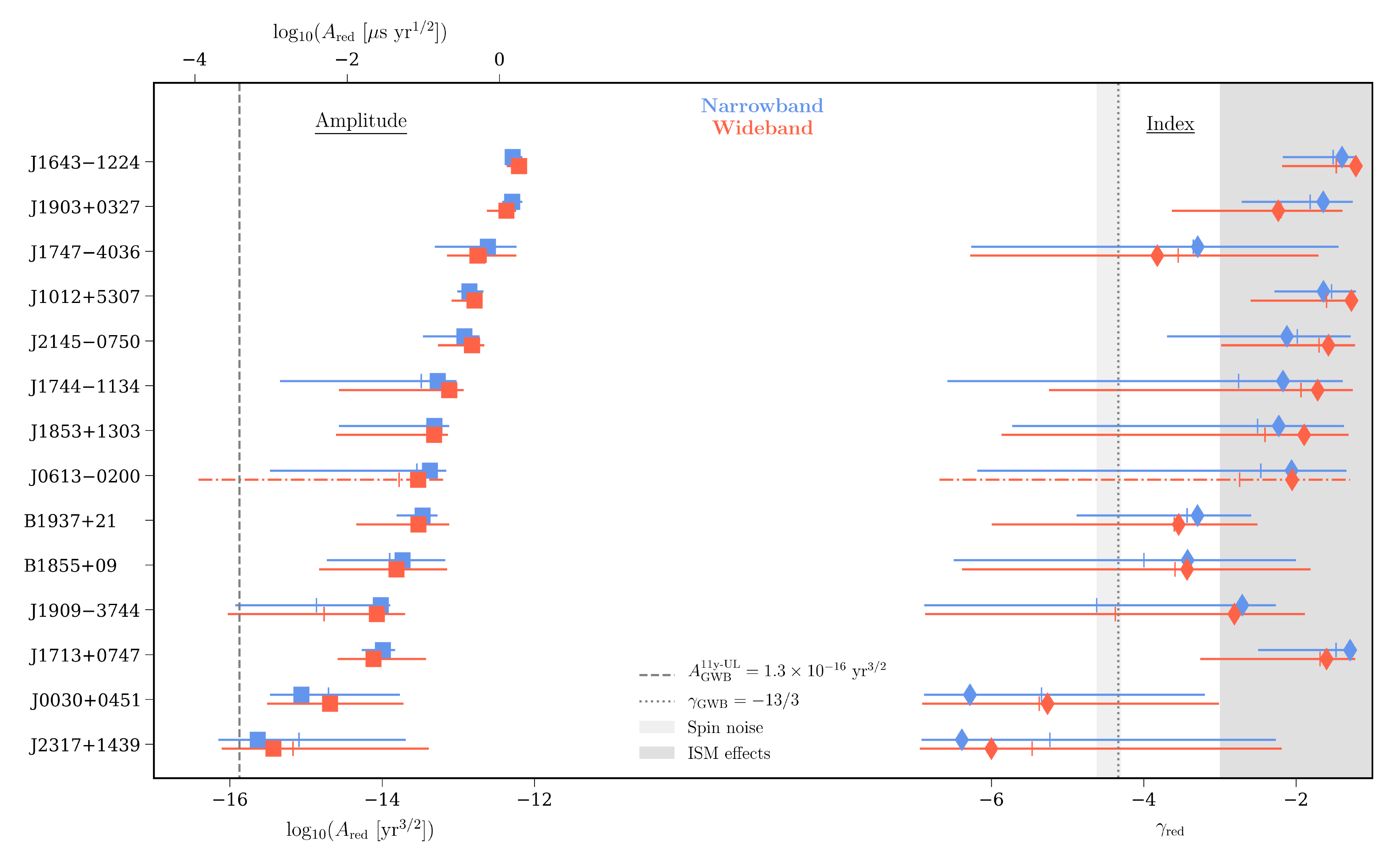}
\caption{\label{fig:rn_compare}
    Comparison of the significantly detected power-law red noise parameters in the two data sets; measurements from the wideband data set (red) are plotted below those from \twelveyr\ (blue).
Pulsars are ordered top-to-bottom by highest-to-lowest red noise amplitude seen in the wideband data set, and the large symbols represent the MAP parameter estimates: squares indicate the logarithm of the amplitude at a frequency of 1\,yr$^{-1}$ (dual units shown), and diamonds represent the power-law index.
The central 95\% of the marginalized posterior distribution for each parameter is shown as a line with a tick indicating the median.
One pulsar (J0613$-$0200) has above-threshold red noise in the narrowband analyses, but not in the wideband analysis; we nevertheless indicate its MAP values and posterior distribution intervals (with dashed-dotted lines) when red noise is included in the modeling.
The apparent correlation between red noise amplitude and index is in part due to the parameterization of referencing the amplitude to a frequency of 1\,yr$^{-1}$.
Unmitigated ISM effects are thought to induce fluctuations with a spectrum having a characteristic index lying within the darker gray region \citep[$\gamma_{\textrm{\scriptsize red}} > -3$, cf.][]{ShannonCordes2017}.
The lighter gray region is the prediction for intrinsic spin noise across a broad pulsar population from \citet{Lam17a}, $\gamma_{\textrm{\scriptsize spin}} = -4.46 \pm 0.16$, although the scatter in the relation is substantial.
We indicate with vertical lines the fiducial index for the stochastic background of gravitational waves and our most recently published 95\% upper limit for its amplitude, from the 11-year data set \citep{Arzoumanian2018b}; this limit accounts for both interpulsar correlations and uncertainties in the solar system ephemeris.
}
\end{center}
\end{figure*}

%% file: text/sec_conclusion.tex

In this paper we have demonstrated the promise of wideband data sets for the purpose of high-precision pulsar timing and GW experiments with PTAs.
Specifically, we reprocessed the pulse profiles from the 47 MSPs comprising the NANOGrav 12.5-year data set using ``wideband'' methods, produced a parallel data set\footnote{Both data sets are available at \href{https://data.nanograv.org/}{data.nanograv.org}.}, and compared the results with those presented in \twelveyr.
Our wideband framework employed an innovative, compact modeling of pulse profile evolution, extracted simultaneous TOA and DM measurements from broadband pulsar observations, and analyzed the data using novel developments to our Bayesian noise models and timing software.

The broad agreement in results spans a variety of metrics, from timing residuals, timing model parameters, and DM time series, to white and red noise model components.
In many of the simplest cases, the wideband timing results are at least on par with their narrowband counterparts, and often times show indications of improvements (e.g. J0931$-$1902, J1125$+$7819, J2043$+$1711).
In other cases, complexities point towards favoring one kind of analysis over the other, or towards an unresolved discrepancy with perhaps interesting results (e.g. J1910$+$1256, J1946$+$3417, J2234$+$0944).

We gain the most assurance from the congruence of results for our most important, best timed pulsars, and from the concurrence in all red noise models.
For example, for J1909$-$3744 and J1713$+$0747, all but one of their timing model parameters agree to within 1$\sigma$ (the last to within 1.5$\sigma$), with very similar parameter uncertainties, and nearly the same form of detected red noise.
In the case of J1713$+$0747, considering that its noise model included 30 parameters -- twelve of which belong to two types of newly implemented parameters -- and that its TOA volume was reduced by a factor of 37, we find this level of agreement remarkable.

Our most significant results include:
\begin{itemize}
    \item{Better than 2$\sigma$ agreement for the large majority (99\%) of timing model parameters (Section~\ref{subsubsec:param_compare}), and better than 3$\sigma$ in all but four of them; 98\% of DMX parameters agree to better than 3$\sigma$ (Section~\ref{subsubsec:dmx_compare}).}
    \item{Consistently detected red noise for 13 of the 14 pulsars in \twelveyr\ (Section~\ref{subsubsec:rn_compare}).}
    \item{Very similar timing residuals and DMX time series (Appendix~\ref{sec:resid}), comparable or improved RMS timing residuals for 31 of the 47 pulsars, with virtually all of them agreeing within a factor of 1.5 (Section~\ref{subsubsec:rms_compare}).}
    \item{Recovery of low S/N profile data (Appendix~\ref{sec:low_snr}), leading to a larger profile data set by 16\% and more precise timing model parameters in several pulsars (Section~\ref{subsec:pulsar_results}).}
    \item{Indications of frequency-dependent DMs in at least four MSPs, and significant variance in the DM measurements in others (Section~\ref{subsec:DM_nu}).}
    \item{Per-observation mean flux density measurements (Section~\ref{subsec:port_results}).}
    \item{Wideband developments to the \texttt{ENTERPRISE} and \texttt{Tempo} software packages (Appendix~\ref{sec:wb_like}).}
    \item{A reduction in the TOA volume of the 12.5-year data set by a factor of 33, and an overall reduction in data volume (once including the DM measurements) by a factor of 16 (Table~\ref{tab:ntoa_nchan}).}
\end{itemize}

The reduction in the data set volume is a particularly important development because PTA experiments are maturing and are well into their second decade of operation; as a result, they have burgeoning facilities, an increasing number of pulsars, growing bandwidths, and, therefore, an exploding number of TOAs.
However, additional developments will need to be made if we are to realize some of the sought-after significant computational speed-ups by moving to wideband data sets.
Nevertheless, wideband techniques offer a number of avenues for tackling problems related to profile evolution, the ISM, and broad-bandwidth observations.

It very well may be that choosing between wideband and narrowband approaches is a pulsar-dependent question, and that some of the advantages that come with retaining frequency-resolved TOAs can aid in making new advances or analyzing wideband measurements.
In all cases, conventional narrowband analyses ought to adopt frequency-dependent profile templates, which would also facilitate a shared framework in which the methods operate.
We note, however, how to optimize a PTA experiment is still an open question, particularly with respect to scheduling and selecting frequencies and bandwidths with which to time individual pulsars \citep[e.g.,][]{Lam18c,Lam18b,Lee14}.
Ultra-wideband receivers and simultaneous multi-band observations are becoming norms in pulsar timing, and along with the anticipated increase in cadence (for which CHIME is the archetype), we anticipate wideband timing techniques to follow closely behind.
In the mean time, we will make new improvements to the wideband strategy, some of which we have already mentioned.
These include incorporating polarization information into the wideband TOA measurement, accounting for additional time-variable effects from the ISM, implementing Gaussian processes to model our DM measurements, and developing the \texttt{PINT} timing software package \citep{Luo21} for full compatibility with wideband data sets.

The final test for our wideband data set will be to analyze it for GWs.
\citet{Arzoumanian20} finds strong evidence for an unidentified common-spectrum stochastic process across pulsars in the narrowband data set, and early analyses indicate we should expect similar results from the wideband data set; we will present that investigation elsewhere.
The NANOGrav 12.5-year data set represents a milestone in part because it lays the groundwork for future wideband data sets, which will meet the challenges posed by PTA experiments.
\\

%% file: text/sec_acknowledgement.tex
{
We thank the telescope operators and all the staff at the Arecibo Observatory and the Green Bank Observatory for the essential role they played in collecting the data for this data release.
We also thank the anonymous referee for their very useful comments that helped to improve the quality of this paper.
The NANOGrav Collaboration dedicates this work to the Arecibo Observatory, its employees and staff, and the many students, teachers, and others who have drawn inspiration from it.
\\\\
\indent The NANOGrav project receives support from National Science Foundation (NSF) Physics Frontiers Center award number 1430284.
The Arecibo Observatory is a facility of the National Science Foundation operated under cooperative agreement (\#AST-1744119) by the University of Central Florida (UCF) in alliance with Universidad Ana G. M\'{e}ndez (UAGM) and Yang Enterprises (YEI), Inc.
The Green Bank Observatory is a facility of the National Science Foundation operated under cooperative agreement by Associated Universities, Inc.
The National Radio Astronomy Observatory is a facility of the National Science Foundation operated under cooperative agreement by Associated Universities, Inc.
Part of this research was carried out at the Jet Propulsion Laboratory, California Institute of Technology, under a contract with the National Aeronautics and Space Administration.
Pulsar research at UBC is supported by an NSERC Discovery Grant and by the Canadian Institute for Advanced Research.
TTP acknowledges support from the MTA-ELTE Extragalactic Astrophysics Research Group, funded by the Hungarian Academy of Sciences (Magyar Tudom\'{a}nyos Akad\'{e}mia), that was used during the development of this research.
WWZ is supported by the CAS Pioneer Hundred Talents Program and the Strategic Priority Research Program of the Chinese Academy of Sciences Grant No. XDB23000000.
}

%% file: text/sec_contribution.tex
{
{\it Author contributions.}
The alphabetical-order author list reflects the broad variety of contributions of authors to the NANOGrav project.
Some specific contributions to this paper, particularly those that overlap with and are in addition to \twelveyr, which made this project possible, are as follows.
TTP was responsible for the research, development, implementation, computation, analysis, and write-up of this work, contributing significantly to all aspects.
PBD and MV developed the formal wideband timing likelihood.
PBD and TTP made necessary developments to \texttt{Tempo}.
MV and TTP made necessary developments to \texttt{ENTERPRISE}.
MED coordinated the development of the primary 12.5-year data set, was its primary lead, co-authored observing proposals, and coordinated the observations at the Arecibo Observatory.
MTL designed the original \texttt{python} notebook analysis pipeline, which was re-purposed by TTP for these analyses.
PBD wrote observing proposals, performed calibration and data reduction, coordinated data management, developed and implemented the methodology used to mitigate the artifact images in the data, and advised closely about this research.
EF wrote observing proposals and assisted in coordination of Arecibo observations.
PBD and KS coordinated GBT observations.
SMR provided computational facilities for this work.
PTB, HTC, JMC, PBD, TD, DLK, MTL, CMFM, DJN and MV contributed directly to portions of the manuscript.  
BJS and JSH contributed to the analysis concerning B1855$+$09.
NGD oversaw and maintained much of the computational infrastructure related and essential to this work, which included regularly updating software and maintaining the servers where data and analysis pipelines are hosted.
The NANOGrav Timing Working Group advised about all stages of the analyses; DJN served as its chair during the development of the 12.5-year data set.
%
%
ZA,
HB,
PTB,
HTC,
MED,
PBD,
TD,
RDF,
ECF,
EF,
PAG,
MLJ,
MAL,
DRL,
RSL,
MAM,
CN,
DJN,
TTP,
SMR,
KS,
IHS,
JKS,
RS,
SJV,
and
WZ
each ran at least 10 sessions and/or 20 hours of observations for this project.
MFA,
KEB,
KC,
RSC,
RLC,
WF,
YG,
DH,
CJ,
KM,
BMXN,
JR,
and
MT
were undergraduate observing-team leaders, supervised by FC, TD, DLK, and JKS.
%
%
PTB, PBD, MED, MTL, JL, MAM, TTP, and KS
developed and refined procedures and computational tools for the timing pipeline.
%
%
%
HB,
PRB,
HTC,
MED,
PBD,
EF,
DCG,
MLJ,
MTL,
MAM,
DJN,
NSP,
TTP,
SMR,
IHS,
and
KS
generated and checked timing solutions for individual pulsars in \twelveyr, which became the starting point for the analyses performed here.
%
%
}

%% file: text/sec_low_snr.tex

Here we justify the empirically determined signal-to-noise ratio (S/N) thresholding applied to the wideband TOA data set described in Section~\ref{subsec:wb_cleaning} and listed in Table~\ref{tab:toa_flags}.
We follow an analogous analysis as Appendix~B of \nineyr\, in which a similar justification was made for excluding all narrowband TOAs with S/N~$ < 8$, a practice that continued in both \elevenyr\ and \twelveyr.
The main reason for excluding these data is that the TOA probability distribution function (PDF) at low S/N becomes very non-Gaussian, gains heavy tails, and approaches a uniform distribution as the noise level becomes comparable to the amplitude of the pulse.

Using the frequency-dependent notation introduced in Section~\ref{subsec:wb_toa_lnlike}, we define the ``per-channel'' S/N in a wideband TOA as
\begin{equation}
\label{eqn:nb_toa_snr}
S_n \equiv a_n T_{n} / \sigma_n,
\end{equation}
where $T_n \equiv \sqrt{\sum_k |t_{nk}|^2}$, $t_{nk}$ is the DFT of the template profile with frequency index $n$ and Fourier harmonic index $k$, $a_n$ is the scaling amplitude, and $\sigma_n$ is the noise level in $t_{nk}$.
This equation is the same form as the conventional, narrowband TOA S/N given in \nineyr\ for each channel's matched-template, but which is subject to the constraint in Equation~\ref{eqn:wb_constraint}.
With this definition in hand, we can immediately write down the joint PDF of a wideband TOA ($\phio$) and DM as a function of S/N, which will be independent of any particular, noisy realization of data.
This PDF has the same form as Equation~13 from \nineyr, but now contains a sum over frequency channels (indexed by $n$),
\begin{equation}
\label{eqn:wb_toa_pdf}
p(\phio, \dm) \propto \textrm{exp} \left (\sum_n \frac{S_n^2}{2} \frac{C_{tt,n}^2(\phi_n)}{T_n^4} \right ),
\end{equation}
where $C_{tt,n}$ is the template autocorrelation as a function of $\phio$ and $\dm$, via $\phi_n$ (Equation~\ref{eqn:wb_constraint}).
$C_{tt,n}$ is normalized such that $C_{tt,n}(\phi_n = 0) = T_n^2$, which explains our definition of the wideband TOA S/N as
\begin{equation}
\label{eqn:wb_toa_snr}
S \equiv \sqrt{\sum_n S_n^2}.
\end{equation}
This is the quantity paired with the wideband TOA flag \texttt{-snr} from Table~\ref{tab:toa_flags}.

We can see that Equation~\ref{eqn:wb_toa_snr} utilizes information that would otherwise be discarded on a per-channel basis by a narrowband TOA threshold of S/N~$< 8$, unless the original data were averaged down to fewer frequency channels to boost each channel's S/N.
As an extreme example, suppose a single subintegration has just enough signal distributed evenly over all frequency channels such that all of its narrowband TOAs woud be removed by thresholding unless all of the profiles are averaged together, thus acquiring a S/N~$> 8$.
By this averaging argument, we should therefore expect that the minimally informative wideband TOA has a ``typical'' per-channel $S_n < 8$, and that its PDF behaves similarly such that it becomes highly non-Gaussian around $S \sim 8$.

However, by the same token, Equations~\ref{eqn:wb_toa_snr}~\&~\ref{eqn:nb_toa_snr} tell us that $S$ will always be biased high, and this bias can be particularly high in the limit of very low S/N.
This is because the maximum-likelihood estimates of $a_n$ are noisy (in particular, they will never be exactly zero), they will be poorly estimated in the low S/N regime, and they enter Equation~\ref{eqn:wb_toa_snr} as $a_n^2$.
That is, even in the absence of any signal, $S$ is strictly positive.
Depending on the statistical properties of $a_n$ in the absence of a signal, $S$ may be more biased when the same bandwidth is divided up into a greater number of frequency channels.
Furthermore, in the low S/N regime, the presence of otherwise low-level RFI will further bias $S$ high.
It is for these reasons that, even though we recover the theoretical behavior of the wideband TOA PDF around $S \sim 8$ (see below), we implemented a larger threshold.
From visual inspection, many of the TOAs between $8 < S < 25$ appeared to be subject to these biases; after implementing the threshold that TOAs with $S < 25$ are cut, only a handful of additional TOAs needed to be manually culled, which were confirmed to be affected by subtle, broadband RFI.

Table~\ref{tab:ntoa_nchan} reveals that, overall, the wideband data set is including a lot of the profiles that were discarded in the narrowband analysis by its threshold of S/N~$< 8$; the number of profiles in the wideband data set is 16\% larger than the number of retained narrowband TOAs.
This hypothesis is supported by pointing out that there are 92,290 TOAs cut from the narrowband data set due to its S/N thresholding, whereas there are only 500 TOAs removed from the wideband data set due to its S/N thresholding.
Taking the number 30 as a representative number of frequency channels per wideband TOA on average, those low S/N wideband TOAs correspond to $\sim$~15,000 profiles.
Without considering any of the other curating described in Section~\ref{subsec:wb_cleaning}, then as a back-of-the-envelope estimate there are $\sim$~77,290 more profiles in the wideband data set that were discarded as TOAs by the narrowband S/N threshold; this is in line with the 16\% difference in Table~\ref{tab:ntoa_nchan}.
The S/N thresholding results in the largest single cut to the narrowband TOA data set by a very large margin (see Table~1 in \twelveyr), and the wideband processing has recovered $\sim$~80\% of these profiles.
It is important to point out, however, that because many of these profiles will have at most a S/N~$\sim 8$ (and will have zero signal in many cases), the impact they have is \textit{much} less than one would expect from having sixteen percent more of typical data.

\input{text/fig_wb_low_snr}

To verify the theoretical expectation for the expected uncertainties on the wideband measurements, we evaluated Equation~\ref{eqn:wb_toa_pdf} as a function of S/N for each of our evolving template models.
For each model, we constructed the frequency-dependent template using a central frequency, bandwidth, and number of channels typical of the receiver in question, for both generations of backend instruments.
Furthermore, we zero-weighted a typical number of random channels for that receiver, to better emulate the measurement uncertainties in the real data set.
For a given model and S/N, we calculated the second and fourth moments of the PDF, related to the variance and kurtosis, respectively, each for the TOA and DM by fixing the other parameter to the maximum-likelihood value.
Figure~\ref{fig:wb_low_snr} shows the results of this analysis for three pulsars that together cover all receiver bands: J1012$+$5307, J1903$+$0327, and J2317$+$1437.
In the figure, we plot R$_\sigma - 1$ and the excess kurtosis.
R$_\sigma$ is the ratio of the PDF's measured standard deviation to the expected 1$\sigma$ parameter uncertainty estimated from the Fisher information of Equation~\ref{eqn:toa_lnlike}, which assumes Gaussian statistics.
An idealized Gaussian-distributed variable will have R$_\sigma = 1$ and an excess kurtosis of zero.
It is evident that below S/N~$\sim$~8, the PDFs deviate largely from the Gaussian expectation, and we also indicate our wideband S/N threshold of 25 in the figure.
These conclusions are bolstered by the Monte Carlo analyses performed in \citet{PDR14}, which demonstrated that the uncertainties are properly estimated (assuming Gaussian noise) down to low S/N.


%% file: text/fig_wb_low_snr.tex
\begin{figure*}[htb!]
\begin{center}
\includegraphics[width=\columnwidth]{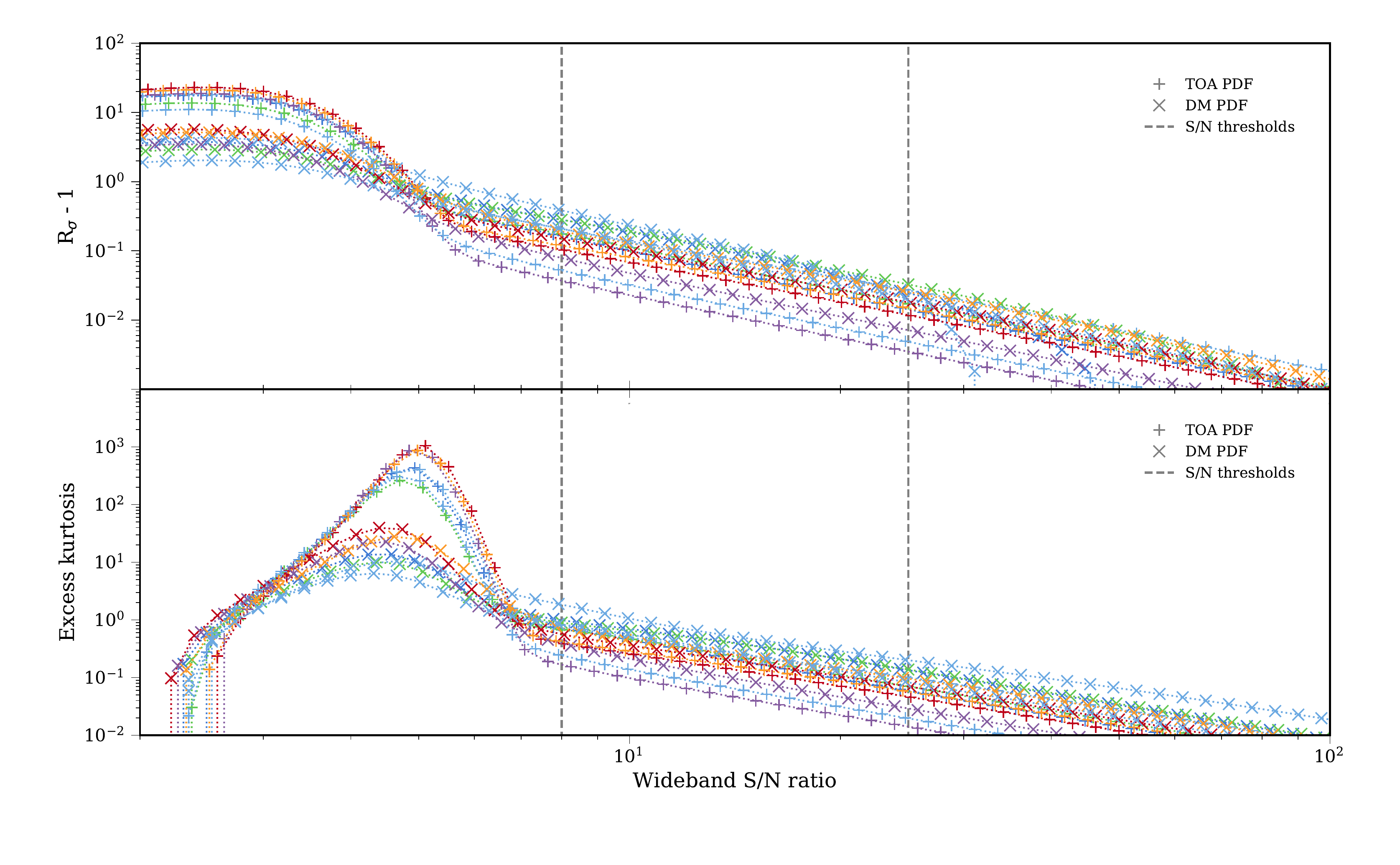}
\caption{\label{fig:wb_low_snr}
Behavior of maximum-likelihood wideband TOA and DM PDFs in the low S/N regime for three pulsars' evolving template models (J1012$+$5307, J1903$+$0327, and J2317$+$1437), which together cover all receivers in the data set.
The colors indicate the receiver as in Figure~\ref{fig:epochs}: 327~MHz (red), 430~MHz (orange), 820~MHz (green), 1.4~GHz (lighter blue for AO, darker blue for the GBT), 2.1~GHz (purple).
The PDFs were calculated to emulate data from PUPPI and GUPPI, but the counterpart ASP and GASP curves overlap almost identically.
The behavior of all other pulsars' profile evolution models is qualitatively the same.
The top panel plots the ratio R$_\sigma$ (minus one), where the ratio is the standard deviation of the evaluated PDF divided by the expected 1$\sigma$ parameter uncertainty.
The zero-covariance TOA reference frequency is used in all calculations (see the Appendix of \citet{PDR14}).
The bottom panel plots the excess kurtosis (i.e., kurtosis minus three), where the kurtosis for a normally distributed random variable is three.
The deviation from the Gaussian expectation (R$_\sigma$ = 1, excess kurtosis = 0) is large below S/N~$\sim$~8, and for reasons described in the text, we make a more conservative threshold at S/N~=~25.
}
\end{center}
\end{figure*}

%% file: text/sec_wb_likelihood.tex
%
In this section we derive our new wideband timing likelihood.
We borrow notation from Appendix~C of \nineyr, which also contains details that we omit here.
The wideband timing residuals $\delta\pmb{t}$ are modeled by deterministic and stochastic components via
\begin{equation}
\label{eqn:toa_resids}
\delta\pmb{t} = M\pmb{\epsilon} + F\pmb{a} + \pmb{n}.
\end{equation}
The product of the timing model design matrix $M$ with small offsets in the timing model parameters $\pmb{\epsilon}$ describes the systematic residuals from subtracting the timing model.
$F$ is the design matrix for the Fourier series decomposition of the red noise, with $\pmb{a}$ being the amplitudes of the Fourier basis functions.
Lastly, $\pmb{n}$ represents the noise remaining in the residuals, which is expected to be uncorrelated in both frequency and time.
This white noise is formally a Gaussian process with likelihood
\begin{equation}
\label{eqn:toa_wn_pdf}
p(\pmb{n}) = \frac{\textrm{exp} \left (-\frac{1}{2}\pmb{n}^TN^{-1}\pmb{n} \right )}{\sqrt{|2\pi N|}},
\end{equation}
where $N$ is an $N_{\textrm{TOA}} \times N_{\textrm{TOA}}$ diagonal data covariance matrix with entries
\begin{equation}
\label{eqn:N_matrix}
N_{ij} = (E_{k(i)}^2 \sigma_i^2 + Q_{k(i)}^2) \, \delta_{ij}.
\end{equation}
The TOA uncertainties $\sigma_i$ are modified by the EFAC and EQUAD parameters $E_k$ and $Q_k$, respectively\footnote{Equation~{\ref{eqn:N_matrix}} is the same formulation as found in \texttt{ENTERPRISE}; in \texttt{Tempo}, however, the EFAC parameter is applied \textit{after} the quadrature sum, meaning a conversion is necessary to obtain the corresponding \texttt{Tempo} EQUAD parameter, $Q_{k(i),\texttt{Tempo}} = Q_{k(i)} / E_{k(i)}$.}.
The label function $k$ indicates the observing system combination of frontend receiver and backend data acquisition instrument for the observation with index $i$ (or $j$), and $\delta_{ij}$ is the Kronecker delta.
As mentioned in Section~\ref{subsec:timing}, because of the nature of wideband TOAs, we do not separately model jitter-like white noise that is completely correlated across simultaneously measured multi-frequency TOAs and completely uncorrelated between epochs (i.e., there are no ECORR parameters).
Any physical effects that would contribute to ECORR in the narrowband analyses will be completely absorbed by EQUAD and the red noise in the wideband analysis.
The red noise is also modeled as a Gaussian process specified by $2n_{\textrm{mode}}$ Fourier basis vectors (the columns of $F$) and by the prior on the $2n_{\textrm{mode}}$ coefficients $\pmb{a}$ (i.e., weights):
\begin{equation}
\label{eqn:a_prior}
p(\pmb{a} \given A_\textrm{red}, \gamma_\textrm{red}) = \frac{\textrm{exp} \left (-\frac{1}{2}\pmb{a}^T\varphi^{-1}\pmb{a} \right )}{\sqrt{|2\pi\varphi|}},
\end{equation}
where $\varphi$ is a $2n_{\textrm{mode}} \times 2n_{\textrm{mode}}$ diagonal matrix with entries $T^{-1}_\textrm{span} P(f_m)$.
Here, $T_\textrm{span}$ is the span of the data set, $P(f_m)$ is the power-law function of Eq.\ \ref{eqn:red_pl}, and $f_m$ are the $n_{\textrm{mode}}$ frequencies of the Fourier components indexed by $m$.

We also model the timing model corrections as a Gaussian process.
We refer to the subset of the timing model offsets $\pmb{\epsilon}$ that describe the piecewise-constant DMX model as $\pmb{\epsilon}^\textrm{\tiny DMX}$.
Except for $\pmb{\epsilon}^\textrm{\tiny DMX}$, the remaining timing model offsets are given uninformative priors; formally, these are zero-mean Gaussian distributions with very large variances.
The novel development we make here is to use the wideband DM measurements to provide a prior for $\pmb{\epsilon}^\textrm{\tiny DMX}$.
Assuming that the $\delta \pmb{t}$ are computed with respect to a fiducial DM that is constant in the data set, and that $\delta \pmb{D}$ is the vector of DM measurements relative to the fiducial DM, then the prior for $\pmb{\epsilon}^\textrm{\tiny DMX}$ can be written as
\begin{equation}
\label{eqn:DMX_prior}
p(\pmb{\epsilon}^{\textrm{\tiny DMX}} \given \delta\pmb{D},\pmb{J}^{\textrm{\tiny DM}},\pmb{E}^{\textrm{\tiny DM}}) =
\frac{\exp
-\frac{1}{2}\left(
(\pmb{\epsilon}^\textrm{\tiny DMX} - \delta \pmb{D} - \pmb{J}^\textrm{\tiny DM})^T
N^{\textrm{\tiny DM}^{-1}}
(\pmb{\epsilon}^\textrm{\tiny DMX} - \delta \pmb{D} - \pmb{J}^\textrm{\tiny DM})
\right)}{\sqrt{|2\pi N^{\textrm{\tiny DM}}|}},
\end{equation}
where $N^{\textrm{\tiny DM}}$ is an $n_{\textrm{DM}} \times n_{\textrm{DM}}~( = n_{\textrm{TOA}} \times n_{\textrm{TOA}})$ diagonal covariance matrix containing the DM measurement errors $\sigma^\textrm{\tiny DM}_i$ scaled by DMEFAC parameters $\pmb{E}^\textrm{\tiny DM} \equiv E^\textrm{\tiny DM}_{k}$,
\begin{equation}
\label{eqn:N_matrix_dm}
N^\textrm{\tiny DM}_{ij} = (E^\textrm{\tiny DM}_{k(i)} \sigma^\textrm{\tiny DM}_i)^2 \, \delta_{ij},
\end{equation}
and where $\pmb{J}^{\textrm{\tiny DM}} \equiv J^\textrm{\tiny DM}_{r(i)}$ are the DMJUMP parameters described in Sections~\ref{subsec:prof_evol}~\&~\ref{subsec:timing}, labeled by receiver $r$.
Altogether, our model of the measurement yields the posterior
\begin{equation}
\label{eqn:wb_post}
p(\pmb{\epsilon},\pmb{a},\pmb{\phi},\pmb{J}^{\textrm{\tiny DM}},\pmb{E}^{\textrm{\tiny DM}} \given \delta\pmb{t},\delta\pmb{D})
\propto
p(\pmb{a} \given A_\textrm{red}, \gamma_\textrm{red})
\times
p(\pmb{\epsilon}^{\textrm{\tiny DMX}} \given \delta\pmb{D},\pmb{J}^{\textrm{\tiny DM}},\pmb{E}^{\textrm{\tiny DM}})
\times
p(\delta\pmb{t} \given \pmb{\epsilon},\pmb{a},\pmb{\phi}),
\end{equation}
where the last term on the right is the usual likelihood for the timing residuals,
\begin{equation}
\label{eqn:wb_like}
p(\delta\pmb{t} \given \pmb{\epsilon},\pmb{a},\pmb{\phi}) = \frac{\textrm{exp} \left (-\frac{1}{2}\pmb{r}^TN^{-1}\pmb{r} \right )}{\sqrt{|2\pi N|}},
\end{equation}
the EFAC and EQUAD parameters comprise the vector $\pmb{\phi}$, and $\pmb{r} = \delta\pmb{t} - M\pmb{\epsilon} - F\pmb{a}$.
The marginalization of the posterior over the timing model parameters proceeds in the same way as described in \nineyr.
Note that the two separate data covariance matrices $N$ and $N^{\textrm{\tiny DM}}$ imply zero covariance between the TOA and DM measurements; as mentioned in Section~\ref{subsec:wb_toa_lnlike}, all wideband TOAs reference a frequency such that the measurement has an estimated zero covariance with its associated DM.
Another way to look at Equation~\ref{eqn:wb_post} is to see the DM measurements as data in a joint likelihood with the wideband TOAs instead of as prior information on the DMX parameters; in either case the formulation will be the same.

We implemented the wideband posterior in the PTA analysis package \texttt{ENTERPRISE}, with which we performed the analyses described in Section~\ref{subsec:timing}.
In \texttt{ENTERPRISE}, the functionality is accessed by using the signal class \texttt{WidebandTimingModel} with a \texttt{Pulsar} object that has wideband TOAs and DM measurements.
Similarly, we implemented the DMX prior in Equation~\ref{eqn:DMX_prior}, which can also be viewed as an additional likelihood component, in the pulsar timing software \texttt{Tempo}.
\texttt{Tempo}'s generalized least squares fit must be used in order to enable the new functionality with wideband TOAs, and the input timing model parameter file must contain the line \texttt{DMDATA 1}, which we have included in the released files.
As a result of these developments, the wideband timing models in this data set come with DMEFAC and DMJUMP parameters, along with the usual EFAC, EQUAD, and red noise parameters.
Our formalism of using the wideband DM measurements currently only works with the DMX model for DM variations.
Extending this to a stochastic, Gaussian process model of DM variations \citep[e.g.,][]{Lentati13} is currently under development (Simon et al. \textit{in prep.}).

%
%
%
%
%

%% file: text/sec_resid.tex

Here we include an appendix of timing residuals and DM variations for each pulsar in our data set, as a complement to the similar appendix in \twelveyr.
Measurements from both wideband and narrowband data sets are plotted for visual comparison.
The predominant backend instrument for a given time period is printed at the top of each plot, with vertical dashed lines indicating times at which the instruments changed.
The color-coding of the timing residuals and wideband DM measurements indicates the receiver used for that observation and is the same as in Figure~\ref{fig:epochs}: 327~MHz (red), 430~MHz (orange), 820~MHz (green), 1.4~GHz (lighter blue for AO, darker blue for the GBT), 2.1~GHz (purple).

\textit{DM variations.}
The top two panels of each figure show the variation in DM for each pulsar.
The black circles in the topmost panel are the DMX model parameters (see Section~\ref{subsec:timing}) and the grey squares are the corresponding DMX parameters modeled in \twelveyr, which may not be visible due to the agreement with their wideband counterparts.
The panel second from the top shows the DM measurements that are paired with each wideband TOA; these measurements are adjusted based on the MAP DMEFAC and DMJUMP parameters (see Appendix~\ref{sec:wb_like}).
The average DM value has been removed from all three time series, and the two panels have the same scale, determined by the maximum DMX deviation in the narrowband data set.
As a result, some of the wideband DM measurements in the second panel fall outside the plotted range (particularly those from the ASP and GASP era), depending on how informative the measurements are to the DMX model.
A number of pulsars show a DM variation that may be a function of frequency; see Section~\ref{subsec:DM_nu} for further discussion.

\textit{Timing residuals.}
The remaining panels contain timing residuals, which are the observed TOAs minus the predicted arrival time from the timing model (see Section~\ref{subsec:timing}).
All residual uncertainties include the white noise model components (i.e., the MAP EFAC and EQUAD parameters; see Appendix~\ref{sec:wb_like}).
Linear and quadratic trends have been subtracted from the plotted timing residuals, as they are completely covariant with the pulsar's rotation frequency and frequency derivative in the timing model, respectively, and hence would be absorbed by these parameters.
The first panel after the DM panels show the averaged \textit{narrowband} timing residuals, which is the same as panel (d) from the analogous residual plot in \twelveyr.
Where red noise is significant in \textit{both} data sets (see Table~\ref{tab:timing_model_summary} and Figure~\ref{fig:rn_compare}), the panel second from the bottom contains the whitened averaged \textit{narrowband} timing residuals, which is the same as panel (e) from \twelveyr.
The panels just below each of these two panels show their \textit{wideband} counterparts, the results of the present work.
The limits in each timing residual panel mirror those from \twelveyr, which permit a useful visual comparison between the data sets.
However, a small number of residuals ($\sim$0.5\%) fall outside of the limits; the averaged narrowband residuals are also plotted in \twelveyr\ such that all data points are visible.
Overall, the agreement in timing residuals is remarkable, though see Sections~\ref{subsec:timing_results}~\&~\ref{subsec:pulsar_results} for further discussion.

%% file: text/fig_summary.tex
\begin{figure*}[p]
\centering
\includegraphics[scale=0.8]{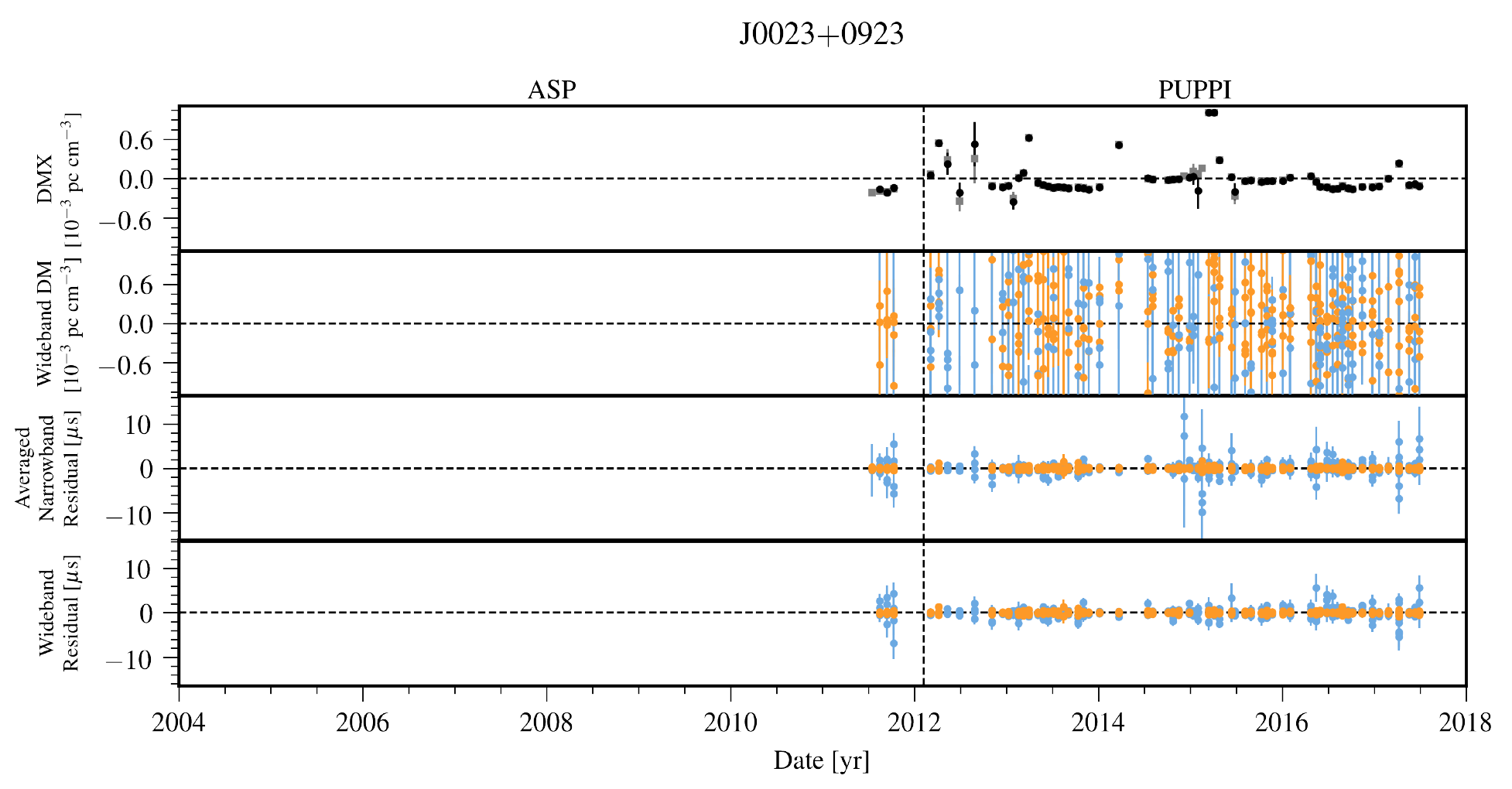}
\caption{Timing residuals and DM variations for J0023$+$0923.  See Appendix~\ref{sec:resid} for details.  DMX model parameters from the wideband (black circles) and narrowband (grey squares) data sets are shown in the top panel.  Colors in the lower panels indicate the receiver for the observation: 430~MHz (Orange) and 1.4~GHz (Light blue).}
\label{fig:summary-J0023+0923}
\end{figure*}

\begin{figure*}[p]
\centering
\includegraphics[scale=0.8]{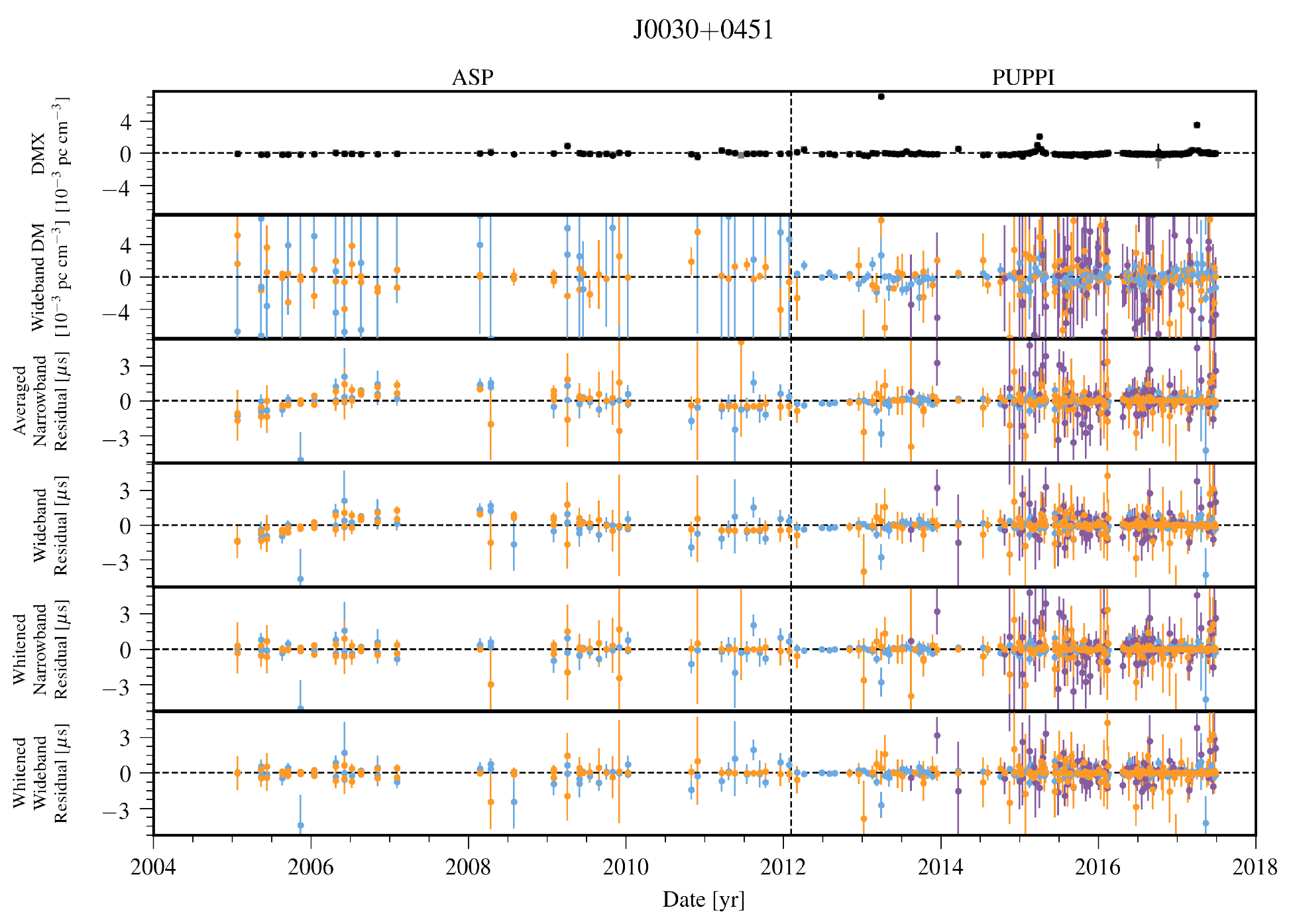}
\caption{Timing residuals and DM variations for J0030$+$0451.  See Appendix~\ref{sec:resid} for details.  DMX model parameters from the wideband (black circles) and narrowband (grey squares) data sets are shown in the top panel.  Colors in the lower panels indicate the receiver for the observation: 430~MHz (Orange), 1.4~GHz (Light blue), and 2.1~GHz (Purple).}
\label{fig:summary-J0030+0451}
\end{figure*}

\begin{figure*}[p]
\centering
\includegraphics[scale=0.8]{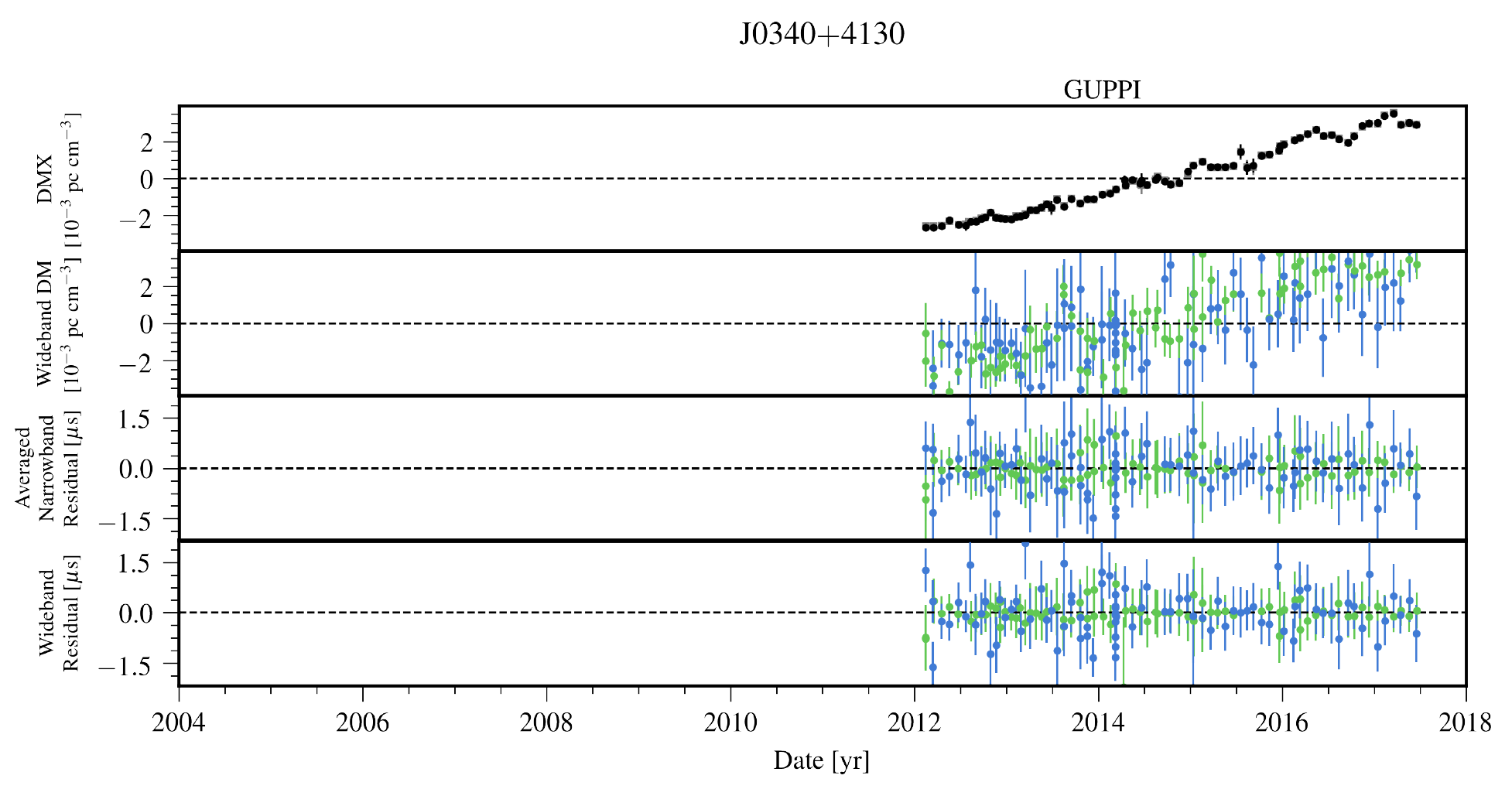}
\caption{Timing residuals and DM variations for J0340$+$4130.  See Appendix~\ref{sec:resid} for details.  DMX model parameters from the wideband (black circles) and narrowband (grey squares) data sets are shown in the top panel.  Colors in the lower panels indicate the receiver for the observation: 820~MHz (Green) and 1.4~GHz (Dark blue).}
\label{fig:summary-J0340+4130}
\end{figure*}

\begin{figure*}[p]
\centering
\includegraphics[scale=0.8]{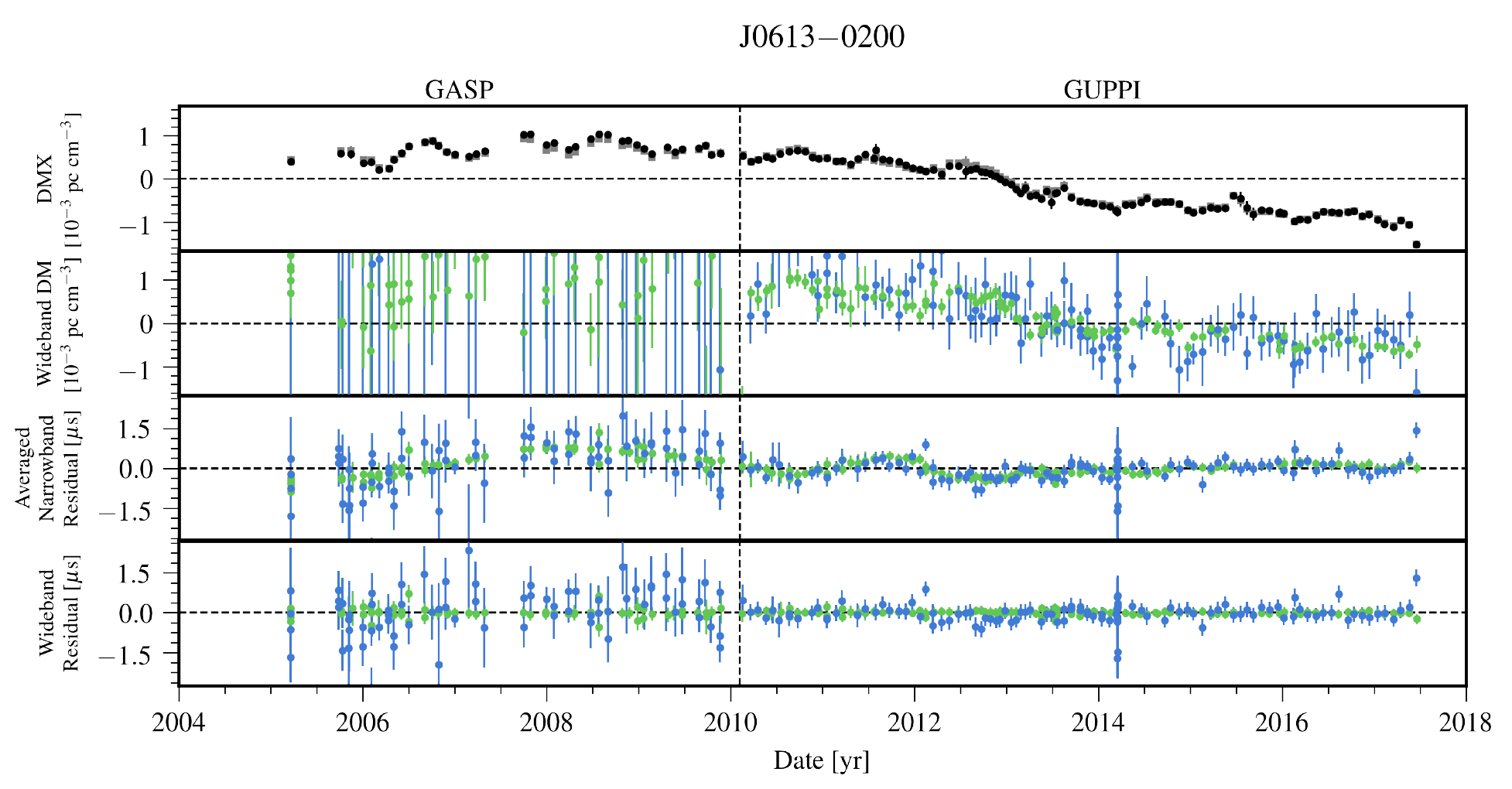}
\caption{Timing residuals and DM variations for J0613$-$0200.  See Appendix~\ref{sec:resid} for details.  DMX model parameters from the wideband (black circles) and narrowband (grey squares) data sets are shown in the top panel.  Colors in the lower panels indicate the receiver for the observation: 820~MHz (Green) and 1.4~GHz (Dark blue).}
\label{fig:summary-J0613-0200}
\end{figure*}

\begin{figure*}[p]
\centering
\includegraphics[scale=0.8]{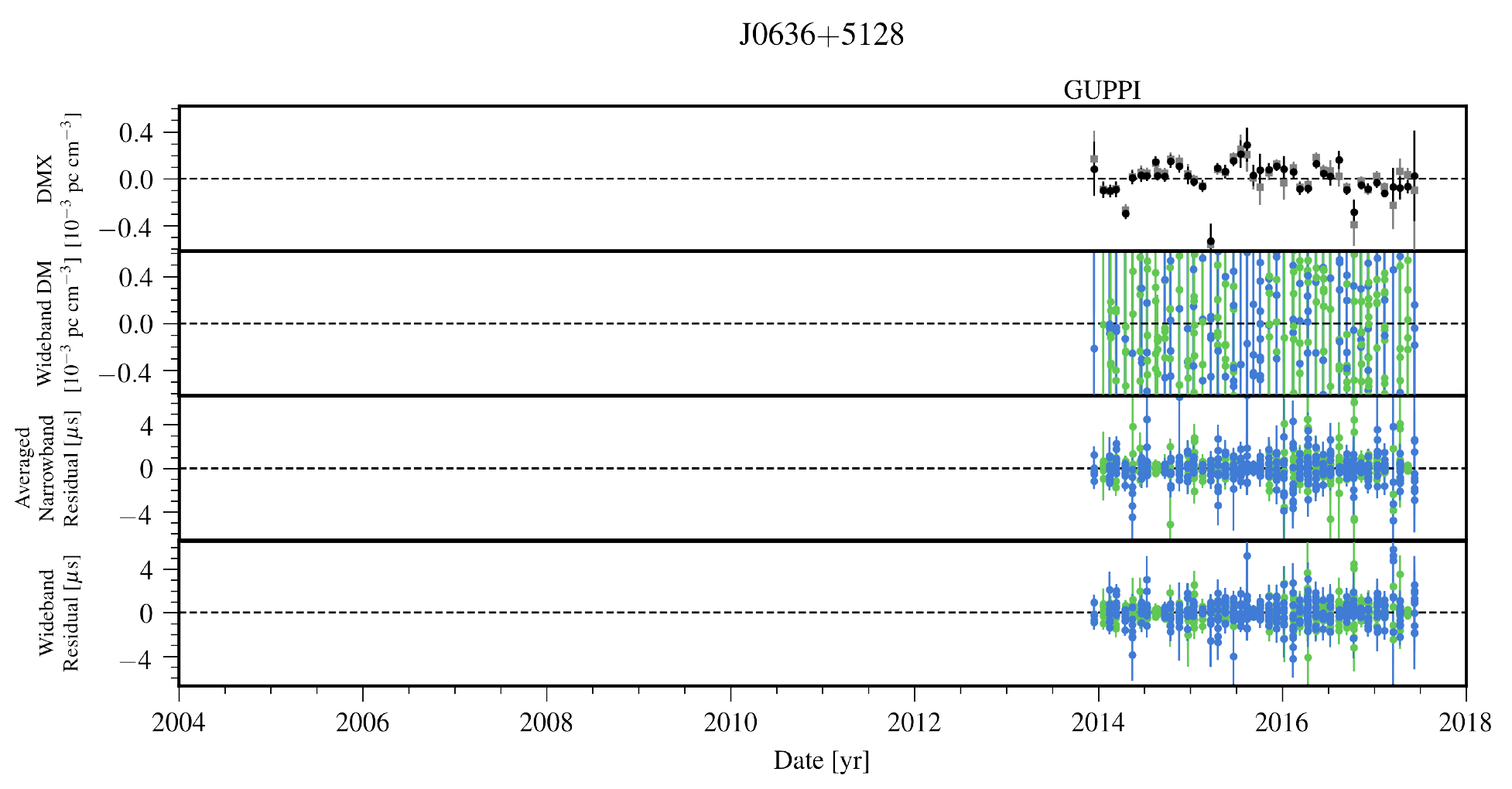}
\caption{Timing residuals and DM variations for J0636$+$5128.  See Appendix~\ref{sec:resid} for details.  DMX model parameters from the wideband (black circles) and narrowband (grey squares) data sets are shown in the top panel.  Colors in the lower panels indicate the receiver for the observation: 820~MHz (Green) and 1.4~GHz (Dark blue).}
\label{fig:summary-J0636+5128}
\end{figure*}

\begin{figure*}[p]
\centering
\includegraphics[scale=0.8]{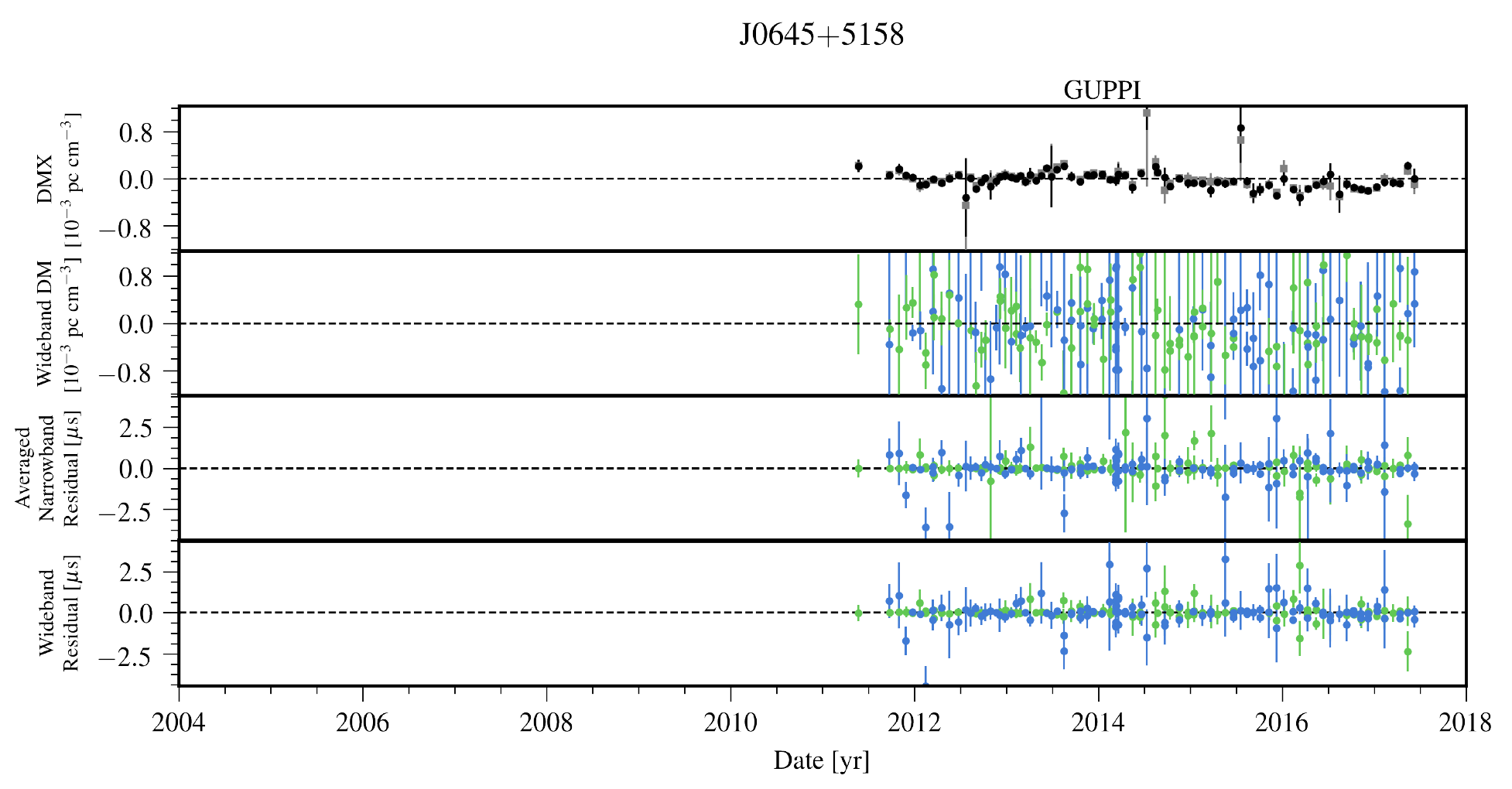}
\caption{Timing residuals and DM variations for J0645$+$5158.  See Appendix~\ref{sec:resid} for details.  DMX model parameters from the wideband (black circles) and narrowband (grey squares) data sets are shown in the top panel.  Colors in the lower panels indicate the receiver for the observation: 820~MHz (Green) and 1.4~GHz (Dark blue).}
\label{fig:summary-J0645+5158}
\end{figure*}

\begin{figure*}[p]
\centering
\includegraphics[scale=0.8]{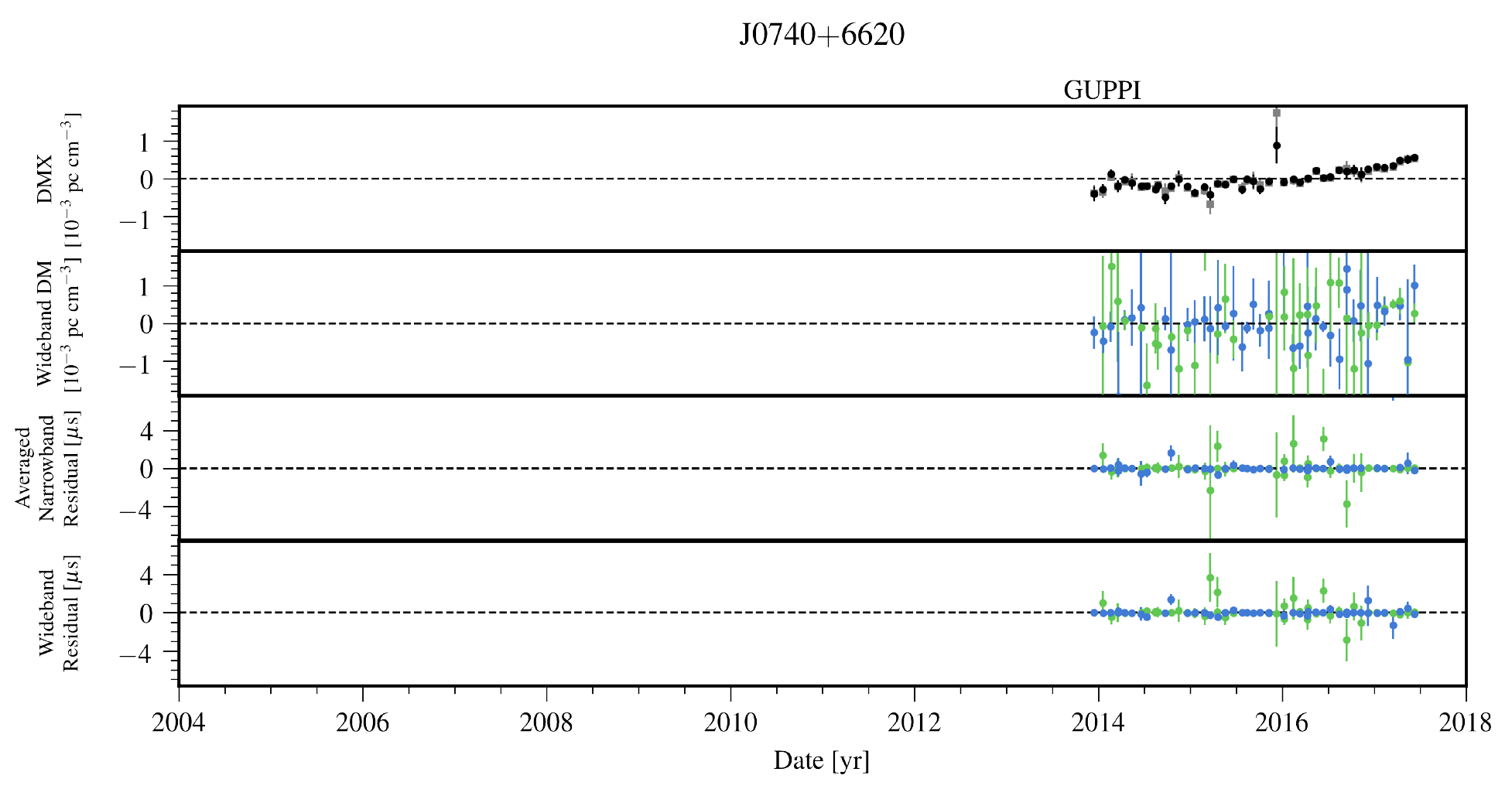}
\caption{Timing residuals and DM variations for J0740$+$6620.  See Appendix~\ref{sec:resid} for details.  DMX model parameters from the wideband (black circles) and narrowband (grey squares) data sets are shown in the top panel.  Colors in the lower panels indicate the receiver for the observation: 820~MHz (Green) and 1.4~GHz (Dark blue).}
\label{fig:summary-J0740+6620}
\end{figure*}

\begin{figure*}[p]
\centering
\includegraphics[scale=0.8]{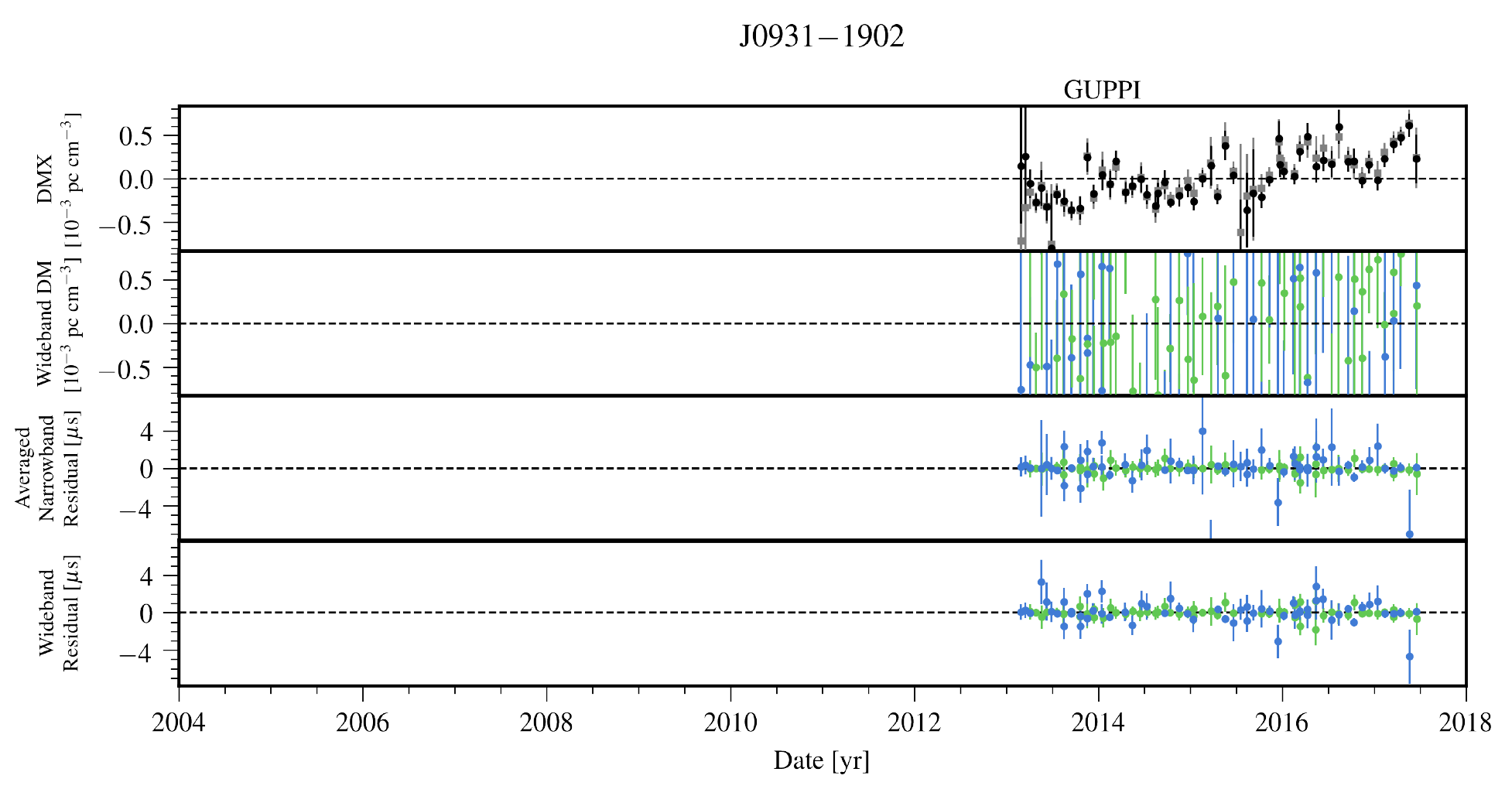}
\caption{Timing residuals and DM variations for J0931$-$1902.  See Appendix~\ref{sec:resid} for details.  DMX model parameters from the wideband (black circles) and narrowband (grey squares) data sets are shown in the top panel.  Colors in the lower panels indicate the receiver for the observation: 820~MHz (Green) and 1.4~GHz (Dark blue).}
\label{fig:summary-J0931-1902}
\end{figure*}

\begin{figure*}[p]
\centering
\includegraphics[scale=0.8]{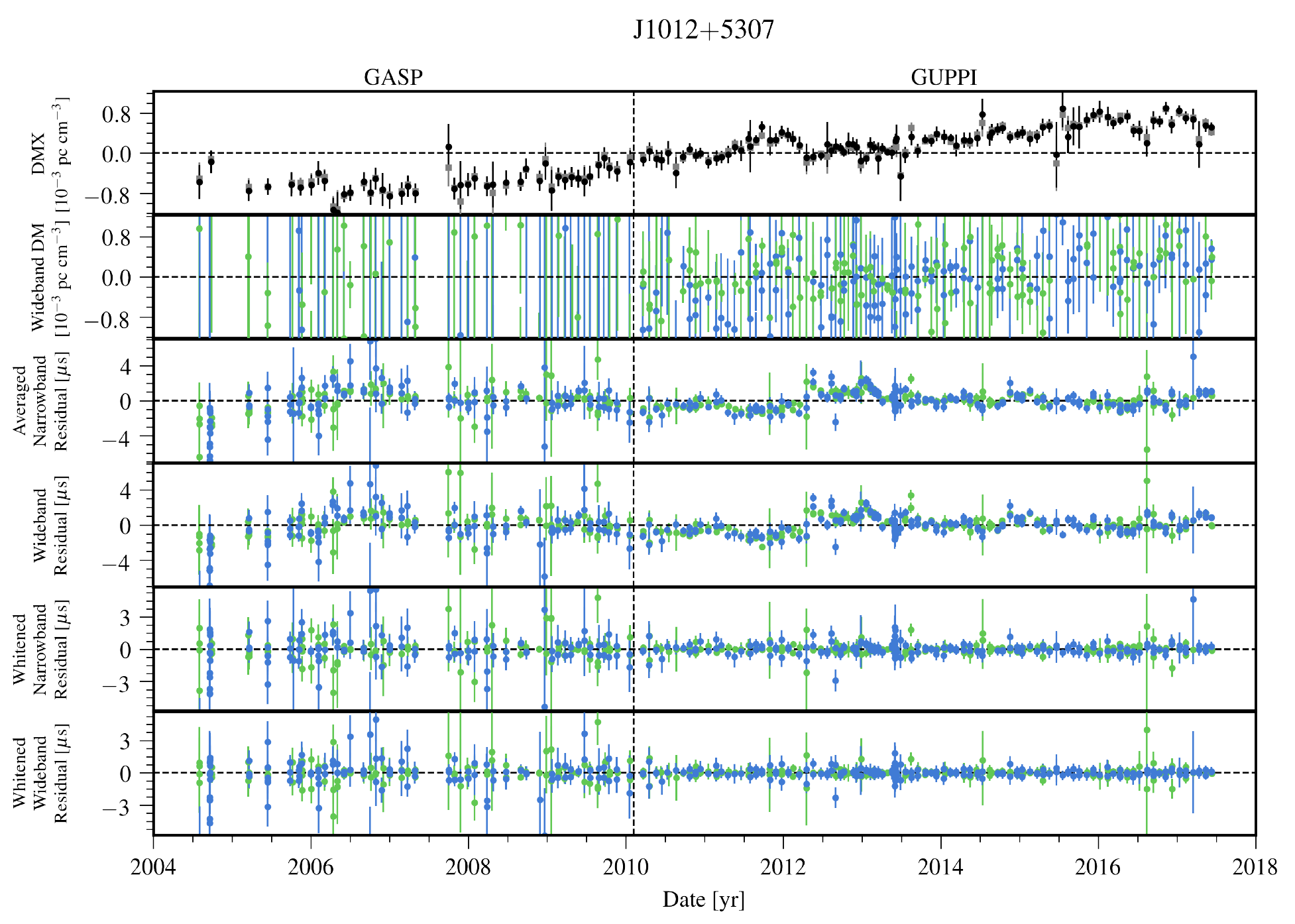}
\caption{Timing residuals and DM variations for J1012$+$5307.  See Appendix~\ref{sec:resid} for details.  DMX model parameters from the wideband (black circles) and narrowband (grey squares) data sets are shown in the top panel.  Colors in the lower panels indicate the receiver for the observation: 820~MHz (Green) and 1.4~GHz (Dark blue).}
\label{fig:summary-J1012+5307}
\end{figure*}

\begin{figure*}[p]
\centering
\includegraphics[scale=0.8]{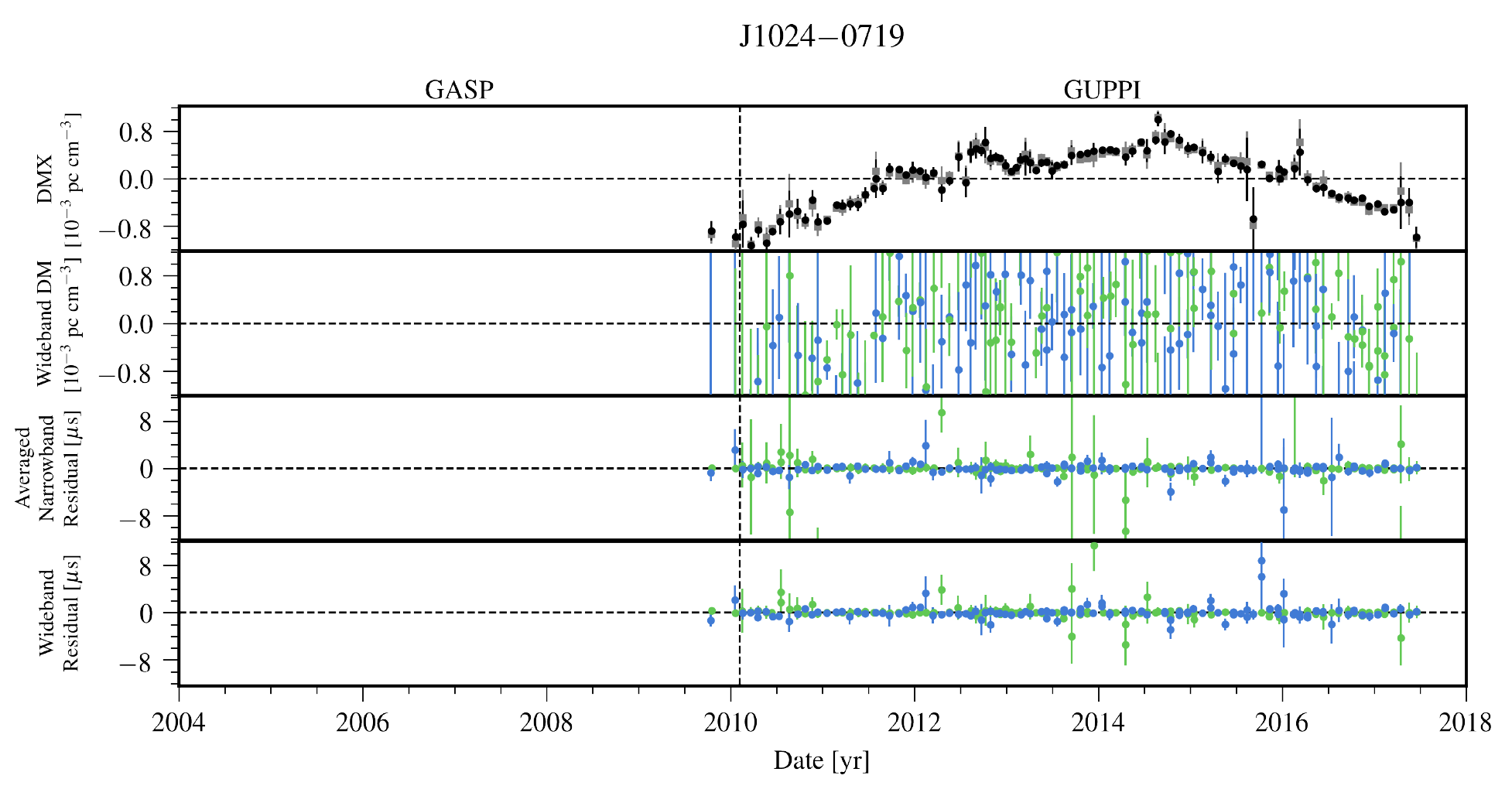}
\caption{Timing residuals and DM variations for J1024$-$0719.  See Appendix~\ref{sec:resid} for details.  DMX model parameters from the wideband (black circles) and narrowband (grey squares) data sets are shown in the top panel.  Colors in the lower panels indicate the receiver for the observation: 820~MHz (Green) and 1.4~GHz (Dark blue).}
\label{fig:summary-J1024-0719}
\end{figure*}

\begin{figure*}[p]
\centering
\includegraphics[scale=0.8]{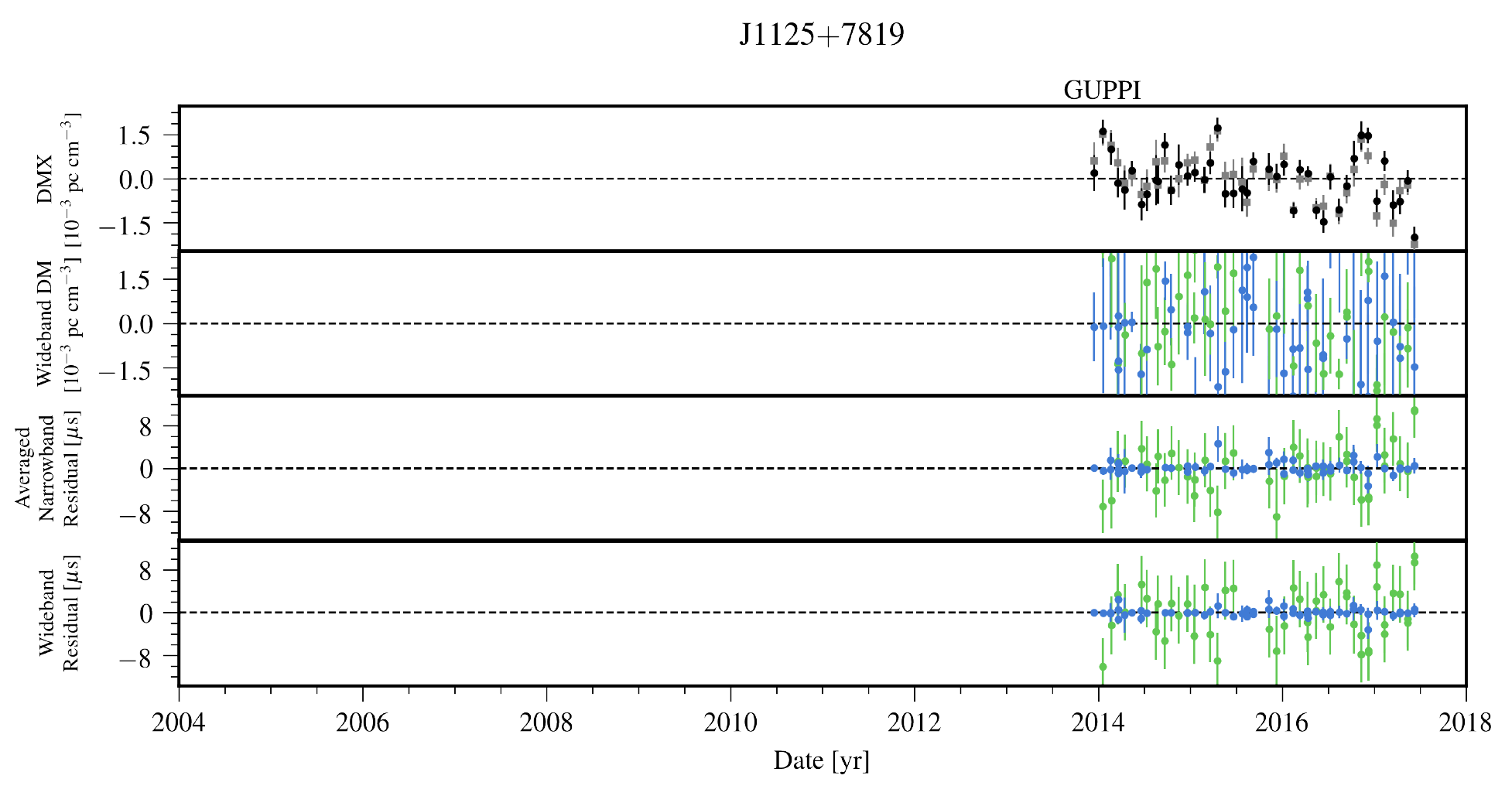}
\caption{Timing residuals and DM variations for J1125$+$7819.  See Appendix~\ref{sec:resid} for details.  DMX model parameters from the wideband (black circles) and narrowband (grey squares) data sets are shown in the top panel.  Colors in the lower panels indicate the receiver for the observation: 820~MHz (Green) and 1.4~GHz (Dark blue).}
\label{fig:summary-J1125+7819}
\end{figure*}

\begin{figure*}[p]
\centering
\includegraphics[scale=0.8]{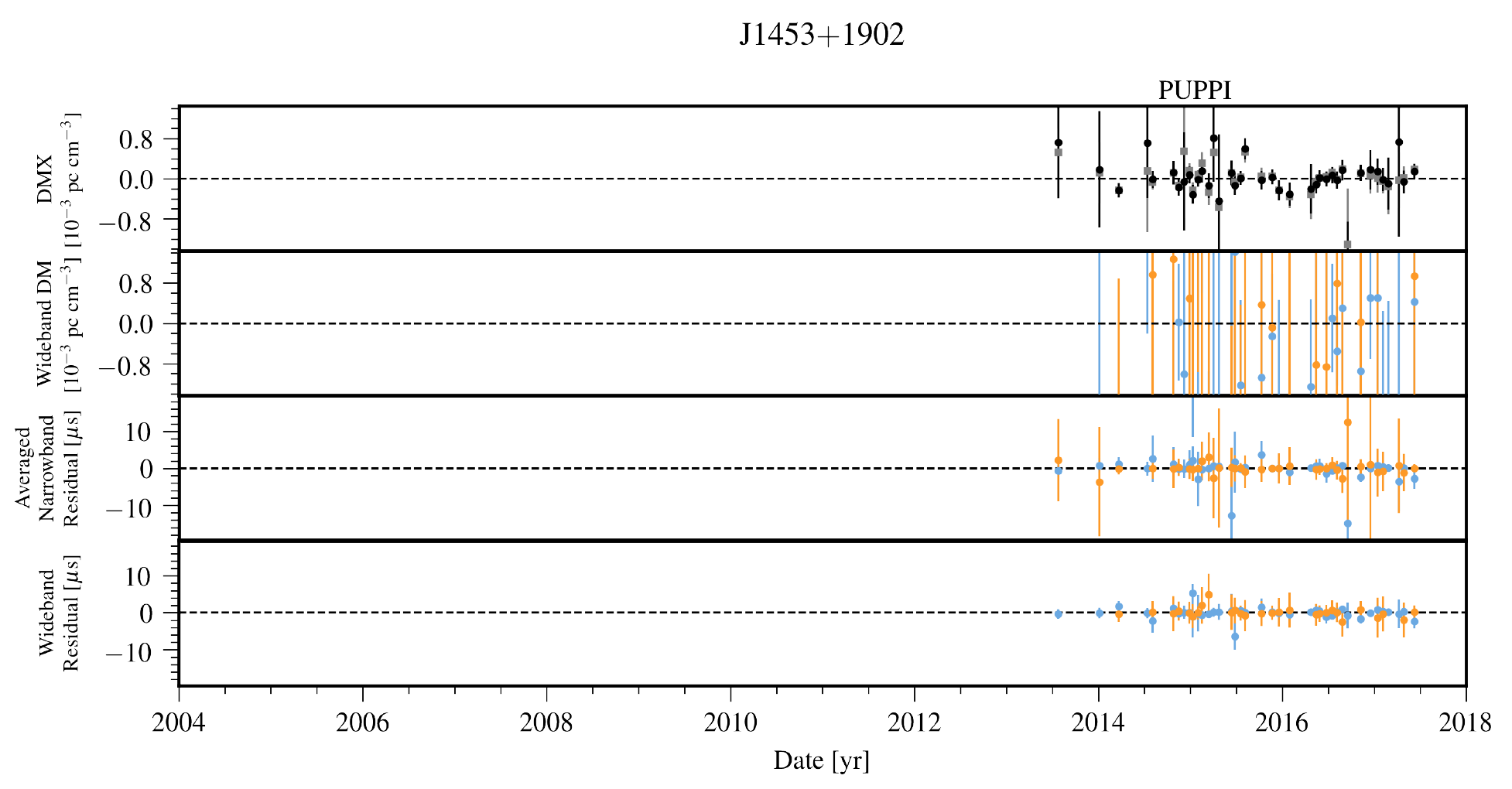}
\caption{Timing residuals and DM variations for J1453$+$1902.  See Appendix~\ref{sec:resid} for details.  DMX model parameters from the wideband (black circles) and narrowband (grey squares) data sets are shown in the top panel.  Colors in the lower panels indicate the receiver for the observation: 430~MHz (Orange) and 1.4~GHz (Light blue).}
\label{fig:summary-J1453+1902}
\end{figure*}

\begin{figure*}[p]
\centering
\includegraphics[scale=0.8]{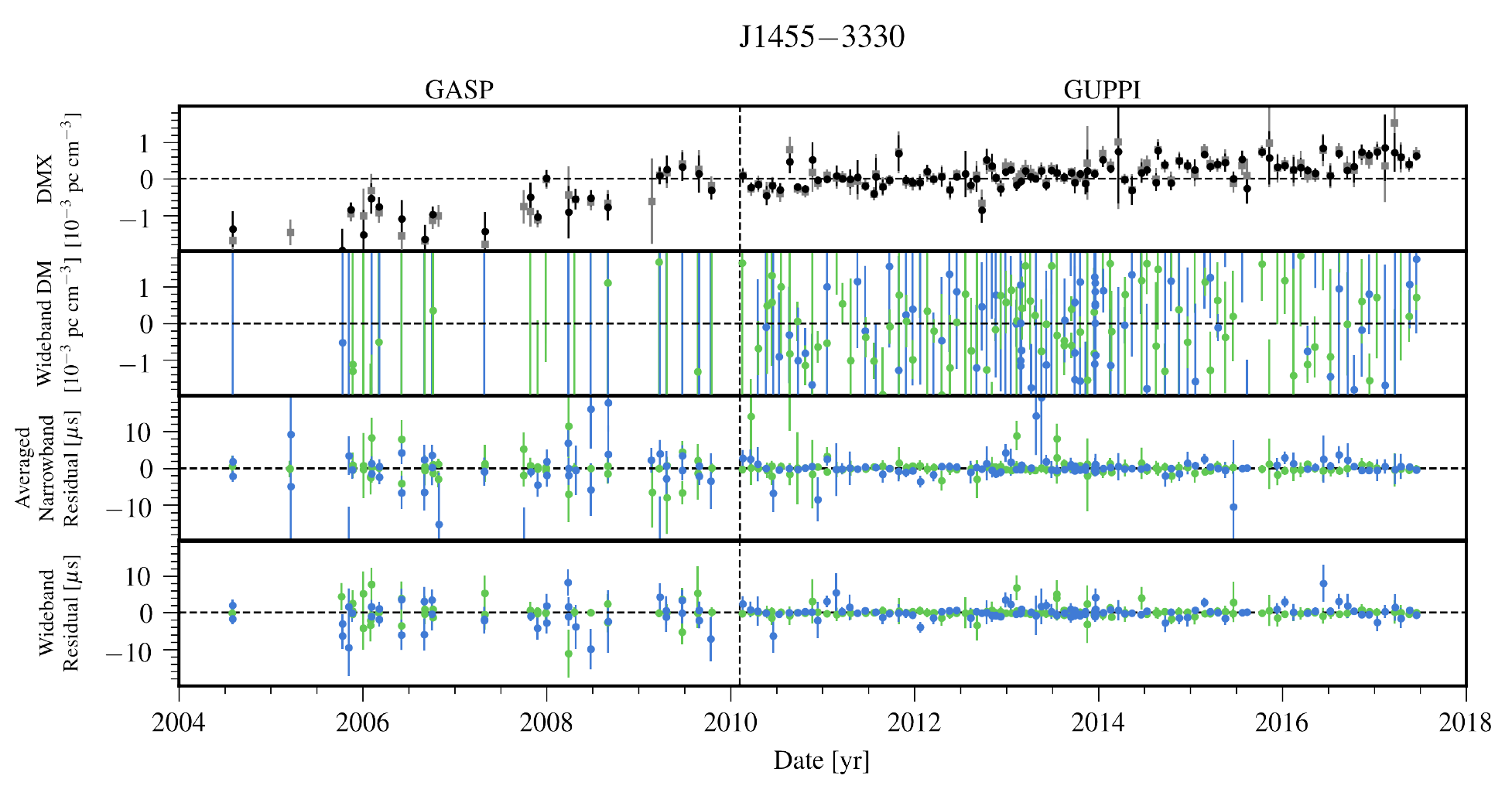}
\caption{Timing residuals and DM variations for J1455$-$3330.  See Appendix~\ref{sec:resid} for details.  DMX model parameters from the wideband (black circles) and narrowband (grey squares) data sets are shown in the top panel.  Colors in the lower panels indicate the receiver for the observation: 820~MHz (Green) and 1.4~GHz (Dark blue).}
\label{fig:summary-J1455-3330}
\end{figure*}

\begin{figure*}[p]
\centering
\includegraphics[scale=0.8]{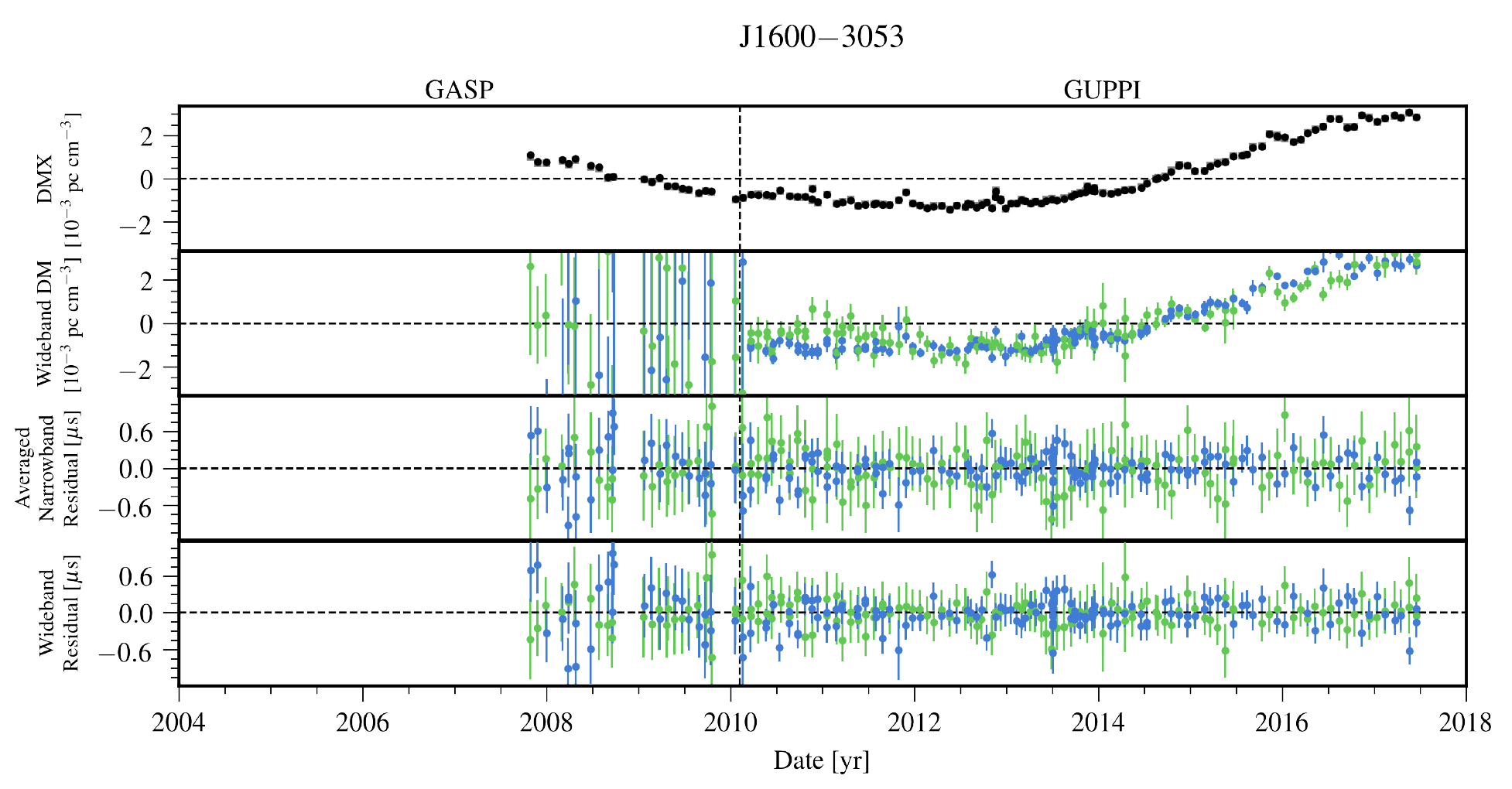}
\caption{Timing residuals and DM variations for J1600$-$3053.  See Appendix~\ref{sec:resid} for details.  DMX model parameters from the wideband (black circles) and narrowband (grey squares) data sets are shown in the top panel.  Colors in the lower panels indicate the receiver for the observation: 820~MHz (Green) and 1.4~GHz (Dark blue).}
\label{fig:summary-J1600-3053}
\end{figure*}

\begin{figure*}[p]
\centering
\includegraphics[scale=0.8]{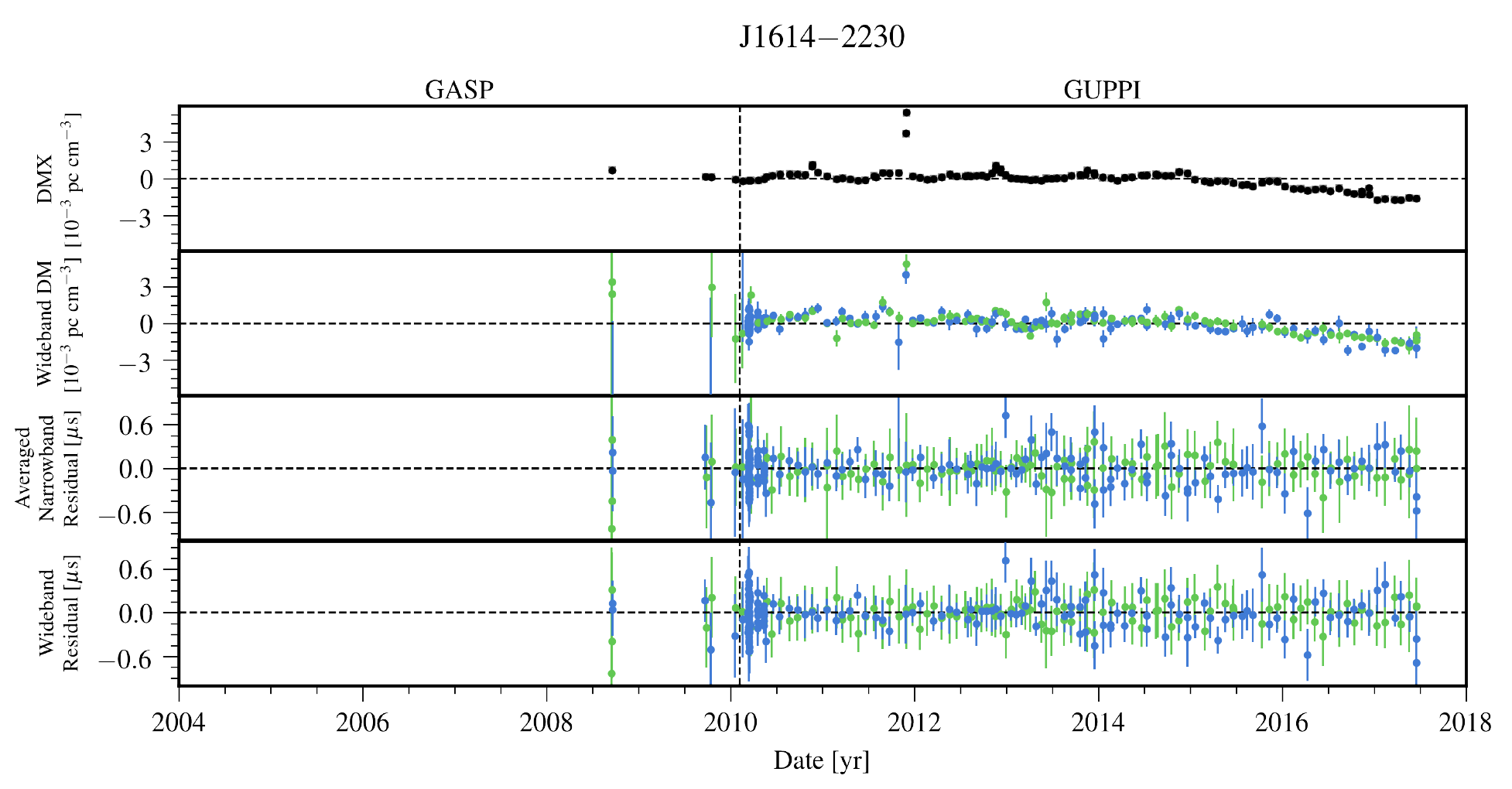}
\caption{Timing residuals and DM variations for J1614$-$2230.  See Appendix~\ref{sec:resid} for details.  DMX model parameters from the wideband (black circles) and narrowband (grey squares) data sets are shown in the top panel.  Colors in the lower panels indicate the receiver for the observation: 820~MHz (Green) and 1.4~GHz (Dark blue).}
\label{fig:summary-J1614-2230}
\end{figure*}

\begin{figure*}[p]
\centering
\includegraphics[scale=0.8]{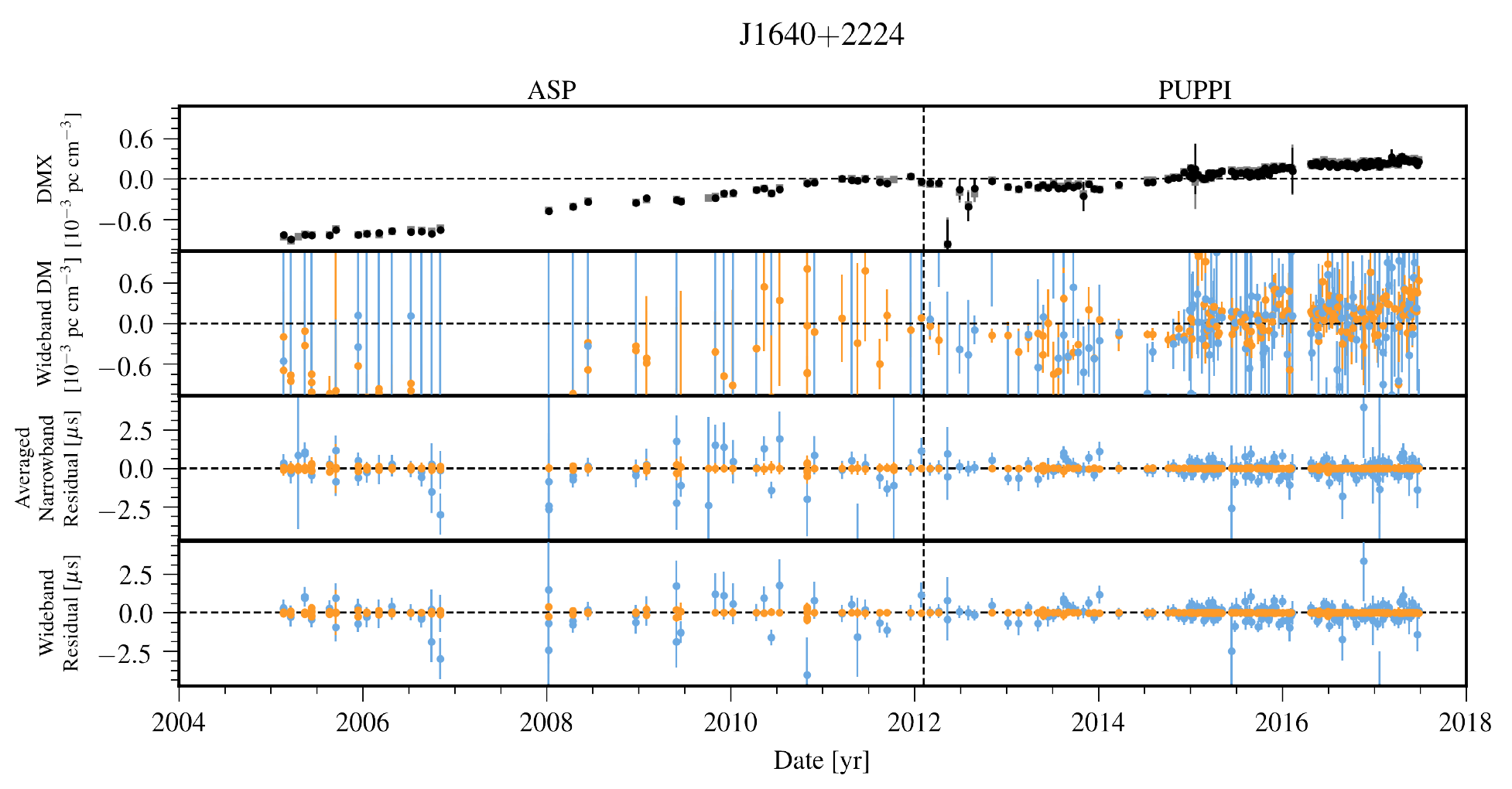}
\caption{Timing residuals and DM variations for J1640$+$2224.  See Appendix~\ref{sec:resid} for details.  DMX model parameters from the wideband (black circles) and narrowband (grey squares) data sets are shown in the top panel.  Colors in the lower panels indicate the receiver for the observation: 430~MHz (Orange) and 1.4~GHz (Light blue).}
\label{fig:summary-J1640+2224}
\end{figure*}

\begin{figure*}[p]
\centering
\includegraphics[scale=0.8]{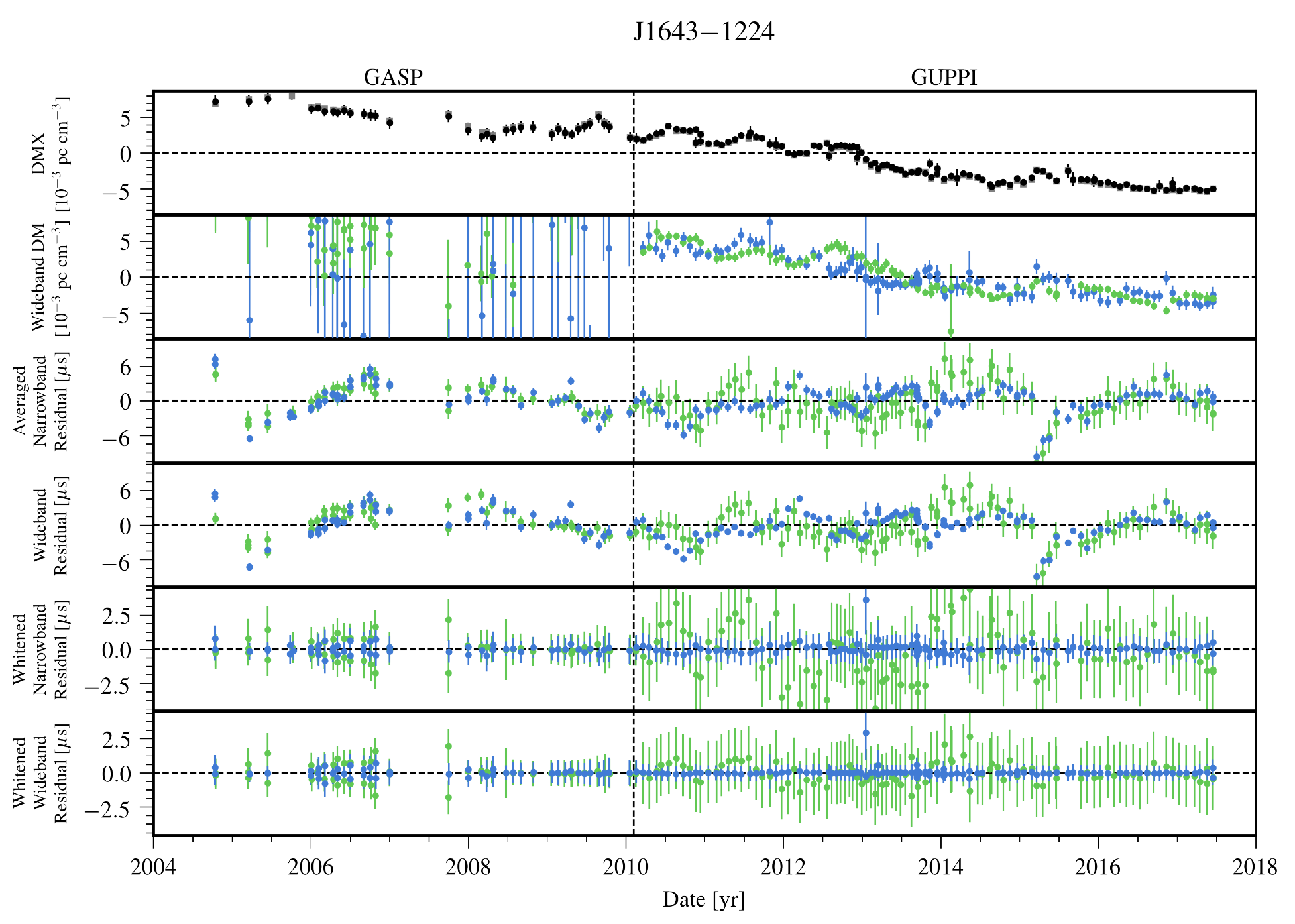}
\caption{Timing residuals and DM variations for J1643$-$1224.  See Appendix~\ref{sec:resid} for details.  DMX model parameters from the wideband (black circles) and narrowband (grey squares) data sets are shown in the top panel.  Colors in the lower panels indicate the receiver for the observation: 820~MHz (Green) and 1.4~GHz (Dark blue).}
\label{fig:summary-J1643-1224}
\end{figure*}

\begin{figure*}[p]
\centering
\includegraphics[scale=0.8]{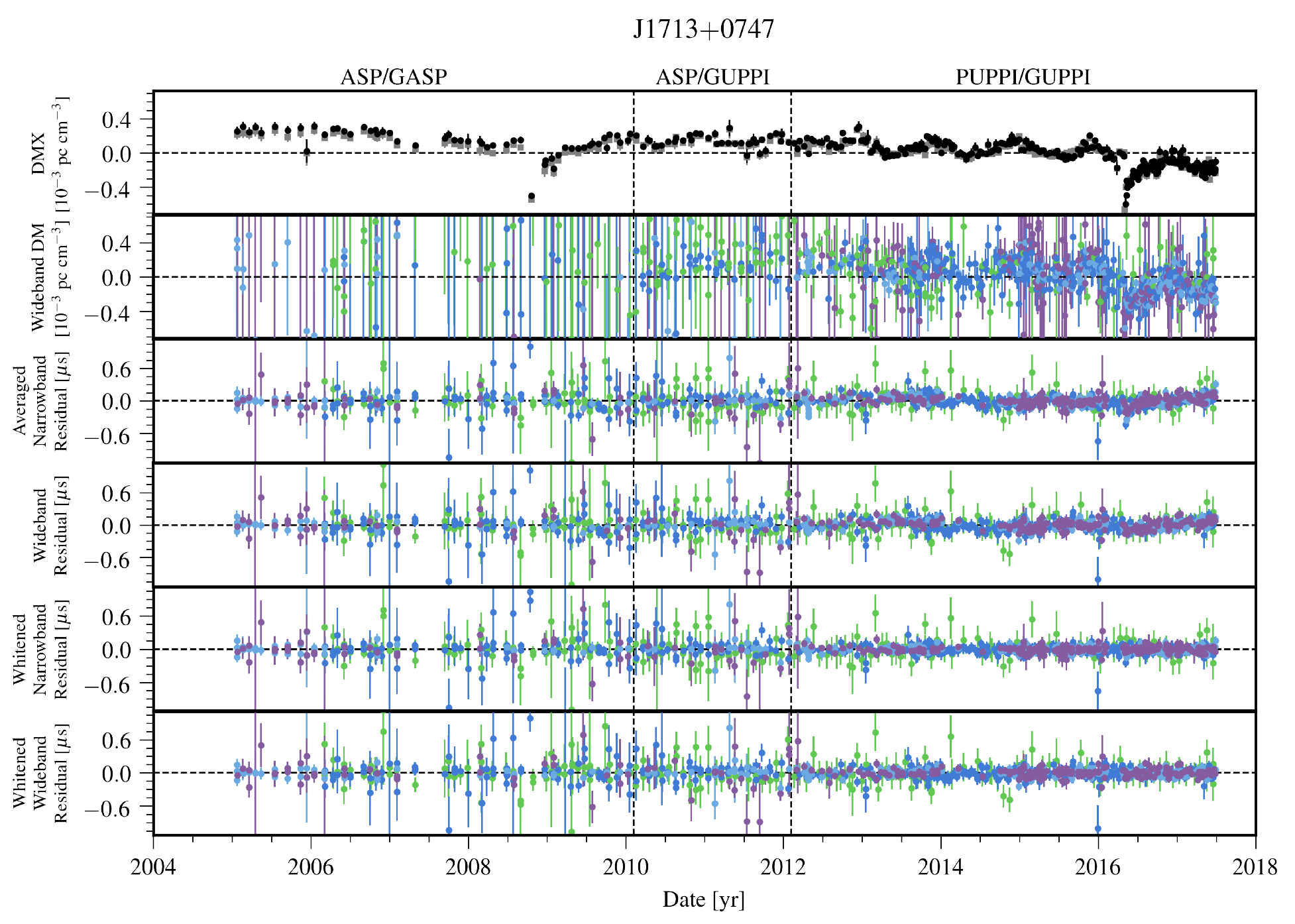}
\caption{Timing residuals and DM variations for J1713$+$0747.  See Appendix~\ref{sec:resid} for details.  DMX model parameters from the wideband (black circles) and narrowband (grey squares) data sets are shown in the top panel.  Colors in the lower panels indicate the receiver for the observation: 820~MHz (Green), 1.4~GHz (Dark blue), 1.4~GHz (Light blue), and 2.1~GHz (Purple).}
\label{fig:summary-J1713+0747}
\end{figure*}

\begin{figure*}[p]
\centering
\includegraphics[scale=0.8]{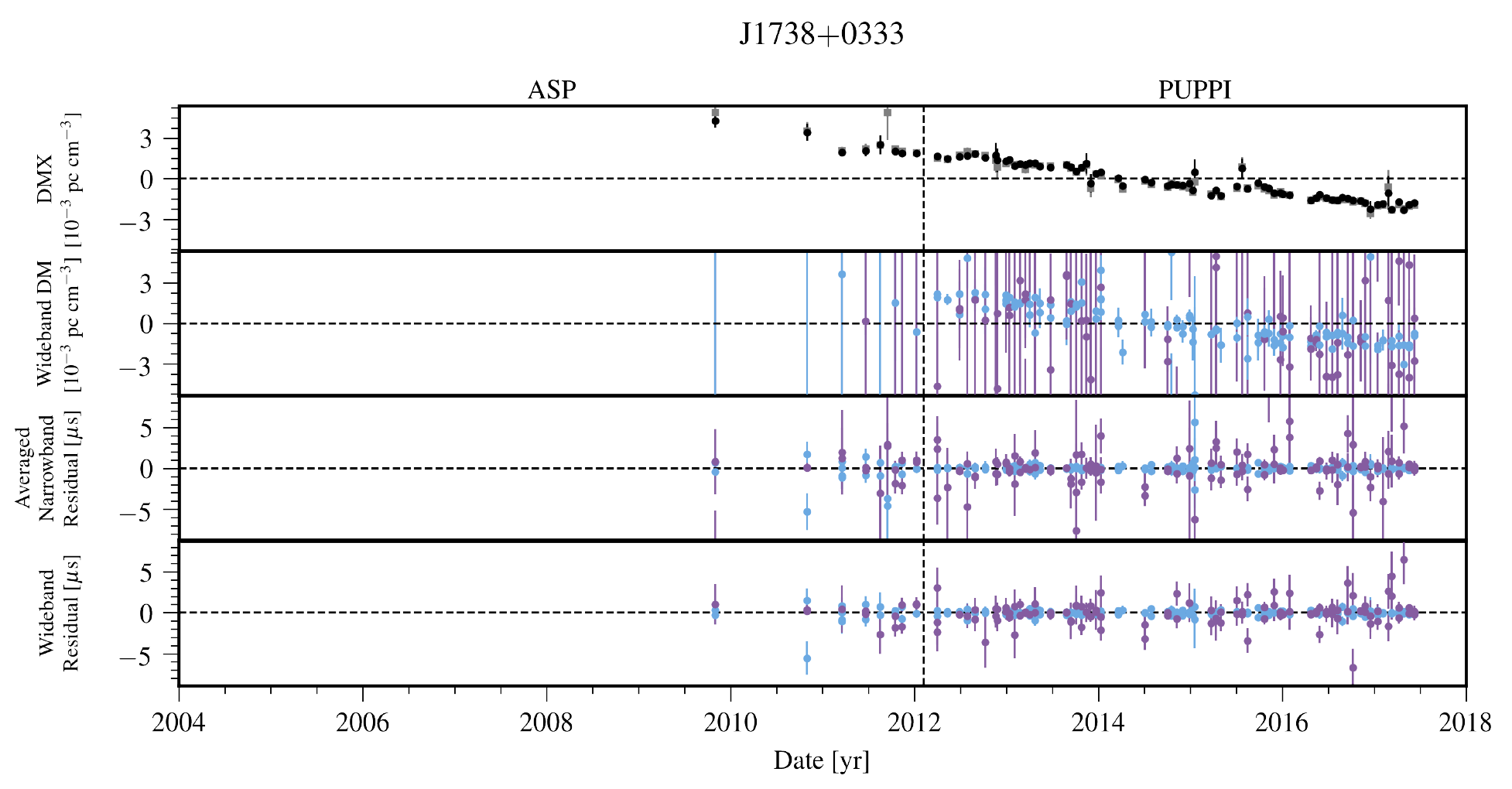}
\caption{Timing residuals and DM variations for J1738$+$0333.  See Appendix~\ref{sec:resid} for details.  DMX model parameters from the wideband (black circles) and narrowband (grey squares) data sets are shown in the top panel.  Colors in the lower panels indicate the receiver for the observation: 1.4~GHz (Light blue) and 2.1~GHz (Purple).}
\label{fig:summary-J1738+0333}
\end{figure*}

\begin{figure*}[p]
\centering
\includegraphics[scale=0.8]{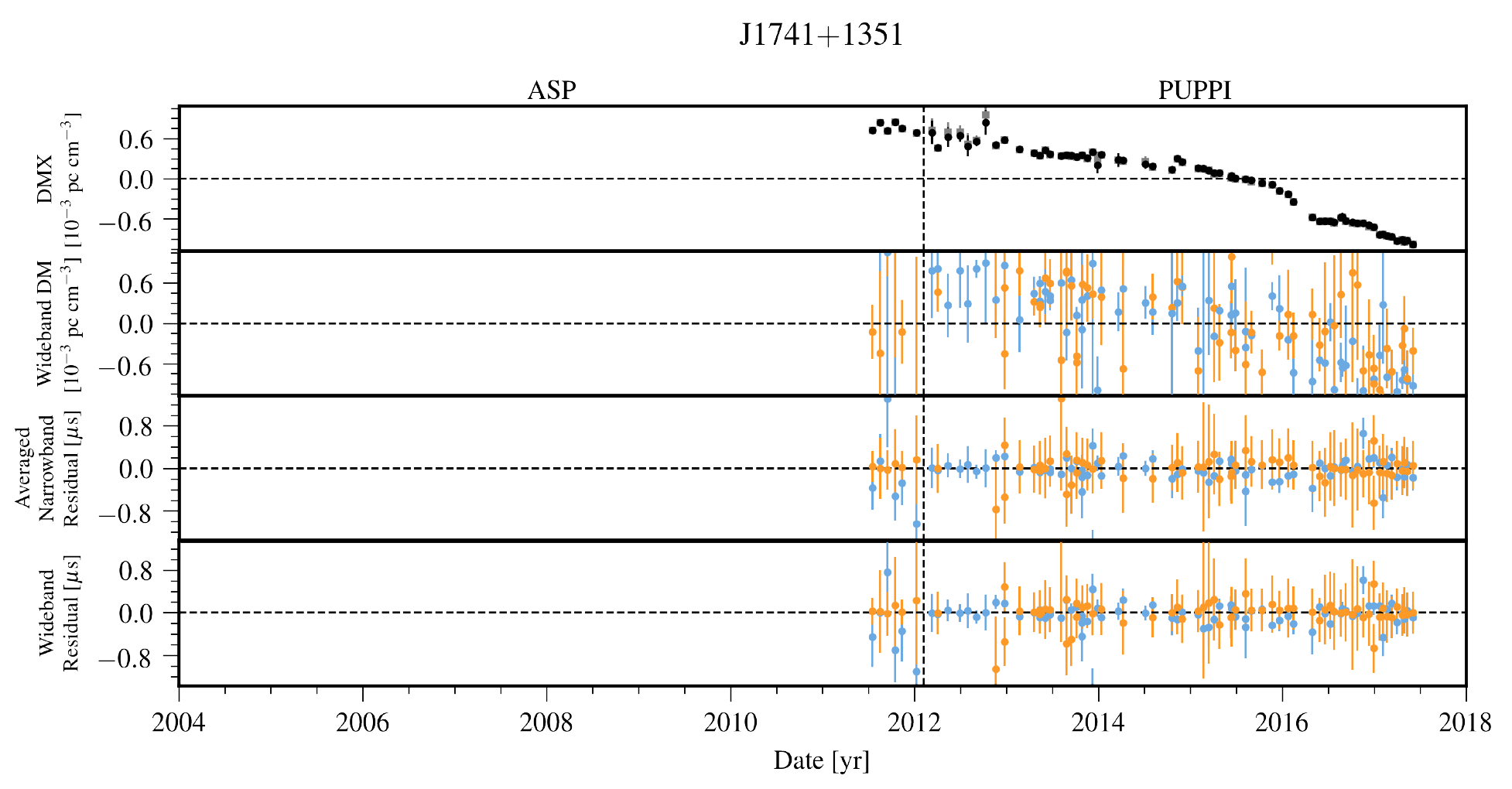}
\caption{Timing residuals and DM variations for J1741$+$1351.  See Appendix~\ref{sec:resid} for details.  DMX model parameters from the wideband (black circles) and narrowband (grey squares) data sets are shown in the top panel.  Colors in the lower panels indicate the receiver for the observation: 430~MHz (Orange) and 1.4~GHz (Light blue).}
\label{fig:summary-J1741+1351}
\end{figure*}

\begin{figure*}[p]
\centering
\includegraphics[scale=0.8]{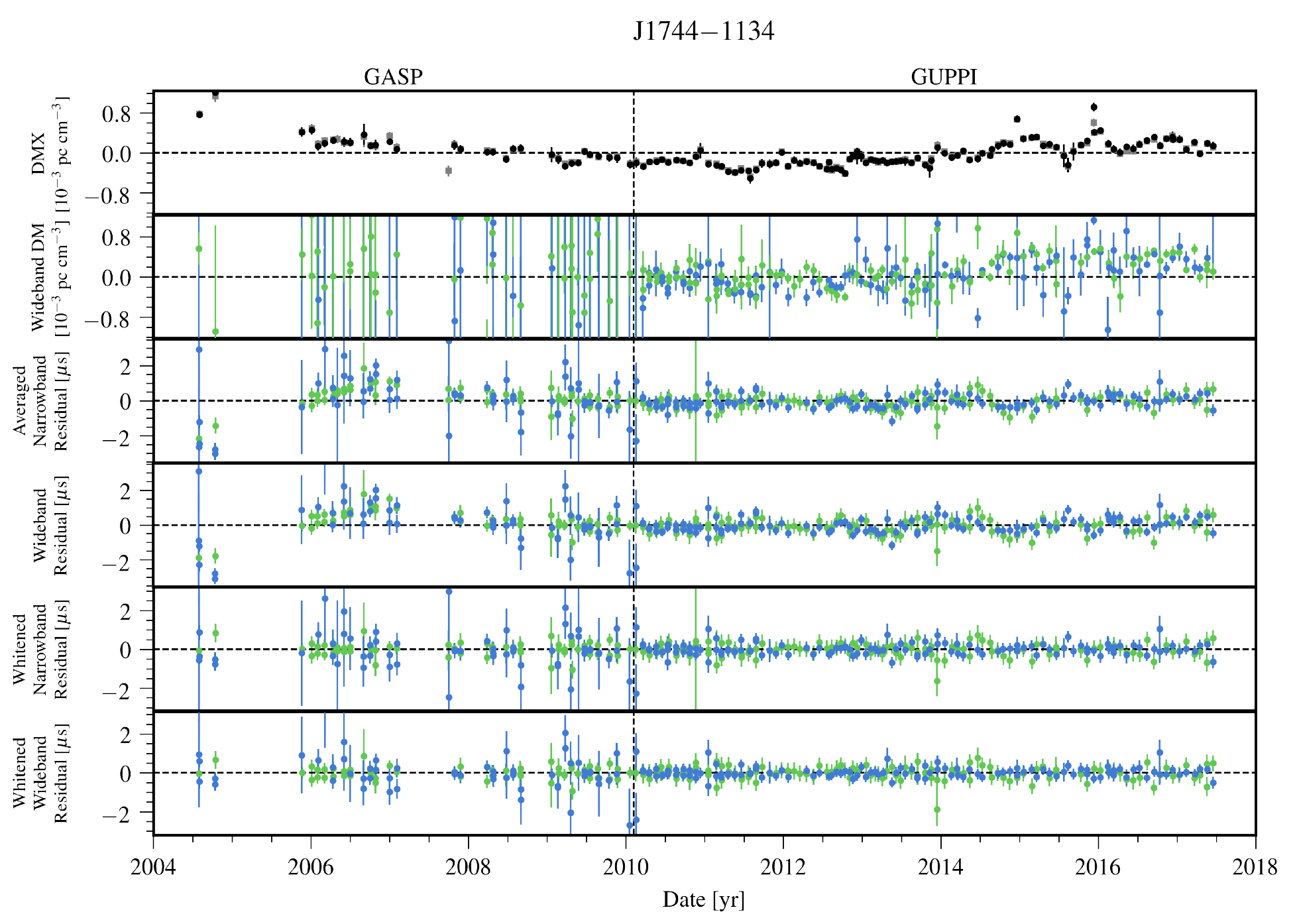}
\caption{Timing residuals and DM variations for J1744$-$1134.  See Appendix~\ref{sec:resid} for details.  DMX model parameters from the wideband (black circles) and narrowband (grey squares) data sets are shown in the top panel.  Colors in the lower panels indicate the receiver for the observation: 820~MHz (Green) and 1.4~GHz (Dark blue).}
\label{fig:summary-J1744-1134}
\end{figure*}

\begin{figure*}[p]
\centering
\includegraphics[scale=0.8]{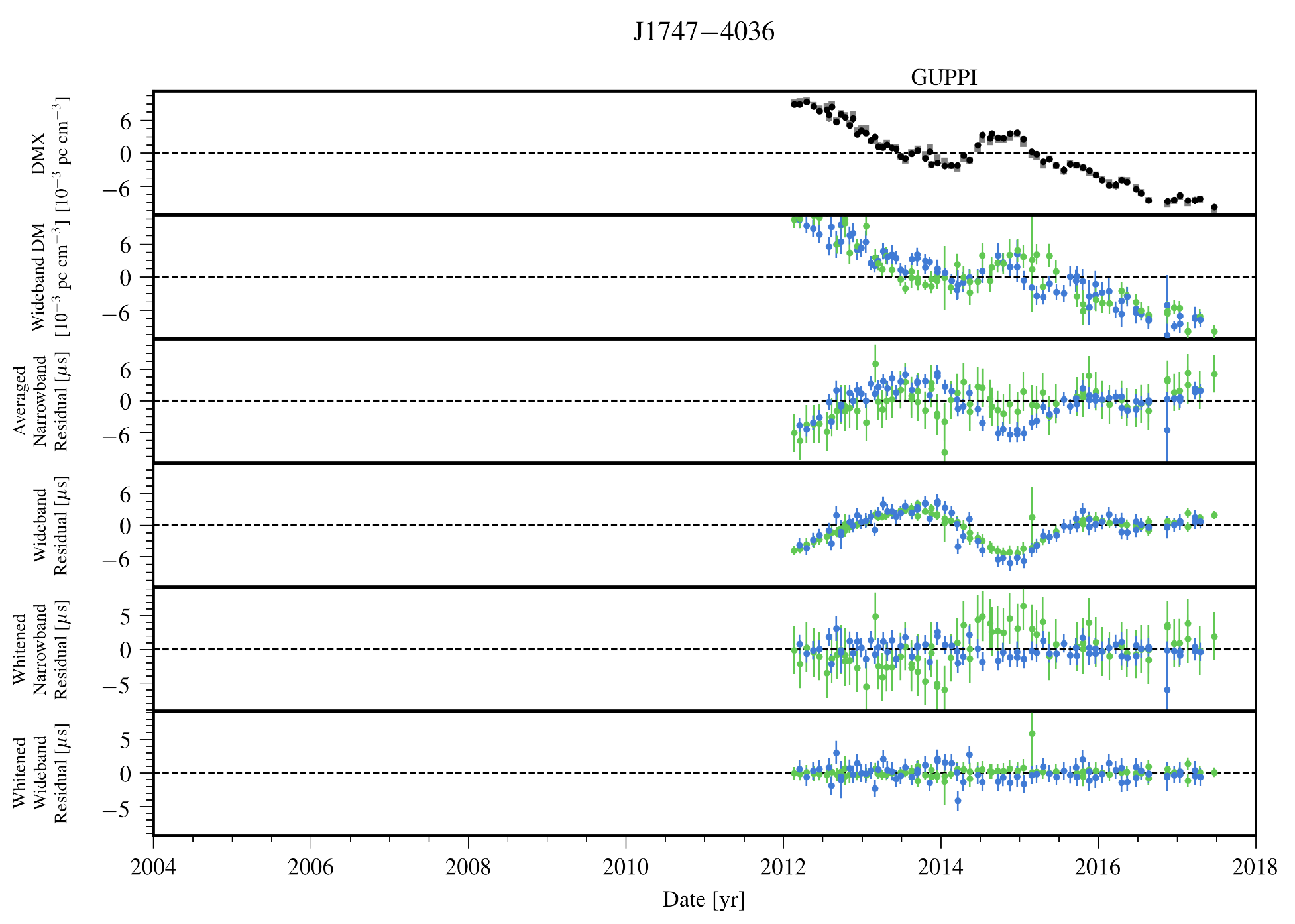}
\caption{Timing residuals and DM variations for J1747$-$4036.  See Appendix~\ref{sec:resid} for details.  DMX model parameters from the wideband (black circles) and narrowband (grey squares) data sets are shown in the top panel.  Colors in the lower panels indicate the receiver for the observation: 820~MHz (Green) and 1.4~GHz (Dark blue).}
\label{fig:summary-J1747-4036}
\end{figure*}

\begin{figure*}[p]
\centering
\includegraphics[scale=0.8]{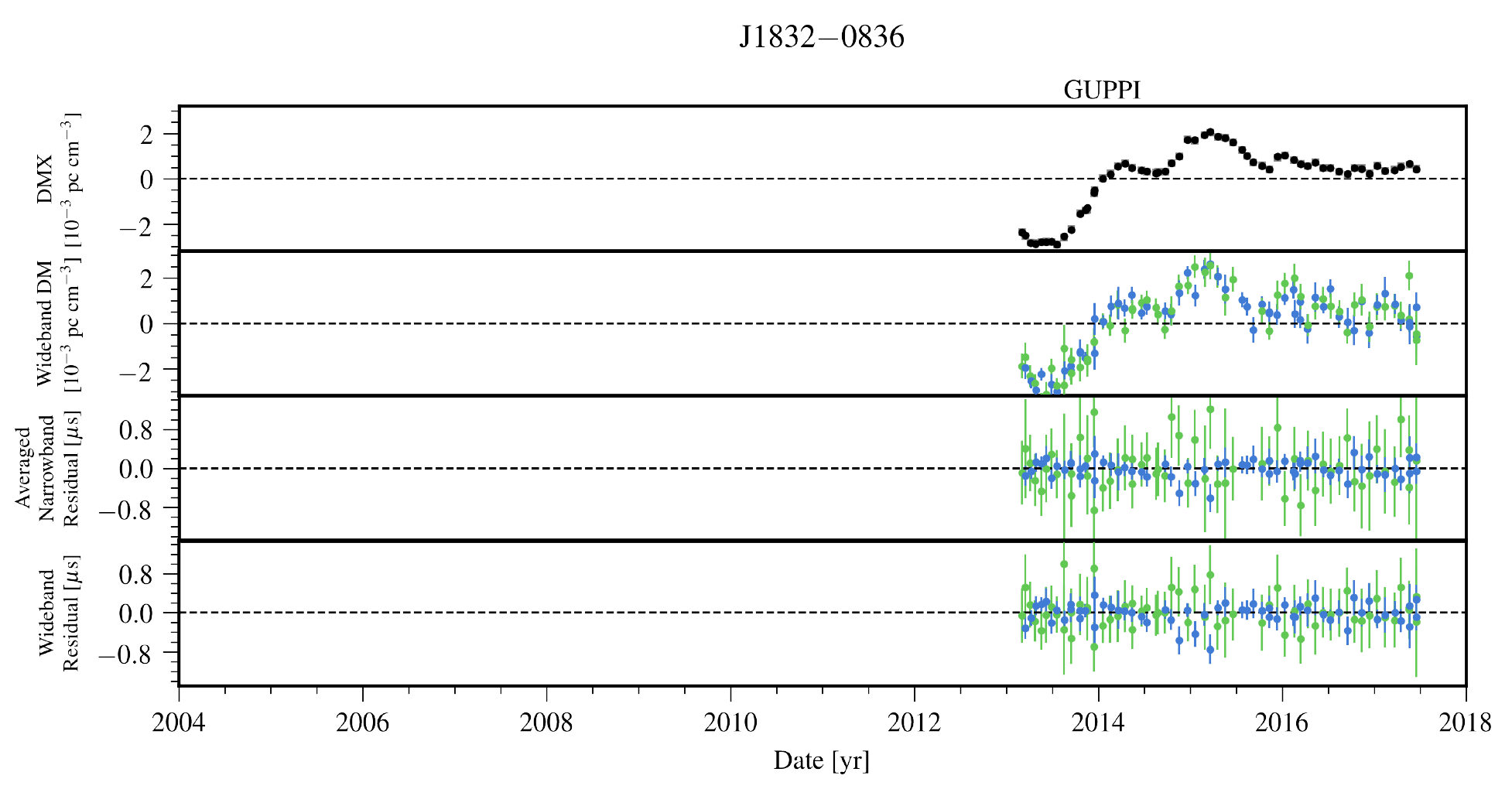}
\caption{Timing residuals and DM variations for J1832$-$0836.  See Appendix~\ref{sec:resid} for details.  DMX model parameters from the wideband (black circles) and narrowband (grey squares) data sets are shown in the top panel.  Colors in the lower panels indicate the receiver for the observation: 820~MHz (Green) and 1.4~GHz (Dark blue).}
\label{fig:summary-J1832-0836}
\end{figure*}

\begin{figure*}[p]
\centering
\includegraphics[scale=0.8]{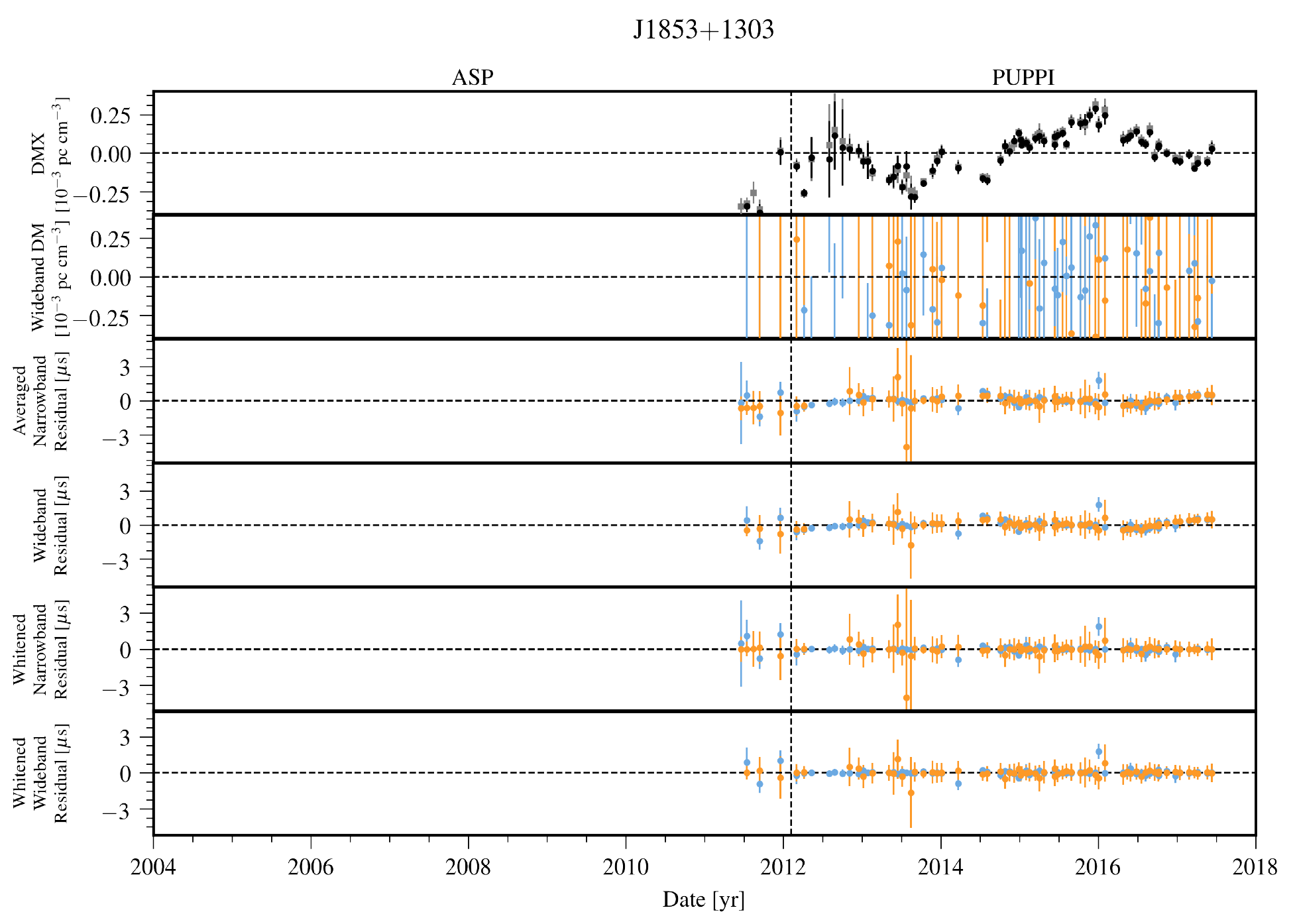}
\caption{Timing residuals and DM variations for J1853$+$1303.  See Appendix~\ref{sec:resid} for details.  DMX model parameters from the wideband (black circles) and narrowband (grey squares) data sets are shown in the top panel.  Colors in the lower panels indicate the receiver for the observation: 430~MHz (Orange) and 1.4~GHz (Light blue).}
\label{fig:summary-J1853+1303}
\end{figure*}

\begin{figure*}[p]
\centering
\includegraphics[scale=0.8]{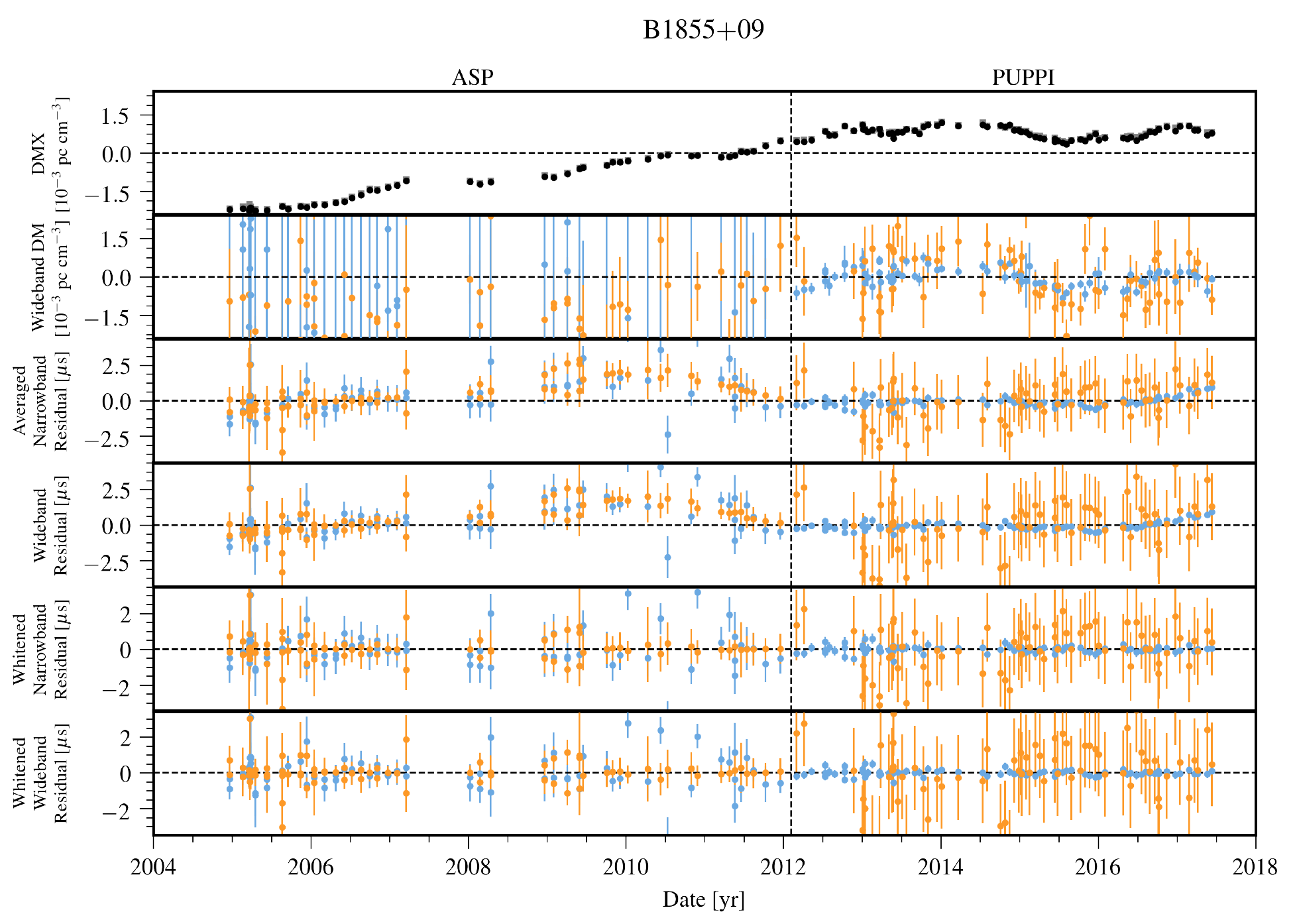}
\caption{Timing residuals and DM variations for B1855$+$09.  See Appendix~\ref{sec:resid} for details.  DMX model parameters from the wideband (black circles) and narrowband (grey squares) data sets are shown in the top panel.  Colors in the lower panels indicate the receiver for the observation: 430~MHz (Orange) and 1.4~GHz (Light blue).}
\label{fig:summary-B1855+09}
\end{figure*}

\begin{figure*}[p]
\centering
\includegraphics[scale=0.8]{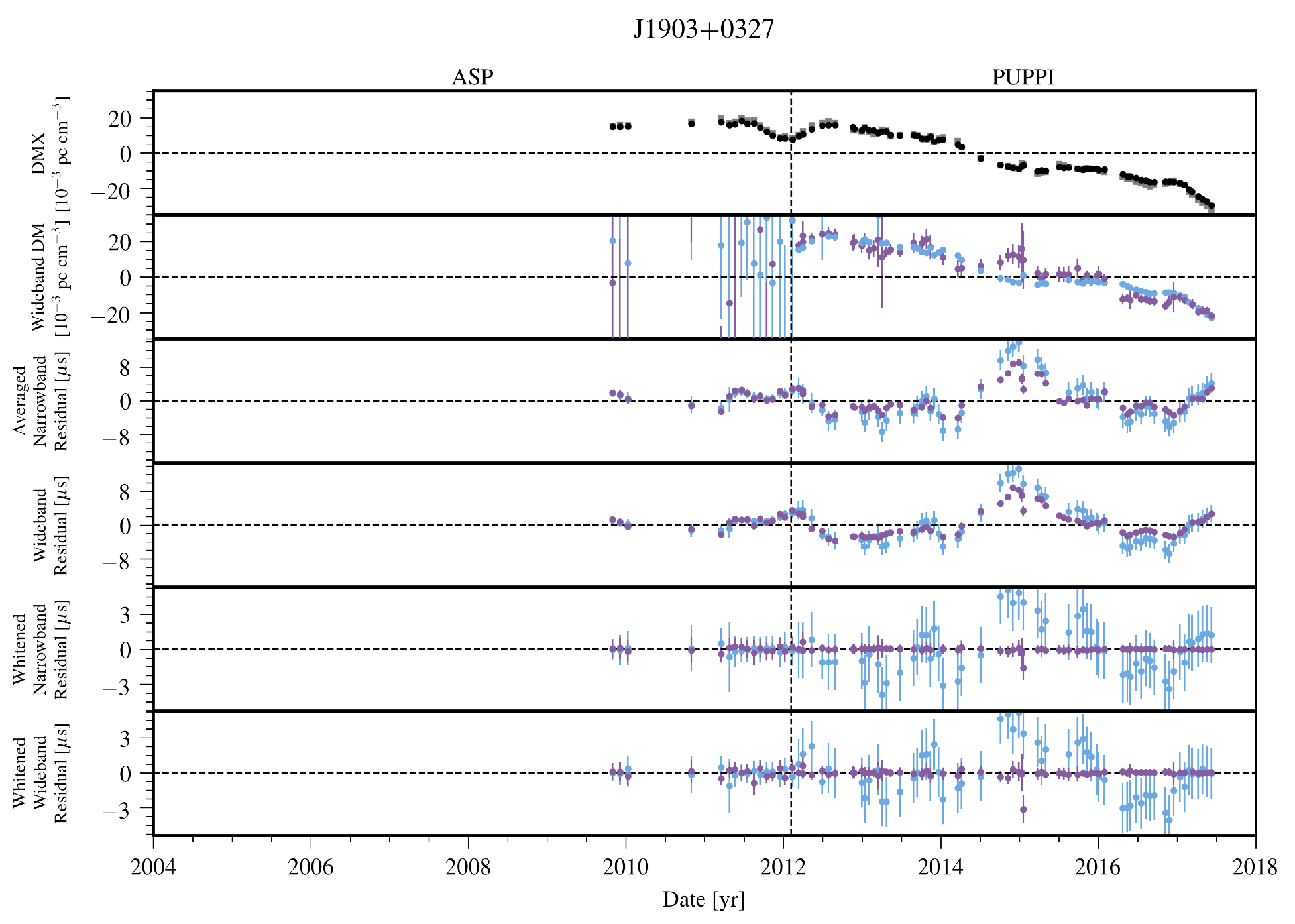}
\caption{Timing residuals and DM variations for J1903$+$0327.  See Appendix~\ref{sec:resid} for details.  DMX model parameters from the wideband (black circles) and narrowband (grey squares) data sets are shown in the top panel.  Colors in the lower panels indicate the receiver for the observation: 1.4~GHz (Light blue) and 2.1~GHz (Purple).}
\label{fig:summary-J1903+0327}
\end{figure*}

\begin{figure*}[p]
\centering
\includegraphics[scale=0.8]{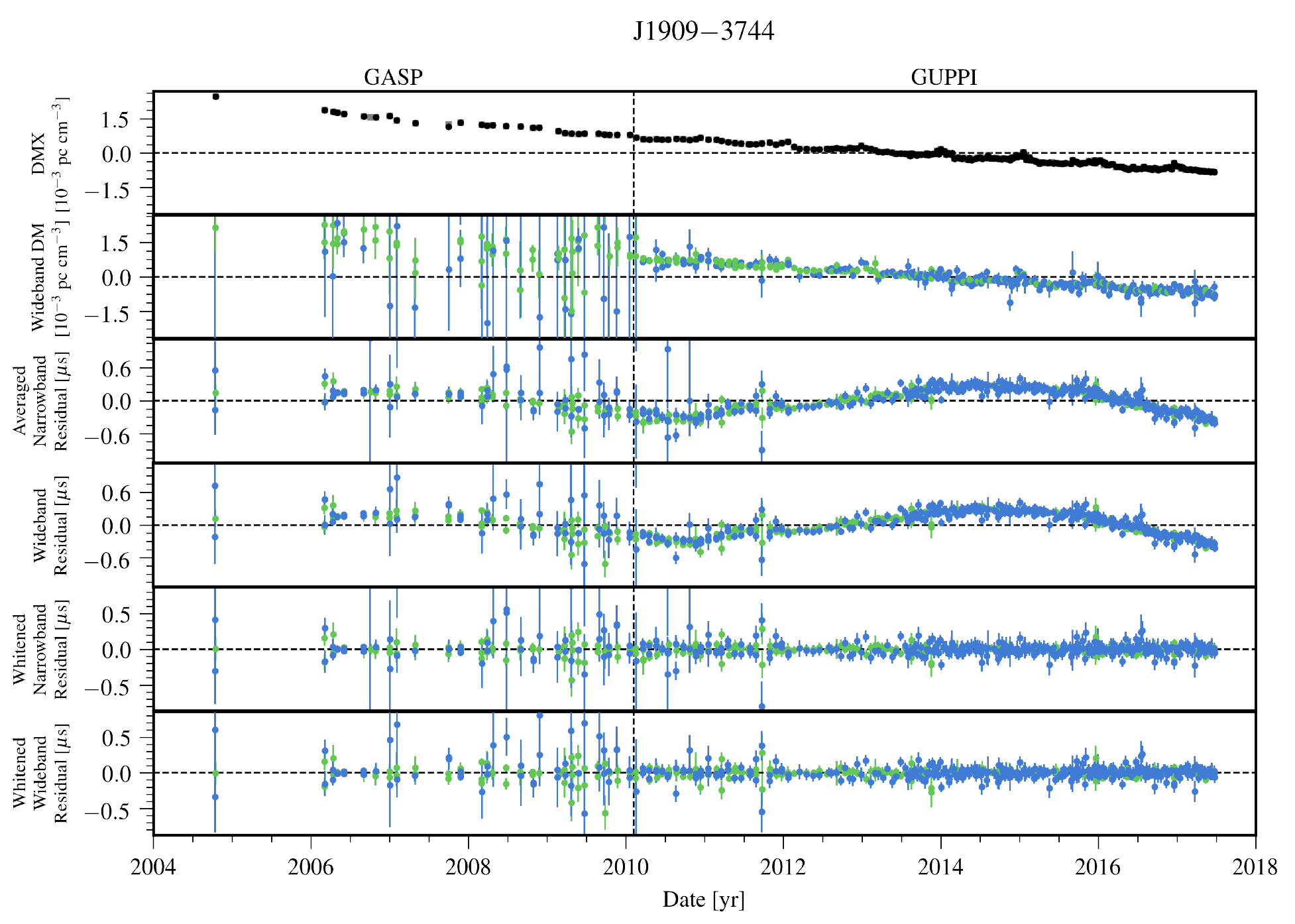}
\caption{Timing residuals and DM variations for J1909$-$3744.  See Appendix~\ref{sec:resid} for details.  DMX model parameters from the wideband (black circles) and narrowband (grey squares) data sets are shown in the top panel.  Colors in the lower panels indicate the receiver for the observation: 820~MHz (Green) and 1.4~GHz (Dark blue).}
\label{fig:summary-J1909-3744}
\end{figure*}

\begin{figure*}[p]
\centering
\includegraphics[scale=0.8]{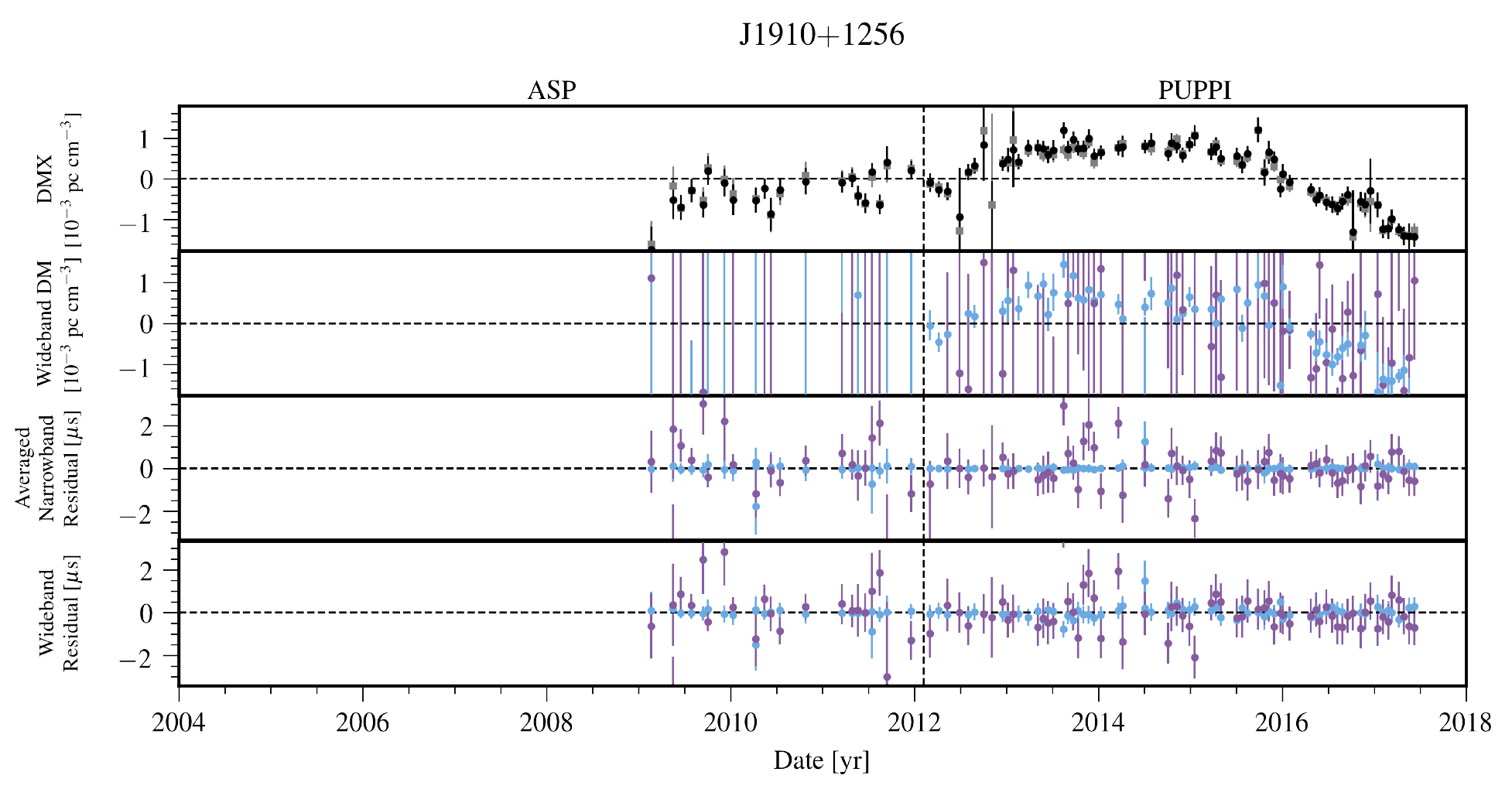}
\caption{Timing residuals and DM variations for J1910$+$1256.  See Appendix~\ref{sec:resid} for details.  DMX model parameters from the wideband (black circles) and narrowband (grey squares) data sets are shown in the top panel.  Colors in the lower panels indicate the receiver for the observation: 1.4~GHz (Light blue) and 2.1~GHz (Purple).}
\label{fig:summary-J1910+1256}
\end{figure*}

\begin{figure*}[p]
\centering
\includegraphics[scale=0.8]{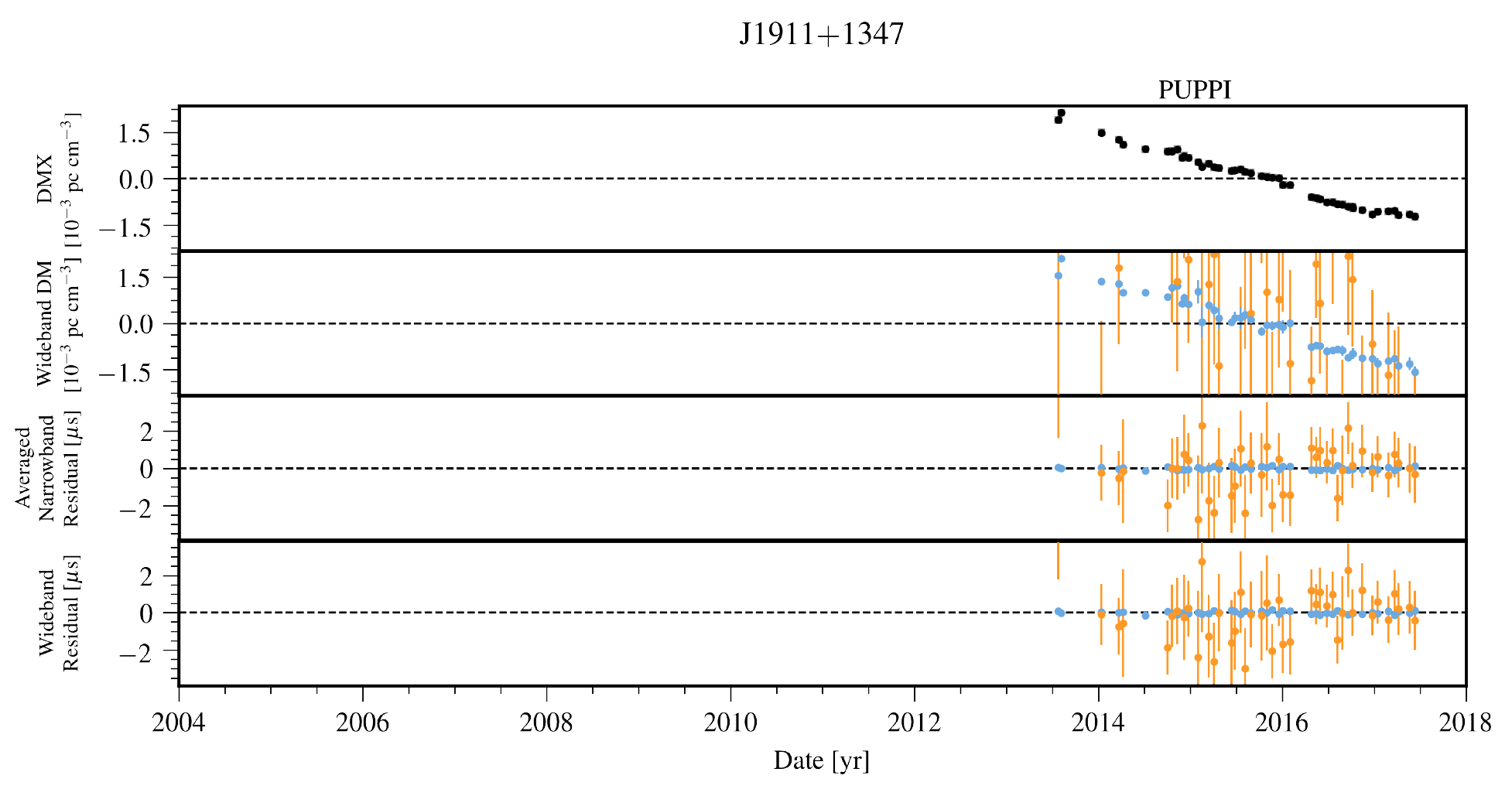}
\caption{Timing residuals and DM variations for J1911$+$1347.  See Appendix~\ref{sec:resid} for details.  DMX model parameters from the wideband (black circles) and narrowband (grey squares) data sets are shown in the top panel.  Colors in the lower panels indicate the receiver for the observation: 430~MHz (Orange) and 1.4~GHz (Light blue).}
\label{fig:summary-J1911+1347}
\end{figure*}

\begin{figure*}[p]
\centering
\includegraphics[scale=0.8]{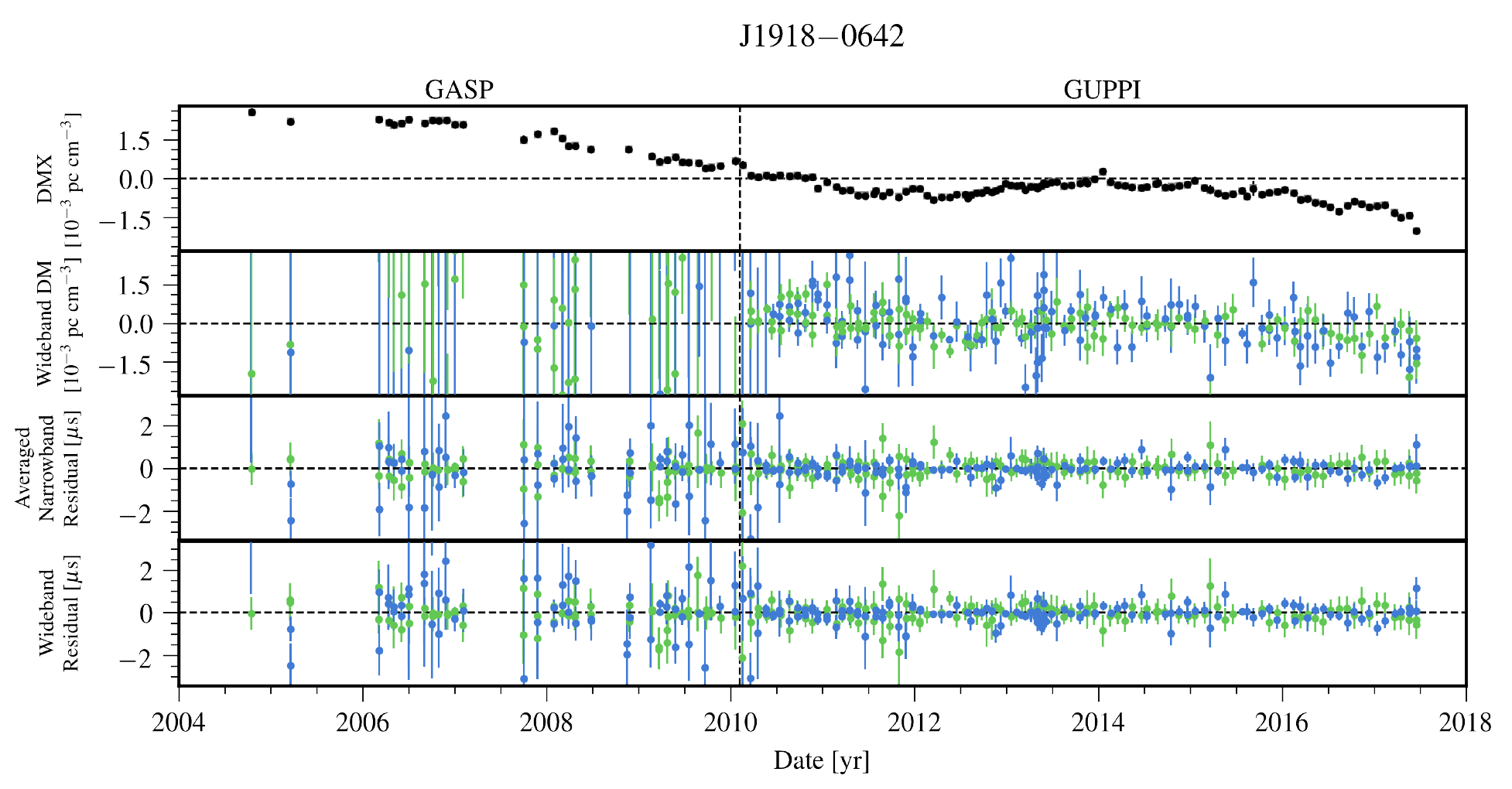}
\caption{Timing residuals and DM variations for J1918$-$0642.  See Appendix~\ref{sec:resid} for details.  DMX model parameters from the wideband (black circles) and narrowband (grey squares) data sets are shown in the top panel.  Colors in the lower panels indicate the receiver for the observation: 820~MHz (Green) and 1.4~GHz (Dark blue).}
\label{fig:summary-J1918-0642}
\end{figure*}

\begin{figure*}[p]
\centering
\includegraphics[scale=0.8]{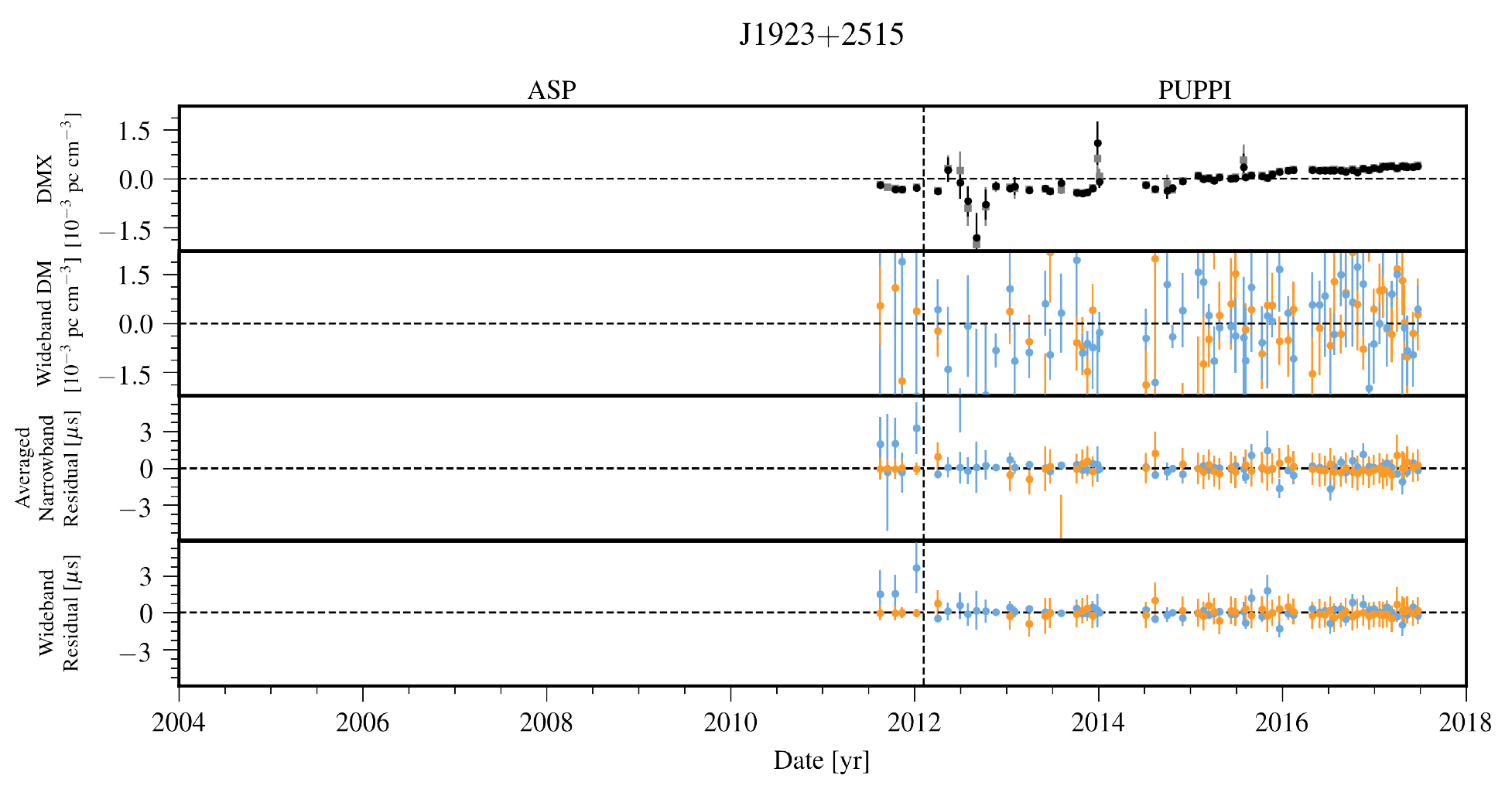}
\caption{Timing residuals and DM variations for J1923$+$2515.  See Appendix~\ref{sec:resid} for details.  DMX model parameters from the wideband (black circles) and narrowband (grey squares) data sets are shown in the top panel.  Colors in the lower panels indicate the receiver for the observation: 430~MHz (Orange) and 1.4~GHz (Light blue).}
\label{fig:summary-J1923+2515}
\end{figure*}

\begin{figure*}[p]
\centering
\includegraphics[scale=0.8]{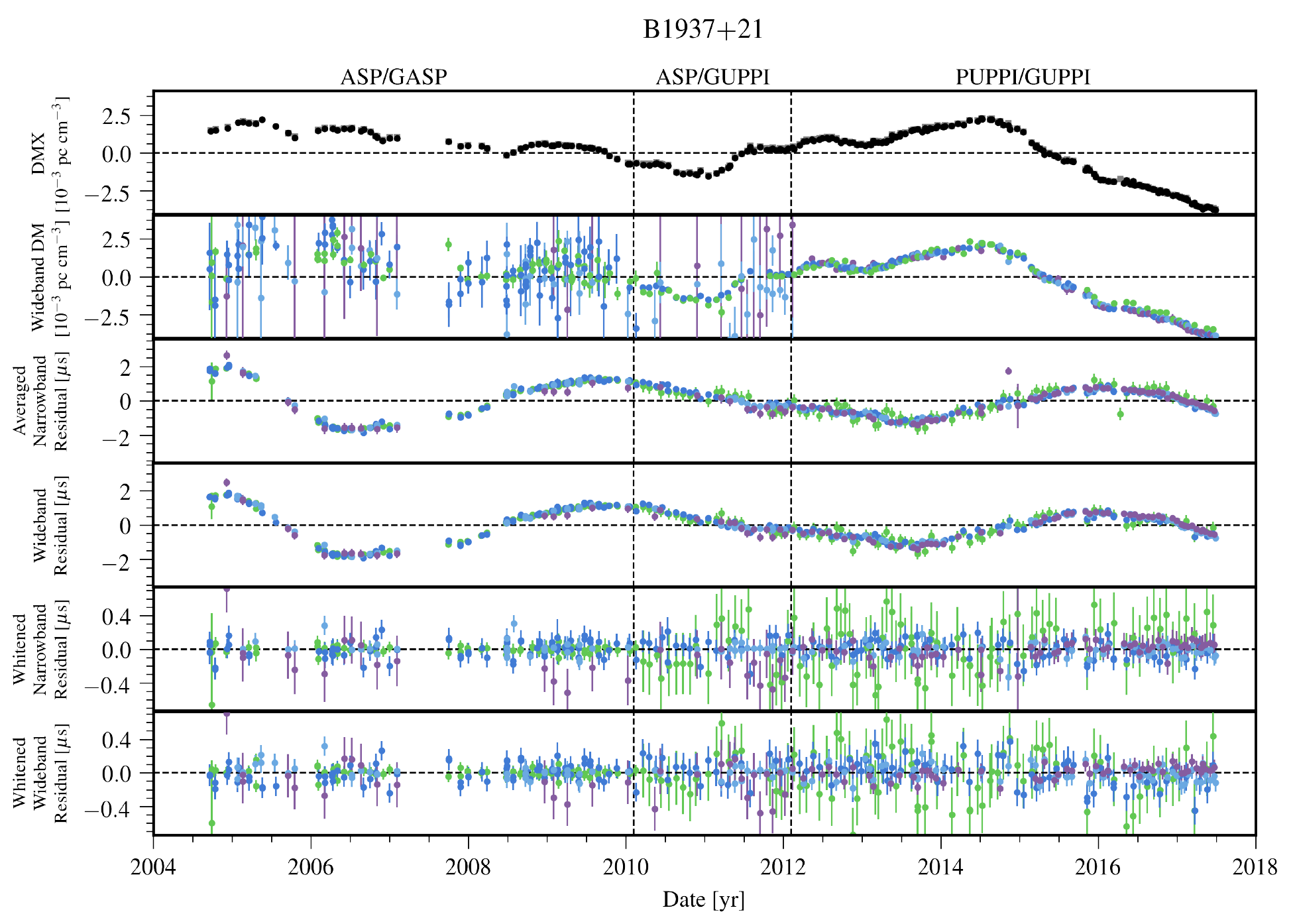}
\caption{Timing residuals and DM variations for B1937$+$21.  See Appendix~\ref{sec:resid} for details.  DMX model parameters from the wideband (black circles) and narrowband (grey squares) data sets are shown in the top panel.  Colors in the lower panels indicate the receiver for the observation: 820~MHz (Green), 1.4~GHz (Dark blue), 1.4~GHz (Light blue), and 2.1~GHz (Purple).}
\label{fig:summary-B1937+21}
\end{figure*}

\begin{figure*}[p]
\centering
\includegraphics[scale=0.8]{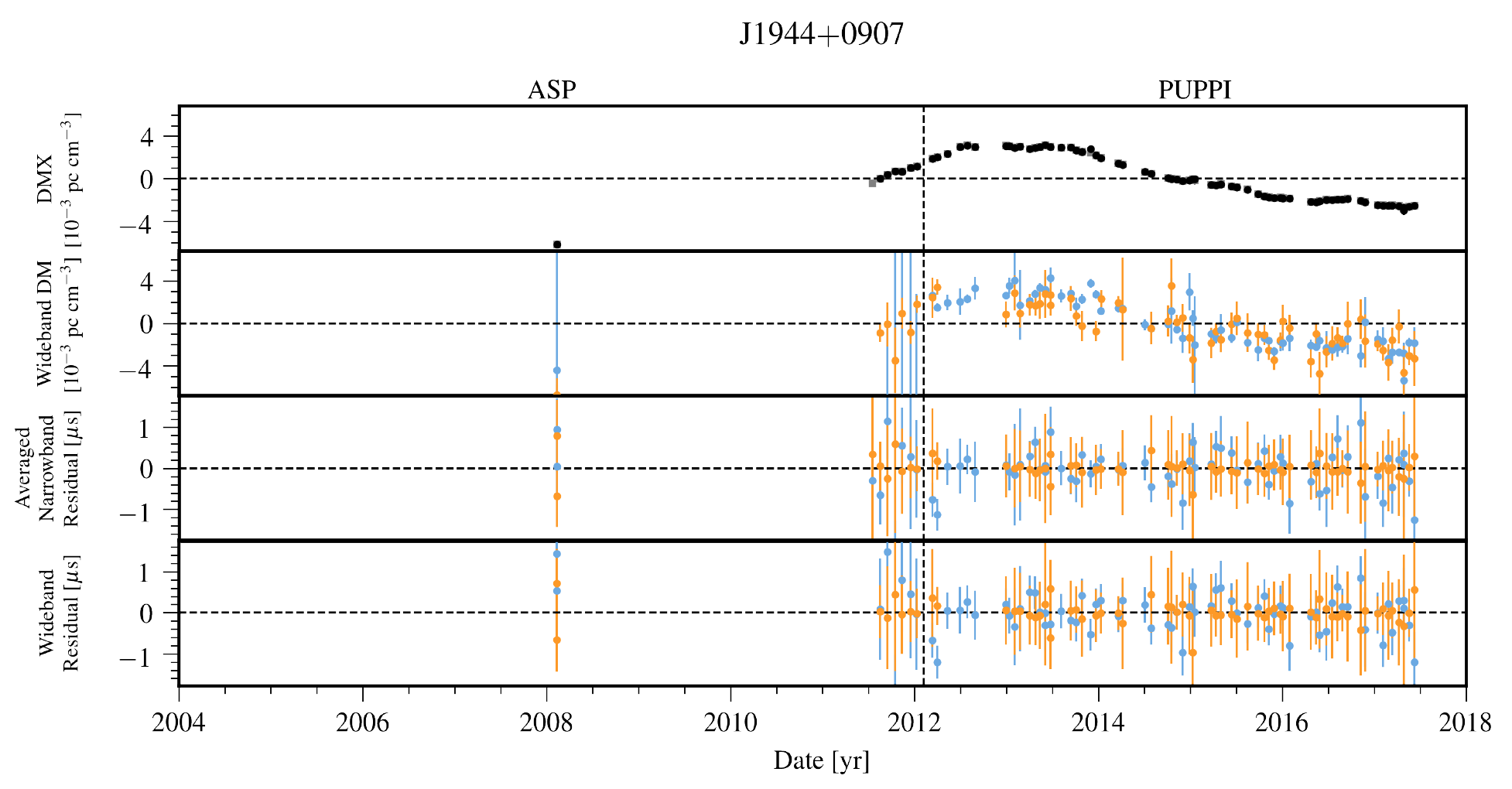}
\caption{Timing residuals and DM variations for J1944$+$0907.  See Appendix~\ref{sec:resid} for details.  DMX model parameters from the wideband (black circles) and narrowband (grey squares) data sets are shown in the top panel.  Colors in the lower panels indicate the receiver for the observation: 430~MHz (Orange) and 1.4~GHz (Light blue).}
\label{fig:summary-J1944+0907}
\end{figure*}

\begin{figure*}[p]
\centering
\includegraphics[scale=0.8]{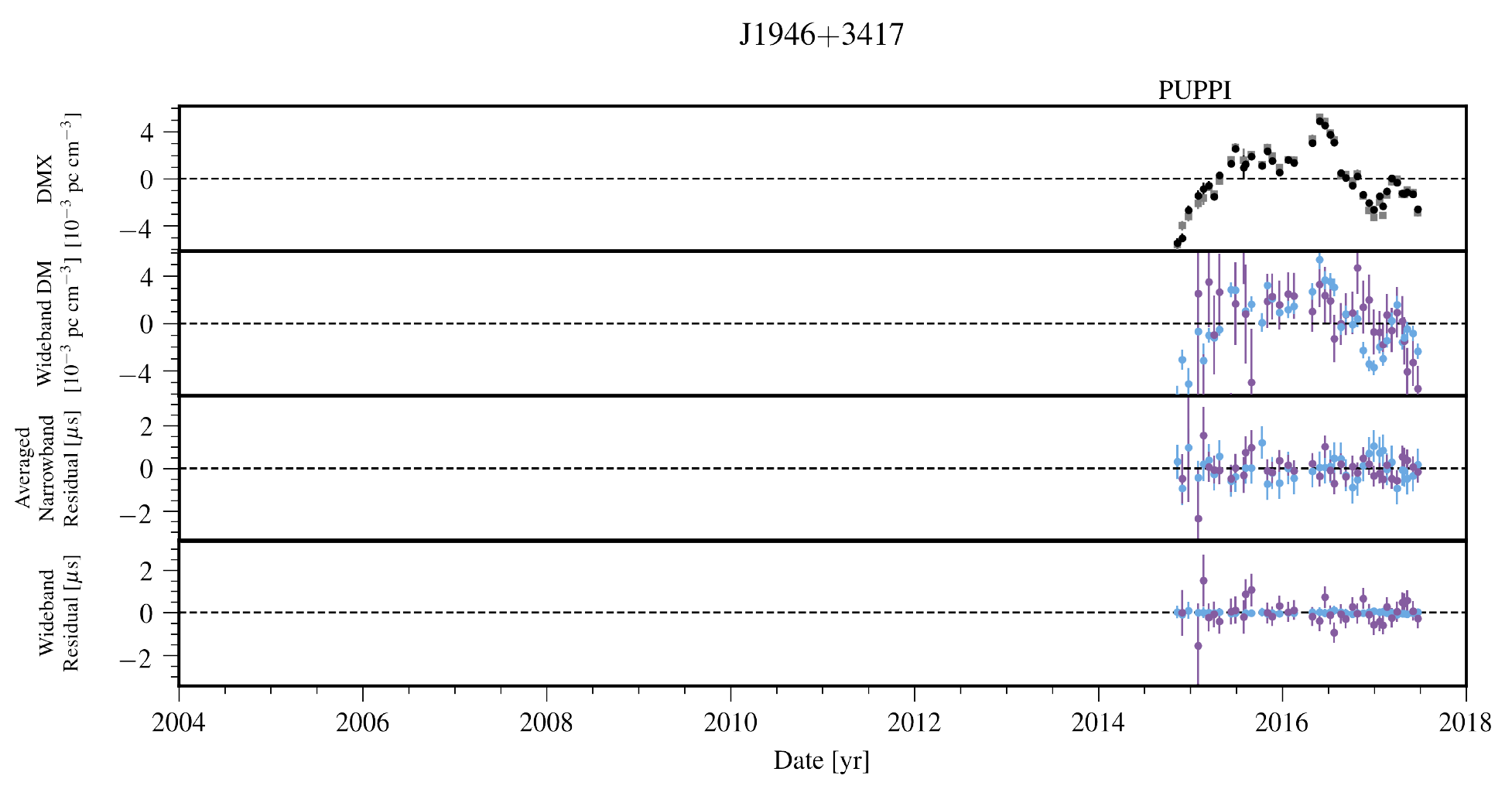}
\caption{Timing residuals and DM variations for J1946$+$3417.  See Appendix~\ref{sec:resid} for details.  DMX model parameters from the wideband (black circles) and narrowband (grey squares) data sets are shown in the top panel.  Colors in the lower panels indicate the receiver for the observation: 1.4~GHz (Light blue) and 2.1~GHz (Purple).}
\label{fig:summary-J1946+3417}
\end{figure*}

\begin{figure*}[p]
\centering
\includegraphics[scale=0.8]{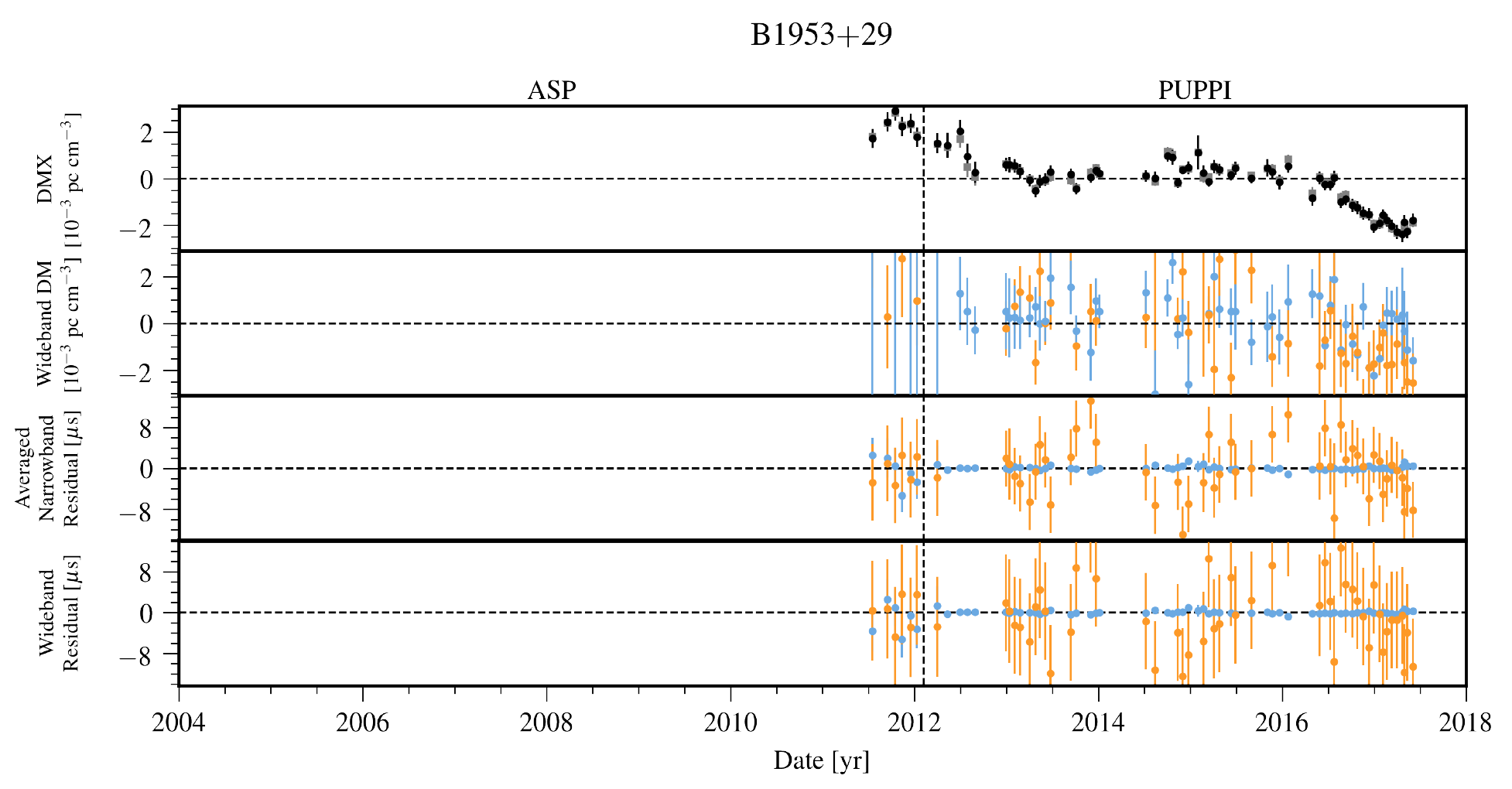}
\caption{Timing residuals and DM variations for B1953$+$29.  See Appendix~\ref{sec:resid} for details.  DMX model parameters from the wideband (black circles) and narrowband (grey squares) data sets are shown in the top panel.  Colors in the lower panels indicate the receiver for the observation: 430~MHz (Orange) and 1.4~GHz (Light blue).}
\label{fig:summary-B1953+29}
\end{figure*}

\begin{figure*}[p]
\centering
\includegraphics[scale=0.8]{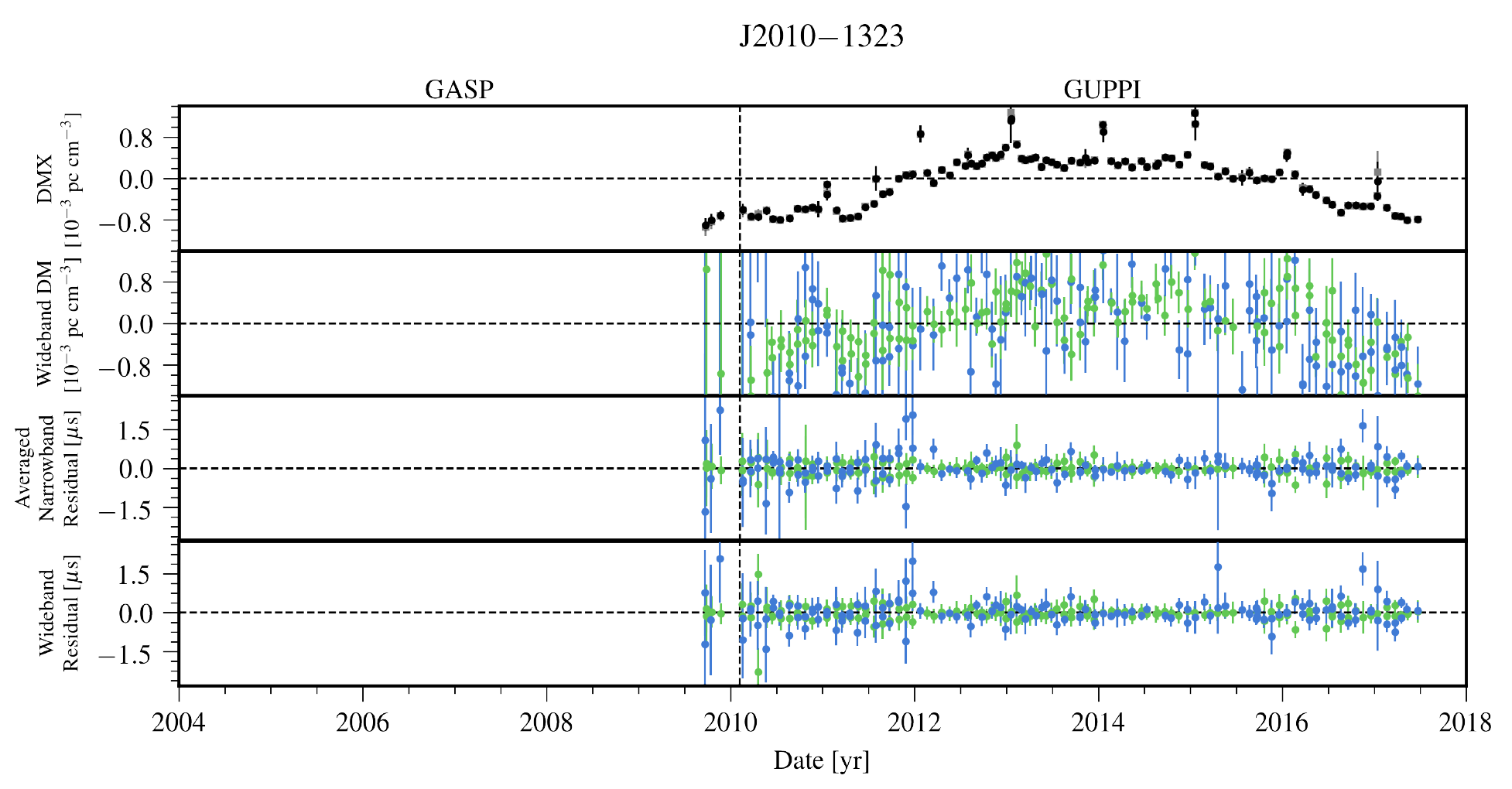}
\caption{Timing residuals and DM variations for J2010$-$1323.  See Appendix~\ref{sec:resid} for details.  DMX model parameters from the wideband (black circles) and narrowband (grey squares) data sets are shown in the top panel.  Colors in the lower panels indicate the receiver for the observation: 820~MHz (Green) and 1.4~GHz (Dark blue).}
\label{fig:summary-J2010-1323}
\end{figure*}

\begin{figure*}[p]
\centering
\includegraphics[scale=0.8]{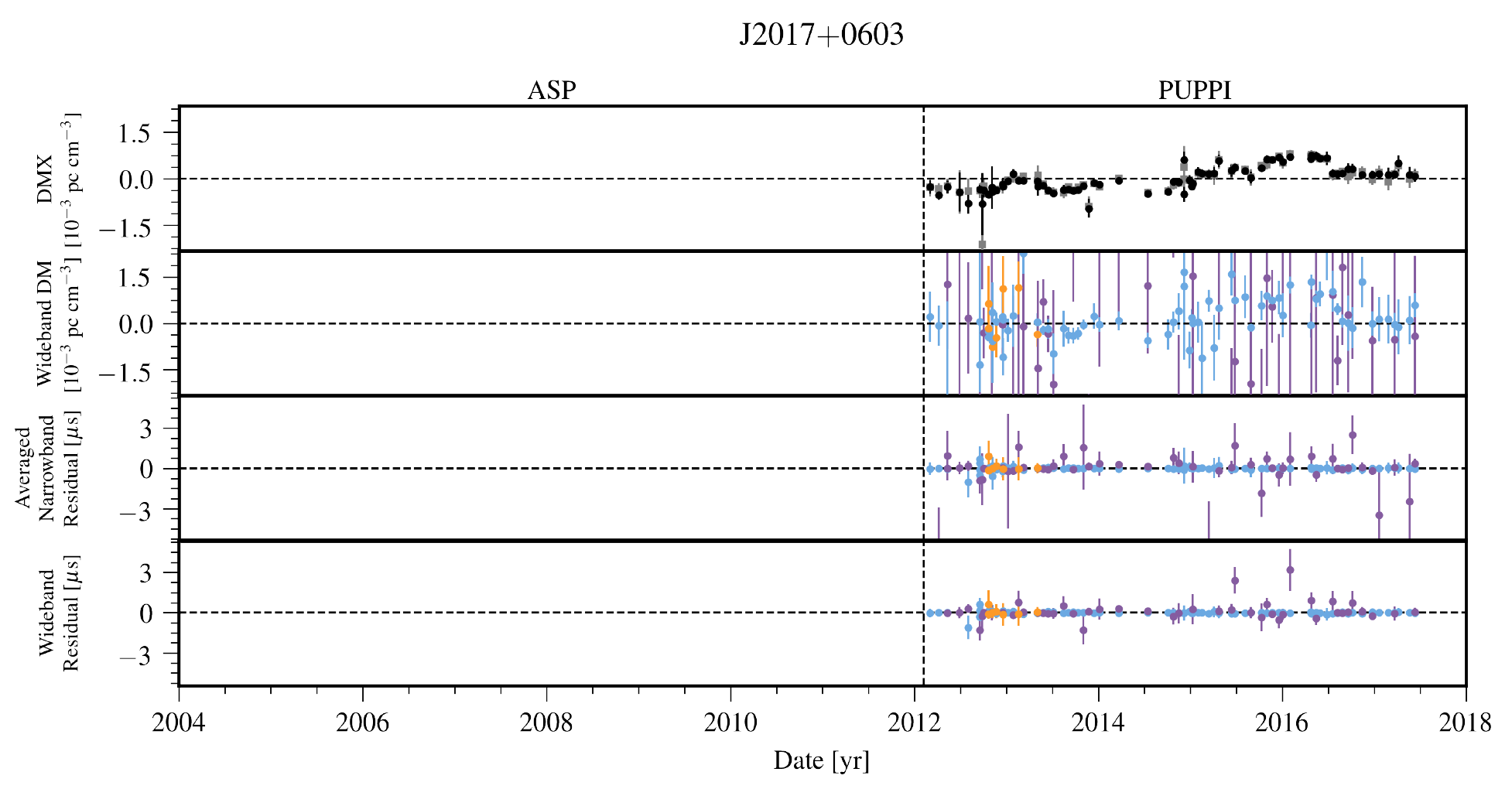}
\caption{Timing residuals and DM variations for J2017$+$0603.  See Appendix~\ref{sec:resid} for details.  DMX model parameters from the wideband (black circles) and narrowband (grey squares) data sets are shown in the top panel.  Colors in the lower panels indicate the receiver for the observation: 430~MHz (Orange), 1.4~GHz (Light blue), and 2.1~GHz (Purple).}
\label{fig:summary-J2017+0603}
\end{figure*}

\begin{figure*}[p]
\centering
\includegraphics[scale=0.8]{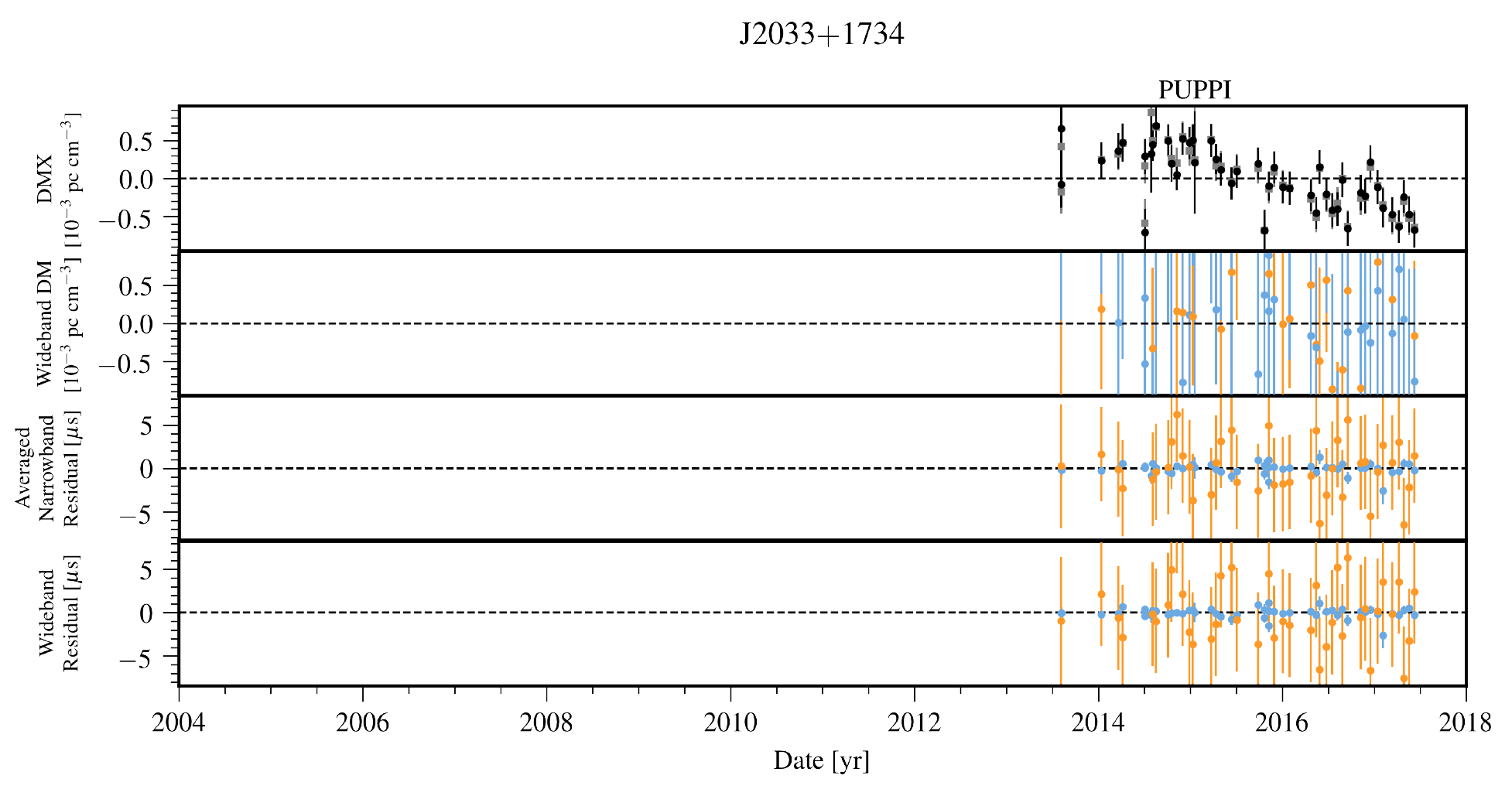}
\caption{Timing residuals and DM variations for J2033$+$1734.  See Appendix~\ref{sec:resid} for details.  DMX model parameters from the wideband (black circles) and narrowband (grey squares) data sets are shown in the top panel.  Colors in the lower panels indicate the receiver for the observation: 430~MHz (Orange) and 1.4~GHz (Light blue).}
\label{fig:summary-J2033+1734}
\end{figure*}

\begin{figure*}[p]
\centering
\includegraphics[scale=0.8]{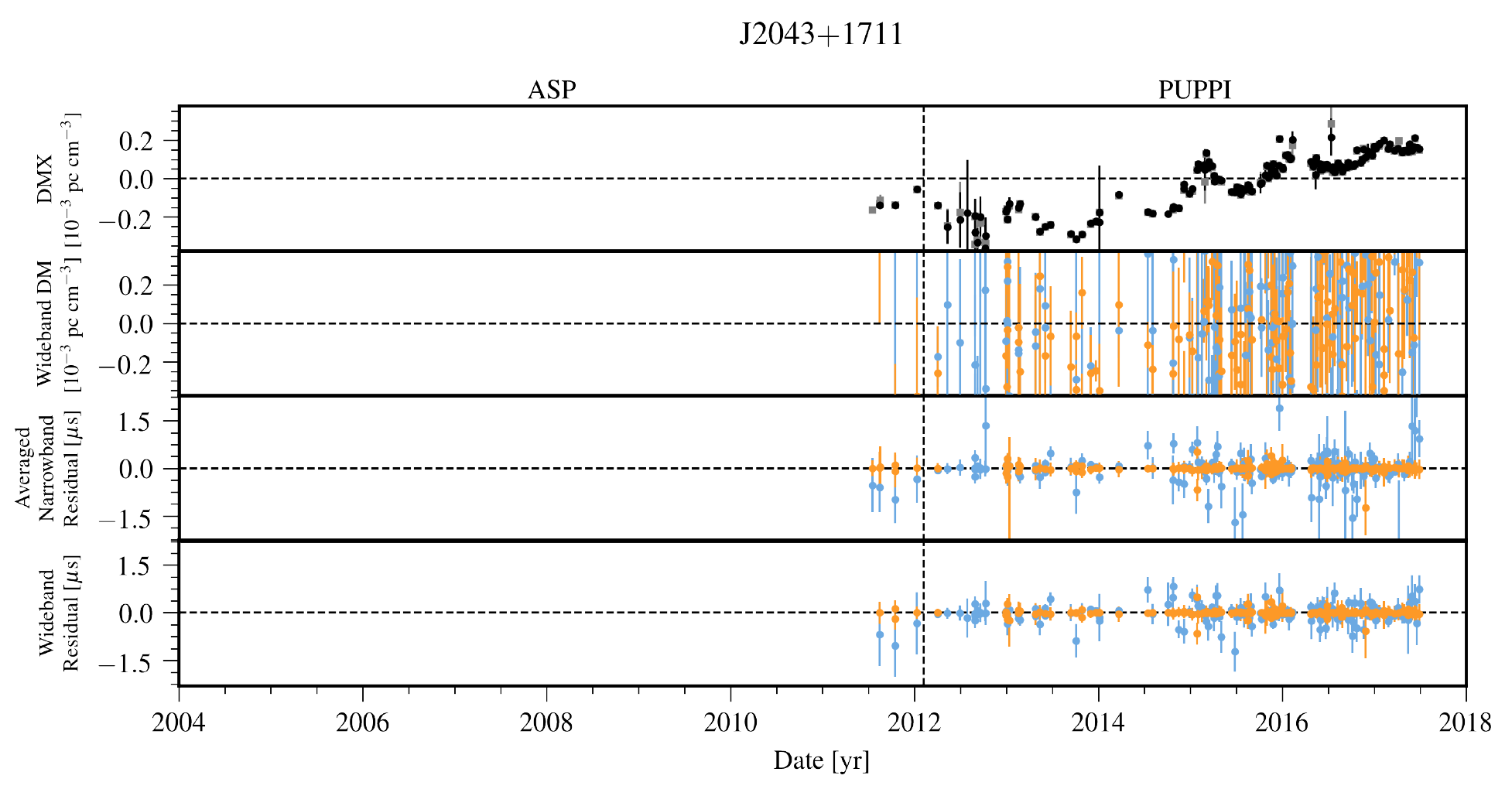}
\caption{Timing residuals and DM variations for J2043$+$1711.  See Appendix~\ref{sec:resid} for details.  DMX model parameters from the wideband (black circles) and narrowband (grey squares) data sets are shown in the top panel.  Colors in the lower panels indicate the receiver for the observation: 430~MHz (Orange) and 1.4~GHz (Light blue).}
\label{fig:summary-J2043+1711}
\end{figure*}

\begin{figure*}[p]
\centering
\includegraphics[scale=0.8]{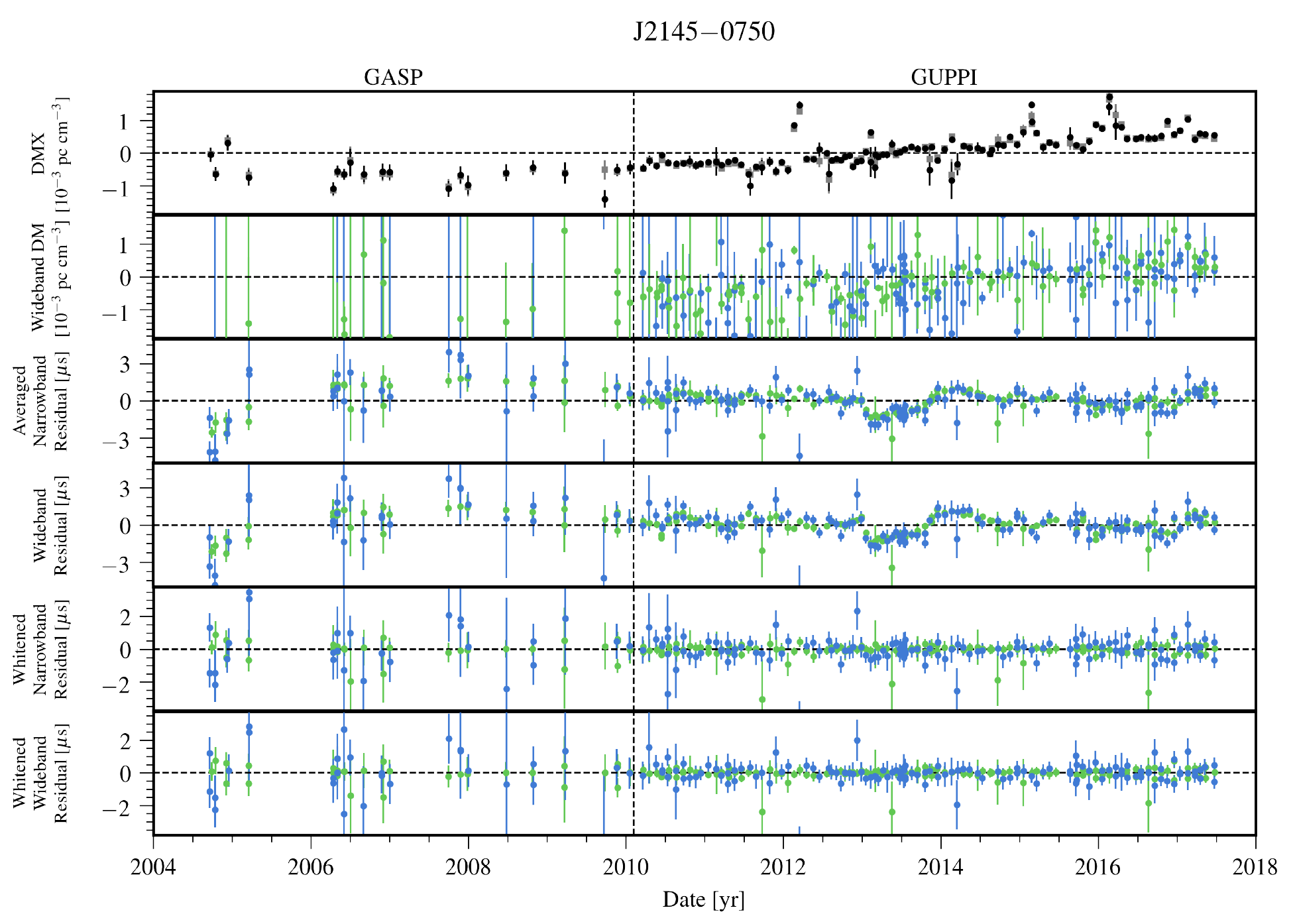}
\caption{Timing residuals and DM variations for J2145$-$0750.  See Appendix~\ref{sec:resid} for details.  DMX model parameters from the wideband (black circles) and narrowband (grey squares) data sets are shown in the top panel.  Colors in the lower panels indicate the receiver for the observation: 820~MHz (Green) and 1.4~GHz (Dark blue).}
\label{fig:summary-J2145-0750}
\end{figure*}

\begin{figure*}[p]
\centering
\includegraphics[scale=0.8]{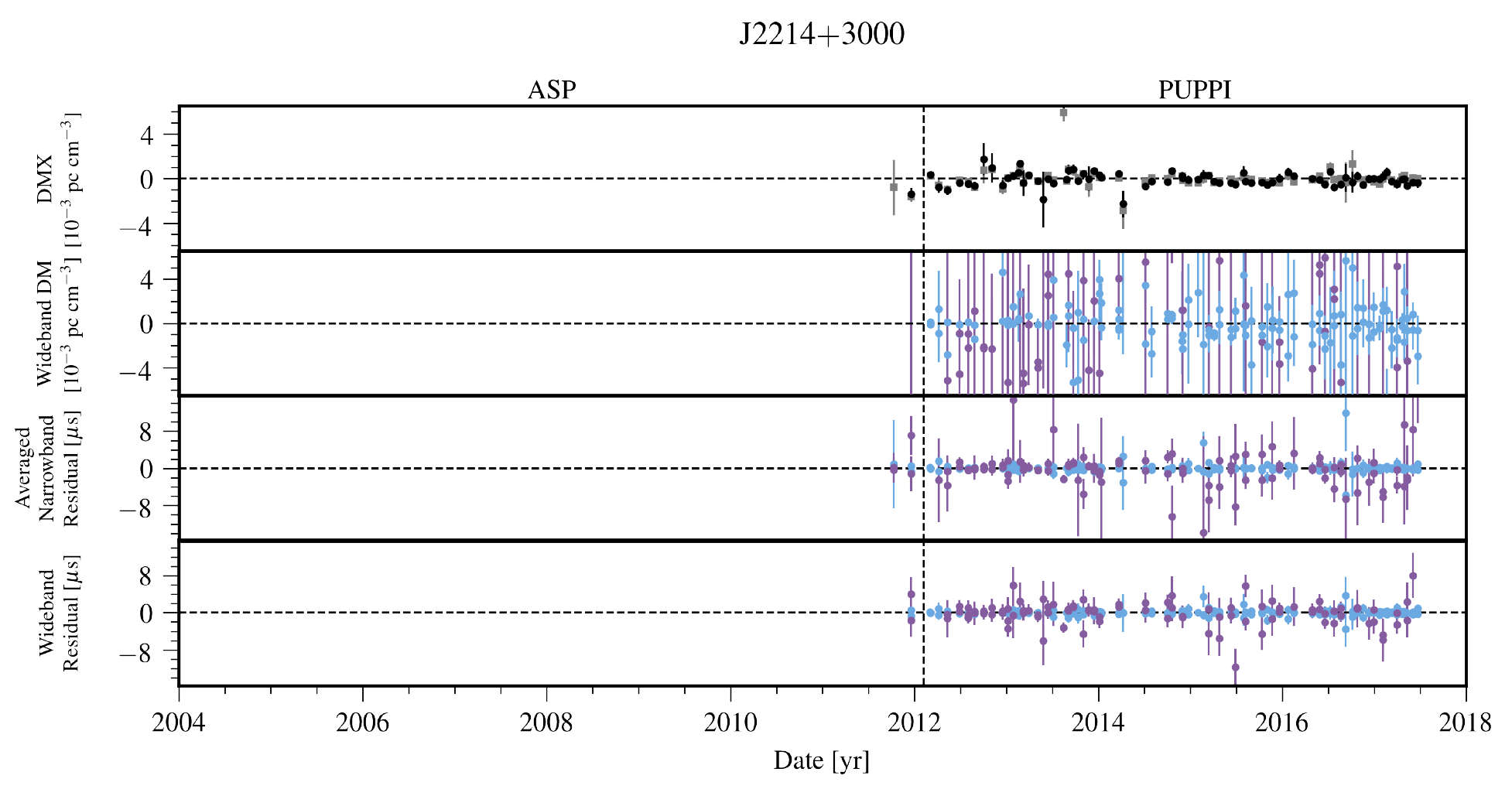}
\caption{Timing residuals and DM variations for J2214$+$3000.  See Appendix~\ref{sec:resid} for details.  DMX model parameters from the wideband (black circles) and narrowband (grey squares) data sets are shown in the top panel.  Colors in the lower panels indicate the receiver for the observation: 1.4~GHz (Light blue) and 2.1~GHz (Purple).}
\label{fig:summary-J2214+3000}
\end{figure*}

\begin{figure*}[p]
\centering
\includegraphics[scale=0.8]{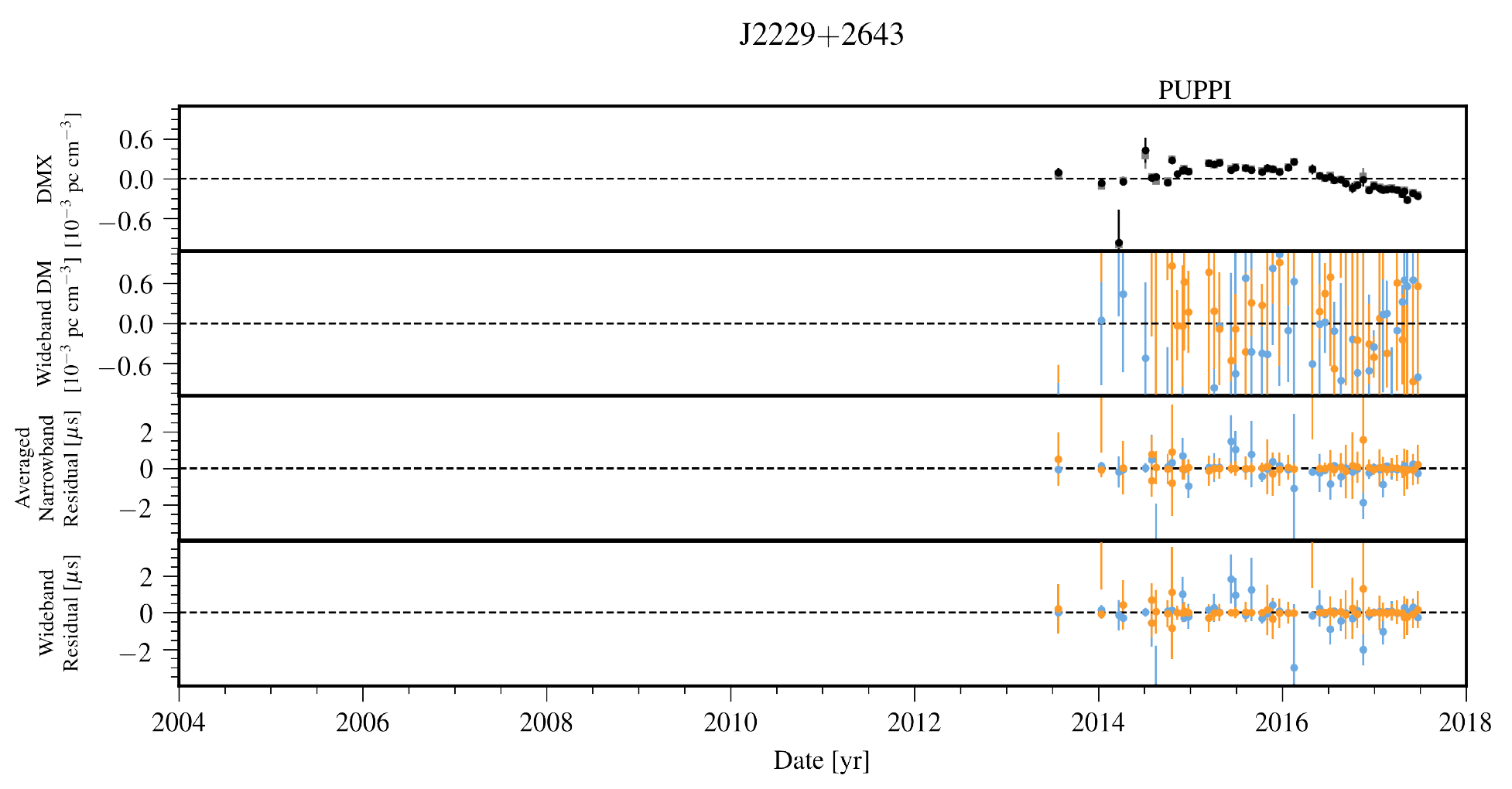}
\caption{Timing residuals and DM variations for J2229$+$2643.  See Appendix~\ref{sec:resid} for details.  DMX model parameters from the wideband (black circles) and narrowband (grey squares) data sets are shown in the top panel.  Colors in the lower panels indicate the receiver for the observation: 430~MHz (Orange) and 1.4~GHz (Light blue).}
\label{fig:summary-J2229+2643}
\end{figure*}

\begin{figure*}[p]
\centering
\includegraphics[scale=0.8]{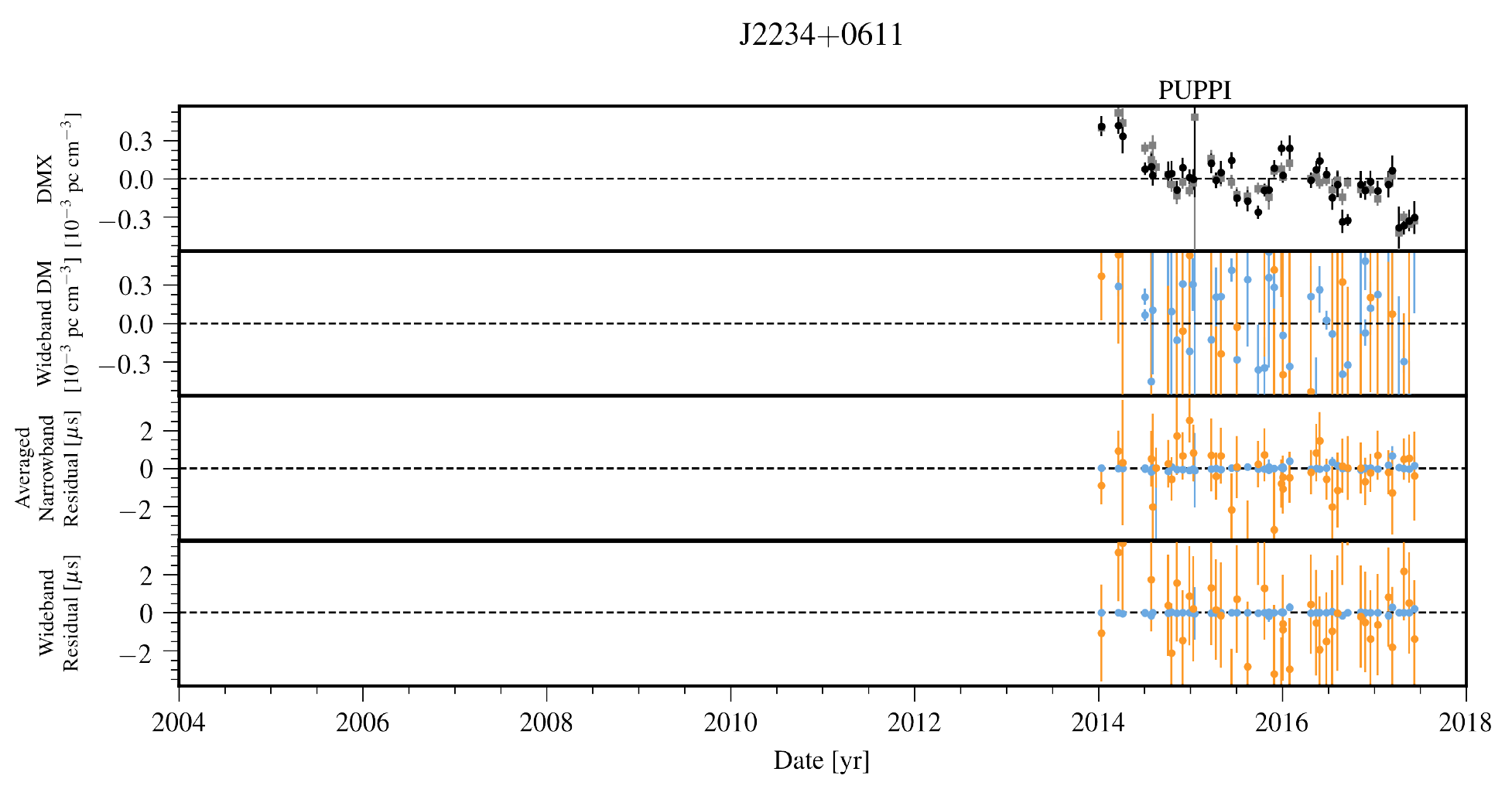}
\caption{Timing residuals and DM variations for J2234$+$0611.  See Appendix~\ref{sec:resid} for details.  DMX model parameters from the wideband (black circles) and narrowband (grey squares) data sets are shown in the top panel.  Colors in the lower panels indicate the receiver for the observation: 430~MHz (Orange) and 1.4~GHz (Light blue).}
\label{fig:summary-J2234+0611}
\end{figure*}

\begin{figure*}[p]
\centering
\includegraphics[scale=0.8]{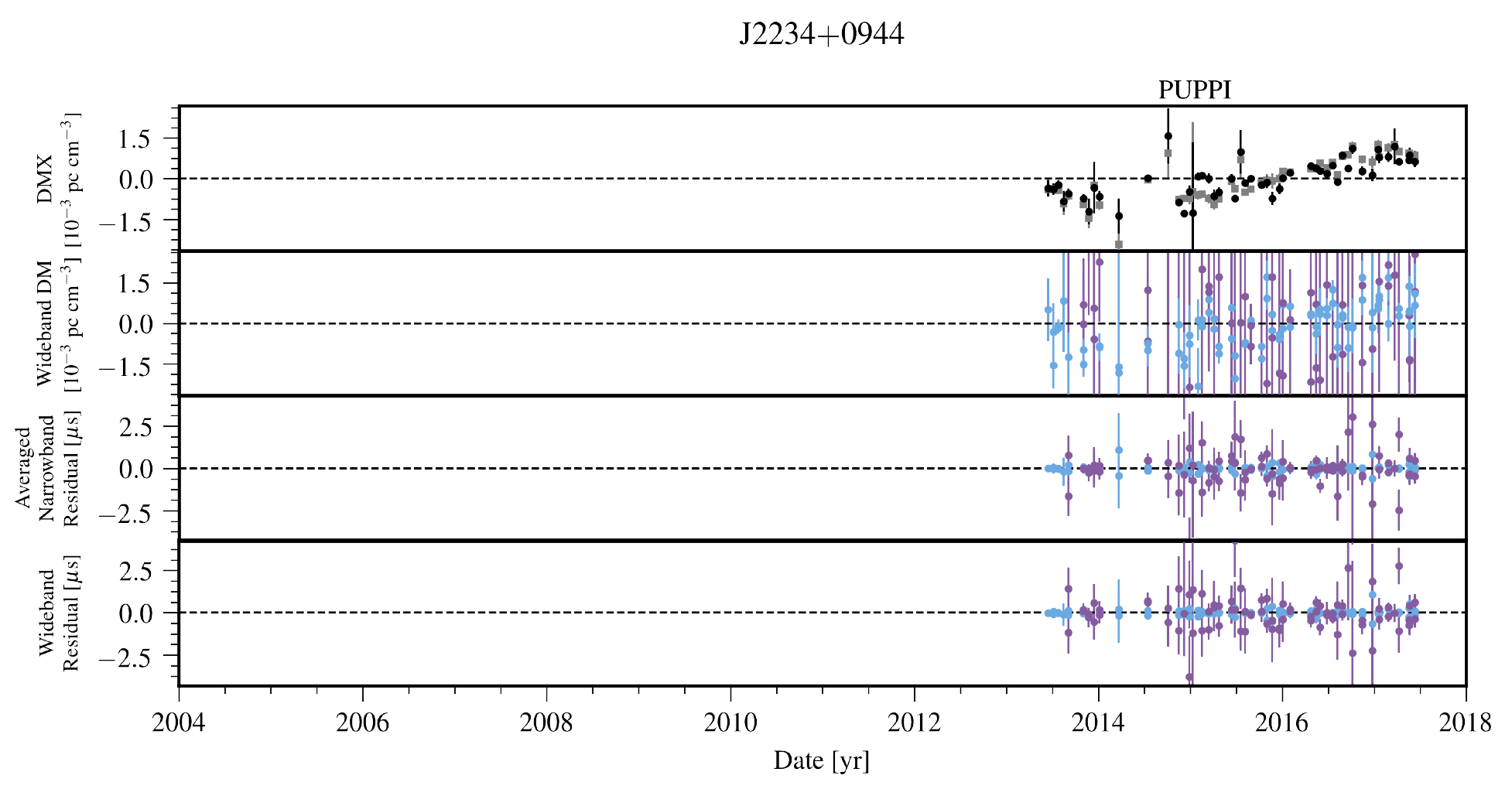}
\caption{Timing residuals and DM variations for J2234$+$0944.  See Appendix~\ref{sec:resid} for details.  DMX model parameters from the wideband (black circles) and narrowband (grey squares) data sets are shown in the top panel.  Colors in the lower panels indicate the receiver for the observation: 1.4~GHz (Light blue) and 2.1~GHz (Purple).}
\label{fig:summary-J2234+0944}
\end{figure*}

\begin{figure*}[p]
\centering
\includegraphics[scale=0.8]{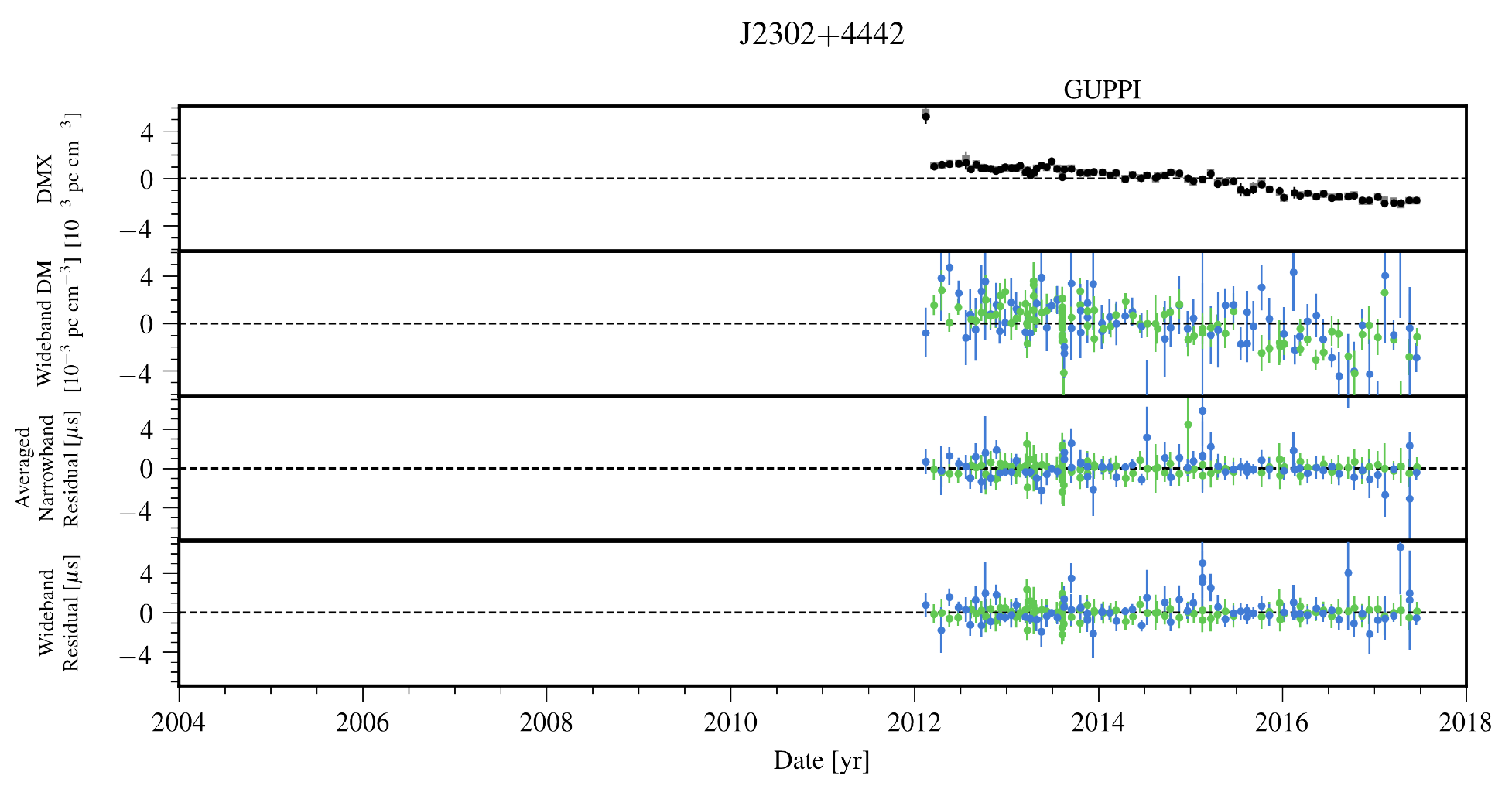}
\caption{Timing residuals and DM variations for J2302$+$4442.  See Appendix~\ref{sec:resid} for details.  DMX model parameters from the wideband (black circles) and narrowband (grey squares) data sets are shown in the top panel.  Colors in the lower panels indicate the receiver for the observation: 820~MHz (Green) and 1.4~GHz (Dark blue).}
\label{fig:summary-J2302+4442}
\end{figure*}

\begin{figure*}[p]
\centering
\includegraphics[scale=0.8]{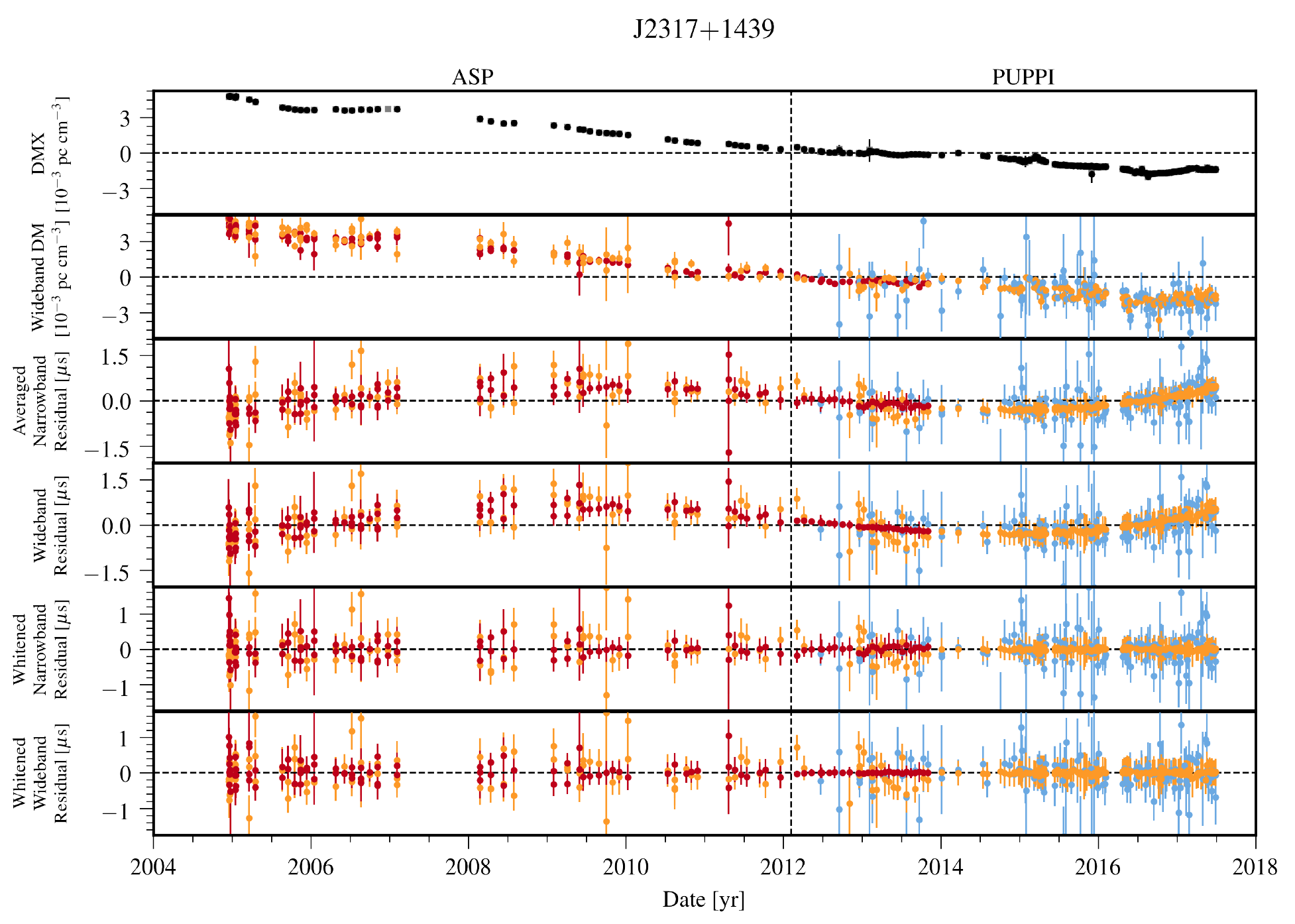}
\caption{Timing residuals and DM variations for J2317$+$1439.  See Appendix~\ref{sec:resid} for details.  DMX model parameters from the wideband (black circles) and narrowband (grey squares) data sets are shown in the top panel.  Colors in the lower panels indicate the receiver for the observation: 327~MHz (Red), 430~MHz (Orange), and 1.4~GHz (Light blue).}
\label{fig:summary-J2317+1439}
\end{figure*}

\begin{figure*}[p]
\centering
\includegraphics[scale=0.8]{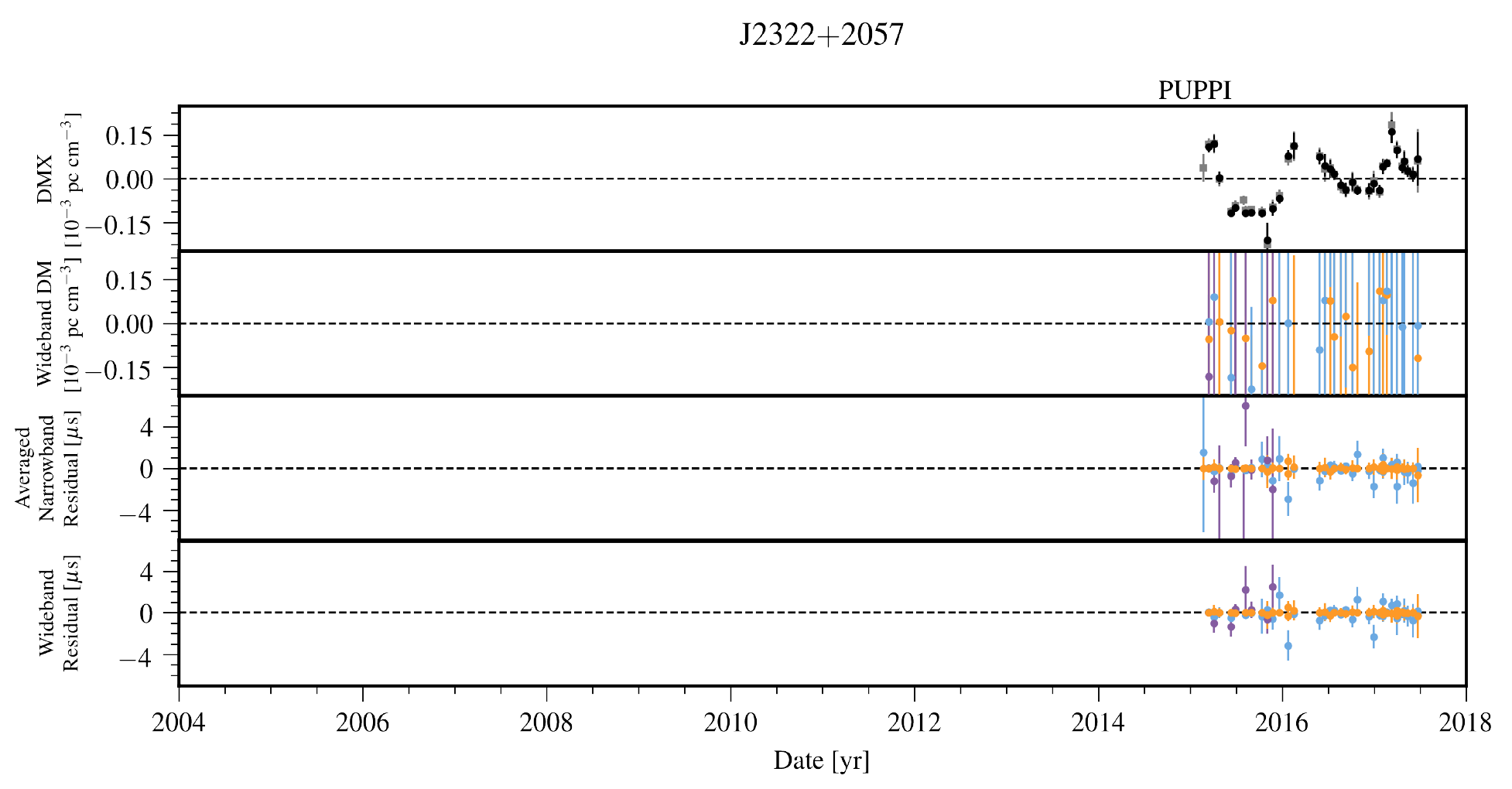}
\caption{Timing residuals and DM variations for J2322$+$2057.  See Appendix~\ref{sec:resid} for details.  DMX model parameters from the wideband (black circles) and narrowband (grey squares) data sets are shown in the top panel.  Colors in the lower panels indicate the receiver for the observation: 430~MHz (Orange), 1.4~GHz (Light blue), and 2.1~GHz (Purple).}
\label{fig:summary-J2322+2057}
\end{figure*}

%% file: nanograv_12y_wb.bbl
\begin{thebibliography}{}
\expandafter\ifx\csname natexlab\endcsname\relax\def\natexlab#1{#1}\fi
\providecommand{\url}[1]{\href{#1}{#1}}
\providecommand{\dodoi}[1]{doi:~\href{http://doi.org/#1}{\nolinkurl{#1}}}
\providecommand{\doeprint}[1]{\href{http://ascl.net/#1}{\nolinkurl{http://ascl.net/#1}}}
\providecommand{\doarXiv}[1]{\href{https://arxiv.org/abs/#1}{\nolinkurl{https://arxiv.org/abs/#1}}}

\bibitem[{{Alam} {et~al.}(2020){Alam}, {Arzoumanian}, {Baker}, {Blumer},
  {Bohler}, {Brazier}, {Brook}, {Burke-Spolaor}, {Caballero}, {Camuccio},
  {Chamberlain}, {Chatterjee}, {Cordes}, {Cornish}, {Crawford}, {Cromartie},
  {DeCesar}, {Demorest}, {Dolch}, {Ellis}, {Ferdman}, {Ferrara}, {Fiore},
  {Fonseca}, {Garcia}, {Garver-Daniels}, {Gentile}, {Good}, {Gusdorff},
  {Halmrast}, {Hazboun}, {Islo}, {Jennings}, {Jessup}, {Jones}, {Kaiser},
  {Kaplan}, {Kelley}, {Shapiro Key}, {Lam}, {Lazio}, {Lorimer}, {Luo}, {Lynch},
  {Madison}, {Maraccini}, {McLaughlin}, {Mingarelli}, {Ng}, {Nguyen}, {Nice},
  {Pennucci}, {Pol}, {Ramette}, {Ransom}, {Ray}, {Shapiro-Albert}, {Siemens},
  {Simon}, {Spiewak}, {Stairs}, {Stinebring}, {Stovall}, {Swiggum}, {Taylor},
  {Tripepi}, {Vallisneri}, {Vigeland}, {Witt}, \& {Zhu}}]{Alam20a}
{Alam}, M.~F., {Arzoumanian}, Z., {Baker}, P.~T., {et~al.} 2020, arXiv
  e-prints, arXiv:2005.06490.
\newblock \doarXiv{2005.06490}

\bibitem[{{Antoniadis}(2014)}]{Antoniadis14}
{Antoniadis}, J. 2014, \apjl, 797, L24, \dodoi{10.1088/2041-8205/797/2/L24}

\bibitem[{{Arzoumanian} {et~al.}(2014){Arzoumanian}, {Brazier},
  {Burke-Spolaor}, {Chamberlin}, {Chatterjee}, {Cordes}, {Demorest}, {Deng},
  {Dolch}, {Ellis}, {Ferdman}, {Garver-Daniels}, {Jenet}, {Jones}, {Kaspi},
  {Koop}, {Lam}, {Lazio}, {Lommen}, {Lorimer}, {Luo}, {Lynch}, {Madison},
  {McLaughlin}, {McWilliams}, {Nice}, {Palliyaguru}, {Pennucci}, {Ransom},
  {Sesana}, {Siemens}, {Stairs}, {Stinebring}, {Stovall}, {Swiggum},
  {Vallisneri}, {van Haasteren}, {Wang}, {Zhu}, \& {NANOGrav
  Collaboration}}]{Arzoumanian2014}
{Arzoumanian}, Z., {Brazier}, A., {Burke-Spolaor}, S., {et~al.} 2014, \apj,
  794, 141, \dodoi{10.1088/0004-637X/794/2/141}

\bibitem[{{Arzoumanian} {et~al.}(2015){Arzoumanian}, {Brazier},
  {Burke-Spolaor}, {Chamberlin}, {Chatterjee}, {Christy}, {Cordes}, {Cornish},
  {Crowter}, {Demorest}, {Dolch}, {Ellis}, {Ferdman}, {Fonseca},
  {Garver-Daniels}, {Gonzalez}, {Jenet}, {Jones}, {Jones}, {Kaspi}, {Koop},
  {Lam}, {Lazio}, {Levin}, {Lommen}, {Lorimer}, {Luo}, {Lynch}, {Madison},
  {McLaughlin}, {McWilliams}, {Nice}, {Palliyaguru}, {Pennucci}, {Ransom},
  {Siemens}, {Stairs}, {Stinebring}, {Stovall}, {Swiggum}, {Vallisneri}, {van
  Haasteren}, {Wang}, \& {Zhu}}]{Arzoumanian2015b}
---. 2015, \apj, 813, 65, \dodoi{10.1088/0004-637X/813/1/65}

\bibitem[{{Arzoumanian} {et~al.}(2018{\natexlab{a}}){Arzoumanian}, {Brazier},
  {Burke-Spolaor}, {Chamberlin}, {Chatterjee}, {Christy}, {Cordes}, {Cornish},
  {Crawford}, {Thankful Cromartie}, {Crowter}, {DeCesar}, {Demorest}, {Dolch},
  {Ellis}, {Ferdman}, {Ferrara}, {Fonseca}, {Garver-Daniels}, {Gentile},
  {Halmrast}, {Huerta}, {Jenet}, {Jessup}, {Jones}, {Jones}, {Kaplan}, {Lam},
  {Lazio}, {Levin}, {Lommen}, {Lorimer}, {Luo}, {Lynch}, {Madison}, {Matthews},
  {McLaughlin}, {McWilliams}, {Mingarelli}, {Ng}, {Nice}, {Pennucci}, {Ransom},
  {Ray}, {Siemens}, {Simon}, {Spiewak}, {Stairs}, {Stinebring}, {Stovall},
  {Swiggum}, {Taylor}, {Vallisneri}, {van Haasteren}, {Vigeland}, {Zhu}, \&
  {NANOGrav Collaboration}}]{Arzoumanian2018a}
---. 2018{\natexlab{a}}, \apjs, 235, 37, \dodoi{10.3847/1538-4365/aab5b0}

\bibitem[{{Arzoumanian} {et~al.}(2018{\natexlab{b}}){Arzoumanian}, {Baker},
  {Brazier}, {Burke-Spolaor}, {Chamberlin}, {Chatterjee}, {Christy}, {Cordes},
  {Cornish}, {Crawford}, {Thankful Cromartie}, {Crowter}, {DeCesar},
  {Demorest}, {Dolch}, {Ellis}, {Ferdman}, {Ferrara}, {Folkner}, {Fonseca},
  {Garver-Daniels}, {Gentile}, {Haas}, {Hazboun}, {Huerta}, {Islo}, {Jones},
  {Jones}, {Kaplan}, {Kaspi}, {Lam}, {Lazio}, {Levin}, {Lommen}, {Lorimer},
  {Luo}, {Lynch}, {Madison}, {McLaughlin}, {McWilliams}, {Mingarelli}, {Ng},
  {Nice}, {Park}, {Pennucci}, {Pol}, {Ransom}, {Ray}, {Rasskazov}, {Siemens},
  {Simon}, {Spiewak}, {Stairs}, {Stinebring}, {Stovall}, {Swiggum}, {Taylor},
  {Vallisneri}, {van Haasteren}, {Vigeland}, {Zhu}, \& {NANOGrav
  Collaboration}}]{Arzoumanian2018b}
{Arzoumanian}, Z., {Baker}, P.~T., {Brazier}, A., {et~al.} 2018{\natexlab{b}},
  \apj, 859, 47, \dodoi{10.3847/1538-4357/aabd3b}

\bibitem[{{Arzoumanian} {et~al.}(2020){Arzoumanian}, {Baker}, {Blumer},
  {Becsy}, {Brazier}, {Brook}, {Burke-Spolaor}, {Chatterjee}, {Chen}, {Cordes},
  {Cornish}, {Crawford}, {Cromartie}, {DeCesar}, {Demorest}, {Dolch}, {Ellis},
  {Ferrara}, {Fiore}, {Fonseca}, {Garver-Daniels}, {Gentile}, {Good},
  {Hazboun}, {Holgado}, {Islo}, {Jennings}, {Jones}, {Kaiser}, {Kaplan},
  {Kelley}, {Shapiro Key}, {Laal}, {Lam}, {Lazio}, {Lorimer}, {Luo}, {Lynch},
  {Madison}, {McLaughlin}, {Mingarelli}, {Ng}, {Nice}, {Pennucci}, {Pol},
  {Ransom}, {Ray}, {Shapiro-Albert}, {Siemens}, {Simon}, {Spiewak}, {Stairs},
  {Stinebring}, {Stovall}, {Sun}, {Swiggum}, {Taylor}, {Turner}, {Vallisneri},
  {Vigeland}, \& {Witt}}]{Arzoumanian20}
{Arzoumanian}, Z., {Baker}, P.~T., {Blumer}, H., {et~al.} 2020, arXiv e-prints,
  arXiv:2009.04496.
\newblock \doarXiv{2009.04496}

\bibitem[{{Bailes} {et~al.}(2016){Bailes}, {Barr}, {Bhat}, {Brink}, {Buchner},
  {Burgay}, {Camilo}, {Champion}, {Hessels}, {Jameson}, {Johnston},
  {Karastergiou}, {Karuppusamy}, {Kaspi}, {Keith}, {Kramer}, {McLaughlin},
  {Moodley}, {Oslowski}, {Possenti}, {Ransom}, {Rasio}, {Sievers}, {Serylak},
  {Stappers}, {Stairs}, {Theureau}, {van Straten}, {Weltevrede}, \&
  {Wex}}]{Bailes16}
{Bailes}, M., {Barr}, E., {Bhat}, N.~D.~R., {et~al.} 2016, 11.
\newblock \doarXiv{1803.07424}

\bibitem[{{Bailes} {et~al.}(2020){Bailes}, {Jameson}, {Abbate}, {Barr}, {Bhat},
  {Bondonneau}, {Burgay}, {Buchner}, {Camilo}, {Champion}, {Cognard},
  {Demorest}, {Freire}, {Gautam}, {Geyer}, {Griessmeier}, {Guillemot}, {Hu},
  {Jankowski}, {Johnston}, {Karastergiou}, {Karuppusamy}, {Kaur}, {Keith},
  {Kramer}, {van Leeuwen}, {Lower}, {Maan}, {McLaughlin}, {Meyers},
  {Os{\l}owski}, {Oswald}, {Parthasarathy}, {Pennucci}, {Posselt}, {Possenti},
  {Ransom}, {Reardon}, {Ridolfi}, {Schollar}, {Serylak}, {Shaifullah},
  {Shamohammadi}, {Shannon}, {Sobey}, {Song}, {Spiewak}, {Stairs}, {Stappers},
  {van Straten}, {Szary}, {Theureau}, {Venkatraman Krishnan}, {Weltevrede},
  {Wex}, {Abbott}, {Adams}, {Burger}, {Gamatham}, {Gouws}, {Horn}, {Hugo},
  {Joubert}, {Manley}, {McAlpine}, {Passmoor}, {Peens-Hough}, {Ramudzuli},
  {Rust}, {Salie}, {Schwardt}, {Siebrits}, {Van Tonder}, {Van Tonder}, \&
  {Welz}}]{Bailes20}
{Bailes}, M., {Jameson}, A., {Abbate}, F., {et~al.} 2020, \pasa, 37, e028,
  \dodoi{10.1017/pasa.2020.19}

\bibitem[{{Bak Nielsen} {et~al.}(2020){Bak Nielsen}, {Janssen}, {Shaifullah},
  {Verbiest}, {Champion}, {Desvignes}, {Guillemot}, {Karuppusamy}, {Kramer},
  {Lyne}, {Possenti}, {Stappers}, {Bassa}, {Cognard}, {Liu}, \&
  {Theureau}}]{BakNielsen20}
{Bak Nielsen}, A.-S., {Janssen}, G.~H., {Shaifullah}, G., {et~al.} 2020,
  \mnras, 494, 2591, \dodoi{10.1093/mnras/staa874}

\bibitem[{{Barr} {et~al.}(2017){Barr}, {Freire}, {Kramer}, {Champion},
  {Berezina}, {Bassa}, {Lyne}, \& {Stappers}}]{Barr17}
{Barr}, E.~D., {Freire}, P.~C.~C., {Kramer}, M., {et~al.} 2017, \mnras, 465,
  1711, \dodoi{10.1093/mnras/stw2947}

\bibitem[{{Brook} {et~al.}(2018){Brook}, {Karastergiou}, {McLaughlin}, {Lam},
  {Arzoumanian}, {Chatterjee}, {Cordes}, {Crowter}, {DeCesar}, {Demorest},
  {Dolch}, {Ellis}, {Ferdman}, {Ferrara}, {Fonseca}, {Gentile}, {Jones},
  {Jones}, {Lazio}, {Levin}, {Lorimer}, {Lynch}, {Ng}, {Nice}, {Pennucci},
  {Ransom}, {Ray}, {Spiewak}, {Stairs}, {Stinebring}, {Stovall}, {Swiggum}, \&
  {Zhu}}]{Brook18}
{Brook}, P.~R., {Karastergiou}, A., {McLaughlin}, M.~A., {et~al.} 2018, \apj,
  868, 122, \dodoi{10.3847/1538-4357/aae9e3}

\bibitem[{{Burke-Spolaor} {et~al.}(2019){Burke-Spolaor}, {Taylor}, {Charisi},
  {Dolch}, {Hazboun}, {Holgado}, {Kelley}, {Lazio}, {Madison}, {McMann},
  {Mingarelli}, {Rasskazov}, {Siemens}, {Simon}, \& {Smith}}]{bs+19}
{Burke-Spolaor}, S., {Taylor}, S.~R., {Charisi}, M., {et~al.} 2019, \aapr, 27,
  5, \dodoi{10.1007/s00159-019-0115-7}

\bibitem[{{Burt} {et~al.}(2011){Burt}, {Lommen}, \& {Finn}}]{Burt2011}
{Burt}, B.~J., {Lommen}, A.~N., \& {Finn}, L.~S. 2011, \apj, 730, 17,
  \dodoi{10.1088/0004-637X/730/1/17}

\bibitem[{{Cordes} {et~al.}(2019){Cordes}, {McLaughlin}, \& {Nanograv
  Collaboration}}]{Cordes19}
{Cordes}, J., {McLaughlin}, M.~A., \& {Nanograv Collaboration}. 2019, \baas,
  51, 447.
\newblock \doarXiv{1903.08653}

\bibitem[{{Cordes} \& {Downs}(1985)}]{CordesDowns85}
{Cordes}, J.~M., \& {Downs}, G.~S. 1985, \apjs, 59, 343, \dodoi{10.1086/191076}

\bibitem[{{Cordes} \& {Shannon}(2010)}]{CordShan10}
{Cordes}, J.~M., \& {Shannon}, R.~M. 2010, ArXiv e-prints.
\newblock \doarXiv{1010.3785}

\bibitem[{{Cordes} {et~al.}(2016){Cordes}, {Shannon}, \&
  {Stinebring}}]{Cordes16}
{Cordes}, J.~M., {Shannon}, R.~M., \& {Stinebring}, D.~R. 2016, \apj, 817, 16,
  \dodoi{10.3847/0004-637X/817/1/16}

\bibitem[{{Cordes} {et~al.}(1990){Cordes}, {Wolszczan}, {Dewey}, {Blaskiewicz},
  \& {Stinebring}}]{Cordes90}
{Cordes}, J.~M., {Wolszczan}, A., {Dewey}, R.~J., {Blaskiewicz}, M., \&
  {Stinebring}, D.~R. 1990, \apj, 349, 245, \dodoi{10.1086/168310}

\bibitem[{{D'Alessandro} {et~al.}(1995){D'Alessandro}, {McCulloch}, {Hamilton},
  \& {Deshpande}}]{DAlessandro95}
{D'Alessandro}, F., {McCulloch}, P.~M., {Hamilton}, P.~A., \& {Deshpande},
  A.~A. 1995, \mnras, 277, 1033, \dodoi{10.1093/mnras/277.3.1033}

\bibitem[{{Demorest}(2007)}]{DemorestPhDT}
{Demorest}, P.~B. 2007, PhD thesis, University of California, Berkeley

\bibitem[{{Demorest}(2018)}]{nanopipe}
---. 2018, {nanopipe: Calibration and data reduction pipeline for pulsar
  timing}, Astrophysics Source Code Library.
\newblock \doeprint{1803.004}

\bibitem[{{Demorest} {et~al.}(2013){Demorest}, {Ferdman}, {Gonzalez}, {Nice},
  {Ransom}, {Stairs}, {Arzoumanian}, {Brazier}, {Burke-Spolaor}, {Chamberlin},
  {Cordes}, {Ellis}, {Finn}, {Freire}, {Giampanis}, {Jenet}, {Kaspi}, {Lazio},
  {Lommen}, {McLaughlin}, {Palliyaguru}, {Perrodin}, {Shannon}, {Siemens},
  {Stinebring}, {Swiggum}, \& {Zhu}}]{Demorest13}
{Demorest}, P.~B., {Ferdman}, R.~D., {Gonzalez}, M.~E., {et~al.} 2013, \apj,
  762, 94, \dodoi{10.1088/0004-637X/762/2/94}

\bibitem[{{Deng} {et~al.}(2020){Deng}, {Gao}, {Li}, \& {Shao}}]{Deng20}
{Deng}, Z.-L., {Gao}, Z.-F., {Li}, X.-D., \& {Shao}, Y. 2020, \apj, 892, 4,
  \dodoi{10.3847/1538-4357/ab76c4}

\bibitem[{{Desvignes} {et~al.}(2016){Desvignes}, {Caballero}, {Lentati},
  {Verbiest}, {Champion}, {Stappers}, {Janssen}, {Lazarus}, {Os{\l}owski},
  {Babak}, {Bassa}, {Brem}, {Burgay}, {Cognard}, {Gair}, {Graikou},
  {Guillemot}, {Hessels}, {Jessner}, {Jordan}, {Karuppusamy}, {Kramer},
  {Lassus}, {Lazaridis}, {Lee}, {Liu}, {Lyne}, {McKee}, {Mingarelli},
  {Perrodin}, {Petiteau}, {Possenti}, {Purver}, {Rosado}, {Sanidas}, {Sesana},
  {Shaifullah}, {Smits}, {Taylor}, {Theureau}, {Tiburzi}, {van Haasteren}, \&
  {Vecchio}}]{Desvignes16}
{Desvignes}, G., {Caballero}, R.~N., {Lentati}, L., {et~al.} 2016, \mnras, 458,
  3341, \dodoi{10.1093/mnras/stw483}

\bibitem[{Dickey(1971)}]{dickey1971}
Dickey, J.~M. 1971, Ann. Math. Statist., 42, 204,
  \dodoi{10.1214/aoms/1177693507}

\bibitem[{{Donner} {et~al.}(2019){Donner}, {Verbiest}, {Tiburzi},
  {Os{\l}owski}, {Michilli}, {Serylak}, {Anderson}, {Horneffer}, {Kramer}, \&
  {Grie{\ss}meier}}]{Donner19}
{Donner}, J.~Y., {Verbiest}, J.~P.~W., {Tiburzi}, C., {et~al.} 2019, \aap, 624,
  A22, \dodoi{10.1051/0004-6361/201834059}

\bibitem[{{Driessen} {et~al.}(2019){Driessen}, {Janssen}, {Bassa}, {Stappers},
  \& {Stinebring}}]{Driessen19}
{Driessen}, L.~N., {Janssen}, G.~H., {Bassa}, C.~G., {Stappers}, B.~W., \&
  {Stinebring}, D.~R. 2019, \mnras, 483, 1224, \dodoi{10.1093/mnras/sty3192}

\bibitem[{{DuPlain} {et~al.}(2008){DuPlain}, {Ransom}, {Demorest}, {Brandt},
  {Ford}, \& {Shelton}}]{DuPlain2008}
{DuPlain}, R., {Ransom}, S., {Demorest}, P., {et~al.} 2008, in \procspie, Vol.
  7019, Advanced Software and Control for Astronomy II, 70191D,
  \dodoi{10.1117/12.790003}

\bibitem[{{Edwards} {et~al.}(2006){Edwards}, {Hobbs}, \&
  {Manchester}}]{Edwards06}
{Edwards}, R.~T., {Hobbs}, G.~B., \& {Manchester}, R.~N. 2006, \mnras, 372,
  1549, \dodoi{10.1111/j.1365-2966.2006.10870.x}

\bibitem[{Ellis \& van Haasteren(2017)}]{ptmcmc}
Ellis, J., \& van Haasteren, R. 2017, jellis18/PTMCMCSampler: Official Release,
  \dodoi{10.5281/zenodo.1037579}

\bibitem[{{Ellis} {et~al.}(2013){Ellis}, {Siemens}, \& {van
  Haasteren}}]{Ellis13a}
{Ellis}, J.~A., {Siemens}, X., \& {van Haasteren}, R. 2013, \apj, 769, 63,
  \dodoi{10.1088/0004-637X/769/1/63}

\bibitem[{{Ellis} {et~al.}(2019){Ellis}, {Vallisneri}, {Taylor}, \&
  {Baker}}]{enterprise}
{Ellis}, J.~A., {Vallisneri}, M., {Taylor}, S.~R., \& {Baker}, P.~T. 2019,
  {ENTERPRISE: Enhanced Numerical Toolbox Enabling a Robust PulsaR Inference
  SuitE}.
\newblock \doeprint{1912.015}

\bibitem[{{Fonseca} {et~al.}(2019){Fonseca}, {Demorest}, {Ransom}, \&
  {Stairs}}]{Fonseca19}
{Fonseca}, E., {Demorest}, P., {Ransom}, S., \& {Stairs}, I. 2019, \baas, 51,
  425.
\newblock \doarXiv{1903.08194}

\bibitem[{{Fonseca} {et~al.}(2016){Fonseca}, {Pennucci}, {Ellis}, {Stairs},
  {Nice}, {Ransom}, {Demorest}, {Arzoumanian}, {Crowter}, {Dolch}, {Ferdman},
  {Gonzalez}, {Jones}, {Jones}, {Lam}, {Levin}, {McLaughlin}, {Stovall},
  {Swiggum}, \& {Zhu}}]{Fonseca2016}
{Fonseca}, E., {Pennucci}, T.~T., {Ellis}, J.~A., {et~al.} 2016, \apj, 832,
  167, \dodoi{10.3847/0004-637X/832/2/167}

\bibitem[{{Ford} {et~al.}(2010){Ford}, {Demorest}, \& {Ransom}}]{Ford2010}
{Ford}, J.~M., {Demorest}, P., \& {Ransom}, S. 2010, in \procspie, Vol. 7740,
  Software and Cyberinfrastructure for Astronomy, 77400A,
  \dodoi{10.1117/12.857666}

\bibitem[{{Foster} \& {Cordes}(1990)}]{FosterCordes90}
{Foster}, R.~S., \& {Cordes}, J.~M. 1990, \apj, 364, 123,
  \dodoi{10.1086/169393}

\bibitem[{{Freire}(2012)}]{Beacon}
{Freire}, P. 2012.
\newblock \url{http://www3.mpifr-bonn.mpg.de/staff/pfreire/BEACON.html}

\bibitem[{{Freire} \& {Tauris}(2014)}]{Freire14}
{Freire}, P. C.~C., \& {Tauris}, T.~M. 2014, \mnras, 438, L86,
  \dodoi{10.1093/mnrasl/slt164}

\bibitem[{{Geyer} \& {Karastergiou}(2016)}]{Geyer16}
{Geyer}, M., \& {Karastergiou}, A. 2016, \mnras, 462, 2587,
  \dodoi{10.1093/mnras/stw1724}

\bibitem[{{Geyer} {et~al.}(2017){Geyer}, {Karastergiou}, {Kondratiev},
  {Zagkouris}, {Kramer}, {Stappers}, {Grie{\ss}meier}, {Hessels}, {Michilli},
  {Pilia}, \& {Sobey}}]{Geyer17}
{Geyer}, M., {Karastergiou}, A., {Kondratiev}, V.~I., {et~al.} 2017, \mnras,
  470, 2659, \dodoi{10.1093/mnras/stx1151}

\bibitem[{{Goulding} {et~al.}(2019){Goulding}, {Pardo}, {Greene}, {Mingarelli},
  {Nyland}, \& {Strauss}}]{Goulding:2019}
{Goulding}, A.~D., {Pardo}, K., {Greene}, J.~E., {et~al.} 2019, \apjl, 879,
  L21, \dodoi{10.3847/2041-8213/ab2a14}

\bibitem[{{Hallinan} {et~al.}(2019){Hallinan}, {Ravi}, {Weinreb}, {Kocz},
  {Huang}, {Woody}, {Lamb}, {D'Addario}, {Catha}, {Law}, {Kulkarni}, {Phinney},
  {Eastwood}, {Bouman}, {McLaughlin}, {Ransom}, {Siemens}, {Cordes}, {Lynch},
  {Kaplan}, {Brazier}, {Bhatnagar}, {Myers}, {Walter}, \&
  {Gaensler}}]{Hallinan19}
{Hallinan}, G., {Ravi}, V., {Weinreb}, S., {et~al.} 2019, in \baas, Vol.~51,
  255.
\newblock \doarXiv{1907.07648}

\bibitem[{{Harding} {et~al.}(1990){Harding}, {Shinbrot}, \&
  {Cordes}}]{Harding90}
{Harding}, A.~K., {Shinbrot}, T., \& {Cordes}, J.~M. 1990, \apj, 353, 588,
  \dodoi{10.1086/168648}

\bibitem[{{Hassall} {et~al.}(2012){Hassall}, {Stappers}, {Hessels}, {Kramer},
  {Alexov}, {Anderson}, {Coenen}, {Karastergiou}, {Keane}, {Kondratiev},
  {Lazaridis}, {van Leeuwen}, {Noutsos}, {Serylak}, {Sobey}, {Verbiest},
  {Weltevrede}, {Zagkouris}, {Fender}, {Wijers}, {B{\"a}hren}, {Bell},
  {Broderick}, {Corbel}, {Daw}, {Dhillon}, {Eisl{\"o}ffel}, {Falcke},
  {Grie{\ss}meier}, {Jonker}, {Law}, {Markoff}, {Miller-Jones}, {Osten}, {Rol},
  {Scaife}, {Scheers}, {Schellart}, {Spreeuw}, {Swinbank}, {ter Veen}, {Wise},
  {Wijnands}, {Wucknitz}, {Zarka}, {Asgekar}, {Bell}, {Bentum}, {Bernardi},
  {Best}, {Bonafede}, {Boonstra}, {Brentjens}, {Brouw}, {Br{\"u}ggen},
  {Butcher}, {Ciardi}, {Garrett}, {Gerbers}, {Gunst}, {van Haarlem}, {Heald},
  {Hoeft}, {Holties}, {de Jong}, {Koopmans}, {Kuniyoshi}, {Kuper}, {Loose},
  {Maat}, {Masters}, {McKean}, {Meulman}, {Mevius}, {Munk}, {Noordam},
  {Orr{\'u}}, {Paas}, {Pandey-Pommier}, {Pandey}, {Pizzo}, {Polatidis},
  {Reich}, {R{\"o}ttgering}, {Sluman}, {Steinmetz}, {Sterks}, {Tagger}, {Tang},
  {Tasse}, {Vermeulen}, {van Weeren}, {Wijnholds}, \& {Yatawatta}}]{Hassall12}
{Hassall}, T.~E., {Stappers}, B.~W., {Hessels}, J.~W.~T., {et~al.} 2012, \aap,
  543, A66, \dodoi{10.1051/0004-6361/201218970}

\bibitem[{{Helfand} {et~al.}(1975){Helfand}, {Manchester}, \&
  {Taylor}}]{Helfand75}
{Helfand}, D.~J., {Manchester}, R.~N., \& {Taylor}, J.~H. 1975, \apj, 198, 661,
  \dodoi{10.1086/153644}

\bibitem[{{Hobbs}(2013)}]{Hobbs13}
{Hobbs}, G. 2013, Classical and Quantum Gravity, 30, 224007,
  \dodoi{10.1088/0264-9381/30/22/224007}

\bibitem[{{Hobbs} {et~al.}(2019){Hobbs}, {Dai}, {Manchester}, {Shannon},
  {Kerr}, {Lee}, \& {Xu}}]{Hobbs19}
{Hobbs}, G., {Dai}, S., {Manchester}, R.~N., {et~al.} 2019, Research in
  Astronomy and Astrophysics, 19, 020, \dodoi{10.1088/1674-4527/19/2/20}

\bibitem[{{Hobbs} \& {Edwards}(2012)}]{tempo2}
{Hobbs}, G., \& {Edwards}, R. 2012, {Tempo2: Pulsar Timing Package}.
\newblock \doeprint{1210.015}

\bibitem[{{Hobbs} {et~al.}(2020){Hobbs}, {Manchester}, {Dunning}, {Jameson},
  {Roberts}, {George}, {Green}, {Tuthill}, {Toomey}, {Kaczmarek}, {Mader},
  {Marquarding}, {Ahmed}, {Amy}, {Bailes}, {Beresford}, {Bhat}, {Bock},
  {Bourne}, {Bowen}, {Brothers}, {Cameron}, {Carretti}, {Carter}, {Castillo},
  {Chekkala}, {Cheng}, {Chung}, {Craig}, {Dai}, {Dawson}, {Dempsey}, {Doherty},
  {Dong}, {Edwards}, {Ergesh}, {Gao}, {Han}, {Hayman}, {Indermuehle},
  {Jeganathan}, {Johnston}, {Kanoniuk}, {Kesteven}, {Kramer}, {Leach},
  {Mcintyre}, {Moss}, {Os{\l}owski}, {Phillips}, {Pope}, {Preisig}, {Price},
  {Reeves}, {Reilly}, {Reynolds}, {Robishaw}, {Roush}, {Ruckley}, {Sadler},
  {Sarkissian}, {Severs}, {Shannon}, {Smart}, {Smith}, {Smith}, {Sobey},
  {Staveley-Smith}, {Tzioumis}, {van Straten}, {Wang}, {Wen}, \&
  {Whiting}}]{Hobbs20}
{Hobbs}, G., {Manchester}, R.~N., {Dunning}, A., {et~al.} 2020, \pasa, 37,
  e012, \dodoi{10.1017/pasa.2020.2}

\bibitem[{{Hotan} {et~al.}(2004){Hotan}, {van Straten}, \&
  {Manchester}}]{Hotan04}
{Hotan}, A.~W., {van Straten}, W., \& {Manchester}, R.~N. 2004, Proc. Astron.
  Soc. Aust., 21, 302, \dodoi{10.1071/AS04022}

\bibitem[{Hunter(2007)}]{matplotlib}
Hunter, J.~D. 2007, Computing In Science \& Engineering, 9, 90,
  \dodoi{10.1109/MCSE.2007.55}

\bibitem[{{Jaffe} \& {Backer}(2003)}]{Jaffe2003}
{Jaffe}, A.~H., \& {Backer}, D.~C. 2003, \apj, 583, 616, \dodoi{10.1086/345443}

\bibitem[{{Jiang} {et~al.}(2015){Jiang}, {Li}, {Dey}, \& {Dey}}]{Jiang15}
{Jiang}, L., {Li}, X.-D., {Dey}, J., \& {Dey}, M. 2015, \apj, 807, 41,
  \dodoi{10.1088/0004-637X/807/1/41}

\bibitem[{{Jones} {et~al.}(2017){Jones}, {McLaughlin}, {Lam}, {Cordes},
  {Levin}, {Chatterjee}, {Arzoumanian}, {Crowter}, {Demorest}, {Dolch},
  {Ellis}, {Ferdman}, {Fonseca}, {Gonzalez}, {Jones}, {Lazio}, {Nice},
  {Pennucci}, {Ransom}, {Stinebring}, {Stairs}, {Stovall}, {Swiggum}, \&
  {Zhu}}]{Jones17}
{Jones}, M.~L., {McLaughlin}, M.~A., {Lam}, M.~T., {et~al.} 2017, \apj, 841,
  125, \dodoi{10.3847/1538-4357/aa73df}

\bibitem[{{Joshi} {et~al.}(2018){Joshi}, {Arumugasamy}, {Bagchi},
  {Bandyopadhyay}, {Basu}, {Dhand a Batra}, {Bethapudi}, {Choudhary}, {De},
  {Dey}, {Gopakumar}, {Gupta}, {Krishnakumar}, {Maan}, {Manoharan}, {Naidu},
  {Nandi}, {Pathak}, {Surnis}, \& {Susobhanan}}]{Joshi18}
{Joshi}, B.~C., {Arumugasamy}, P., {Bagchi}, M., {et~al.} 2018, Journal of
  Astrophysics and Astronomy, 39, 51, \dodoi{10.1007/s12036-018-9549-y}

\bibitem[{{Keith} {et~al.}(2013){Keith}, {Coles}, {Shannon}, {Hobbs},
  {Manchester}, {Bailes}, {Bhat}, {Burke-Spolaor}, {Champion}, {Chaudhary},
  {Hotan}, {Khoo}, {Kocz}, {Os{\l}owski}, {Ravi}, {Reynolds}, {Sarkissian},
  {van Straten}, \& {Yardley}}]{Keith13}
{Keith}, M.~J., {Coles}, W., {Shannon}, R.~M., {et~al.} 2013, \mnras, 429,
  2161, \dodoi{10.1093/mnras/sts486}

\bibitem[{{Kelley} {et~al.}(2019){Kelley}, {Charisi}, {Burke-Spolaor}, {Simon},
  {Blecha}, {Bogdanovic}, {Colpi}, {Comerford}, {D'Orazio}, {Dotti},
  {Eracleous}, {Graham}, {Greene}, {Haiman}, {Holley-Bockelmann}, {Kara},
  {Kelly}, {Komossa}, {Larson}, {Liu}, {Ma}, {Noble}, {Paschalidis}, {Rafikov},
  {Ravi}, {Runnoe}, {Sesana}, {Stern}, {Strauss}, {U}, {Volonteri}, \&
  {Nanograv Collaboration}}]{Kelley19}
{Kelley}, L., {Charisi}, M., {Burke-Spolaor}, S., {et~al.} 2019, \baas, 51,
  490.
\newblock \doarXiv{1903.07644}

\bibitem[{{Kerr} {et~al.}(2020){Kerr}, {Reardon}, {Hobbs}, {Shannon},
  {Manchester}, {Dai}, {Russell}, {Zhang}, {van Straten}, {Os{\l}owski},
  {Parthasarathy}, {Spiewak}, {Bailes}, {Bhat}, {Cameron}, {Coles}, {Dempsey},
  {Deng}, {Goncharov}, {Kaczmarek}, {Keith}, {Lasky}, {Lower}, {Preisig},
  {Sarkissian}, {Toomey}, {Wang}, {Wang}, {Zhang}, \& {Zhu}}]{Kerr20}
{Kerr}, M., {Reardon}, D.~J., {Hobbs}, G., {et~al.} 2020, \pasa, 37, e020,
  \dodoi{10.1017/pasa.2020.11}

\bibitem[{{Kramer} \& {Champion}(2013)}]{Kramer13}
{Kramer}, M., \& {Champion}, D.~J. 2013, Classical and Quantum Gravity, 30,
  224009, \dodoi{10.1088/0264-9381/30/22/224009}

\bibitem[{Kurosawa {et~al.}(2001)Kurosawa, Kobayashi, Maruyama, Sugawara, \&
  Kobayashi}]{Kurosawa01}
Kurosawa, N., Kobayashi, H., Maruyama, K., Sugawara, H., \& Kobayashi, K. 2001,
  Circuits and Systems I: Fundamental Theory and Applications, IEEE
  Transactions on, 48, 261 , \dodoi{10.1109/81.915383}

\bibitem[{{Lam}(2017)}]{pypulse}
{Lam}, M.~T. 2017, {PyPulse: PSRFITS handler}.
\newblock \doeprint{1706.011}

\bibitem[{{Lam}(2018)}]{Lam18c}
---. 2018, \apj, 868, 33, \dodoi{10.3847/1538-4357/aae533}

\bibitem[{{Lam} {et~al.}(2016{\natexlab{a}}){Lam}, {Cordes}, {Chatterjee},
  {Jones}, {McLaughlin}, \& {Armstrong}}]{Lam2016}
{Lam}, M.~T., {Cordes}, J.~M., {Chatterjee}, S., {et~al.} 2016{\natexlab{a}},
  \apj, 821, 66, \dodoi{10.3847/0004-637X/821/1/66}

\bibitem[{{Lam} {et~al.}(2020){Lam}, {Lazio}, {Dolch}, {Jones}, {McLaughlin},
  {Stinebring}, \& {Surnis}}]{Lam19}
{Lam}, M.~T., {Lazio}, T.~J.~W., {Dolch}, T., {et~al.} 2020, \apj, 892, 89,
  \dodoi{10.3847/1538-4357/ab7b6b}

\bibitem[{{Lam} {et~al.}(2018{\natexlab{a}}){Lam}, {McLaughlin}, {Cordes},
  {Chatterjee}, \& {Lazio}}]{Lam18b}
{Lam}, M.~T., {McLaughlin}, M.~A., {Cordes}, J.~M., {Chatterjee}, S., \&
  {Lazio}, T.~J.~W. 2018{\natexlab{a}}, \apj, 861, 12,
  \dodoi{10.3847/1538-4357/aac48d}

\bibitem[{{Lam} {et~al.}(2016{\natexlab{b}}){Lam}, {Cordes}, {Chatterjee},
  {Arzoumanian}, {Crowter}, {Demorest}, {Dolch}, {Ellis}, {Ferdman}, {Fonseca},
  {Gonzalez}, {Jones}, {Jones}, {Levin}, {Madison}, {McLaughlin}, {Nice},
  {Pennucci}, {Ransom}, {Siemens}, {Stairs}, {Stovall}, {Swiggum}, \&
  {Zhu}}]{Lam16}
{Lam}, M.~T., {Cordes}, J.~M., {Chatterjee}, S., {et~al.} 2016{\natexlab{b}},
  \apj, 819, 155, \dodoi{10.3847/0004-637X/819/2/155}

\bibitem[{{Lam} {et~al.}(2017){Lam}, {Cordes}, {Chatterjee}, {Arzoumanian},
  {Crowter}, {Demorest}, {Dolch}, {Ellis}, {Ferdman}, {Fonseca}, {Gonzalez},
  {Jones}, {Jones}, {Levin}, {Madison}, {McLaughlin}, {Nice}, {Pennucci},
  {Ransom}, {Shannon}, {Siemens}, {Stairs}, {Stovall}, {Swiggum}, \&
  {Zhu}}]{Lam17a}
---. 2017, \apj, 834, 35, \dodoi{10.3847/1538-4357/834/1/35}

\bibitem[{{Lam} {et~al.}(2018{\natexlab{b}}){Lam}, {Ellis}, {Grillo}, {Jones},
  {Hazboun}, {Brook}, {Turner}, {Chatterjee}, {Cordes}, {Lazio}, {DeCesar},
  {Arzoumanian}, {Blumer}, {Cromartie}, {Demorest}, {Dolch}, {Ferdman},
  {Ferrara}, {Fonseca}, {Garver-Daniels}, {Gentile}, {Gupta}, {Lorimer},
  {Lynch}, {Madison}, {McLaughlin}, {Ng}, {Nice}, {Pennucci}, {Ransom},
  {Spiewak}, {Stairs}, {Stinebring}, {Stovall}, {Swiggum}, {Vigeland}, \&
  {Zhu}}]{Lam2018}
{Lam}, M.~T., {Ellis}, J.~A., {Grillo}, G., {et~al.} 2018{\natexlab{b}}, \apj,
  861, 132, \dodoi{10.3847/1538-4357/aac770}

\bibitem[{{Lam} {et~al.}(2019){Lam}, {McLaughlin}, {Arzoumanian}, {Blumer},
  {Brook}, {Cromartie}, {Demorest}, {DeCesar}, {Dolch}, {Ellis}, {Ferdman},
  {Ferrara}, {Fonseca}, {Garver-Daniels}, {Gentile}, {Jones}, {Lorimer},
  {Lynch}, {Ng}, {Nice}, {Pennucci}, {Ransom}, {Spiewak}, {Stairs}, {Stovall},
  {Swiggum}, {Vigeland}, \& {Zhu}}]{Lam19b}
{Lam}, M.~T., {McLaughlin}, M.~A., {Arzoumanian}, Z., {et~al.} 2019, \apj, 872,
  193, \dodoi{10.3847/1538-4357/ab01cd}

\bibitem[{{Lee}(2016)}]{Lee16}
{Lee}, K.~J. 2016, Astronomical Society of the Pacific Conference Series, Vol.
  502, {Prospects of Gravitational Wave Detection Using Pulsar Timing Array for
  Chinese Future Telescopes}, ed. L.~{Qain} \& D.~{Li}, 19

\bibitem[{{Lee} {et~al.}(2014){Lee}, {Bassa}, {Janssen}, {Karuppusamy},
  {Kramer}, {Liu}, {Perrodin}, {Smits}, {Stappers}, {van Haasteren}, \&
  {Lentati}}]{Lee14}
{Lee}, K.~J., {Bassa}, C.~G., {Janssen}, G.~H., {et~al.} 2014, \mnras, 441,
  2831, \dodoi{10.1093/mnras/stu664}

\bibitem[{{Lentati} {et~al.}(2013{\natexlab{a}}){Lentati}, {Alexander}, \&
  {Hobson}}]{Lentati13}
{Lentati}, L., {Alexander}, P., \& {Hobson}, M.~P. 2013{\natexlab{a}}, ArXiv
  e-prints.
\newblock \doarXiv{1312.2403}

\bibitem[{{Lentati} {et~al.}(2015){Lentati}, {Alexander}, \&
  {Hobson}}]{Lentati15a}
---. 2015, \mnras, 447, 2159, \dodoi{10.1093/mnras/stu2611}

\bibitem[{{Lentati} {et~al.}(2014){Lentati}, {Alexander}, {Hobson}, {Feroz},
  {van Haasteren}, {Lee}, \& {Shannon}}]{Lentati14}
{Lentati}, L., {Alexander}, P., {Hobson}, M.~P., {et~al.} 2014, \mnras, 437,
  3004, \dodoi{10.1093/mnras/stt2122}

\bibitem[{{Lentati} {et~al.}(2013{\natexlab{b}}){Lentati}, {Alexander},
  {Hobson}, {Taylor}, {Gair}, {Balan}, \& {van Haasteren}}]{Lentati13b}
---. 2013{\natexlab{b}}, \prd, 87, 104021, \dodoi{10.1103/PhysRevD.87.104021}

\bibitem[{{Lentati} {et~al.}(2017{\natexlab{a}}){Lentati}, {Kerr}, {Dai},
  {Shannon}, {Hobbs}, \& {Os{\l}owski}}]{lkd+17}
{Lentati}, L., {Kerr}, M., {Dai}, S., {et~al.} 2017{\natexlab{a}}, \mnras, 468,
  1474, \dodoi{10.1093/mnras/stx580}

\bibitem[{{Lentati} {et~al.}(2017{\natexlab{b}}){Lentati}, {Kerr}, {Dai},
  {Hobson}, {Shannon}, {Hobbs}, {Bailes}, {Bhat}, {Burke-Spolaor}, {Coles},
  {Dempsey}, {Lasky}, {Levin}, {Manchester}, {Os{\l}owski}, {Ravi}, {Reardon},
  {Rosado}, {Spiewak}, {van Straten}, {Toomey}, {Wang}, {Wen}, {You}, \&
  {Zhu}}]{Lentati17a}
---. 2017{\natexlab{b}}, \mnras, 466, 3706, \dodoi{10.1093/mnras/stw3359}

\bibitem[{{Levin} {et~al.}(2016){Levin}, {McLaughlin}, {Jones}, {Cordes},
  {Stinebring}, {Chatterjee}, {Dolch}, {Lam}, {Lazio}, {Palliyaguru},
  {Arzoumanian}, {Crowter}, {Demorest}, {Ellis}, {Ferdman}, {Fonseca},
  {Gonzalez}, {Jones}, {Nice}, {Pennucci}, {Ransom}, {Stairs}, {Stovall},
  {Swiggum}, \& {Zhu}}]{Levin16}
{Levin}, L., {McLaughlin}, M.~A., {Jones}, G., {et~al.} 2016, \apj, 818, 166,
  \dodoi{10.3847/0004-637X/818/2/166}

\bibitem[{{Liu} {et~al.}(2014){Liu}, {Desvignes}, {Cognard}, {Stappers},
  {Verbiest}, {Lee}, {Champion}, {Kramer}, {Freire}, \& {Karuppusamy}}]{Liu14}
{Liu}, K., {Desvignes}, G., {Cognard}, I., {et~al.} 2014, \mnras, 443, 3752,
  \dodoi{10.1093/mnras/stu1420}

\bibitem[{{Lommen} \& {Demorest}(2013)}]{Lommen13}
{Lommen}, A.~N., \& {Demorest}, P. 2013, Classical and Quantum Gravity, 30,
  224001, \dodoi{10.1088/0264-9381/30/22/224001}

\bibitem[{{Lorimer} \& {Kramer}(2005)}]{L&K05}
{Lorimer}, D.~R., \& {Kramer}, M. 2005, {Handbook of Pulsar Astronomy}, ed.
  R.~{Ellis}, J.~{Huchra}, S.~{Kahn}, G.~{Rieke}, \& P.~B. {Stetson} (The Press
  Syndicate of the University of Cambridge)

\bibitem[{{Luo} {et~al.}(2019){Luo}, {Ransom}, {Demorest}, {van Haasteren},
  {Ray}, {Stovall}, {Bachetti}, {Archibald}, {Kerr}, {Colen}, \&
  {Jenet}}]{PINT}
{Luo}, J., {Ransom}, S., {Demorest}, P., {et~al.} 2019, {PINT: High-precision
  pulsar timing analysis package}.
\newblock \doeprint{1902.007}

\bibitem[{{Luo} {et~al.}(2020){Luo}, {Ransom}, {Demorest}, {Ray}, {Archibald},
  {Kerr}, {Jennings}, {Bachetti}, {van Haasteren}, {Champagne}, {Colen},
  {Phillips}, {Zimmerman}, {Stovall}, {Lam}, \& {Jenet}}]{Luo21}
---. 2020, arXiv e-prints, arXiv:2012.00074.
\newblock \doarXiv{2012.00074}

\bibitem[{{Lynch} {et~al.}(2019){Lynch}, {Brook}, {Chatterjee}, {Dolch},
  {Kramer}, {Lam}, {Lewand owska}, {McLaughlin}, {Pol}, \& {Stairs}}]{Lynch19}
{Lynch}, R., {Brook}, P., {Chatterjee}, S., {et~al.} 2019, \baas, 51, 461

\bibitem[{{Maitia} {et~al.}(2003){Maitia}, {Lestrade}, \& {Cognard}}]{Maitia03}
{Maitia}, V., {Lestrade}, J.-F., \& {Cognard}, I. 2003, \apj, 582, 972,
  \dodoi{10.1086/344816}

\bibitem[{{Manchester} \& {IPTA}(2013)}]{Manchester13b}
{Manchester}, R.~N., \& {IPTA}. 2013, Classical and Quantum Gravity, 30,
  224010, \dodoi{10.1088/0264-9381/30/22/224010}

\bibitem[{{McKinnon} {et~al.}(2019){McKinnon}, {Beasley}, {Murphy}, {Selina},
  {Farnsworth}, \& {Walter}}]{Mckinnon19}
{McKinnon}, M., {Beasley}, A., {Murphy}, E., {et~al.} 2019, in \baas, Vol.~51,
  81

\bibitem[{{Mingarelli}(2019)}]{Mingarelli2019}
{Mingarelli}, C. M.~F. 2019, Nature Astronomy, 3, 8,
  \dodoi{10.1038/s41550-018-0666-y}

\bibitem[{{Ng}(2018)}]{Ng17}
{Ng}, C. 2018, in IAU Symposium, Vol. 337, Pulsar Astrophysics the Next Fifty
  Years, ed. P.~{Weltevrede}, B.~B.~P. {Perera}, L.~L. {Preston}, \&
  S.~{Sanidas}, 179--182, \dodoi{10.1017/S1743921317010638}

\bibitem[{{Nice} {et~al.}(2015){Nice}, {Demorest}, {Stairs}, {Manchester},
  {Taylor}, {Peters}, {Weisberg}, {Irwin}, {Wex}, \& {Huang}}]{tempo}
{Nice}, D., {Demorest}, P., {Stairs}, I., {et~al.} 2015, {Tempo: Pulsar timing
  data analysis}.
\newblock \doeprint{1509.002}

\bibitem[{{Ocker} {et~al.}(2020){Ocker}, {Cordes}, \& {Chatterjee}}]{Ocker20}
{Ocker}, S.~K., {Cordes}, J.~M., \& {Chatterjee}, S. 2020, \apj, 897, 124,
  \dodoi{10.3847/1538-4357/ab98f9}

\bibitem[{{Os{\l}owski} {et~al.}(2013){Os{\l}owski}, {van Straten}, {Demorest},
  \& {Bailes}}]{Oslowski13}
{Os{\l}owski}, S., {van Straten}, W., {Demorest}, P., \& {Bailes}, M. 2013,
  \mnras, 430, 416, \dodoi{10.1093/mnras/sts662}

\bibitem[{{Os{\l}owski} {et~al.}(2011){Os{\l}owski}, {van Straten}, {Hobbs},
  {Bailes}, \& {Demorest}}]{Oslowski11}
{Os{\l}owski}, S., {van Straten}, W., {Hobbs}, G.~B., {Bailes}, M., \&
  {Demorest}, P. 2011, \mnras, 418, 1258,
  \dodoi{10.1111/j.1365-2966.2011.19578.x}

\bibitem[{{Pennucci}(2015)}]{PennucciPhDT}
{Pennucci}, T.~T. 2015, PhD thesis, University of Virginia.
\newblock \url{http://libra.virginia.edu/catalog/libra-oa:9384}

\bibitem[{{Pennucci}(2019)}]{Pennucci19}
---. 2019, \apj, 871, 34, \dodoi{10.3847/1538-4357/aaf6ef}

\bibitem[{{Pennucci} {et~al.}(2014){Pennucci}, {Demorest}, \& {Ransom}}]{PDR14}
{Pennucci}, T.~T., {Demorest}, P.~B., \& {Ransom}, S.~M. 2014, \apj, 790, 93,
  \dodoi{10.1088/0004-637X/790/2/93}

\bibitem[{{Pennucci} {et~al.}(2016){Pennucci}, {Demorest}, \&
  {Ransom}}]{pulseportraiture}
---. 2016, {Pulse Portraiture: Pulsar Timing}, Astrophysics Source Code
  Library.
\newblock \doeprint{1606.013}

\bibitem[{{Pennucci} {et~al.}(2015){Pennucci}, {Possenti}, {Esposito}, {Rea},
  {Haggard}, {Baganoff}, {Burgay}, {Coti Zelati}, {Israel}, \&
  {Minter}}]{Pennucci15}
{Pennucci}, T.~T., {Possenti}, A., {Esposito}, P., {et~al.} 2015, \apj, 808,
  81, \dodoi{10.1088/0004-637X/808/1/81}

\bibitem[{{Perera} {et~al.}(2019){Perera}, {DeCesar}, {Demorest}, {Kerr},
  {Lentati}, {Nice}, {Os{\l}owski}, {Ransom}, {Keith}, {Arzoumanian}, {Bailes},
  {Baker}, {Bassa}, {Bhat}, {Brazier}, {Burgay}, {Burke-Spolaor}, {Caballero},
  {Champion}, {Chatterjee}, {Chen}, {Cognard}, {Cordes}, {Crowter}, {Dai},
  {Desvignes}, {Dolch}, {Ferdman}, {Ferrara}, {Fonseca}, {Goldstein},
  {Graikou}, {Guillemot}, {Hazboun}, {Hobbs}, {Hu}, {Islo}, {Janssen},
  {Karuppusamy}, {Kramer}, {Lam}, {Lee}, {Liu}, {Luo}, {Lyne}, {Manchester},
  {McKee}, {McLaughlin}, {Mingarelli}, {Parthasarathy}, {Pennucci}, {Perrodin},
  {Possenti}, {Reardon}, {Russell}, {Sanidas}, {Sesana}, {Shaifullah},
  {Shannon}, {Siemens}, {Simon}, {Spiewak}, {Stairs}, {Stappers}, {Swiggum},
  {Taylor}, {Theureau}, {Tiburzi}, {Vallisneri}, {Vecchio}, {Wang}, {Zhang},
  {Zhang}, {Zhu}, \& {Zhu}}]{Perera19}
{Perera}, B.~B.~P., {DeCesar}, M.~E., {Demorest}, P.~B., {et~al.} 2019, \mnras,
  490, 4666, \dodoi{10.1093/mnras/stz2857}

\bibitem[{{Phinney}(2001)}]{Phinney2001}
{Phinney}, E.~S. 2001, arXiv e-prints, astro.
\newblock \doarXiv{astro-ph/0108028}

\bibitem[{{Ramachandran} {et~al.}(2006){Ramachandran}, {Demorest}, {Backer},
  {Cognard}, \& {Lommen}}]{Ramachandran06}
{Ramachandran}, R., {Demorest}, P., {Backer}, D.~C., {Cognard}, I., \&
  {Lommen}, A. 2006, \apj, 645, 303, \dodoi{10.1086/500634}

\bibitem[{{Ransom} {et~al.}(2019){Ransom}, {Brazier}, {Chatterjee}, {Cohen},
  {Cordes}, {DeCesar}, {Demorest}, {Hazboun}, {Lam}, {Lynch}, {McLaughlin},
  {Ransom}, {Siemens}, {Taylor}, \& {Vigeland}}]{Ransom19}
{Ransom}, S., {Brazier}, A., {Chatterjee}, S., {et~al.} 2019, in \baas,
  Vol.~51, 195.
\newblock \doarXiv{1908.05356}

\bibitem[{{Rickett} \& {Lyne}(1990)}]{Rickett90}
{Rickett}, B.~J., \& {Lyne}, A.~G. 1990, \mnras, 244, 68

\bibitem[{{Rosado} {et~al.}(2015){Rosado}, {Sesana}, \&
  {Gair}}]{RosadoSesanaGair:2015}
{Rosado}, P.~A., {Sesana}, A., \& {Gair}, J. 2015, \mnras, 451, 2417,
  \dodoi{10.1093/mnras/stv1098}

\bibitem[{{Shannon} \& {Cordes}(2017)}]{ShannonCordes2017}
{Shannon}, R.~M., \& {Cordes}, J.~M. 2017, \mnras, 464, 2075,
  \dodoi{10.1093/mnras/stw2449}

\bibitem[{{Shannon} {et~al.}(2014){Shannon}, {Os{\l}owski}, {Dai}, {Bailes},
  {Hobbs}, {Manchester}, {van Straten}, {Raithel}, {Ravi}, {Toomey}, {Bhat},
  {Burke-Spolaor}, {Coles}, {Keith}, {Kerr}, {Levin}, {Sarkissian}, {Wang},
  {Wen}, \& {Zhu}}]{Shannon14}
{Shannon}, R.~M., {Os{\l}owski}, S., {Dai}, S., {et~al.} 2014, \mnras, 443,
  1463, \dodoi{10.1093/mnras/stu1213}

\bibitem[{{Shannon} {et~al.}(2016){Shannon}, {Lentati}, {Kerr}, {Bailes},
  {Bhat}, {Coles}, {Dai}, {Dempsey}, {Hobbs}, {Keith}, {Lasky}, {Levin},
  {Manchester}, {Os{\l}owski}, {Ravi}, {Reardon}, {Rosado}, {Spiewak}, {van
  Straten}, {Toomey}, {Wang}, {Wen}, {You}, \& {Zhu}}]{slk+16}
{Shannon}, R.~M., {Lentati}, L.~T., {Kerr}, M., {et~al.} 2016, \apjl, 828, L1,
  \dodoi{10.3847/2041-8205/828/1/L1}

\bibitem[{{Siemens} {et~al.}(2013){Siemens}, {Ellis}, {Jenet}, \&
  {Romano}}]{Siemens2013}
{Siemens}, X., {Ellis}, J., {Jenet}, F., \& {Romano}, J.~D. 2013, Classical and
  Quantum Gravity, 30, 224015, \dodoi{10.1088/0264-9381/30/22/224015}

\bibitem[{{Siemens} {et~al.}(2019){Siemens}, {Hazboun}, {Baker},
  {Burke-Spolaor}, {Madison}, {Mingarelli}, {Simon}, \& {Smith}}]{Siemens19}
{Siemens}, X., {Hazboun}, J., {Baker}, P.~T., {et~al.} 2019, \baas, 51, 437

\bibitem[{{Stinebring} {et~al.}(2019){Stinebring}, {Chatterjee}, {Clark},
  {Cordes}, {Dolch}, {Heiles}, {Hill}, {Jones}, {Kaspi}, {Lam}, {Lazio},
  {Madison}, {McLaughlin}, {McClure-Griffiths}, {Palliyaguru}, {Rickett}, \&
  {Surnis}}]{Stinebring19}
{Stinebring}, D.~R., {Chatterjee}, S., {Clark}, S.~E., {et~al.} 2019, \baas,
  51, 492.
\newblock \doarXiv{1903.07370}

\bibitem[{{Stovall} {et~al.}(2019){Stovall}, {Freire}, {Antoniadis}, {Bagchi},
  {Deneva}, {Garver-Daniels}, {Martinez}, {McLaughlin}, {Arzoumanian},
  {Blumer}, {Brook}, {Cromartie}, {Demorest}, {DeCesar}, {Dolch}, {Ellis},
  {Ferdman}, {Ferrara}, {Fonseca}, {Gentile}, {Jones}, {Lam}, {Lorimer},
  {Lynch}, {Ng}, {Nice}, {Pennucci}, {Ransom}, {Spiewak}, {Stairs}, {Swiggum},
  {Vigeland}, \& {Zhu}}]{Stovall2019}
{Stovall}, K., {Freire}, P.~C.~C., {Antoniadis}, J., {et~al.} 2019, \apj, 870,
  74, \dodoi{10.3847/1538-4357/aaf37d}

\bibitem[{{Susobhanan} {et~al.}(2020){Susobhanan}, {Maan}, {Joshi}, {Prabu},
  {Desai}, {Gupta}, {Gopakumar}, {Dhanda Batra}, {Choudhary}, {Surnis}, {Dey},
  {Singha}, {Nobleson}, {Bagchi}, {Basu}, {Bethapudi}, {De}, {Girgaonkar},
  {Krishnakumar}, {Manoharan}, {Naidu}, {Pathak}, \& {Chaitanya
  Susarla}}]{Susobhanan20}
{Susobhanan}, A., {Maan}, Y., {Joshi}, B.~C., {et~al.} 2020, arXiv e-prints,
  arXiv:2007.02930.
\newblock \doarXiv{2007.02930}

\bibitem[{{Taylor}(1992)}]{Taylor92}
{Taylor}, J.~H. 1992, Royal Society of London Philosophical Transactions Series
  A, 341, 117, \dodoi{10.1098/rsta.1992.0088}

\bibitem[{{Taylor} {et~al.}(2019){Taylor}, {Burke-Spolaor}, {Baker}, {Charisi},
  {Islo}, {Kelley}, {Madison}, {Simon}, {Vigeland}, \& {Nanograv
  Collaboration}}]{Taylor19}
{Taylor}, S., {Burke-Spolaor}, S., {Baker}, P.~T., {et~al.} 2019, \baas, 51,
  336.
\newblock \doarXiv{1903.08183}

\bibitem[{{Taylor} {et~al.}(2016){Taylor}, {Vallisneri}, {Ellis}, {Mingarelli},
  {Lazio}, \& {van Haasteren}}]{Taylor2016}
{Taylor}, S.~R., {Vallisneri}, M., {Ellis}, J.~A., {et~al.} 2016, \apjl, 819,
  L6, \dodoi{10.3847/2041-8205/819/1/L6}

\bibitem[{{Torres} {et~al.}(2017){Torres}, {Ji}, {Li}, {Papitto}, {Rea}, {de
  O{\~n}a Wilhelmi}, \& {Zhang}}]{Torres17}
{Torres}, D.~F., {Ji}, L., {Li}, J., {et~al.} 2017, \apj, 836, 68,
  \dodoi{10.3847/1538-4357/836/1/68}

\bibitem[{{Vallisneri}(2020)}]{libstempo}
{Vallisneri}, M. 2020, {libstempo: Python wrapper for Tempo2}.
\newblock \doeprint{2002.017}

\bibitem[{{Vallisneri} \& {van Haasteren}(2017)}]{Vallisneri2017}
{Vallisneri}, M., \& {van Haasteren}, R. 2017, \mnras, 466, 4954,
  \dodoi{10.1093/mnras/stx069}

\bibitem[{{van Haasteren} \& {Vallisneri}(2014)}]{vhV14}
{van Haasteren}, R., \& {Vallisneri}, M. 2014, \prd, 90, 104012,
  \dodoi{10.1103/PhysRevD.90.104012}

\bibitem[{{van Haasteren} \& {Vallisneri}(2015)}]{vHV15}
---. 2015, \mnras, 446, 1170, \dodoi{10.1093/mnras/stu2157}

\bibitem[{{van Straten}(2006)}]{vS06}
{van Straten}, W. 2006, \apj, 642, 1004, \dodoi{10.1086/501001}

\bibitem[{{van Straten}(2013)}]{vS13}
---. 2013, \apjs, 204, 13, \dodoi{10.1088/0067-0049/204/1/13}

\bibitem[{{van Straten} {et~al.}(2011){van Straten}, {Demorest}, {Khoo},
  {Keith}, {Hotan}, \& {et al.}}]{vS11}
{van Straten}, W., {Demorest}, P., {Khoo}, J., {et~al.} 2011, {PSRCHIVE:
  Development Library for the Analysis of Pulsar Astronomical Data}.
\newblock \doeprint{1105.014}

\bibitem[{{Verbiest} \& {Shaifullah}(2018)}]{Verbiest18}
{Verbiest}, J.~P.~W., \& {Shaifullah}, G.~M. 2018, Classical and Quantum
  Gravity, 35, 133001, \dodoi{10.1088/1361-6382/aac412}

\bibitem[{{Vigeland} {et~al.}(2018){Vigeland}, {Deller}, {Kaplan}, {Istrate},
  {Stappers}, \& {Tauris}}]{Vigeland18}
{Vigeland}, S.~J., {Deller}, A.~T., {Kaplan}, D.~L., {et~al.} 2018, \apj, 855,
  122, \dodoi{10.3847/1538-4357/aaaa73}

\end{thebibliography}
